\documentclass[showpacs,prd,amsfonts,amsmath,twocolumn,aps,preprintnumbers,letterpaper]{revtex4}
\usepackage{amsmath,amssymb}
\usepackage{slashed}
\usepackage{graphicx}
\usepackage{tikz}
\usepackage{tabularx}
\usepackage{epsfig}
\usepackage{bm}
\usepackage[top=1.0in,bottom=1.0in,left=1.0in,right=1.0in]{geometry}
\usepackage[caption=false]{subfig}
\usepackage[pdftex, pdfborder= 0 0 0, citecolor=blue, urlcolor=blue, linkcolor=blue, colorlinks=true, bookmarksopen=true]{hyperref}
\usepackage{booktabs}

\newcommand{\be}{\begin{equation}}
\newcommand{\ee}{\end{equation}}
\newcommand{\bear}{\begin{eqnarray}}
\newcommand{\eear}{\end{eqnarray}}
\newcommand{\ba}{\begin{array}}
\newcommand{\ea}{\end{array}}

\def\be{\begin{eqnarray}}
\def\ee{\end{eqnarray}}
\def\bea{\be}
\def\eea{\ee}

\def\roughly#1{\mathrel{\raise.3ex\hbox{$#1$\kern-.75em%
\lower1ex\hbox{$\sim$}}}}

\def\abs#1{{\left| #1 \right|}}

  \long\def\comment#1{ }

  \newcommand{\Tr}{{\rm Tr}}

  \newcommand{\beq}{\begin{eqnarray}}
  \newcommand{\eeq}{\end{eqnarray}}
  
 \def\simge{\mathrel{%
   \rlap{\raise 0.511ex \hbox{$>$}}{\lower 0.511ex \hbox{$\sim$}}}}
\def\simle{\mathrel{
   \rlap{\raise 0.511ex \hbox{$<$}}{\lower 0.511ex \hbox{$\sim$}}}}

\begin{document}


\title{Quark and gluon GPDs at finite skewness from strings in holographic QCD: 
evolved and compared with experiment}

\author{Kiminad A. Mamo}
\email{kmamo@anl.gov}
\affiliation{Physics Division, Argonne National Laboratory, Argonne, Illinois 60439, USA
}

\author{Ismail Zahed}
\email{ismail.zahed@stonybrook.edu}
\affiliation{Center for Nuclear Theory, Department of Physics and Astronomy, Stony Brook University, Stony Brook, New York 11794-3800, USA}

\date{\today}

\begin{abstract}

We present a framework for constructing Generalized Parton Distributions (GPDs) using holographic QCD in the large $N_c$ limit, with a focus on low-x and finite skewness. Our approach utilizes holographic amplitudes for exclusive electroproduction processes to extract the spin-j conformal (Gegenbauer) moments of GPDs, which are then evolved to higher resolution scales using QCD evolution equations. Our evolved GPDs (reconstructed from their evolved conformal moments) are applied to analyze the electroproduction of $\rho^{0,+}$ and $\phi$ mesons, and we account for non-perturbative contributions in the $s+u$\,-\,channel using holographic QCD. The results compare well with the existing experimental data. Our GPDs provide detailed information about partonic distributions and are useful for future experimental studies and global data analyses. 

\end{abstract}

\maketitle

%

\section{Introduction}
\label{Introduction}
The General Parton Distributions (GPDs) provide a comprehensive framework for addressing the longitudinal momentum and spatial distributions of partons in hadrons. They consolidate the physical content of form factors, parton distribution amplitudes (DA), and parton distribution functions (PDFs) into a single framework. A number of current and future electron machines, such as the EIC and EIcC~\cite{AbdulKhalek:2021gbh,Anderle:2021wcy}, will be dedicated to measuring them.

The GPDs capture several invariants of the off-forward and non-local quark or gluon bilinears. In this work, we will focus on the unpolarized quark and gluon GPDs $H(x, \eta,t)$ and quark axial GPDs $\widetilde{H}(x, \eta,t)$ in the large $N_c$ limit. For the quarks, they can be combined into valence (isovector) and singlet GPDs. GPDs are a function of the parton longitudinal momentum fraction $x$, the skewness $\eta\sim\xi=-\Delta^+/2P^+$, and the momentum transfer $\Delta^2=-t$. $\Delta^+$ and $P^+$ are the light front momentum transfer and averaged momentum of the in-out protons, respectively.

The GPDs are characterized by two distinct kinematical regions for fixed Mandelstam $t$ and positive skewness: the DGLAP regions for $|x|>\xi$ and the ERBL region for $|x|<\xi$. In the DGLAP region with positive (negative) $x$, the GPDs correspond to removing a quark (antiquark) with momentum $k$ and re-inserting it with momentum $k+\Delta$. In the ERBL region, the GPDs correspond to emitting a meson-like quark-antiquark pair with momentum $-\Delta$. As a result, the forward limit with zero skewness of the GPDs coincides with the quark and gluon PDFs, while the x-integrated GPDs with finite skewness coincide with form factors.

GPDs play an important role in exclusive processes and are at the cornerstone of hadronic tomography. Deeply Virtual Compton Scattering (DVCS) is one of the processes suggested for the empirical extraction of the GPDs~\cite{Ji:1996nm,Radyushkin:1997ki}. Assuming factorization, the invariant Compton amplitudes are usually expressed as integral transforms of the leading twist-2 quark operators in the form of generalized sum rules. The extraction of the quark GPDs requires their inversion, usually using a perturbative analysis of the pertinent moments. This is a non-trivial deconvolution problem~\cite{Bertone:2021yyz}.

Holographic QCD provides a non-perturbative approach to a variety of scattering processes in QCD in the double limit of a large number of colors and strong $^\prime$t Hooft gauge coupling. It is a proposal following on the AdS/CFT or gauge/gravity duality established in string theory~\cite{Nastase:2015wjb}. In short, a strongly coupled gauge theory in four dimensions is dual to weakly coupled string theory in higher dimensions. The original correspondence holds for type IIB superstring theory in $AdS_5\times S_5$, but is commonly assumed to hold for a string theory in a general background.

In this work, we propose to use holographic QCD with a soft wall to evaluate electroproduction of photons and mesons by combining the leading s- and t-channel exchanges at finite skewness. The t-channel exchanges are dominated by open and closed string exchanges in the double limit, and they are dominant in the Regge kinematics.

A key outcome from this construction is exclusive amplitudes with a well-defined dependence on all kinematical variables. This yields well-defined expressions for the initial quark and gluon GPDs in the ERBL region to evolve at higher resolution with the help of perturbative renormalization group equations. The interface between the perturbative and non-perturbative and stringy aspects of QCD will be shown to set in for a gauge coupling $\alpha_s(\mu_0)\sim \frac 13$ at a resolution $\mu_0\sim 1.8$ GeV.

In our analysis, the deconvolution problem is altogether bypassed through the use of Gegenbauer moments for pertinent GPDs. The holographic construction allows their identification for any value of the conformal spin-j in the large number of colors limit. Their spin-j resummation in the GPDs can be carried explicitly in the ERBL region in the Regge limit. The extension to the DGLAP region can be sought using a re-organized Gegenbauer expansion. Once evolved, the results can be used in any exclusive process. They also should prove useful for the extraction of the GPDs in global data analyses. For completeness, we note the holographic GPDs analysis in~\cite{Nishio:2014rya}.

To our knowledge, the holographic framework provides the most economical way to encode the stringy properties of QCD at low resolution, in line with the dual Veneziano amplitudes and empirical Regge phenomenology. In the double limit of large $N_c$ and strong $^\prime$t Hooft coupling, the holographic amplitudes capture the key aspects used in QCD dispersive analyses, such as crossing symmetry, unitarity, and spectral densities, all while enforcing the essential symmetries of QCD with very few stringy parameters.

The paper is structured as follows: In section~\ref{sec_GPDHOLO}, we use large $N_c$ arguments to extract the conformal moments of GPDs from the holographic exclusive amplitudes for electroproduction of mesons at low-x and low resolution $\mu_0$ by matching the holographic exclusive amplitudes with the exclusive amplitudes in QCD which are based on factorization theorems. This matching is done explicitly for the electroproduction of mesons, photons (DVCS), a pair of pions, and a neutral pion. In section~\ref{RGEVOLUTION1}, the extracted conformal moments of GPDs will be evoloved, to higher resolution $\mu$, using Renormalization Group Equations (RGEs). The evolved GPDs will also be reconstructed using their evolved conformal moments. In section~\ref{sec_RHO0}, we use the evolved singlet quark and gluon GPDs to analyze in detail the electroproduction of neutral $\rho^0$. The total and differential cross-sections are derived and compared to the available data for a wide range of energies $\sqrt{s}$ and momentum transfer. In section~\ref{sec_PHI}, we extend the analysis to the electroproduction of $\phi$, using the evolved gluon GPDs. The results are compared to some existing data. In section~\ref{sec_RHO+}, we make use of the evolved non-singlet (valence) quark GPDs to derive the total and differential cross-sections for the electroproduction of charged $\rho^+$. The results are also compared to existing data, mostly at low $\sqrt{s}$. Our conclusions are in Section~\ref{sec_CON}.







\begin{widetext}

\section{Conformal moments of GPDs from t-channel string exchange in AdS/QCD}~\label{sec_GPDHOLO}

In this paper, we investigate the use of General Parton Distributions (GPDs) in exclusive processes, such as leptoproduction of photons and hadrons, where factorization is assumed to work. GPDs capture the off-forward partonic content of a hadron, and their importance lies in their ability to combine the physical content of form factors, parton distribution amplitudes (DA), and parton distribution functions (PDFs) in a single framework. We focus on the unpolarized quark and gluon GPDs $H(x, \eta,t)$, and quark axial GPDs $\widetilde H(x, \eta,t)$, in the large $N_c$ limit.

We propose a method to determine the matrix elements that define GPDs at some initial soft renormalization scale $\mu = \mu_0$, which carries information on confinement and chiral symmetry breaking, and then evolve them using the Renormalization Group Equation (RGE) to higher $\mu$. Our proposed method involves using holographic QCD, a dual string approach in AdS, in the double limit of large $N_c$ and strong 't Hooft coupling. We identify the twist-2 quark and gluon GPD conformal moments at finite skewness $\eta \neq 0$ and an initial renormalization scale $\mu_0 \sim Q_0$ using various exclusive t-channel open and closed string exchange electroproduction amplitudes in soft-wall holographic QCD, as detailed in Appendix~\ref{ACTION11}. The closed strings in soft-wall AdS are dual to spin-j glueballs in QCD, while the open strings attached to bulk filling $N_{f}$\,D9-branes in soft-wall AdS are dual to spin-j vector and axial mesons.

We match various exclusive electroproduction amplitudes in holographic QCD to various electroproduction amplitudes in QCD, written in terms of quark and gluon GPDs based on the leading-order factorization theorems, by matching their respective large-$N_{c}$ dependence at fixed 't Hooft coupling $\lambda_s = g_s^2 N_c$. We determine the large-$N_c$ dependence of the exclusive processes in the holographic side by using the bulk gravitational constant $2\kappa^2 \sim {1}/{N_c^2}$ and the bulk D9-brane coupling constant $g_5^2 \sim {1}/{N_c}$. The matching of exclusive electroproduction amplitudes is illustrated in Figs.~\ref{fig_HVM2}\,-\,\ref{fig_PI2}.

We consider only tree-level Witten diagrams in AdS and the leading-order exclusive electroproduction amplitudes in QCD, as they are the only diagrams that contribute in the large-$N_{c}$ limit. The rules for the tree-level Witten diagrams stem from the SUGRA fields and their couplings in bulk, which are detailed in Appendix~\ref{ACTION11}. These rules enable us to evaluate electroproduction of photons and mesons in holographic QCD, combining the leading s- and t-channel exchanges at finite skewness, and resulting in exclusive amplitudes with well-defined dependence on all kinematical variables.

In summary, this section presents the proposed method for determining GPDs using holographic QCD in the large $N_c$ limit, and summarizes the key results of this paper. The relevant notations and kinematics are summarized in Appendix~\ref{sec:notations}, and Appendix~\ref{KINEMATICS}, respectively. The matrix elements that define GPD are detailed in the Appendix~\ref{sec_PQCD}. The RGE used to evolve the matrix elements to higher $\mu$ is detailed in Appendix~\ref{RGEVOLUTION}. Furthermore, the exclusive electroproduction amplitudes in QCD are summarized in Appendix~\ref{PQCDAMPLITUDES}, which should be compared with the corresponding holographic amplitudes derived in Appendix~\ref{ACTION11} where we have illustrated the comparisons in Figs.~\ref{fig_HVM2}\,-\,\ref{fig_PI2}.


 \subsection{Electroproduction of vector mesons}
The leading order electroproduction amplitude for vector mesons (with quark DA and gluon GPD) is of order $\frac{g_s^2}{N_c}$ as illustrated in  Fig.~\ref{fig_HVM2} left. 
In the large-$N_c$ limit, it is of order  $\frac{g_s^2N_c}{N_c^2}=\frac{\lambda_s}{N_c^2}\,.$  This matches  the $\frac{1}{N_c^2}$ dependence for the holographic vector meson electroproduction amplitude, using the corresponding tree-level Witten diagram illustrated in Fig.~\ref{fig_HVM2} right. More specifically, the $N_c$ dependence of the electroproduction of vector mesons written in terms of Gegenbauer moments,


\begin{figure*}
\subfloat[\label{fig_HVM2a}]{%
  \includegraphics[height=5.5cm,width=.45\linewidth]{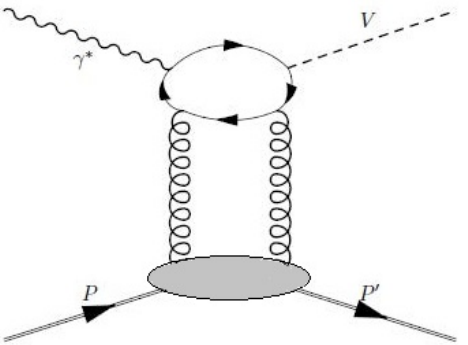}%
}\hfill
\subfloat[\label{fig_HVM2b}]{%
  \includegraphics[height=5.5cm,width=.45\linewidth]{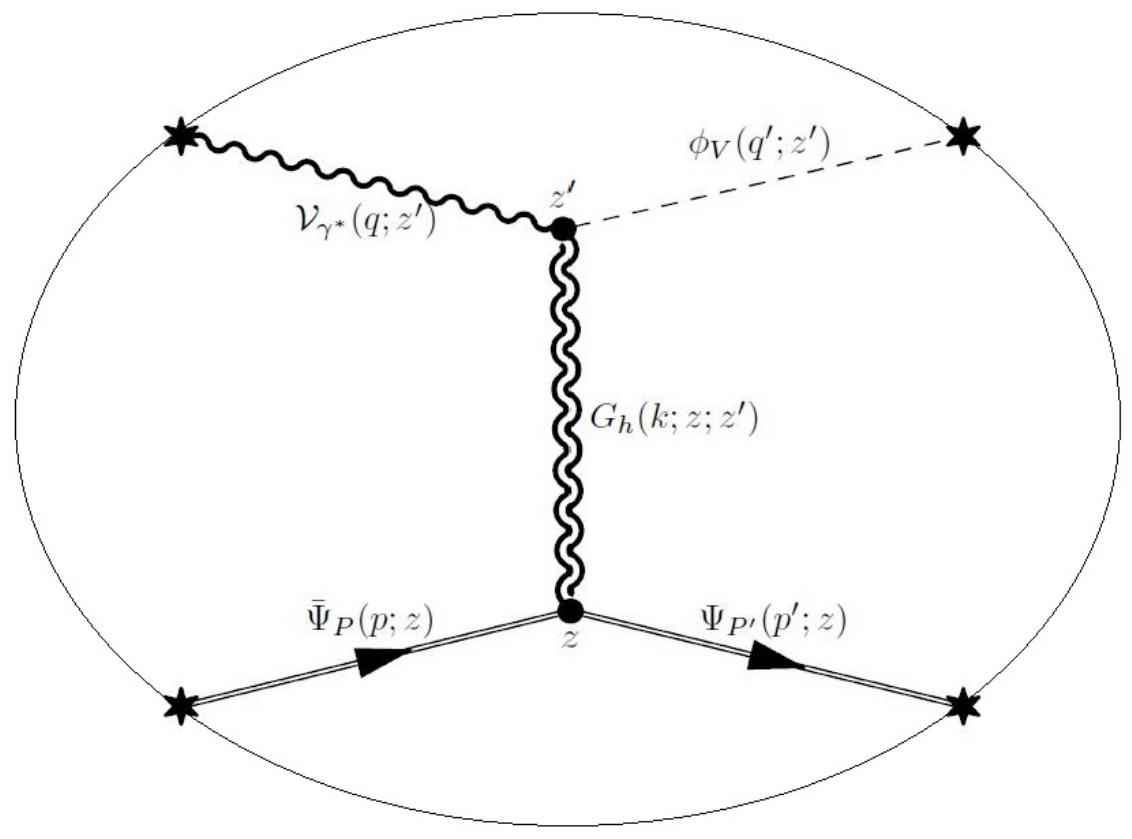}%
}
\caption{electroproduction of a vector meson probing the $\mathbf{gluon~GPD}$: (a) leading QCD contribution in the Regge limit; (b) leading Witten diagram in the large-$N_c$ limit.}
 \label{fig_HVM2}
\end{figure*}

\bea\label{gluonFinal2new}
{\cal A}^{LL}_{\gamma^{*} p\rightarrow  V p} (s,t,Q_0,\epsilon_{L},\epsilon^{\prime}_{L})
&\approx &e\times\frac{1}{4}\times\frac{1}{N_c^2}\times \frac{1}{Q_0}
\times \lambda_{s}(\mu^2)\times\left[\int_0^1 dz\, \sum_{q} e_q \frac{\Phi^q(z)}{z(1-z)}\right]\times \frac{1}{m_N}
\times\bar u(p_2)u(p_1)\nonumber\\
&\times &\sum_{j = 2}^{\infty}
\frac{1}{\xi^{j}}\times\frac{1}{N_{j-2} ( \frac{5}{2})}\times
\left[\int_{0}^{\xi}\frac{dx}{\xi}\, \left(1-\frac{x^2}{\xi^2}\right)\times C_{j-2}^{5/2}\left(\frac{x}{\xi}\right)\right]\times\mathbb{F}^g_{j} (\xi, t ; \mu_0^2)\,,\nonumber\\ 
\eea
for even $j=2,4,...$, matches  the holographic result 
\bea\label{spin-j-2new}
{\cal A}^{LL}_{\gamma^{*} p\rightarrow  V p} (s,t,Q_0,\epsilon_{L},\epsilon^{\prime}_{L})&\sim &e\times\frac{1}{g_5}\times g_5\times 2\kappa^2\times\frac{1}{2}\times\frac{1}{Q_0}\times\left[f_{V}\right]\times\frac{1}{m_N}
\times\bar u(p_2)u(p_1)\nonumber\\
&\times & \sum_{j=2}^{\infty}\,\frac{1}{\xi^j}\times \left[\frac{1}{\Gamma(\Delta_{g}(j)-2)}\times\left[\mathcal{A}(j,\tau,\Delta_{g}(j), t)+\mathcal{D}_{\eta}(j,\tau,\Delta_g(j),t)\right]\right]\,,\nonumber\\
\eea
for even $j=2,4,...$, with the details of the derivation given in  Appendix \ref{ACTION11}.
The 5-dimensional bulk gravitational coupling is $2\kappa^2={8\pi^2}/{N_c^2}$, and $g_5 \sim 1/\sqrt{N_c}$.

A comparison of (\ref{gluonFinal2new}) and (\ref{spin-j-2new}), allows the extraction of 
the Gegenbauer (conformal) moments of the gluon GPDs 
\bea\label{gluonGgrMoments2}
\mathbb{F}^g_{j} (\eta\sim\xi, t ; \mu_0^2)\propto \frac{1}{\Gamma(\Delta_{g}(j)-2)}\times\left[\mathcal{A}(j,\tau,\Delta_{g}(j), t)+\mathcal{D}_{\eta}(j,\tau,\Delta_g(j),t)\right]\,,\nonumber\\
\eea
with $\eta\sim \xi$ following from $\sqrt{s}\gg  -t$,  with the details of the  kinematics in Appendix~\ref{KINEMATICS}. 
The anomalous dimension of the spin-j conformal gluon operator at $\mu=\mu_0\sim Q_0$, and even $j=2,4,...$ is  $$\Delta_g(j)=2+\sqrt{2\sqrt{\lambda}(j-j_{0g})}\,,$$ with $j_{0g}=2-2/\sqrt{\lambda}$, and the 't Hooft coupling of QCD at $\mu=\mu_0$ given by $$\lambda\equiv\lambda_s(\mu_0)=4\pi\alpha_s(\mu_0)N_c\rightarrow 11.243\,.$$ 
Below we suggest the matching scale $\mu_0=1.808$ GeV (see section~\ref{sec_MATCH}). In addition, we have set the Mandelstam variable  $t=-K^2$, and defined the $\eta$-independent gluonic spin-j (with even $j=2,4,...$) form factors of the proton (with twist $\tau$) as
\bea\label{Ajj}
&&\mathcal{A}(j,\tau,\Delta_{g}(j), -t=K^2) = \nonumber\\ 
&&\frac{2^{1-\Delta_{g}(j) }}{\Gamma(\tau)} \Bigg( (\tau -1) \Gamma \left(\frac{j-2}{2}+\tau -\frac{\Delta_{g}(j) }{2}+1\right) \Gamma \left(\frac{j-2}{2}+\frac{\Delta_{g}(j) }{2}+\tau -1\right)\nonumber\\
&\times & \, _2\tilde{F}_1\left(\frac{a_K}{2}-\frac{\Delta_{g}(j) }{2}+2,\frac{j-2}{2}+\tau -\frac{\Delta_{g}(j) }{2}+1;\frac{a_K}{2}+\frac{j-2}{2}+\tau +1;-1\right)\nonumber\\
&&+\Gamma \left(\frac{j-2}{2}+\tau -\frac{\Delta_{g}(j) }{2}+2\right) \Gamma \left(\frac{j-2}{2}+\frac{\Delta_{g}(j) }{2}+\tau \right)\nonumber\\
&\times & _2\tilde{F}_1\left(\frac{a_K}{2}-\frac{\Delta_{g}(j) }{2}+2,\frac{j-2}{2}+\tau -\frac{\Delta_{g}(j) }{2}+2;\frac{a_K}{2}+\frac{j-2}{2}+\tau +2;-1\right)\Bigg)\,,\nonumber\\
\eea
where $_2\tilde{F}_1$ is the regularized hypergeometric function, with $a_K=K^2/4\tilde{\kappa}_T^2$. The skewness or
$\eta$-dependent spin-j $\mathcal{D}_{\eta}$-terms are given by
\bea\label{DKj22}
\mathcal{D}_{\eta}(j,\tau,\Delta_g(j),-t=K^2)&=&\left(\hat{d}_{j}(\eta,-K^2)-1\right)\times \left[\mathcal{A}(j,\tau,\Delta_g(j),K,\tilde{\kappa}_T)-\mathcal{A}_S(j,\tau,\Delta_g(j),K,\tilde{\kappa}_S)\right]\,,\nonumber\\
\eea
where
\bea
\label{ASj22}
\mathcal{A}_S(j,\tau,\Delta_g(j),K,\tilde{\kappa}_S)&\equiv&\mathcal{A}(j,\tau,\Delta_g(j),K,\tilde{\kappa}_T\rightarrow \tilde{\kappa}_S)\,.
\eea
and
\be\label{etapoly232}
\hat{d}_{j}(\eta,-K^2)=\, _2F_1\left(-\frac{j}{2},\frac{1-j}{2};\frac{1}{2}-j;\frac{4 m_N^2}{K^2}\times\eta ^2\right)\,.
\ee
Note that the A-form factor for the spin-2 exchange is given by  $$A(t)=\mathcal{A}(j=2,\tau,\Delta_{g}(j=2)=4, -t=K^2)\,,$$
and the D-form factor for the same spin-2 exchange (the spin-2 D-term) is given by 
$$\eta^2D(t)=\mathcal{D}_{\eta}(j=2,\tau,\Delta_g(j=2)=4,-t=K^2)\,.$$ Both form factors play an important role in the holographic
electroproduction of heavy mesons near treshold~\cite{Mamo:2019mka,Mamo:2021tzd,Mamo:2022eui} (and references therein).
\subsection{Electroproduction of pion pair}
The electroproduction amplitude for pair of mesons (with gluon DA and quark GPDs) at leading-order is order of $\frac{g_s^2}{N_c}$. Therefore, in the large-$N_c$ limit, it is of  order $\frac{g_s^2N_c}{N_c^2}=\frac{\lambda_s}{N_c^2}\,,$ (see Fig.~\ref{fig_PIPI2} left). We find the same $\frac{1}{N_c^2}$ dependence for the holographic meson pair electroproduction amplitude, using the corresponding tree-level Witten diagram (see Fig.~\ref{fig_PIPI2} right), and compare the $N_c$ dependence of the electroproduction of pair mesons written in terms of Gegenbauer momnents (\ref{pairGPD22new}), and the holographic one (\ref{LLmesonPairj2new}) below


\begin{figure*}
\subfloat[\label{fig_PIPI2a}]{%
  \includegraphics[height=5.5cm,width=.45\linewidth]{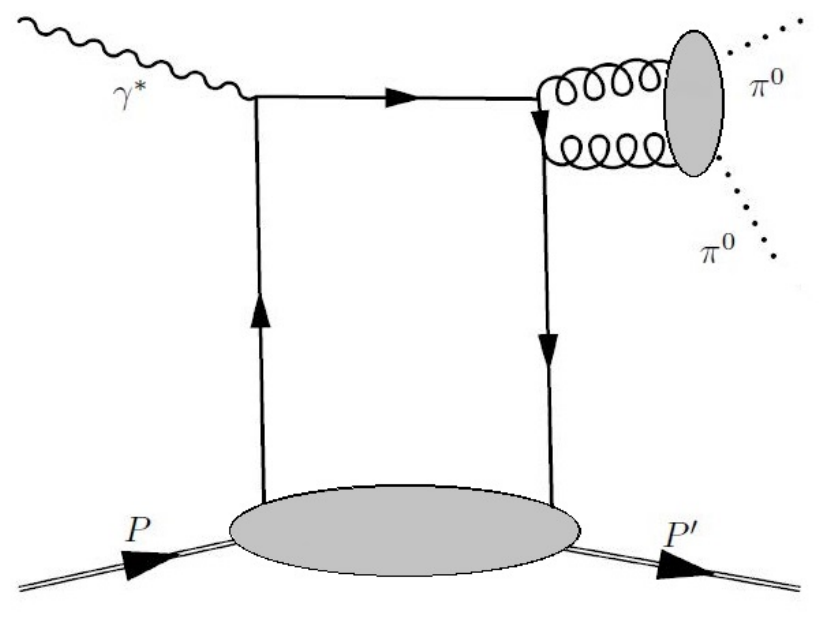}%
}\hfill
\subfloat[\label{fig_PIPI2b}]{%
  \includegraphics[height=5.5cm,width=.45\linewidth]{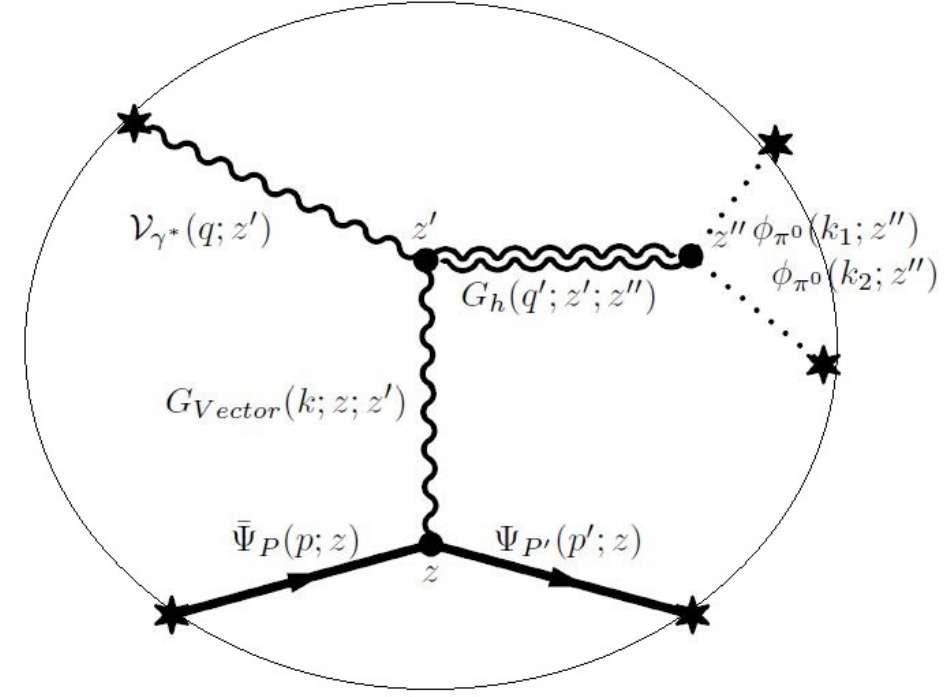}%
}
\caption{electroproduction of double pions probing the $\bf{non-singlet~(valence)~vector~quark~GPDs}$ : (a) leading QCD contribution in the Regge limit; (b) leading Witten diagram in the large-$N_c$ limit.}
\label{fig_PIPI2}
\end{figure*}

\bea\label{pairGPD22new}
{\cal A}^{LL}_{\gamma^* p\rightarrow  \pi\pi p} (s,t,Q_0,\epsilon_{L};m_{\pi\pi})
&\approx &
e\times\lambda_{s}(\mu_0)\times \frac{4}{N_c^2}\times \frac{1}{Q_0}
\times\left[\int_0^1 dz\, \frac{\Phi^g(z)}{z(1-z)}\right]\times\frac{1}{2m_N}\times\bar u(p_2)u(p_1) \nonumber\\
&\times &\sum_{j = 1}^{\infty}\frac{1}{\xi^{j}}
\times\frac{1}{N_{j-1}(\frac{3}{2})}
\times\left[\int_{0}^{\xi}\frac{dx}{\xi}\,C_{j-1}^{3/2}\left(\frac{x}{\xi}\right)
\right]\times\sum_{q} Q_q\,\mathbb{F}^q_{j(valence)} (\eta\sim\xi, t ; \mu_0^2)  
\,,\nonumber\\
\eea
for odd $j=1,3,..$, and
 \bea 
\label{LLmesonPairj2new}
{\cal A}^{LL}_{\gamma^* p\rightarrow  \pi\pi p} (s,t,Q_0,\epsilon_{L};m_{\pi\pi})&\sim & e\times\frac{1}{g_5}\times g_5\times 2\kappa^2\times \frac{1}{4}\times\frac{1}{Q_0}\times \left[A_{gluon}^{\pi}(m_{\pi\pi})\right]\times \frac{1}{2m_N}
\times\bar u(p_2)u(p_1)\nonumber\\
&\times &\sum_{j=1}^{\infty}\,\frac{1}{\xi^{j}}\times\frac{1}{\Gamma(\Delta_{q}(j)-2)}\times\left[\mathcal{F}_1(j,\tau,\Delta_{q}(j), t)+\mathcal{D}_{q\eta}(j,\tau,\Delta_{q}(j), t)\right]\,,\nonumber\\
\eea                    
for odd $j=1,3,....$, with the 5-dimensional bulk gravitational coupling $2\kappa^2={8\pi^2}/{N_c^2}$, and $g_5 \sim 1/\sqrt{N_c}$. 
The detailed derivation of (\ref{LLmesonPairj2new}) is given in Appendix~\ref{ACTION11}. 
Comparing (\ref{pairGPD22new}) to (\ref{LLmesonPairj2new}), we identify the conformal moments
\bea\label{valenceGgbrMoments2updown}
\mathbb{F}^q_{j(valence)} (\eta\sim\xi, t ; \mu_0^2) &\propto& \frac{1}{\Gamma(\Delta_{q}(j)-2)}\times \left[\mathcal{F}_1(j,\tau,\Delta_{q}(j), t)+\mathcal{D}_{q\eta}(j,\tau,\Delta_{q}(j), t)\right]\,.
\eea
The anomalous dimension of the spin-j conformal valence quark operator at $\mu=\mu_0\sim Q_0$, and odd $j=1,3,...$ is given by $$\Delta_{q}(j)=2+\sqrt{\sqrt{\lambda}(j-j_{0q})}\,,$$ with $j_{0q}=1-1/\sqrt{\lambda}$, and the 't Hooft coupling at $\mu=\mu_0$ fixed as  $$\lambda\equiv\lambda_s(\mu_0)=4\pi\alpha_s(\mu_0)N_c=11.243\,.$$ The  $t$-dependent valence quark spin-j (with odd $j=1,3,5,7,...$) form factors of the proton with twist $\tau=d-s$, is
\bea
\mathcal{F}_1(j,\tau,\Delta_{q}(j), -t=K^2) &\propto& \mathcal{A}(j\rightarrow j+1,\tau,\Delta_{g}\rightarrow\Delta_{q},t)\nonumber\\
\eea
and the quark spin-j skewness dependent $\mathcal{D}_{q\eta}$-terms are
\bea\label{spinjFFD2quark2}
\mathcal{D}_{q\eta}(j,\tau,\Delta_{q}(j),t)
&\propto&\mathcal{D}_{\eta}(j\rightarrow j+1,\tau,\Delta_{g}\rightarrow\Delta_{q},t;\tilde{\kappa}_T\rightarrow\frac{1}{2}\times\tilde{\kappa}_V)\,,
\eea
where $\mathcal{A}$ and $\mathcal{D}_{\eta}$ are given by (\ref{Ajj}) and (\ref{DKj22}), respectively, for odd $j=1,3,5,7,...$.


Note that the Dirac electromagnetic form factor (ignoring the Pauli contribution) for the spin-1 exchange is given by  $$F_1(t)=\mathcal{F}_1(j=1,\tau,\Delta_{q}(j=1)=3, -t=K^2)\,.$$
\subsection{DVCS amplitude}
The DVCS amplitude in QCD at leading-order is order of $N_c^0$ (see Fig.~\ref{fig_DVCS2} left), the NLO correction (due to the gluon GPD contribution) 
is order of $g_s^2=\frac{g_s^2N_c}{N_c}=\frac{\lambda_s}{N_c}\,$, and can be ignored in the large-$N_c$ and small 't Hooft coupling limit. Using the corresponding tree-level Witten diagrams (see Fig.~\ref{fig_DVCS2} right), we find the same $N_c^0$ dependence for the holographic DVCS amplitudes in the large-$N_c$ limit (compare the $N_c$ dependence of the axial part of DVCS written in terms of Gegenbauer moments (\ref{cont2Tw2Parton222new}) and the holographic DVCS (\ref{LTDVCSaxialH2new}) below, 


\begin{figure*}
\subfloat[\label{fig_DVCS2a}]{%
  \includegraphics[height=5.5cm,width=.45\linewidth]{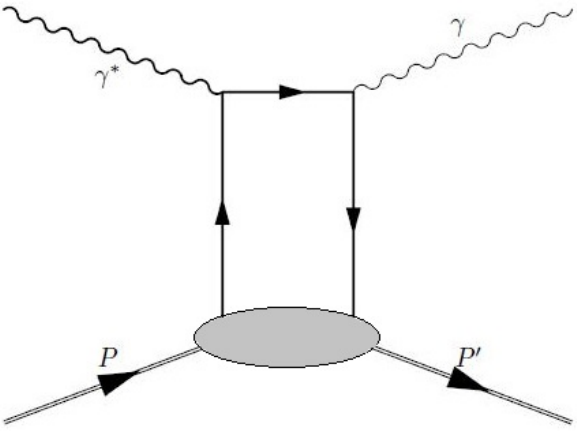}%
}\hfill
\subfloat[\label{fig_DVCS2b}]{%
  \includegraphics[height=5.5cm,width=.45\linewidth]{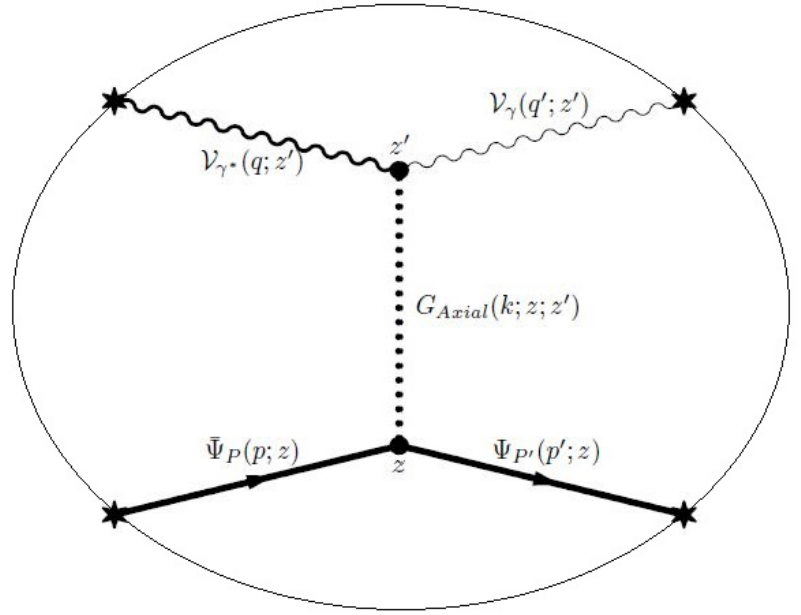}%
}
\caption{Deeply virtual Compton scattering (DVCS) probing the $\bf{singlet~axial~quark~GPDs}$ : (a) leading QCD contribution in the Regge limit; (b) leading Witten diagram in the large-$N_c$ limit.}
\label{fig_DVCS2}
\end{figure*}

\begin{eqnarray}
\label{cont2Tw2Parton222new}
{\cal A}^{LT}_{\gamma^{*} p\rightarrow  \gamma p} (s,t,Q_0,\epsilon_{L},\epsilon^{\prime}_{T})
&\approx &\!\!\!
\epsilon_{L\mu}\epsilon_{T\nu}^{\prime*}\times N_c^0\times i \varepsilon^{\mu \nu \rho \sigma} \tilde q_\sigma p_\rho
\times \frac{1}{p^{+}}
\times \bar u(p_2)\gamma^{+}\gamma_5 u(p_1)\nonumber\\
&\times &\frac{4}{Q_0^2}\times\sum_{j=1}^{\infty}\,\frac{1}{\xi^{j-1}} 
\times\frac{1}{N_{j-1}(\frac{3}{2})}\times\left[\int_{0}^{\xi}\frac{dx}{\xi}\,C_{j-1}^{3/2}\left(\frac{x}{\xi}\right)\right]\times\sum_{q}Q_q^2\,\tilde{\mathbb{F}}^q_{j(singlet)} (\eta\sim\xi, t ; \mu_0^2)\,,\nonumber\\
\end{eqnarray}
for odd $j=1,3,...$, and
\bea 
\label{LTDVCSaxialH2new}
{\cal A}^{LT}_{\gamma^{*} p\rightarrow  \gamma p} (s,t,Q_0,\epsilon_{L},\epsilon^{\prime}_{T})&=&\frac{1}{g_5}\times\frac{1}{g_5}\times g_5\times g_5^3\kappa_{cs}\times\frac{1}{\tilde{\kappa}_{N}^2}\epsilon_{L\mu}\epsilon_{T\nu}^{\prime*}\times i \varepsilon^{\mu \nu \rho \sigma} \tilde q_\sigma
\times\bar u(p_2)\gamma_{\rho}\gamma_5 u(p_1)\nonumber\\
&\times &\sum_{j=1}^{\infty}\,\frac{1}{\xi^{j-1}}\times \frac{1}{\Gamma(\Delta_{q}(j)-2)}\times\left[\mathcal{F}_{A}(j,\tau,\Delta_{q}(j), t)+\mathcal{D}_{A\eta}(j,\tau,\Delta_{q}(j), t)\right]\,,\nonumber\\
\eea
for odd $j=1,3,...$ with $g_5\sim 1/\sqrt{N_{c}}$ and the Chern-Simons coupling $\kappa_{cs}\sim N_c$. 
The detailed derivation of (\ref{LTDVCSaxialH2new}) is given in Appendix~\ref{ACTION11}. 
The matching of (\ref{cont2Tw2Parton222new}) to (\ref{LTDVCSaxialH2new}) in leading order in $1/N_c$, yields

\bea\label{GgbrMomentsSingletAxial2}
\tilde{\mathbb{F}}^q_{j(singlet)} (\eta\sim\xi, t ; \mu_0^2)&\propto & \frac{1}{\Gamma(\Delta_{q}(j)-2)}\times\left[\mathcal{F}_{A}(j,\tau,\Delta_{q}(j), t)+\mathcal{D}_{A\eta}(j,\tau,\Delta_{q}(j), t)\right]\,,\nonumber\\
\eea 
where the spin-j axial form factors $\mathcal{F}_{A}(j,\tau,\Delta_{q}(j), t)$ and $\mathcal{D}_{A\eta}(j,\tau,\Delta_{q}(j), t)$ are given by (\ref{FAj1}) and (\ref{FAetaKj22}), respectively, for odd $j=1,3,...$. The anomalous dimension of the spin-j conformal singlet axial quark operator at $\mu=\mu_0\sim Q_0$, and odd $j=1,3,...$ is given by $$\Delta_{q}(j)=2+\sqrt{\sqrt{\lambda}(j-j_{0q})}\,,$$ with $j_{0q}=1-1/\sqrt{\lambda}$.  

Note that the axial electromagnetic form factor for the spin-1 exchange is given by  $$F_A(t)=\mathcal{F}_{A}(j=1,\tau=3,\Delta_{q}(j=1)=3, -t=K^2)\,.$$ 
 \subsection{Electroproduction of neutral pion}
Finally, the electroproduction amplitude for mesons (with quark DAs and quark GPDs) at leading-order is order of $\frac{C_Fg_s^2}{N_c}$ with $C_F=\frac{N_c^2-1}{2N_c}$ (see Fig.~\ref{fig_PI2} left). Therefore, the electroproduction amplitude for mesons (with quark DAs and quark GPDs) at leading-order is of order $\frac{g_s^2N_c}{2N_c}=\frac{\lambda_s}{2N_c}\,,$ using $C_{F}\sim \frac{N_c}{2}$ in the large-$N_c$ limit. We find the same $\frac{1}{N_c}$ dependence for the holographic meson electroproduction amplitude using the corresponding tree-level Witten diagram (see Fig.~\ref{fig_PI2} right), and compare the $N_c$ dependence of the electroproduction of $\pi^{0}$ written in terms of Gegenbauer momnents (\ref{pi0GPD22new}) and the holographic one (\ref{LLmesonAxialj2new}) below, 


\begin{figure*}
\subfloat[\label{fig_PI2a}]{%
  \includegraphics[height=5.5cm,width=.45\linewidth]{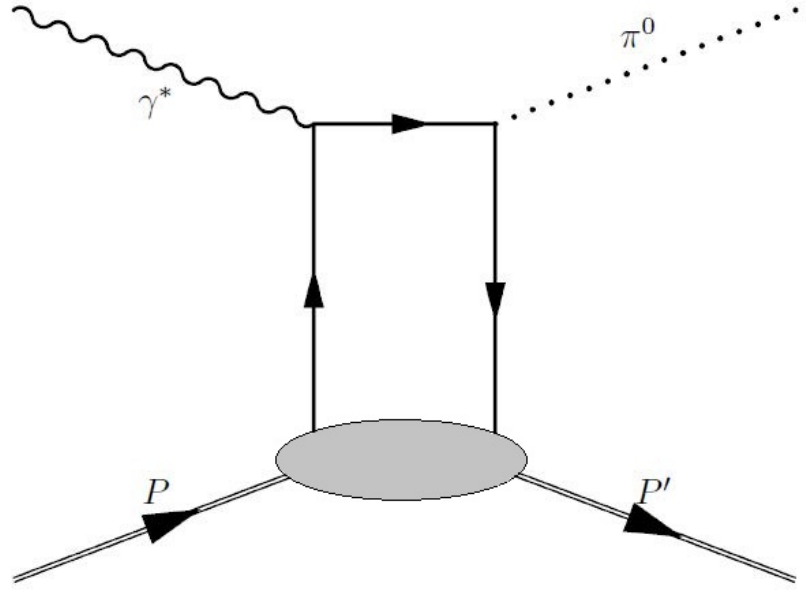}%
}\hfill
\subfloat[\label{fig_PI2b}]{%
  \includegraphics[height=5.5cm,width=.45\linewidth]{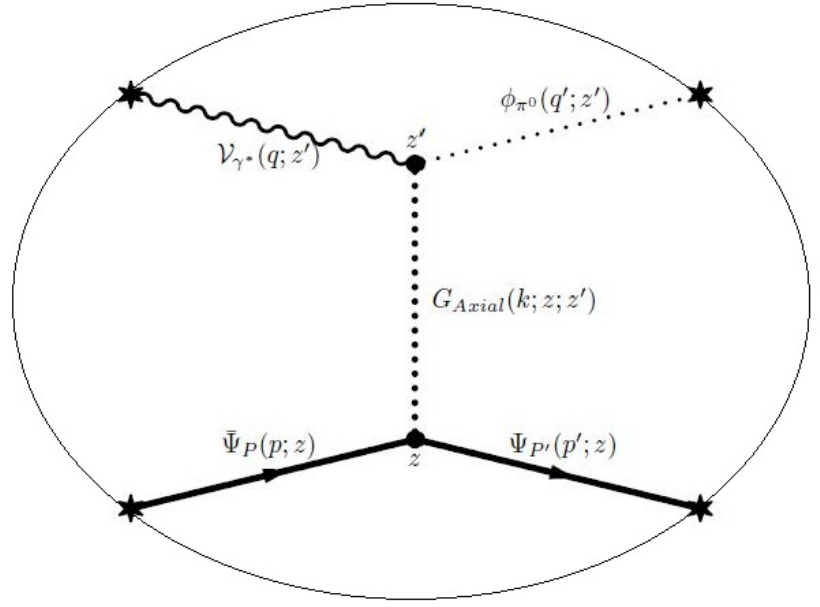}%
}
\caption{electroproduction of neutral pion probing the $\bf{non-singlet~(valence)~axial~quark~GPDs}$ : (a) leading QCD contribution in the Regge limit; (b) leading Witten diagram in the large-$N_c$ limit.}
 \label{fig_PI2}
\end{figure*}

\begin{eqnarray}\label{pi0GPD22new}
{\cal A}^{LL}_{\gamma^{*} p\rightarrow  \pi^{0} p} (s,t,Q_0,\epsilon_{L},\epsilon_{L}^{\prime}) 
&\,\approx\,& e\times {1\over 2}\times \frac{1}{N_c}\times\lambda_s(\mu_0)\times{1\over {Q_0}}\times\left[ \, \int _{0}^{1}dz{{\Phi_{\pi}(z)}\over z}\right] \times
\frac{1}{p^{+}}\times u(p_2)\gamma^{+}\,\gamma_5 u(p_1)\times{-1 \over {\sqrt 2}}\nonumber\\
&\times &\sum_{j = 2}^{\infty}
\frac{1}{\xi^{j}}\times\frac{1}{N_{j-1} ( \frac{3}{2})}
\times\left[\int_{0}^\xi \frac{dx}{\xi}\,\frac{x}{\xi}\,C_{j-1}^{3/2}\left(\frac{x}{\xi}\right)\right]\nonumber\\
&\times &\left(Q_u \ \widetilde{\mathbb{F}}^u_{j(valence)} (\eta\sim\xi, t ; \mu_0^2) \,-\, Q_d \ \widetilde{\mathbb{F}}^d_{j(valence)} (\eta\sim\xi, t ; \mu_0^2)\right)\,,\nonumber\\
\end{eqnarray}
for even $j=2,4,...$, and
\bea 
\label{LLmesonAxialj2new}
{\cal A}^{LL}_{\gamma^{*} p\rightarrow  \pi^{0} p} (s,t,Q_0,\epsilon_{L},\epsilon_{L}^{\prime})&=& e\times\frac{1}{g_5}\times g_5^2\times g_5\times \left[f_{\pi}\right]\times\frac{1}{2}\times\frac{1}{Q_0}\times\frac{1}{p^+}\times\bar u(p_2)\gamma^{+}\,\gamma_5 u(p_1)\times \frac{-1}{\sqrt{2}}\nonumber\\
&&\times \sum_{j=2}^{\infty}\,\frac{1}{\xi^{j}}\times \frac{1}{\Gamma(\Delta_{q}(j)-2)}\nonumber\\
&\times& \mathcal{F}_{A}(j,\tau,\Delta_{q}(j), t)\times\mathcal{O}\left(1/N_c\right)\rightarrow 0\,,
\eea
for even $j=2,4,...$, with $g_5 \sim 1/\sqrt{N_c}$. 
The details of the derivation of (\ref{LLmesonAxialj2new}) can be found in Appendix~\ref{ACTION11}.
Note that the holographic amplitude (\ref{LLmesonAxialj2new}) is $1/N_c$ suppressed since there is no direct coupling between axial-axial-vector mesons at tree level. Similarly, comparing (\ref{pi0GPD22new}) to (\ref{LLmesonAxialj2new}), the Gegenbauer (conformal) moments of the valence axial GPDs at $\mu=\mu_0$ vanish
\bea\label{GgbrMomentsValenceAxial2}
\widetilde{\mathbb{F}}^q_{j(valence)} (\eta\sim\xi, t ; \mu_0^2) &=& 0\,.\nonumber\\
\eea

\section{RG Evolution of Quark and Gluon GPDs in QCD}~\label{RGEVOLUTION1}
In the inclusive DIS process, the hadron's partonic content varies with the probing virtual photon's $Q^2$. In massless QCD, a scale-free theory, this leads to scaling violations or logarithmic deviations from free partons. DIS structure functions are convolutions of perturbatively determined coefficient functions with fixed $Q^2/\mu^2$ and relevant PDFs at resolution $\mu$, due to factorization. The PDFs' evolution and mixing are determined by DGLAP, reflecting the RG flow independence of the original structure functions.

Contrasting with PDFs (diagonal matrix elements of gauge invariant bilocals), GPDs are off-diagonal (double parton distributions) and follow the exclusive kernels' evolution described by ERBL. The ERBL evolution can be recast into a DGLAP evolution when expressed in terms of hadronic DA moments. Similarly, GPDs' evolution can be transformed into a DGLAP-type evolution when expressed in terms of Gegenbauer (conformal) moments, as detailed in~\cite{Belitsky:2005qn}. 

We provide a concise summary of the central RG evolution results for conformal (Gegenbauer) moments derived from the holographic construction at the initial resolution $\mu_0$. See Appendix \ref{RGEVOLUTION} for detailed derivations. The evolved GPDs, reconstructed from the evolved conformal moments, will be utilized in the electroproduction analysis of $\rho^{0}$, $\phi$, and $\rho^{+}$ vector mesons, and compared with experimental data in sections \ref{sec_RHO0}, \ref{sec_PHI}, and \ref{sec_RHO+}, respectively.

\subsection{RG evolution of non-singlet (valence) quark GPDs}

The leading-order RG evolution of Gegenbauer (conformal) moments of valence (non-singlet) vector quark GPDs is given by (see Eq.4.244 in \cite{Belitsky:2005qn})
\begin{equation}
\label{EvolutionConfMomGPDsValence}
\mathbb{F}^q_{j(valence)} (\eta, t ; \mu^2)
=
\mathbb{F}^q_{j(valence)} (\eta, t  ; \mu_0^2)
\times\left(
\frac{\alpha_s (\mu_0^2)}{\alpha_s (\mu^2)}
\right)^{\gamma_{(0)j-1}^{qq;\text{V};\text{NS}}/\beta_0}
\end{equation}
for odd $j=1,3,...$, where the input valence quark GPDs $\mathbb{F}^q_{j(valence)} (\eta, t  ; \mu_0^2)$ are given in (\ref{valenceGgbrMoments2updown}). Hence, the evolved valence quark GPDs $H_{valence}^q (x, \eta, t ; \mu^2)$ are given by
\begin{equation}
\label{2ExpansionGPDeigenfunctionsH}
H_{valence}^q (x, \eta, t ; \mu^2)
=
\frac{1}{\eta} \sum_{j = 1}^\infty
\frac{w \big( \frac{x}{\eta} | \frac{3}{2} \big)}{\eta^{j-1} \, N_{j-1} ( \frac{3}{2})}
C_{j-1}^{3/2}\left(\frac{x}{\eta}\right)
\times\mathbb{F}^q_{j(valence)} (\eta, t ; \mu^2)
\, ,
\end{equation}
for odd $j=1,3,...$. Here the weight and normalization factors are
$$
w (x | \nu) = (1 - x^2)^{\nu - 1/2}
\, , \qquad
N_{j-1} (\nu) = 2^{1 - 2 \nu}
\frac{
{\mit\Gamma}^2 (1/2) {\mit\Gamma} (2 \nu + (j-1))
}{
{\mit\Gamma}^2 (\nu) {\mit\Gamma} ((j-1) + 1) (\nu + (j-1))
}
\, ,
$$
and $C_{j}^{\nu}(x)$ are the Gegenbauer polynomials.

\subsection{RG evolution of singlet quark and gluon GPDs}
The singlet quark and gluon GPDs mix under evolution. Also as we noted earlier, the initial holographic
input for the singlet quark evolution vanishes in leading order in $1/N_c$, simplifying the initial data. 

The evolved conformal moments of singlet quark $\mathbb{F}^q_{j(singlet)}(\eta, t ; \mu^2)$ and gluon $\mathbb{F}^g_j(\eta, t ; \mu^2)$ GPDs are given by

\bea\label{evolvedGgnbrMomentsSingletQuark}
&&\sum_q \mathbb{F}^q_{j(singlet)}(\eta, t ; \mu^2)\nonumber\\
&=&
\frac{1}{3}\times\frac{2\left(\gamma^{qq}_{(0)j-1} - \gamma^-_{j-1}\right)}{\gamma^+_{j-1} - \gamma^-_{j-1}}\times \mathbb{F}^+_j(\eta, t;\mu_0^2)\times\left(
\frac{\alpha_s (\mu^2_0)}{\alpha_s (\mu^2)}
\right)^{\gamma^+_{j-1} / \beta_0} \nonumber\\
&+& \frac{1}{3}\times\frac{-2\left(\gamma^{qq}_{(0)j-1} - \gamma^+_{j-1}\right)}{\gamma^+_{j-1} - \gamma^-_{j-1}}\times \mathbb{F}^-_j(\eta, t;\mu_0^2)\times\left(
\frac{\alpha_s (\mu^2_0)}{\alpha_s (\mu^2)}
\right)^{\gamma^-_{j-1} / \beta_0}\,,
\eea
\bea\label{evolvedGgnbrMomentsGluon}
&&\mathbb{F}^g_j(\eta, t ; \mu^2)\nonumber\\
&=&2\times\frac{2\left(\gamma^{gq}_{(0)j-1}\right)}{\gamma^+_{j-1} - \gamma^-_{j-1}}\times \mathbb{F}^+_j(\eta, t;\mu_0^2)\times\left(
\frac{\alpha_s (\mu^2_0)}{\alpha_s (\mu^2)}
\right)^{\gamma^+_{j-1} / \beta_0} \nonumber\\ 
&+& 2\times\frac{-2\left(\gamma^{gq}_{(0)j-1}\right)}{\gamma^+_{j-1} - \gamma^-_{j-1}}\times \mathbb{F}^-_j(\eta, t;\mu_0^2)\times\left(
\frac{\alpha_s (\mu^2_0)}{\alpha_s (\mu^2)}
\right)^{\gamma^-_{j-1} / \beta_0}\,,
\eea
where
\begin{equation}\label{cqcg}
F^{\pm}_j(\eta, t ; \mu_0^2) = 
\frac{1}{4}
\frac{\gamma^{qg}_{(0)j-1}}{\gamma^{qq}_{(0)j-1} - \gamma^{\mp}_{j-1}} \mathbb{F}^g_j(\eta, t ; \mu_0^2)
\, .
\end{equation}
For two flavor (up (u) and down (d)) quarks, we will also assume that $\sum_q \mathbb{F}^q_{j(singlet)}(\eta, t ; \mu^2)\propto\mathbb{F}^{u+d}_{j(singlet)}(\eta, t ; \mu^2)\propto\mathbb{F}^{u-d}_{j(singlet)}(\eta, t ; \mu^2)$. Also note that the input gluon GPD $\mathbb{F}^g_j(\eta, t ; \mu_0^2)$ is given in (\ref{gluonGgrMoments2}).

Finally, in terms of the
evolved conformal (Gegenbauer) moments (\ref{evolvedGgnbrMomentsSingletQuark}-\ref{evolvedGgnbrMomentsGluon}),  the evolved singlet quark and gluon GPDs are given by
\bea
\sum_q F_{singlet}^q (x, \eta, t;\mu^2) &=& 
\frac{1}{\eta}\sum_{j=2}^{\infty}\frac{w \left(\frac{x}{\eta} | \frac{3}{2} \right)}{\eta^{j-1} N_{j-1} \left(\frac{3}{2}\right)}
C_{j-1}^{3/2}\left(\frac{x}{\eta}\right)\times\sum_q \mathbb{F}^q_{j(singlet)}(\eta, t ; \mu^2)\,,
\nonumber\\
&=&\frac{1}{\eta}\times
\left(1-\frac{x^2}{\eta^2}\right)\times\sum_{j=2}^{\infty}\frac{1}{\eta^{j-1}}\times\frac{1}{N_{j-1}\left(\frac{3}{2}\right)}
\times C_{j-1}^{3/2}\left(\frac{x}{\eta}\right)\times\sum_q \mathbb{F}^q_{j(singlet)}(\eta, t ; \mu^2)\,,\nonumber\\
\eea
and
\bea
F^g (x, \eta, t;\mu^2)&=&\frac{1}{\eta}\sum_{j=2}^{\infty}\frac{w \left(\frac{x}{\eta} | \frac{5}{2} \right)}{\eta^{j-2} N_{j-2} \left(\frac{5}{2}\right)}
C_{j-2}^{5/2}\left(\frac{x}{\eta}\right)\times\mathbb{F}^g_{j}(\eta, t ; \mu^2)\,,\nonumber\\
&=&\frac{1}{\eta}\times \left(1-\frac{x^2}{\eta^2}\right)^2\times \sum_{j=2}^{\infty}\frac{1}{\eta^{j-2}}\times \frac{1}{N_{j-2} \left(\frac{5}{2}\right)}
\times C_{j-2}^{5/2}\left(\frac{x}{\eta}\right)\times\mathbb{F}^g_{j}(\eta, t ; \mu^2)\,.
\eea

\section{Electroproduction of longitudinal $\rho^0$ meson with evolved singlet quark and gluon GPDs: a comparison to experiment}~\label{sec_RHO0}

We now apply the preceding results to the electroproduction of vector mesons as illustrated in Fig.~\ref{fig_HVM2}. In this section, we will
focus mostly on the electroproduction of longitudinal neutral $\rho^0$, and the comparison to the available detailed data for this process.
The extension to the charged $\rho^+$ production, and the $\phi$ mesons, with comparison to data will be discussed in the next sections. Clearly,
most of our analysis applies to heavier meson production such as $J/\Psi$ and $\Upsilon$ with minor kinematical changes.

\subsection{Evolved quark GPD contribution for $\rho^0$}~\label{RHO-1}

To summarize, the  electroproduction of neutral rho meson ($\rho^0$) in terms of quark DAs and quark GPDs is given by 
(see Eqs.~219-221 in \cite{Goeke:2001tz}) 
\begin{eqnarray}\label{rho0GPD11}
&&{\cal A}^{LL(quark)}_{\gamma^{*} p\rightarrow  \rho^{0} p} (s,t,Q,\epsilon_{L},\epsilon_{L}^{\prime}) =\nonumber\\ 
&&-e\times {C_{F}\over N_c}\times{1 \over 2}\times (4\pi \alpha _{s}(\mu^2))\times{1\over {Q}}\times\left[ \, \int _{0}^{1}dz{{\Phi_{\rho}(z)}\over z}\right] 
\times
\frac{1}{p^{+}}\left\{ A_{\rho^0\,p} 
\,h^{+}
\,+\, B_{\rho^0\,p} 
\,e^{+} 
\right\} \, , 
\end{eqnarray}
where $\Phi_{\rho}(z)$ is the neutral rho meson DA,
and
\bea
A_{\rho^0 \, p} \,&=&\, \int_{-1}^1 dx \; 
{-1 \over {\sqrt 2}} \, 
{\left(Q_u \ H^u(x,\eta, t;\mu^2) \,-\, Q_d \ H^d(x,\eta, t;\mu^2)\right)} 
\; \left\{ {{1} \over {\xi - x - i \epsilon}}
- {{1} \over {\xi + x -i \epsilon}} \right\}\nonumber\\
B_{\rho^0 \, p} \,&=&\, \int_{-1}^1 dx \;
{-1 \over {\sqrt 2}} \, 
{\left(Q_u \ E^u(x,\eta, t;\mu^2) \,-\, Q_d \ E^d(x,\eta, t;\mu^2)\right)} 
\;\left\{ {{1} \over {\xi - x - i \epsilon}}
- {{1} \over {\xi + x - i \epsilon}} \right\}\,.
\eea
Using $C_{F}=\frac{N_c^2-1}{2N_c}$, we can rewrite (\ref{rho0GPD11}) as 
\begin{eqnarray}\label{rho0GPD222}
&&{\cal A}^{LL(quark)}_{\gamma^{*} p\rightarrow  \rho^{0} p} (s,t,Q,\epsilon_{L},\epsilon_{L}^{\prime}) =\nonumber\\
&& e\times {1\over 2}\times{1 \over 2}\times\left(1-\frac{1}{N_c^2}\right)\times 4\pi\times\alpha_s(\mu^2)
\times {1\over {Q}}\times\left[ \, \int _{0}^{1}dz{{\Phi_{\rho}(z)}\over z}\right] \times
\frac{1}{p^{+}}\times u(p_2)\gamma^{+} u(p_1)\times{1 \over {\sqrt 2}}\nonumber\\
&&\times  \int_{0}^1 dx \; 
{\left(Q_u \ H_{singlet}^u(x,\eta, t;\mu^2) \,-\, Q_d \  H_{singlet}^d(x,\eta, t;\mu^2)\right)} 
\times 
\left\{ {{1} \over {\xi - x - i \epsilon}}
- {{1} \over {\xi + x -i \epsilon}} \right\}\,,\nonumber\\
\end{eqnarray}
where we have defined $H_{singlet}^q(x,\eta, t;\mu^2)=H^q(x,\eta, t;\mu^2)-H^q(-x,\eta, t;\mu^2)$, and we have ignored the Pauli contribution for $-t\ll 4m_N^2$. Using the conformal expansion of the singlet quark GPDs in the ERBL region, we can write the amplitude (\ref{rho0GPD222}) in terms of the conformal moments as
\begin{eqnarray}\label{rho0GPD22EVOLVEDHEnergy22}
&&{\cal A}^{LL(quark)}_{\gamma^{*} p\rightarrow  \rho^{0} p} (s,t,Q,\epsilon_{L},\epsilon_{L}^{\prime})\nonumber\\
&=& e\times {1\over 2}\times{1 \over 2}\times\left(1-\frac{1}{N_c^2}\right)\times 4\pi\alpha_s(\mu)\times{1\over {Q}}\times\left[ \, \int _{0}^{1}dz{{\Phi_{\rho}(z)}\over z}\right] \times
\frac{1}{2m_{N}}\times \bar u(p_2)u(p_1)\times{1 \over {\sqrt 2}}\nonumber\\
&\times & \Bigg[\int_{0}^1 dx\,\frac{2x}{\xi^2 - x^2}\times\left(Q_u \ H_{singlet}^u(x,\eta, t;\mu^2) \,-\, Q_d \ H_{singlet}^d(x,\eta, t;\mu^2)\right)\nonumber\\
&+&i\pi\left(Q_u \ H_{singlet}^u(\xi,\eta, t;\mu^2) \,-\, Q_d \ H_{singlet}^d(\xi,\eta, t;\mu^2)\right)\nonumber\\
&-&i\pi\left(Q_u \ H_{singlet}^u(-\xi,\eta, t;\mu^2) \,-\, Q_d \ H_{singlet}^d(-\xi,\eta, t;\mu^2)\right)\Bigg]\,,\nonumber\\
&\,=\,& e\times {1\over 2}\times{1 \over 2}\times\left(1-\frac{1}{N_c^2}\right)\times 4\pi\alpha_s(\mu)\times{1\over {Q}}\times\left[ \, \int _{0}^{1}dz{{\Phi_{\rho}(z)}\over z}\right] \times
\frac{1}{2m_N}\times \bar u(p_2)u(p_1)\times{1 \over {\sqrt 2}}\nonumber\\
&\times &\Bigg[\frac{2}{\xi}\times\sum_{j = 2}^{\infty}
\frac{1}{\eta^{j-1}}\times\frac{1}{N_{j-1} ( \frac{3}{2})}
\times\left[\int_{0}^\eta \frac{dx}{\eta}\,\frac{\frac{x}{\xi}\left(1 - \frac{x^2}{\eta^2}\right)}{1 - \frac{x^2}{\xi^2}}\times C_{j-1}^{3/2}\left(\frac{x}{\eta}\right)\right]\nonumber\\
&\times &\left(Q_u \ \mathbb{F}^u_{j(singlet)} (\eta, t ; \mu^2) \,-\, Q_d \ \mathbb{F}^d_{j(singlet)} (\eta, t ; \mu^2)\right)\nonumber\\
&+&i\pi\left(Q_u \ H_{singlet}^u(\xi,\eta, t;\mu^2) \,-\, Q_d \ H_{singlet}^d(\xi,\eta, t;\mu^2)\right)\nonumber\\
&-&i\pi\left(Q_u \ H_{singlet}^u(-\xi,\eta, t;\mu^2) \,-\, Q_d \ H_{singlet}^d(-\xi,\eta, t;\mu^2)\right)\Bigg]\,,\nonumber\\
&\,\approx\,& e\times {1\over 2}\times\left(1-\frac{1}{N_c^2}\right)\times 4\pi\alpha_s(\mu)\times{1\over {Q}}\times\left[ \, \int _{0}^{1}dz{{\Phi_{\rho}(z)}\over z}\right] \times
\frac{1}{2m_{N}}\times \bar u(p_2)u(p_1)\times{1 \over {\sqrt 2}}\nonumber\\
&\times &
\sum_{j=2}^{\infty}\frac{1}{\xi^{j}}\times\frac{1}{N_{j-1} ( \frac{3}{2})}
\times\left[\int_{0}^\xi \frac{dx}{\xi}\,\frac{x}{\xi}\,C_{j-1}^{3/2}\left(\frac{x}{\xi}\right)\right]\nonumber\\
&\times &\left(Q_u \ \mathbb{F}^u_{j(singlet)} (\eta\sim\xi, t ; \mu^2) \,-\, Q_d \ \mathbb{F}^d_{j(singlet)} (\eta\sim\xi, t ; \mu^2)\right)\,,\nonumber\\
\end{eqnarray}
where in the last line we have used $\eta\sim\xi$, and dropped the $i\pi H$ contributions in the Regge limit. We can also rewrite the amplitude(\ref{rho0GPD22EVOLVEDHEnergy22}) in a more compact form
\bea
\label{rho0GPD22EVOLVEDHEnergy2}
&&{\cal A}^{LL(quark)}_{\gamma^{*} p\rightarrow  \rho^{0} p} (s,t,Q,\epsilon_{L},\epsilon_{L}^{\prime})\approx  e\times f_V\times\alpha_s(\mu)\times{1\over {Q}}\nonumber\\
&&\times \sum_{j=2}^{\infty}\frac{1}{\xi^{j}}\times\mathcal{N}_{q}(j)\times \left(Q_u \ \mathbb{F}^u_{j(singlet)} (\xi, t ; \mu^2) \,-\, Q_d \ \mathbb{F}^d_{j(singlet)} (\xi, t ; \mu^2)\right)
\times  \frac{1}{2\sqrt{2} m_{N}}\times \bar u(p_2)u(p_1)\,,\nonumber\\
\eea
where we have defined 
\bea
\mathcal{N}_{q}(j) &\equiv &  {1\over 2}\times\left(1-\frac{1}{N_c^2}\right)\times 4\pi\times
\frac{1}{N_{j-1} ( \frac{3}{2})}
\times I_q(j)\,,
\eea
and the integrals (with $\tilde{x}=x/\xi$.)
\bea
f_V =\int _{0}^{1}dz{{\Phi_{\rho}(z)}\over z}\,,\qquad\qquad
I_{q}(j)=\left[\int_{0}^1 d\tilde{x}\,\tilde{x}\,C_{j-1}^{3/2}\left(\tilde{x}\right)\right]\,.
\eea
\\
\\
\\
{\bf Resummation by j-contour}
\\

\begin{figure*}
  \includegraphics[height=6cm,width=.6\linewidth]{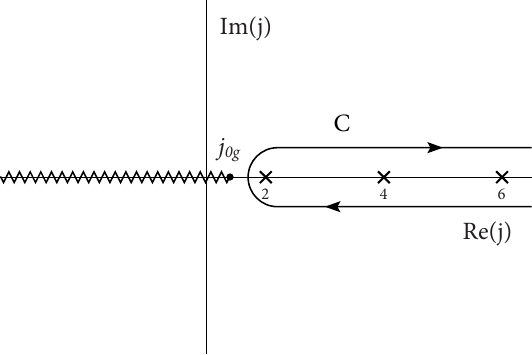}%
\caption{Illustration of the integration contour $\mathbb C$ in the complex j-plane used for computing the contour integral (\ref{rho0GPD22EVOLVEDHEnergy3INPUTquark}), with a branch cut displayed for $\text{Re(j)} \leq j_{0g}=2-2/\sqrt{\lambda}$.}
\label{contourgluon}
\end{figure*}

Following  \cite{Brower:2007xg}, we rewrite the sum over even $j=2,4,...$ in (\ref{rho0GPD22EVOLVEDHEnergy2}) as a contour integral in the complex j-plane 
\bea
\label{rho0GPD22EVOLVEDHEnergy3INPUTquark}
{\cal A}^{LL(quark)}_{\gamma^{*} p\rightarrow  \rho^{0} p} (s,t,Q,\epsilon_{L},\epsilon_{L}^{\prime})&\approx& -\int_{\mathbb C}\frac{dj}{4i}\frac{1+e^{-i\pi j}}{{\rm sin}\pi j}
\times\frac{1}{\xi^{j}}\times\frac{1}{\Gamma(\Delta_{g}(j)-2)}\times\mathcal{N}_{q}(j)\nonumber\\
&\times&\left(Q_u \ \widehat{\mathbb{F}}^u_{j(singlet)} (\xi, t ; \mu^2 ) \,-\, Q_d \ \widehat{\mathbb{F}}^d_{j(singlet)} (\xi, t ; \mu^2 )\right)\nonumber\\
&\times & e\times f_V\times\alpha_s(\mu)\times{1\over {Q}}\times \frac{1}{2\sqrt{2} m_{N}}\times \bar u(p_2)u(p_1)\,,
\eea
where the contour $\mathbb C $, shown in Fig.~\ref{contourgluon}, is at the right most of the branch point of $$\frac{1}{\Gamma(\Delta_{g}(j)-2)}=\frac{1}{\Gamma\left(\sqrt{2}\lambda^{1/4}(j-j_{0g})^{1/2}\right)}$$ and enclosing the poles at even $j=2,4,...$. Here we have defined
\bea
\widehat{\mathbb{F}}^q_{j(singlet)} (\xi, t ; \mu^2 )=\Gamma(\Delta_{g}(j)-2)\times\mathbb{F}^q_{j(singlet)} (\xi, t ; \mu^2 ) 
\eea
in order to reveal the branch point at $j=j_{0g}=2-2/\sqrt{\lambda}$ coming from our holographic input conformal (Gegenbauer) moments of the gluon GPD at $\mu=\mu_0\sim Q_0$. 

We evaluate the amplitude (\ref{rho0GPD22EVOLVEDHEnergy3INPUTquark}) by wrapping the j-plane contour $\mathbb C $ to the left
\bea
\label{rho0GPD22EVOLVEDHEnergy4INPUTquark}
{\cal A}^{LL(quark)}_{\gamma^{*} p\rightarrow  \rho^{0} p} (s,t,Q,\epsilon_{L},\epsilon_{L}^{\prime})&\approx& -\int_{-\infty}^{j_{0g}}\frac{dj}{2i}\frac{1+e^{-i\pi j}}{{\rm sin}\pi j}
\times\frac{1}{\xi^{j}}\times\text{Im}\left[\frac{1}{\Gamma(iy)}\right]\nonumber\\
&\times & \left(Q_u \ \widehat{\mathbb{F}}^u_{j(singlet)} (\xi, t ; \mu^2) \,-\, Q_d \ \widehat{\mathbb{F}}^d_{j(singlet)} (\xi, t ; \mu^2)\right)\nonumber\\
&\times &\mathcal{N}_{q}(j)\times e\times f_V\times\alpha_s(\mu)\times{1\over {Q}}\times\frac{1}{2\sqrt{2} m_{N}}\times \bar u(p_2)u(p_1)\,,
\eea
with $iy = i\sqrt{2\sqrt{\lambda}(j_{0g}-j)}\,.$ For, $y\rightarrow 0$, we may approximate $\frac{1}{\Gamma(iy)}\approx iy\,e^{i\gamma_E y}$ with the Euler constant $\gamma_E\approx 0.577216$, and write $\text{Im}\left[\frac{1}{\Gamma(iy)}\right]\approx y\,\text{cos}(\gamma_E y)\approx \text{sin}(y)$ for $y\rightarrow 0$ or $j\rightarrow j_{0g}$. Therefore, for $j\rightarrow j_{0g}$, we find
\bea
\label{rho0GPD22EVOLVEDHEnergy5INPUTquark}
{\cal A}^{LL(quark)}_{\gamma^{*} p\rightarrow  \rho^{0} p} (s,t,Q,\epsilon_{L},\epsilon_{L}^{\prime})&\approx&\left(Q_u\widehat{\mathbb{F}}^u_{j_0(singlet)} (\xi, t ; \mu^2)-Q_d\widehat{\mathbb{F}}^d_{j_0(singlet)} (\xi, t ; \mu^2)\right)\nonumber\\
&\times &\mathcal{N}_{q}(j_{0g})\times e\,f_V\,\alpha_s(\mu)\,{1\over {Q}}\,\frac{1}{2\sqrt{2} m_{N}}\,\bar u(p_2)u(p_1)\nonumber\\
&\times & -\int_{-\infty}^{j_{0g}}\frac{dj}{2}\frac{1+e^{-i\pi j}}{{\rm sin}\pi j}
\times\frac{1}{\xi^{j}}\times \text{sin}\left(\sqrt{2\sqrt{\lambda}(j_{0g}-j)}\right)\,.
\eea
\\
\\
{\bf Real and Imaginary parts}
\\
We separate the amplitude (\ref{rho0GPD22EVOLVEDHEnergy5INPUTquark}) into its real and imaginary parts $${\cal A}^{LL(quark)}_{\gamma^{*} p\rightarrow  \rho^{0} p} (s,t,Q,\epsilon_{L},\epsilon_{L}^{\prime})\equiv{\cal A}^{LL(quark)}=\text{Re}\left[{\cal A}^{LL(quark)}\right]+i\text{Im}\left[{\cal A}^{LL(quark)}\right]\,,$$ with
\bea\label{REIMrho0GPD22EVOLVEDHEnergy4}
\text{Re}\left[{\cal A}^{LL(quark)}\right]&=& \mathcal{N}_{q}(j_{0g})\times \left(Q_u \ \widehat{\mathbb{F}}^u_{j_{0q}(singlet)} (\xi, t ; \mu^2) \,-\, Q_d \ \widehat{\mathbb{F}}^d_{j_{0q}(singlet)} (\xi, t ; \mu^2)\right)\nonumber\\
&\times & e\,f_V\,\alpha_s(\mu)\,{1\over {Q}}\,\frac{1}{2\sqrt{2} m_{N}}\,\bar u(p_2)u(p_1) \times -\int_{-\infty}^{j_{0g}}\frac{dj}{2}\frac{1+\text{cos}\pi j}{{\rm sin}\pi j}
\times\frac{1}{\xi^{j}}\times \text{sin}\left(\sqrt{2\sqrt{\lambda}(j_{0g}-j)}\right)\,,\nonumber\\
\text{Im}\left[{\cal A}^{LL(quark)}\right] &=&\mathcal{N}_{q}(j_{0q})\times \left(Q_u \ \widehat{\mathbb{F}}^u_{j_{0q}(singlet)} (\xi, t ; \mu^2) \,-\, Q_d \ \widehat{\mathbb{F}}^d_{j_{0q}(singlet)} (\xi, t ; \mu^2)\right)\nonumber\\
&\times & e\,f_V\,\alpha_s(\mu)\,{1\over {Q}}\,\frac{1}{2\sqrt{2} m_{N}}\,\bar u(p_2)u(p_1) \times\int_{-\infty}^{j_{0g}}\frac{dj}{2}\,\frac{1}{\xi^{j}}\times \text{sin}\left(\sqrt{2\sqrt{\lambda}(j_{0g}-j)}\right)\,.\nonumber\\
\eea
With the change of integration variable $y^2=2\sqrt{\lambda}(j_{0g}-j)$, we turn the imaginary part into a Gaussian integral that can be evaluated exactly (with $\tilde{\tau}=\text{log}(1/\xi)$)
\be\label{IMrho0GPD22EVOLVEDHEnergy4}
\text{Im}\left[{\cal A}^{LL(quark)}\right]&=&\mathcal{N}_{q}(j_{0g})\times \left(Q_u \ \widehat{\mathbb{F}}^u_{j_{0q}(singlet)} (\xi, t ; \mu^2) \,-\, Q_d \ \widehat{\mathbb{F}}^d_{j_{0q}(singlet)} (\xi, t ; \mu^2)\right)\nonumber\\
&\times & e\,f_V\,\alpha_s(\mu)\,{1\over {Q}}\,\frac{1}{2\sqrt{2} m_{N}}\,\bar u(p_2)u(p_1) \times\frac{1}{\xi^{j_{0g}}}\times\int_{-\infty}^{\infty}\frac{dy}{4\pi i\sqrt{\lambda}}\,y\,e^{-\tilde{\tau}y^2/2\sqrt{\lambda}}\,e^{iy}\,,\nonumber\\
&=&\mathcal{N}_{q}(j_{0g})\times \left(Q_u \ \widehat{\mathbb{F}}^u_{j_{0g}(singlet)} (\xi, t ; \mu^2) \,-\, Q_d \ \widehat{\mathbb{F}}^d_{j_{0q}(singlet)} (\xi, t ; \mu^2)\right)\nonumber\\
&\times & e\,f_V\,\alpha_s(\mu)\,{1\over {Q}}\,\frac{1}{2\sqrt{2} m_{N}}\,\bar u(p_2)u(p_1) \times\frac{1}{\xi^{j_{0g}}}\times (\sqrt{\lambda}/2\pi)^{1/2}\times \frac{e^{-\sqrt{\lambda}/2\tilde{\tau}}}{\tilde{\tau}^{3/2}}\,.
\ee
Note the emergence of $1/\xi^{j_{0g}}$ in the imaginary part of the amplitude in (\ref{IMrho0GPD22EVOLVEDHEnergy4}) with $j_{0g}=2-2/\sqrt{\lambda}$.
This follows from the transmutation of the graviton with spin $j=2$, to a Pomeron with spin $j_{0g}$, after Reggeization.

We now compute the real part of the amplitude $\text{Re}\left[{\cal A}^{LL(quark)}\right]$,  by first expanding its prefactor near $j=2$ 
\bea
\frac{1+\text{cos}\pi j}{{\rm sin}\pi j}\simeq \frac{2}{\pi(j-2)}+\mathcal{O}(j-2)\,,
\eea
and establishing the identity $\partial_{\tilde{\tau}}\left[e^{-2\tilde{\tau}}\text{Re}\left[{\cal A}^{LL(quark)}\right]\right]=-(2/\pi)e^{-2\tilde{\tau}}\text{Im}\left[{\cal A}^{LL(quark)}\right]$ which gives us an approximation for the real part
\bea\label{RErho0GPD22EVOLVEDHEnergy4}
\text{Re}\left[{\cal A}^{LL(quark)}\right] &\simeq& \mathcal{N}_{q}(j_{0g})\times \left(Q_u \ \widehat{\mathbb{F}}^u_{j_{0q}(singlet)} (\xi, t ; \mu^2) \,-\, Q_d \ \widehat{\mathbb{F}}^d_{j_{0g}(singlet)} (\xi, t ; \mu^2)\right)\nonumber\\
&\times & e\,f_V\,\alpha_s(\mu)\,{1\over {Q}}\,\frac{1}{2\sqrt{2} m_{N}}\,\bar u(p_2)u(p_1)\times (\sqrt{\lambda}/2\pi)^{1/2}\times \frac{1}{\xi^{2}}\times\int_{\tilde{\tau}}^{\infty}d\tilde{\tau}^{\prime}\,\frac{2e^{-2\tilde{\tau}^{\prime}/\sqrt{\lambda}-\sqrt{\lambda}/2\tilde{\tau}^{\prime}}}{\pi\tilde{\tau}^{\prime 3/2}}\,.\nonumber\\
\eea
Small corrections in the order of $\mathcal{O}(\text{Im}\left[{\cal A}^{LL(quark)}\right]/\text{log}(1/\xi))$ and $\mathcal{O}(\text{Im}\left[{\cal A}^{LL(quark)}\right]/\sqrt{\lambda})$ to (\ref{RErho0GPD22EVOLVEDHEnergy4}) can be computed in a standard perturbation series,  but can be ignored for $\xi\rightarrow 0$ and fixed large $\sqrt{\lambda}$. 
\\
\\
{\bf small-$\xi$ Regime}
\\
We then compute the integral in (\ref{RErho0GPD22EVOLVEDHEnergy4}) in the small-$\xi$ regime, i.e., $\tilde{\tau}=\text{log}(1/\xi)\rightarrow\infty$ at fixed large $\sqrt{\lambda}$, where the integral is dominated by its end point, and can be approximated by
\be
\int_{\tilde{\tau}}^{\infty}d\tilde{\tau}^{\prime}\,\frac{2e^{-2\tilde{\tau}^{\prime}/\sqrt{\lambda}-\sqrt{\lambda}/2\tilde{\tau}^{\prime}}}{\pi\tilde{\tau}^{\prime 3/2}}=(\sqrt{\lambda}/\pi)\times e^{-2\tilde{\tau}/\sqrt{\lambda}}\times\frac{e^{-\sqrt{\lambda}/2\tilde{\tau}}}{\tilde{\tau}^{3/2}}\left(1+\mathcal{O}(\sqrt{\lambda}/\tilde{\tau})\right)\,.
\ee 
Finally, combining this approximation for $\text{Re}\left[{\cal A}^{LL(quark)}\right]$ with the exact result for $\text{Im}\left[{\cal A}^{LL(quark)}\right]$ (\ref{IMrho0GPD22EVOLVEDHEnergy4}), we find the complex amplitude with evolved singlet quark GPDs (for input gluon GPDs part) 
\bea\label{fullquarkEVOLVEDquarkINPUT}
{\cal A}^{LL(quark)}_{\gamma^{*} p\rightarrow  \rho^{0} p} (s,t,Q,\epsilon_{L},\epsilon_{L}^{\prime})&\simeq&  \Bigg(Q_u\widehat{\mathbb{F}}^u_{j_{0g}(singlet)} (\xi, t ; \mu^2)- Q_d \widehat{\mathbb{F}}^d_{j_{0g}(singlet)} (\xi, t ; \mu^2)\Bigg)\nonumber\\
&\times& \Bigg[(\sqrt{\lambda}/2\pi)^{1/2}\times \frac{1}{\xi^{2}}\times(\sqrt{\lambda}/\pi)\times e^{-2\tilde{\tau}/\sqrt{\lambda}}\times\frac{e^{-\sqrt{\lambda}/2\tilde{\tau}}}{\tilde{\tau}^{3/2}}\nonumber\\
&+& i\times \frac{1}{\xi^{j_{0g}}}\times (\sqrt{\lambda}/2\pi)^{1/2}\times \frac{e^{-\sqrt{\lambda}/2\tilde{\tau}}}{\tilde{\tau}^{3/2}}\Bigg]\times\mathcal{N}_{q}(j_{0g})e\,f_V\,\alpha_s(\mu)\,{1\over {Q}}\,\frac{1}{2\sqrt{2} m_{N}}\,\bar u(p_2)u(p_1)\,,\nonumber\\  
&\simeq&  \Bigg(Q_u\widehat{\mathbb{F}}^u_{j_{0g}(singlet)} (\xi, t ; \mu^2)-Q_d\widehat{\mathbb{F}}^d_{j_{0g}(singlet)} (\xi, t ; \mu^2)\Bigg)\nonumber\\
&\times& \mathcal{N}_{q}(j_{0g})\times e\,f_V\,\alpha_s(\mu)\,{1\over {Q}}\,\frac{1}{2\sqrt{2} m_{N}}\,\bar u(p_2)u(p_1)\nonumber\\ 
&\times& \left[\frac{(\sqrt{\lambda}/\pi)}{\xi^{2-2/\sqrt{\lambda}}}+\frac{i}{\xi^{2-2/\sqrt{\lambda}}}\right]\times(\sqrt{\lambda}/2\pi)^{1/2}\times \frac{e^{-\sqrt{\lambda}/2\tilde{\tau}}}{\tilde{\tau}^{3/2}}\,. 
\eea
Therefore, the full complex amplitude due to the evolved singlet quark GPDs is given by (\ref{fullquarkEVOLVEDquarkINPUT})
\bea\label{FULLquarkSinglet}
{\cal A}^{LL(quark)}_{\gamma^{*} p\rightarrow  \rho^{0} p} (s,t,Q,\epsilon_{L},\epsilon_{L}^{\prime})
&\propto&\Bigg(Q_u\widehat{\mathbb{F}}^u_{j_{0g}(singlet)} (\xi, t ; \mu^2)-Q_d\widehat{\mathbb{F}}^d_{j_{0g}(singlet)} (\xi, t ; \mu^2)\Bigg)\nonumber\\
&\times &f_V\,\alpha_s(\mu)\,{1\over {Q}}\,\frac{1}{2\sqrt{2} m_{N}}\,\bar u(p_2)u(p_1)\times \left[\frac{(\sqrt{\lambda}/\pi)}{\xi^{2-2/\sqrt{\lambda}}}+\frac{i}{\xi^{2-2/\sqrt{\lambda}}}\right]\,,
\eea
where we have approximated $\mathcal{N}_{q}(j_{0g})\times e\times(\sqrt{\lambda}/2\pi)^{1/2}\times \frac{e^{-\sqrt{\lambda}/2\tilde{\tau}}}{\tilde{\tau}^{3/2}}$ to be a constant since it varies very slowly with $s$.

\subsection{Evolved gluon GPD contribution for $\rho^0$}~\label{RHO-2}
In addition, the electroproduction of neutral rho meson ($\rho^0$) in terms of quark DAs and gluon GPDs is given by (see Eq.282 in \cite{Diehl:2003ny})
\begin{eqnarray}
  \label{meson-22}
&&\mathcal{A}^{LL(gluon)}_{\gamma^{*} p\rightarrow  \rho^{0} p} (s,t,Q,\epsilon_{L},\epsilon_{L}^{\prime})=\nonumber\\
&&e\times\frac{1}{8}\times\frac{1}{N_c}\times 4\pi \times \alpha_s(\mu^2) \times \frac{1}{Q}
\times \left[\int_0^1 dz\, \sum_{q} e_q \frac{\Phi^q(z)}{z(1-z)}\right] \times
\int_{-1}^1 dx\, \frac{F^g(x,\eta, t;\mu^2)}{x}  
\left\{\frac{1}{\xi - x -i\epsilon} - \frac{1}{\xi + x -i\epsilon} \right\}\,.\nonumber\\
\end{eqnarray}
On the light front, 
$$h^+=\bar u(p_2)\gamma^{+}u(p_1)\approx \frac{p^+}{2m_N}\times \bar u(p_2)u(p_1),$$
and ignoring the Pauli contribution $E^g(x,\eta, t;\mu^2)$ to $F^g(x,\eta, t;\mu^2)$ for $-t \ll 4m_N^2$, we have
\bea
F^g(x,\eta, t;\mu^2)\approx\frac{h^{+}}{p^{+}}H^g(x,\eta, t;\mu^2)= H^g(x,\eta, t;\mu^2)\times \frac{1}{2m_N}\times \bar u(p_2)u(p_1)\,,
\eea
which simplifies (\ref{meson-22}) 
\bea\label{gluonFinal22}
\mathcal{A}^{LL(gluon)}_{\gamma^{*} p\rightarrow  \rho^{0} p} (s,t,Q,\epsilon_{L},\epsilon_{L}^{\prime})&=& e\times\frac{1}{8}\times\frac{1}{N_c}\times \frac{1}{Q}
\times 4\pi \times \alpha_{s}(\mu^2) \times \frac{1}{2m_N}
\times\bar u(p_2)u(p_1)\nonumber\\
&\times&\left[\int_0^1 dz\, \sum_{q} e_q \frac{\Phi^q(z)}{z(1-z)}\right]\times
\int_{0}^1 dx\, \frac{H^g(x,\eta, t;\mu^2)}{x}\times\left\{\frac{1}{\xi - x -i\epsilon} - \frac{1}{\xi + x -i\epsilon} \right\}  
\,.\nonumber\\
\eea

Using the conformal expansion of gluon GPDs, the amplitude (\ref{gluonFinal22}) can be written in terms of the conformal moments of gluon GPDs as
\bea\label{gluonFinal2EVOLVEDHEnergy}
&&\mathcal{A}^{LL(gluon)}_{\gamma^{*} p\rightarrow  \rho^{0} p} (s,t,Q,\epsilon_{L},\epsilon_{L}^{\prime})\nonumber\\
&=& e\times\frac{1}{4}\times\frac{1}{N_c}\times \frac{1}{Q}
\times \alpha_{s}(\mu^2)\times\left[\int_0^1 dz\, \sum_{q} e_q \frac{\Phi^q(z)}{z(1-z)}\right]\times \Bigg[\frac{1}{\xi^2} \times
\int_{0}^1 dx\, \frac{H^g(x,\eta, t;\mu^2)}{(1-\frac{x^2}{\xi^2})}\nonumber\\  
&+&i\pi H^g(\xi,\eta, t;\mu^2)-i\pi H^g(-\xi,\eta, t;\mu^2)\Bigg]\times \frac{1}{m_N}
\times\bar u(p_2)u(p_1)\,,\nonumber\\
&=& e\times\frac{1}{4}\times\frac{1}{N_c}\times \frac{1}{Q}
\times \alpha_{s}(\mu^2)\times\left[\int_0^1 dz\, \sum_{q} e_q \frac{\Phi^q(z)}{z(1-z)}\right]\times \frac{1}{m_N}
\times\bar u(p_2)u(p_1)\nonumber\\
&\times & \Bigg[\frac{1}{\xi^2}\times\sum_{j = 2}^{\infty}
\frac{1}{\eta^{j-2}}\times\frac{1}{N_{j-2} ( \frac{5}{2})}\times
\left[\int_{0}^{\eta}\frac{dx}{\eta}\, \frac{\left(1-\frac{x^2}{\eta^2}\right)^2}{(1-\frac{x^2}{\xi^2})}\times C_{j-2}^{5/2}\left(\frac{x}{\eta}\right)\right]\times\mathbb{F}^g_{j} (\eta, t ; \mu^2)\nonumber\\ 
&+&i\pi H^g(\xi,\eta, t;\mu^2)-i\pi H^g(-\xi,\eta, t;\mu^2)\Bigg]\nonumber\\
&\approx & e\times\frac{1}{4}\times\frac{1}{N_c}\times \frac{1}{Q}
\times \alpha_{s}(\mu^2)\times\left[\int_0^1 dz\, \sum_{q} e_q \frac{\Phi^q(z)}{z(1-z)}\right]\times \frac{1}{m_N}
\times\bar u(p_2)u(p_1)\nonumber\\
&\times &
\sum_{j=2}^{\infty}\frac{1}{\xi^{j}}\times\frac{1}{N_{j-2} ( \frac{5}{2})}\times
\left[\int_{0}^{\xi}\frac{dx}{\xi}\, \left(1-\frac{x^2}{\xi^2}\right)\times C_{j-2}^{5/2}\left(\frac{x}{\xi}\right)\right]\times\mathbb{F}^g_{j} (\xi, t ; \mu^2)\,,\nonumber\\ 
\eea
in the last line we have used $\eta\sim\xi$. Note that the  $i\pi H^g$ contribution vanishes for $\xi=\pm \eta$. We can rewrite (\ref{gluonFinal2EVOLVEDHEnergy}) more compactly as 
\bea
\label{gluonFinal2EVOLVEDHEnergy2}
\mathcal{A}^{LL(gluon)}_{\gamma^{*} p\rightarrow  \rho^{0} p} (s,t,Q,\epsilon_{L},\epsilon_{L}^{\prime})&\approx &e\times f_V^{\prime}\times \alpha_{s}(\mu^2)\times {1\over {Q}}\times\sum_{j=2}^{\infty}\frac{1}{\xi^j}\times\mathcal{N}_g(j)\times\mathbb{F}^g_{j} (\xi, t ; \mu^2)\times\frac{1}{2\sqrt{2}m_N}
\times\bar u(p_2)u(p_1)\,,\nonumber\\ 
\eea
where we have defined 
\bea
\mathcal{N}_g(j) &\equiv &\frac{1}{4}\times \frac{1}{N_c}\times 2\sqrt{2}\times\frac{1}{N_{j-2}( \frac{5}{2})}\times I_{g}(j)\,,
\eea
and the integrals ($\tilde{x}=x/\xi$)
\bea
f_V^{\prime} =\int_0^1 dz\, \sum_{q} e_q \frac{\Phi^q(z)}{z(1-z)}\,,\qquad\qquad
I_{g}(j)=\int_{0}^{1}d\tilde{x}\, \left(1-\tilde{x}^2\right)C_{j-2}^{5/2}\left(\tilde{x}\right)\,.
\eea
The resummation by j-contour for the gluon GPD contribution (\ref{gluonFinal2EVOLVEDHEnergy2}) is similar to the resummation for the singlet quark GPDs, and the detailed derivation is given in Appendix~\ref{gluonsum}. Here we only provide the final answer.

After the resummation by j-contour, the full complex amplitude due to the evolved gluon GPD is given by (\ref{REIMgluonFinal2EVOLVEDHEnergy6INPUTgluon})  i.e.,
\bea\label{FULLgluon}
{\cal A}^{LL(gluon)}_{\gamma^{*} p\rightarrow  \rho^{0} p} (s,t,Q,\epsilon_{L},\epsilon_{L}^{\prime})&\propto&\widehat{\mathbb{F}}^g_{j_{0g}} (\xi, t ; \mu^2)\times e\,f_V^{\prime}\,\alpha_s(\mu)\,{1\over {Q}}\,\frac{1}{2\sqrt{2} m_{N}}\,\bar u(p_2)u(p_1)\nonumber\\  
&\times& \frac{1}{\xi^{2-2/\sqrt{\lambda}}}\times\left[(\sqrt{\lambda}/\pi)+i\right]\,,
\eea
with $j_{0g}=2-2/\sqrt{\lambda}$, and $\tilde{\tau}=\text{log}(1/\xi)$. We have also approximated $\mathcal{N}_{g}(j_{0g})\times e\times(\sqrt{\lambda}/2\pi)^{1/2}\times \frac{e^{-\sqrt{\lambda}/2\tilde{\tau}}}{\tilde{\tau}^{3/2}}$ to be a constant (that will be fixed by experimental data) since it varies slowly with $s$.

\subsection{Total quark and gluon GPDs contribution to the electroproduction of $\rho^0$}
To summarize, we have found the singlet quark GPD contribution to the $\rho^0$ amplitude to be given by
\bea\label{FULLquarkSinglet2}
{\cal A}^{LL(quark)}_{\gamma^{*} p\rightarrow  \rho^{0} p} (s,t,Q,\epsilon_{L},\epsilon_{L}^{\prime})&\propto&
\Bigg(Q_u\widehat{\widehat{\mathbb{F}}}^u_{j_{0g}(singlet)} (\xi, t ; \mu^2)-Q_d\widehat{\widehat{\mathbb{F}}}^d_{j_{0g}(singlet)} (\xi, t ; \mu^2)\Bigg)\nonumber\\
&\times &f_V\,\alpha_s(\mu)\,{1\over {Q}}\,\frac{1}{2\sqrt{2} m_{N}}\,\bar u(p_2)u(p_1)\times \left[\frac{(\sqrt{\lambda}/\pi)}{\xi^{2-2/\sqrt{\lambda}}}+\frac{i}{\xi^{2-2/\sqrt{\lambda}}}\right]\nonumber\\
&\propto &{\cal A}^{quark}_{\rho^{0}} (s,t,Q)\times\frac{1}{2\sqrt{2} m_{N}}\,\bar u(p_2)u(p_1)\,,
\eea
where we have normalized the singlet quark conformal moments using the lattice data from \cite{LHPC:2007blg} as
\bea
\widehat{\widehat{\mathbb{F}}}^{u+d}_{j(singlet)} (\xi, t ; \mu^2)&=&0.526\times\frac{\widehat{\mathbb{F}}^{u+d}_{j(singlet)} (\xi, t ; \mu^2)}{\widehat{\mathbb{F}}^{u+d}_{j=2(singlet)} (\xi=0, t=0 ; \mu^2=4~\text{GeV}^2)}\\
\widehat{\widehat{\mathbb{F}}}^{u-d}_{j(singlet)} (\xi, t ; \mu^2)&=&0.210\times\frac{\widehat{\mathbb{F}}^{u-d}_{j(singlet)} (\xi, t ; \mu^2)}{\widehat{\mathbb{F}}^{u-d}_{j=2(singlet)} (\xi=0, t=0 ; \mu^2=4~\text{GeV}^2)}\,,
\eea
with $\widehat{\mathbb{F}}^{u\pm d}_{j(singlet)} (\xi, t ; \mu^2)=\Gamma(\Delta_{g}(j)-2)\times\mathbb{F}^{u\pm d}_{j(singlet)} (\xi, t ; \mu^2)$ which is given in (\ref{evolvedGgnbrMomentsSingletQuark}). Note that we have used the average value of $A_2^{u+d}(-t=0)=\widehat{\widehat{\mathbb{F}}}^{u+d}_{j=2(singlet)} (\xi=0, t=0 ; \mu^2=4~\text{GeV}^2)=0.526$ from Table XV and Table XVI of \cite{LHPC:2007blg}. We have also used the average value of $A_2^{u-d}(-t=0)=\widehat{\widehat{\mathbb{F}}}^{u-d}_{j=2(singlet)} (\xi=0, t=0 ; \mu^2=4~\text{GeV}^2)=0.210$ from Table IX and Table X of \cite{LHPC:2007blg}. Also note that we have 
\bea
\widehat{\widehat{\mathbb{F}}}^{u}_{j(singlet)} (\xi, t ; \mu^2)&=&\frac{1}{2}\times\left(\widehat{\widehat{\mathbb{F}}}^{u+d}_{j(singlet)} (\xi, t ; \mu^2)+\widehat{\widehat{\mathbb{F}}}^{u-d}_{j(singlet)} (\xi, t ; \mu^2)\right)\\
\widehat{\widehat{\mathbb{F}}}^{d}_{j(singlet)} (\xi, t ; \mu^2)&=&\frac{1}{2}\times\left(\widehat{\widehat{\mathbb{F}}}^{u+d}_{j(singlet)} (\xi, t ; \mu^2)-\widehat{\widehat{\mathbb{F}}}^{u-d}_{j(singlet)} (\xi, t ; \mu^2)\right)\,.
\eea

We have also found the gluon GPD contribution to the $\rho^0$ amplitude to be
\bea\label{FULLgluon2}
{\cal A}^{LL(gluon)}_{\gamma^{*} p\rightarrow  \rho^{0} p} (s,t,Q,\epsilon_{L},\epsilon_{L}^{\prime})&\propto&\widehat{\widehat{\mathbb{F}}}^g_{j_{0g}} (\xi, t ; \mu^2)\times f_V^{\prime}\,\alpha_s(\mu)\,{1\over {Q}}\,\frac{1}{2\sqrt{2} m_{N}}\,\bar u(p_2)u(p_1)\nonumber\\  
&\times &\left[\frac{(\sqrt{\lambda}/\pi)}{\xi^{2-2/\sqrt{\lambda}}}+\frac{i}{\xi^{2-2/\sqrt{\lambda}}}\right]\nonumber\\
&\propto&{\cal A}^{gluon}_{\rho^{0}} (s,t,Q)\times\frac{1}{2\sqrt{2} m_{N}}\,\bar u(p_2)u(p_1)\,,
\eea
with $j_{0g}=2-2/\sqrt{\lambda}$, $\tilde{\tau}=\text{log}(1/\xi)$, $f_V^{\prime}\approx f_{V}$, and we have normalized the gluonic conformal moments using the lattice data from \cite{LHPC:2007blg} as
\bea
\widehat{\widehat{\mathbb{F}}}^g_{j} (\xi, t ; \mu^2)=0.474\times\frac{\widehat{\mathbb{F}}^g_{j} (\xi, t ; \mu^2)}{\widehat{\mathbb{F}}^g_{j=2} (\xi=0, t=0 ; \mu^2=4~\text{Gev}^2)}
\eea
where $\widehat{\mathbb{F}}^g_{j} (\xi, t ; \mu^2)=\Gamma(\Delta_{g}(j)-2)\times\mathbb{F}^g_{j} (\xi, t ; \mu^2)$ is given in (\ref{evolvedGgnbrMomentsGluon}). Note that we have used the fact that $A_2^{g}(-t=0)=\widehat{\widehat{\mathbb{F}}}^g_{j=2} (\xi=0, t=0 ; \mu^2=4~\text{Gev}^2)=1-A_2^{u+d}(-t=0)=1-\widehat{\widehat{\mathbb{F}}}^{u+d}_{j=2(singlet)} (\xi=0, t=0 ; \mu^2=4~\text{GeV}^2)=1-0.526=0.474$ from \cite{LHPC:2007blg}. 

\subsection{s+u-channel contribution to the electroproduction of $\rho^0$}
In this section, we first compute the $s+u$-channel contribution to DVCS, shown in Fig.~\ref{SWITTEN1}, and generalize the result to determine the s+u-channel contribution to the electroproduction of $\rho^0$, shown in Fig.~\ref{SWITTEN2}. See~(\ref{s-channelRho}) for the final answer to $\rho^0$ electroproduction.

\begin{figure*}
\subfloat[\label{SWITTEN1}]{%
  \includegraphics[height=5.5cm,width=.45\linewidth]{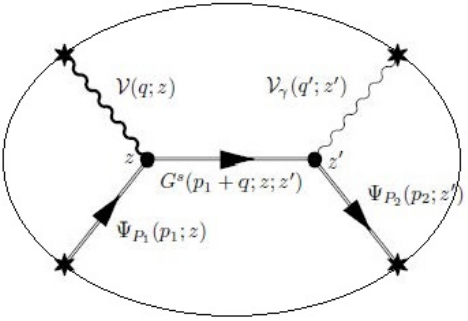}%
}\hfill
\subfloat[\label{SWITTEN2}]{%
  \includegraphics[height=5.5cm,width=.45\linewidth]{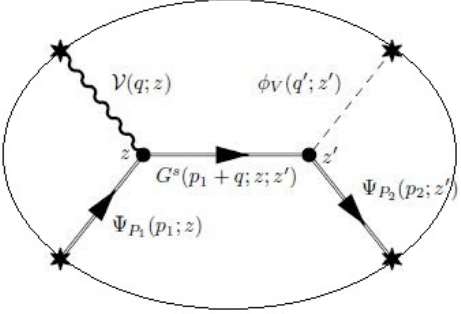}%
}
\caption{(a) Holographic $\textbf{s-channel}$ contribution to the electroproduction of a photon ($\textbf{DVCS}$). (b) Holographic $\textbf{s-channel}$ contribution to the electroproduction of a $\textbf{vector meson}$.}
 \label{SWITTEN}
\end{figure*}

The holographic DVCS amplitude shown in Fig.~\ref{SWITTEN1} involves the bulk Dirac fields

\bea
&&\Psi_{1p_1/2p_1}(z)=\left(\psi_R(z)P_{\pm}+\psi_L(z)P_{\mp}\right)\times u(p_1)\nonumber\\
&&\bar\Psi_{1p_2/2p_2}(z^\prime)=\bar u(p_2)\times\left(\psi_R(z^\prime)P_{\mp}+\psi_L(z^\prime)P_{\pm}\right)\nonumber\\
&&\Psi_{1p/2p}(z)\equiv \Psi_{1/2}(p,z;n=0)\nonumber\\
&&\psi_{R/L}(z)\equiv \psi_{R/L}(z;n=0)\,.\nonumber\\
\eea
To proceed, we define 
 
\bea
\tilde{G}^{s}(p_1+q,M_{n_x})=iG^{s}(p_1+q,M_{n_x})=\frac{i(p\hspace{-6pt}\slash_{1} +q\hspace{-6pt}\slash + M_{n_x})}{(p_1+q)^2-M_{n_x} ^{2}+i\epsilon }\nonumber\\
\eea
in terms of which the Dirac contribution to the s-channel is

\begin{eqnarray}
&&i\epsilon_{\mu}\epsilon_{\nu}^{\prime*} T_{s(\text{Dirac})}^{\mu\nu}(s,t;Q^2,Q^{\prime 2}) = \sum_{n_x}i\epsilon_{\mu}\epsilon_{\nu}^{\prime*}T_{s(\text{Dirac})}^{\mu\nu}(n_x,s,t;Q^2,Q^{\prime 2})\nonumber\\
&&i\epsilon_{\mu}\epsilon_{\nu}^{\prime*}T_{s(\text{Dirac})}^{\mu\nu}(n_x,s,t;Q^2,Q^{\prime 2}) = \int dz \sqrt{g} e^{- \tilde\kappa^2 z^2} \,\frac{z}{R}\int dz^{\prime
}\sqrt{g^\prime} e^{- \tilde\kappa^2 z^{\prime 2}}\,\frac{z^\prime}{R}\nonumber\\
&&\times \bar\Psi_{1p_2}(z^\prime)\times\left(-i\gamma^\nu\right)\times\Big(\frac{1}{3}V_\nu^{0}(q^\prime z^\prime)T^0+ V_\nu^{3}(q^\prime z^\prime)T^3\Big)\times\Big(\psi_R(z^\prime;n_x)P_{+}+ \psi_L(z^\prime;n_x)P_{-}\Big) \nonumber\\
&&\times \tilde{G}^{s}(p_1+q,M_{n_x})\times \Big(\psi_R(z;n_x)P_{-}+ \psi_L(z;n_x)P_{+}\Big)\times\left(-i\gamma^{\mu}\right)\times\Big(\frac{1}{3}V_\mu^{0}(qz)T^0+ V_\mu^{3}(qz)T^3\Big)\Psi_{1p_1}(z)\nonumber\\
&+&\int dz \sqrt{g} e^{- \tilde\kappa^2 z^2} \,\frac{z}{R}\int dz^{\prime
}\sqrt{g^\prime} e^{- \tilde\kappa^2 z^{\prime 2}} \,\frac{z^\prime}{R}\nonumber\\
&&\times \bar\Psi_{2p_2}(z^\prime)\times\left(-i\gamma^\nu\right)\times\Big(\frac{1}{3}V_\nu^{0}(q^\prime z^\prime)T^0+ V_\nu^{3}(q^\prime z^\prime)T^3\Big)\times\Big(\psi_R(z^\prime;n_x)P_{-}+ \psi_L(z^\prime;n_x)P_{+}\Big) \nonumber\\
&&\times \tilde{G}^{s}(p_1+q,M_{n_x})\times \Big(\psi_R(z;n_x)P_{+}+ \psi_L(z;n_x)P_{-}\Big)\times\left(-i\gamma^{\mu}\right)\times\Big(\frac{1}{3}V_\mu^{0}(qz)T^0+ V_\mu^{3}(qz)T^3\Big)\Psi_{2p_1}(z)\,,\nonumber\\
\end{eqnarray}
Setting  $q^2=-Q^2$, we have $\frac{1}{3}V_\mu^{0}(qz)T^0+ V_\mu^{3}(qz)T^3=\epsilon_{\mu}\times\frac{1}{2}\mathcal{V}(Qz)$ for the proton,
 we have
 
\bea
&&i\epsilon_{\mu}\epsilon_{\nu}^{\prime*}T_{s(\text{Dirac})}^{\mu\nu}(n_x,s,t;Q^2,Q^{\prime 2})=\int dz \sqrt{g} e^{- \tilde\kappa^2 z^2} \,\frac{z}{R}\int dz^{\prime
}\sqrt{g^\prime} e^{- \tilde\kappa^2 z^{\prime 2} }\,\frac{z^\prime}{R}\nonumber\\
&&\times \bar u(p_2)\times\left(-i\gamma^\nu\right)\times\epsilon_{\nu}^{\prime*}\mathcal{V}(Q^\prime z^\prime)\times\frac{1}{2}\Big(\psi_R(z^\prime)\psi_R(z^\prime;n_x)P_{+}+ \psi_L(z^\prime)\psi_L(z^\prime;n_x)P_{-}\Big) \nonumber\\
&&\times\tilde{G}^{s}(p_1+q,M_{n_x})\times\left(-i\gamma^{\mu}\right)\times \frac{1}{2}\Big(\psi_R(z;n_x)\psi_R(z)P_{+}+ \psi_L(z;n_x)\psi_L(z)P_{-}\Big)\times\epsilon_{\mu}\mathcal{V}(Qz)\times u(p_1)\nonumber\\
&+&\int dz \sqrt{g} e^{- \tilde\kappa^2 z^2} \,\frac{z}{R}\int dz^{\prime
}\sqrt{g^\prime} e^{- \tilde\kappa^2 z^{\prime 2}} \,\frac{z^\prime}{R}\nonumber\\
&&\times \bar u(p_2)\times\left(-i\gamma^\nu\right)\times\epsilon_{\nu}^{\prime*}\mathcal{V}(Q^\prime z^\prime)\times\frac{1}{2}\Big(\psi_R(z^\prime)\psi_R(z^\prime;n_x)P_{-}+ \psi_L(z^\prime)\psi_L(z^\prime;n_x)P_{+}\Big) \nonumber\\
&&\times\tilde{G}^{s}(p_1+q,M_{n_x})\times\left(-i\gamma^{\mu}\right)\times \frac{1}{2}\Big(\psi_R(z;n_x)\psi_R(z)P_{-}+ \psi_L(z;n_x)\psi_L(z)P_{+}\Big)\times\epsilon_{\mu}\mathcal{V}(Qz)\times u(p_1)\nonumber\\
&=&\bar u(p_2)\times\epsilon_{\nu}^{\prime*}\times\frac{1}{2}\Big(\mathcal{I}_{R}(Q^\prime,n_x)P_{-}+ \mathcal{I}_{L}(Q^\prime,n_x)P_{+}\Big)\times\left(-i\gamma^\nu\right)\times\tilde{G}^{s}(p_1+q,M_{n_x})\times\left(-i\gamma^{\mu}\right)\nonumber\\
&\times & \frac{1}{2}\Big(\mathcal{I}_{R}(Q,n_x)P_{+}+ \mathcal{I}_{L}(Q,n_x)P_{-}\Big)\times\epsilon_{\mu}\times u(p_1)\nonumber\\
&+&\bar u(p_2)\times\epsilon_{\nu}^{\prime*}\times\frac{1}{2}\Big(\mathcal{I}_{R}(Q^\prime,n_x)P_{+}+ \mathcal{I}_{L}(Q^\prime,n_x)P_{-}\Big)\times\left(-i\gamma^\nu\right)\times\tilde{G}^{s}(p_1+q,M_{n_x})\times\left(-i\gamma^{\mu}\right)\nonumber\\
&\times & \frac{1}{2}\Big(\mathcal{I}_{R}(Q,n_x)P_{-}+ \mathcal{I}_{L}(Q,n_x)P_{+}\Big)\times\epsilon_{\mu}\times u(p_1)\nonumber\\
\eea
or more explicitly

\bea
=&&\bar u(p_2)\times\epsilon_{\nu}^{\prime*}\times\left(-i\gamma^\nu\right)\times\,iG_{1}^{s}(p_1+q,M_{n_x})\times\left(-i\gamma^{\mu}\right)\times\epsilon_{\mu}
\nonumber\\&&\times\frac{1}{4}\Big(\mathcal{I}_{R}(Q^\prime,n_x)\mathcal{I}_{R}(Q,n_x)P_{+}+ \mathcal{I}_{L}(Q^\prime,n_x)\mathcal{I}_{L}(Q,n_x)P_{-}\Big)\times u(p_1)\nonumber\\
&+&\bar u(p_2)\times\epsilon_{\nu}^{\prime*}\times\left(-i\gamma^\nu\right)\times\,iG_{2}^{s}(p_1+q,M_{n_x})\times\left(-i\gamma^{\mu}\right)\times\epsilon_{\mu}
\nonumber\\&&\times\frac{1}{4}\Big(\mathcal{I}_{R}(Q^\prime,n_x)\mathcal{I}_{L}(Q,n_x)P_{-}+ \mathcal{I}_{L}(Q^\prime,n_x)\mathcal{I}_{R}(Q,n_x)P_{+}\Big)\times u(p_1)\nonumber\\
&+&\bar u(p_2)\times\epsilon_{\nu}^{\prime*}\times\left(-i\gamma^\nu\right)\times\,iG_{1}^{s}(p_1+q,M_{n_x})\times\left(-i\gamma^{\mu}\right)\times\epsilon_{\mu}
\nonumber\\&&\times\frac{1}{4}\Big(\mathcal{I}_{R}(Q^\prime,n_x)\mathcal{I}_{R}(Q,n_x)P_{-}+ \mathcal{I}_{L}(Q^\prime,n_x)\mathcal{I}_{L}(Q,n_x)P_{+}\Big)\times u(p_1)\nonumber\\
&+&\bar u(p_2)\times\epsilon_{\nu}^{\prime*}\times\left(-i\gamma^\nu\right)\times\,iG_{2}^{s}(p_1+q,M_{n_x})\times\left(-i\gamma^{\mu}\right)\times\epsilon_{\mu}
\nonumber\\&&\times\frac{1}{4}\Big(\mathcal{I}_{R}(Q^\prime,n_x)\mathcal{I}_{L}(Q,n_x)P_{+}+ \mathcal{I}_{L}(Q^\prime,n_x)\mathcal{I}_{R}(Q,n_x)P_{-}\Big)\times u(p_1)\nonumber\\
&=&\bar u(p_2)\times\epsilon_{\nu}^{\prime*}\times\left(-i\gamma^\nu\right)\times\,iG_{1}^{s}(p_1+q,M_{n_x})\times\left(-i\gamma^{\mu}\right)\times\epsilon_{\mu}
\nonumber\\&&\times\frac{1}{4}\Big(\mathcal{I}_{R}(Q^\prime,n_x)\mathcal{I}_{R}(Q,n_x)+ \mathcal{I}_{L}(Q^\prime,n_x)\mathcal{I}_{L}(Q,n_x)\Big)\times u(p_1)\nonumber\\
&+&\bar u(p_2)\times\epsilon_{\nu}^{\prime*}\times\left(-i\gamma^\nu\right)\times\,iG_{2}^{s}(p_1+q,M_{n_x})\times\left(-i\gamma^{\mu}\right)\times\epsilon_{\mu}
\nonumber\\&&\times\frac{1}{4}\Big(\mathcal{I}_{R}(Q^\prime,n_x)\mathcal{I}_{L}(Q,n_x)+ \mathcal{I}_{L}(Q^\prime,n_x)\mathcal{I}_{R}(Q,n_x)\Big)\times u(p_1)\,,\nonumber\\
\eea
where we have used the identities $P_{\pm}^2=P_{\pm}$, $P_{\pm}P_{\mp}=0$, and $\gamma^{\mu}P_{\pm}=P_{\mp}\gamma^{\mu}$, and we have defined
\be
\tilde{G}^{s}(p_1+q,M_{n_x})=\frac{i(p\hspace{-6pt}\slash_{1} +q\hspace{-6pt}\slash + M_{n_x})}{(p_1+q)^2-M_{n_x} ^{2}+i\epsilon }=\,iG_{1}^{s}(p_1+q,M_{n_x})\,+\,iG_2^{s}(p_1+q,M_{n_x})
\ee
with 
\bea
G_{1}^{s}(p_1+q,M_{n_x})&=&\frac{(p\hspace{-6pt}\slash_{1} +q\hspace{-6pt}\slash)}{(p_1+q)^2-M_{n_x} ^{2}+i\epsilon }\nonumber\\
G_{2}^{s}(p_1+q,M_{n_x})&=&\frac{M_{n_x}}{(p_1+q)^2-M_{n_x} ^{2}+i\epsilon }\,,
\eea
we have also dropped the longitudinal contributions by assuming $\epsilon\cdot q=\epsilon^{\prime}\cdot q^{\prime}=0$.

Finally, after some simplification, we find 
\bea
&&\epsilon_{\mu}\epsilon_{\nu}^{\prime*}T_{s(\text{Dirac})}^{\mu\nu}(s,t;Q^2,Q^{\prime 2})\nonumber\\
&&= -\sum_{n_x}\bigg[\bar u(p_2)\times\epsilon_{\nu}^{\prime*}\times\gamma^\nu\times G_{1}^{s}(p_1+q,M_{n_x})\times\gamma^{\mu}\times\epsilon_{\mu}
\nonumber\\&&\times\frac{1}{2}\Big(\mathcal{I}_{R}(Q^\prime,n_x)\mathcal{I}_{R}(Q,n_x)+ \mathcal{I}_{L}(Q^\prime,n_x)\mathcal{I}_{L}(Q,n_x)\Big)\times u(p_1)\nonumber\\
&&\qquad\qquad+\bar u(p_2)\times\epsilon_{\nu}^{\prime*}\times\gamma^\nu\times G_{2}^{s}(p_1+q,M_{n_x})\times\gamma^{\mu}\times\epsilon_{\mu}
\nonumber\\&&\times\frac{1}{2}\Big(\mathcal{I}_{R}(Q^\prime,n_x)\mathcal{I}_{L}(Q,n_x)+ \mathcal{I}_{L}(Q^\prime,n_x)\mathcal{I}_{R}(Q,n_x)\Big)\times u(p_1)\bigg]\,.\nonumber\\
\eea
for transverse $\epsilon, \epsilon^\prime$ polarizations.
For $Q^{\prime 2}=0$, the sum is dominated by the ground state since $\mathcal{I}_{R/L}(Q^\prime=0,n_x)=\delta_{0,n_x}$, and we find

\bea
T_{s(\text{Dirac})}^{\mu\nu}(s,t;Q^2) &=& -\bar u(p_2)\times\gamma^\nu\times G_{1}^{s}(p_1+q,M_{0})\times\gamma^{\mu}\times\frac{1}{2}\Big(\mathcal{I}_{R}(Q,0)+\mathcal{I}_{L}(Q,0)\Big)\times u(p_1)\nonumber\\
&&+\bar u(p_2)\times\gamma^\nu\times G_{2}^{s}(p_1+q,M_{0})\times\gamma^{\mu}\times\frac{1}{2}\Big(\mathcal{I}_{L}(Q,0)+\mathcal{I}_{R}(Q,0)\Big)\times u(p_1)\,,\nonumber\\
&=&-C_1(Q)\times\bar u(p_2)\times\gamma^\nu\times G^{s}(p_1+q,M_{0})\times\gamma^{\mu}\times u(p_1)\,.\nonumber\\
\eea
Including the Pauli contribution in similar fashion, we find

\bea
\epsilon_{\mu}\epsilon_{\nu}^{\prime*} T_{s}^{\mu\nu}(s,t;Q^2)&=&-\epsilon_{\mu}\epsilon_{\nu}^{\prime*}\times\bar u(p_2)\times\gamma^\nu\times G^{s}(p_1+q,M_{0})\times\gamma^{\mu}\times\left(F_1(Q) - F_2(Q) \frac{\slashed q}{2M_0}\right)\times u(p_1)\,,\nonumber\\
\eea
where we have also ignored the longitudinal contribution (by imposing $\epsilon\cdot q=0$) in order to write 
$$\epsilon_{\mu}\times\sigma^{\mu\nu}q_\nu=\epsilon_{\mu}\times i\left(\gamma^{\mu}\gamma^{\nu}-\eta^{\mu\nu}\right)q_{\nu}=\epsilon_{\mu}\times i\gamma^{\mu}\slashed{q}$$
The holographic proton Dirac and Pauli form factors $F_{1,2}(Q)$ are given in (\ref{F1PQ}).
The u-channel contribution follows by inspection

\bea
\epsilon_{\mu}\epsilon_{\nu}^{\prime*}T_{u}^{\mu\nu}(s,t;Q^2)&=&-\epsilon_{\mu}\epsilon_{\nu}^{\prime*}\times \bar u(p_2)\times\gamma^\mu\times\left(F_1^{P}(Q) - F_2^{P}(Q) \frac{\slashed{q}}{2M_0}\right)\times G^{u}(p_1-q,M_{0})\times\gamma^{\nu}\times u(p_1)\,.\nonumber\\
\eea

\subsubsection{High-energy limit}\label{high energy limit}

In the high energy limit $(s,u\gg M_0^2)$ we can ignore the target mass, and $G^{s,u}$ simplify

\bea
G^{s}(p_1+q,M_{0})&=&\frac{p\hspace{-6pt}\slash_{1} +q\hspace{-6pt}\slash + M_{0}}{(p_1+q)^2-M_{0} ^{2}+i\epsilon }\approx \frac{p\hspace{-6pt}\slash_{1} +q\hspace{-6pt}\slash}{(p_1+q)^2+i\epsilon }\,,\nonumber\\
G^{u}(p_1-q,M_{0})&=&\frac{p\hspace{-6pt}\slash_{1} -q\hspace{-6pt}\slash + M_{0}}{(p_1-q)^2-M_{0} ^{2}+i\epsilon }\approx \frac{p\hspace{-6pt}\slash_{1} -q\hspace{-6pt}\slash}{(p_1-q)^2+i\epsilon }\,.
\eea
Using the light-cone kinematics~\cite{Belitsky:2005qn}
\bea
\frac{1}{(p_1+q)^2+i\epsilon }\approx\frac{1}{(p \cdot \tilde q)}
\frac{1}{1 - \xi + i\epsilon}\,,
\eea
with $\xi$ the longitudinal momentum transfer or skewness, and the spin reduction (for $p_{\perp}=p_{x,y}=0$)~\cite{Belitsky:2005qn}

\begin{eqnarray*}
\gamma^\mu G^{s}(p_1+q,M_{0}) \gamma^\nu
\simeq
\frac{1}{(p \cdot \tilde q) (1 - \xi + i\epsilon)}
\Bigg[
S^{\mu\nu;\rho\sigma}
\left(
\frac{1}{2}(p \cdot \tilde q) n_\sigma +\frac{1}{2}(1 - \xi) p_\sigma
\right) \gamma_\rho
+
i \varepsilon^{\mu\nu\rho\sigma}
\left(\frac{1}{2} p_\sigma + \tilde q_\sigma \right) \gamma_\rho \gamma^5
\Bigg]
\, .
\end{eqnarray*}
we have for the $s+u$-contribution

\begin{eqnarray}
\label{DVCSUS}
T^{\mu\nu}_{s+u}&=&-\bar{u}(p_2)\times\left[\gamma^\mu G^{s}(p_1+q,M_{0}) \gamma^\nu +\gamma^\nu G^{u}(p_1-q,M_{0}) \gamma^\mu\right]\times\left(F_1^{P}(Q) - F_2^{P}(Q) \frac{q\hspace{-6pt}\slash}{2M_0}\right)\times u(p_1)\nonumber\\
&=&\!\!\!
\frac{1}{2 p \cdot \tilde q}
\,
C^{p[-]}_{(0)} (\xi)
\left\{
(p \cdot \tilde q) S^{\mu \nu; \rho \sigma} n_\sigma F^p_\rho(Q)
+
i \varepsilon^{\mu \nu \rho \sigma} p_\sigma
\, \widetilde F^p_\rho(Q)
\right\} +
\frac{1}{2 p \cdot \tilde q}
C^{p[+]}_{(0)} (\xi)
\left\{2 i \varepsilon^{\mu \nu \rho \sigma} \tilde q_\sigma
\widetilde F^p_\rho(Q) 
\right\}\nonumber\\
\, ,
\end{eqnarray}

We have defined the spin structure 
$$S^{\mu \nu; \rho \sigma}\equiv\eta^{\mu\rho} \eta^{\nu\sigma}+\eta^{\mu\sigma} \eta^{\nu\rho}-\eta^{\mu\nu} \eta^{\rho\sigma}$$
and the sum of the ratios

\bea
\label{ProtonCoeffFunct}
C^{p[\pm]}_{(0)} (\xi) =
\frac{1}{\xi - 1 - i\epsilon} \pm \frac{1}{\xi + 1 - i \epsilon}
\eea
The vector and axial form factors are

\begin{eqnarray}
\label{VectorTw2Nucleon}
F^p_\rho(Q)=
p_\rho\left( \frac{h^+}{p^+}F_1^{P}(Q) + \frac{e^+}{p^+}F_2^{P}(Q)\right)=p_\rho\times F^p(Q)\, ,\nonumber\\
\widetilde{F}^p_\rho(Q)=
 p_\rho\left(
\frac{\widetilde{h}^+}{p^+}F_1^{P}(Q) +
\frac{\widetilde{e}^+}{p^+}F_2^{P}(Q)\right)=p_\rho\times\widetilde{F}^p(Q)\, .\nonumber\\
\end{eqnarray}
respectively, with  the Dirac bilinears

\begin{eqnarray}
\label{DiracBilinearsSpinOneHalf}
e^\mu &=& \frac{t^{\nu\mu} \tilde q_\nu}{2M_{0}}
\, , \qquad
\widetilde e^\mu = \frac{\widetilde{t}^{\nu\mu} \tilde q_\nu}{2M_{0}}
\, .
\end{eqnarray}
and  the nucleon matrix elements

\begin{equation}
\label{Def-ForFac}
\begin{array}{ll}
b = \bar u (p_2) u (p_1)
\, , \qquad
&
\widetilde b = \bar u (p_2) \gamma^5 u (p_1)
\, , \\
h^{\mu} = \bar u (p_2) \gamma^\mu u (p_1)
\, , \qquad
&
\widetilde{h}^\mu = \bar u (p_2) \gamma^\mu \gamma^5 u (p_1)
\, , \\
t^{\mu\nu} = \bar u (p_2) i \sigma^{\mu\nu} u (p_1)
\, , \qquad
&
\widetilde{t}^{\mu\nu} = \bar u (p_2) i \sigma^{\mu\nu} \gamma^5 u (p_1)
\, ,
\end{array}
\end{equation}

\subsubsection{TT and LL s+u-channel holographic DVCS amplitude}

Contracting the $s+u$-channel holographic amplitude $T_{s+u}^{\mu\nu}$ with the transverse polariztion tensors $\epsilon_{T\mu}$ and $\epsilon_{T\nu}^{\prime*}$, we find

\bea
\label{cont1Tw2ProtonTT}
\epsilon_{T\mu}\epsilon_{T\nu}^{\prime*} T_{s+u}^{\mu\nu}(s,t;Q^2)=-\epsilon_{T}\cdot\epsilon_{T}^{\prime}\times\left[ \frac{1}{2}
C^{p[-]}_{(0)} (\xi)
\times F^p(Q)\right]\,.\nonumber\\
\eea

Similarly, contracting the $s+u$-channel holographic amplitude $T_{s+u}^{\mu\nu}$ with the longitudinal polarization tensors $\epsilon_{L\mu}$ and $\epsilon_{L\nu}^{\prime*}$, we find

\bea
\label{cont1Tw2ProtonLL}
\epsilon_{L\mu}\epsilon_{L\nu}^{\prime*} T_{s+u}^{\mu\nu}(s,t;Q^2)=-\epsilon_{L}\cdot\epsilon_{L}^{\prime}\times\left[ \frac{1}{2}
C^{p[-]}_{(0)} (\xi)
\times F^p(Q)\right]\,.\nonumber\\
\eea

We will also use the approximation  
\be
h^+=\bar u(p_2)\gamma^{+}u(p_1)\approx \frac{p^+}{2m_N}\times \bar u(p_2)u(p_1)\,,
\ee
and our holographic electromagnetic form factors $F_{1,2}^{P}(Q)$ 
\bea\label{F1PQ}
F_{1}^{P}(Q)&=&\left(\frac{a_Q}{2}+\tau_N \right)\times B(\tau_N ,a_Q+1)+0.448\times \frac{a_Q (a_Q (\tau_N -1)-1)}{(a_Q+\tau_N ) (a_Q+\tau_N +1)}\times B(\tau_N -1,a_Q+1)\,,\nonumber\\
F_{2}^{P}(Q)&=&1.793\times \tau_N \times B(\tau_N ,a_Q+1)\,,\nonumber\\
F_{1}^{N}(Q)&=&-0.478\times \frac{a_Q (a_Q (\tau_N -1)-1)}{(a_Q+\tau_N +1) (a_Q+\tau_N )}\times B(\tau_N -1,a_Q+1)\,,\nonumber\\
F_{2}^{N}(Q)&=&-1.913\times \tau_N \times B(\tau_N ,a_Q+1)\,,\nonumber\\
\eea
where $a_Q=Q^2/4\kappa_N^2$ with $Q^2=-q^2$.

\subsubsection{Generalization of s+u-channel contribution to DVCS to the electroproduction of $\rho^0$}\label{srhofinal}
The holographic s-channel contribution to the electroproduction of $\rho^0$ follows from the Witten diagram shown in Fig.~\ref{SWITTEN2} which is readily obtained from the  
results (\ref{cont1Tw2ProtonLL}\,-\,\ref{F1PQ}) for the s-channel DVCS amplitude, shown in Fig.~\ref{SWITTEN1},
by assuming that the bulk wave function of the rho vector meson is independent of $z$, i.e., $\phi_{V}(z)\approx g_5\times\frac{f_V}{M_V}$) 
\bea\label{s-channelRho}
{\cal A}^{LL(s+u-channel)}_{\gamma^{*} p\rightarrow  \rho^{0} p} (s,t,Q,\epsilon_{L},\epsilon_{L}^{\prime})&\propto&e\times g_5\times\left[\epsilon_{L\mu}\epsilon_{L\nu}^{\prime*} T_{s+u}^{\mu\nu}(s,t;Q^2)\right]\times g_5\times\frac{f_V}{M_V}  \nonumber\\
&\propto&\frac{Q}{M_V}\times F_1^{P}(Q)\times \frac{f_V}{M_V}\times\frac{1}{2\sqrt{2} m_{N}}\,\bar u(p_2)u(p_1)\times\frac{1}{1-\xi^2}\,,\nonumber\\
&\propto&{\cal A}^{s+u}_{\rho^{0}} (s,t,Q)\times\frac{1}{2\sqrt{2} m_{N}}\,\bar u(p_2)u(p_1)\,,\nonumber\\
\eea
where in the last line we have absorbed $e\times\sqrt{2}\times g_5^2$ into the proportionality constant that will be fixed by experimental data. We have also approximated $\epsilon_{L}\cdot\epsilon_{L}^{\prime}\approx \frac{Q}{M_V}$ in the high energy limit.

\subsection{Differential cross section for $\rho^{0}$}

Finally, combining the singlet quark (\ref{FULLquarkSinglet2}) and gluon (\ref{FULLgluon2}) GPD contributions, along with the s-channel contribution,
we have the total longitudinal differential cross section (including the real s-channel background amplitude (\ref{s-channelRho})) for electroproduction of $\rho^{0}$ meson 
\bea\label{totaldcs}
\frac{d\sigma_{L}(\rho^0)}{dt}&=&\frac{1}{16\pi s^2}\times \frac{1}{2}\sum_{\rm spin} \Big|\mathcal{A}_{\gamma^{*} p\rightarrow  \rho^{0} p}^{LL(total)}\Big|^2\nonumber\\
&=&\frac{1}{16\pi s^2}\times \frac{1}{2}\sum_{\rm spin} \Big|\mathcal{N}_{s}\times \mathcal{A}_{\gamma^{*} p\rightarrow  \rho^{0} p}^{LL(s+u-channel)}+\mathcal{N}_{t}\times\left(\mathcal{A}_{\gamma^{*} p\rightarrow  \rho^{0} p}^{LL(quark)}+\mathcal{A}_{\gamma^{*} p\rightarrow  \rho^{0} p}^{LL(gluon)}\right)\Big|^2 \nonumber\\
&=&\frac{1}{16\pi s^2}\times \frac{1}{2}\times\Big|\mathcal{N}_{s}\times \mathcal{A}_{\rho^{0}}^{s+u}+\mathcal{N}_{t}\times\left(\mathcal{A}_{\rho^{0}}^{quark}+\mathcal{A}_{\rho^{0}}^{gluon}\right)\Big|^2\,,\nonumber\\
\eea
where
\bea
{\cal A}^{s+u}_{\rho^{0}} (s,t,Q)&=&\frac{Q}{M_V}\times F_1^{P}(Q)\times \frac{f_V}{M_V}\times\frac{1}{1-\xi^2}\,,\nonumber\\
\mathcal{A}_{\rho^{0}}^{quark} (s,t,Q)&=&\Bigg(Q_u\widehat{\widehat{\mathbb{F}}}^u_{j_{0g}(singlet)} (\xi, t ; \mu^2)-Q_d\widehat{\widehat{\mathbb{F}}}^d_{j_{0g}(singlet)} (\xi, t ; \mu^2)\Bigg)\times f_V\,\alpha_s(\mu)\,{1\over {Q}}\times \left[\frac{(\sqrt{\lambda}/\pi)}{\xi^{2-2/\sqrt{\lambda}}}+\frac{i}{\xi^{2-2/\sqrt{\lambda}}}\right]\,,\nonumber\\
\mathcal{A}_{\rho^{0}}^{gluon} (s,t,Q)&=&\widehat{\widehat{\mathbb{F}}}^g_{j_{0g}} (\xi, t ; \mu^2)\times f_V^{\prime}\,\alpha_s(\mu)\,{1\over {Q}}\times\left[\frac{(\sqrt{\lambda}/\pi)}{\xi^{2-2/\sqrt{\lambda}}}+\frac{i}{\xi^{2-2/\sqrt{\lambda}}}\right]\,.
\eea
Note that, in the last line of (\ref{totaldcs}), we have summed over the spin as
\bea
\sum_{s,s'}\bar u_{s'}(p_2)u_s(p_1)\bar u_s(p_1)u_{s'}(p_2)&=& \Tr\Big(\sum_{s.s'}u_{s'}(p_2)\bar u_{s'}(p_2)u_s(p_1)\bar u_s(p_1)\Big)\nonumber\\
&=&\frac 14 \Tr\Big(\big(\gamma_\mu p_2^\mu+m_N\big)\big(\gamma_\mu p_1^\mu+m_N\big)\Big)=8m_N^2\times\left(1+\frac{K^2}{4m_N^2}\right)\approx 8m_N^2\,,\nonumber\\
\eea
for $\frac{K^2}{4m_N^2}\ll1$ or $p_1\sim p_2$ which we are assuming in the high energy limit $s\gg -t$.

\subsection{Total cross section for $\rho^{0}$}
Using (\ref{totaldcs}), we can compute the total cross section by employing the optical theorem as
\be\label{totalcs}
\sigma_{L}(\rho^0)=\left(\frac{16\pi\,\frac{d\sigma_{L}(\rho^0)}{dt}}{1+\rho}\right)^{1/2}_{t=0}\,,
\ee
where $\rho$ is the ratio of the real and imaginary part of our total complex amplitude $\mathcal{A}_{\gamma^{*} p\rightarrow  \rho^{0} p}^{LL(total)}$ defined in (\ref{totaldcs}). 

We have compared our total cross section (\ref{totalcs}) 
to experimental data in Figs. \ref{fig_FIVEE}\,-\,\ref{fig_EIGHT}. We have also compared our total differential cross section (\ref{totaldcs}) 
to experimental data in Fig. \ref{fig_ELEVENTH}. Note that throughout this paper, we use the 't Hooft coupling constant $\lambda=11.243$, mass scales $\tilde{\kappa}_N=\tilde{\kappa}_V=0.402$~GeV, and  $\tilde{\kappa}_T=0.388$~GeV.

\begin{table}[h!]
\centering
\renewcommand{\arraystretch}{1.5}
\begin{tabular}{
 l 
 c 
 c 
 c 
 c 
 l 
 l 
}
\toprule
Holographic QCD $(\text{This Work})$ & $\lambda$ & $\tau$ & $\tilde{\kappa}_{T,S}(\text{GeV})$ & $\tilde{\kappa}_{N,V}(\text{GeV})$ & $\mathcal{N}_{t}(\mu b\times\text{GeV}^2)$ & $\mathcal{N}_{s}(\mu b\times\text{GeV}^2)$ \\
\midrule
$\sigma_{L}(\rho^0)$ vs $W$ or $Q^2$ & & & & & & \\
Figs. \ref{fig_FIVEE}\,-\,\ref{fig_EIGHT} $(\textcolor{blue}{blue})$ & 11.243 & 3 & 0.388 & 0.402 & 1.687 &  0 \\
Figs. \ref{fig_FIVEE}\,-\,\ref{fig_EIGHT} $(\textcolor{red}{red})$ & 11.243 & 3 & 0.388 & 0.402 & 1.687 & -131.503 \\
\midrule
$\frac{d\sigma_{L}(\rho^0)}{dt}$ vs $-t$ & & & & & & \\
Figs. \ref{fig_ELEVENTHa} $(\textcolor{blue}{blue})$ & 11.243 & 3 & 0.388 & 0.402 & 6.319 &  0 \\
Figs. \ref{fig_ELEVENTHa} $(\textcolor{red}{red})$ & 11.243 & 3 & 0.388 & 0.402 & 6.319 & -5.662$\times 10^{-7}$ \\
Figs. \ref{fig_ELEVENTHb} $(\textcolor{blue}{blue})$ & 11.243 & 3 & 0.388 & 0.402 & 194.649 & 0 \\
Figs. \ref{fig_ELEVENTHb} $(\textcolor{red}{red})$ & 11.243 & 3 & 0.388 & 0.402 & 194.649 & -5.454$\times 10^{-7}$ \\
Figs. \ref{fig_ELEVENTHc} $(\textcolor{blue}{blue})$ & 11.243 & 3 & 0.388 & 0.402 & 125.468 & 0 \\
Figs. \ref{fig_ELEVENTHc} $(\textcolor{red}{red})$ & 11.243 & 3 & 0.388 & 0.402 & 125.468 & -196.937 \\
Figs. \ref{fig_ELEVENTHd} $(\textcolor{blue}{blue})$ & 11.243 & 3 & 0.388 & 0.402 & 1206.470 & 0 \\
Figs. \ref{fig_ELEVENTHd} $(\textcolor{red}{red})$ & 11.243 & 3 & 0.388 & 0.402 & 1206.470 & -838.781 \\
Figs. \ref{fig_ELEVENTHe} $(\textcolor{blue}{blue})$ & 11.243 & 3 & 0.388 & 0.402 & 342.628 & 0 \\
Figs. \ref{fig_ELEVENTHe} $(\textcolor{red}{red})$ & 11.243 & 3 & 0.388 & 0.402 & 342.628 & -767.554 \\
\midrule
$\sigma_{L}(\phi)$ vs $W$ or $Q^2$ & & & & & & \\
Figs. \ref{fig_12phi}\,-\,\ref{fig_13phi} $(\textcolor{blue}{blue})$ & 11.243 & 3 & 0.388 & 0.402 & 0.200 & 0 \\
\midrule
$\sigma_{L}(\rho^{+})$ vs $W$ & & & & & & \\
Figs. \ref{fig_14rhoplus} $(\textcolor{blue}{blue})$ & 11.243 & 3 & 0.388 & 0.402 & 724.513 & 0 \\
\bottomrule
\end{tabular}
\caption{Summary of the values we used for the pre-determined parameters ($\lambda=11.243$, $\tau=3$, $\tilde{\kappa}_{T,S}=0.388~\text{GeV}$, $\tilde{\kappa}_{N,V}=0.402~\text{GeV}$) of holographic QCD with a soft wall, and the overall normalization constants $\mathcal{N}_{t,s}$ which are fitted to the pertinent cross-section data, according to the corresponding (colored) figures for the neutral rho ($\rho^0$), phi ($\phi$), and charged rho ($\rho^+$) meson production.}\label{table11}
\end{table}

\begin{figure*}
\subfloat[\label{fig_1pt6} $Q^2=1.6-1.9$~GeV$^2$]{%
  \includegraphics[height=4cm,width=.49\linewidth]{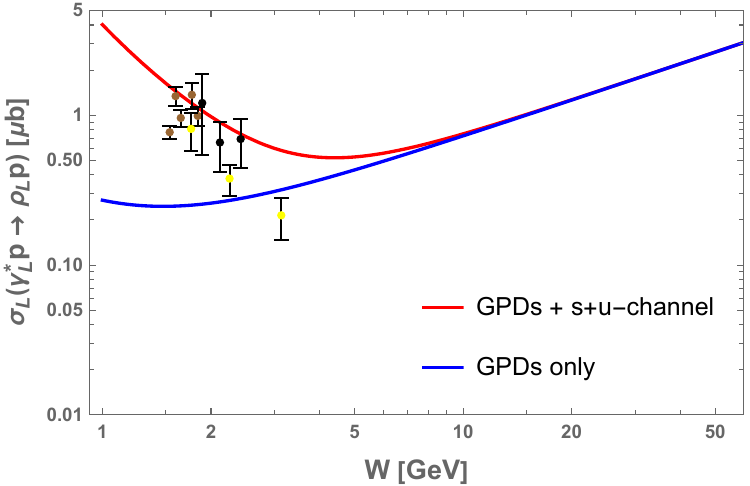}%
}\hfill
\subfloat[\label{fig_1pt9} $Q^2=1.9-2.2$~GeV$^2$]{%
  \includegraphics[height=4cm,width=.49\linewidth]{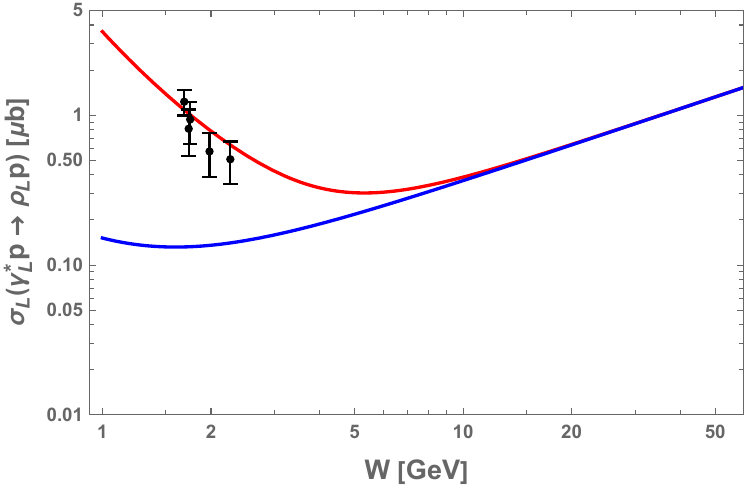}%
}\hfill
\subfloat[\label{fig_2pt2} $Q^2=2.2-2.5$~GeV$^2$]{%
  \includegraphics[height=4cm,width=.49\linewidth]{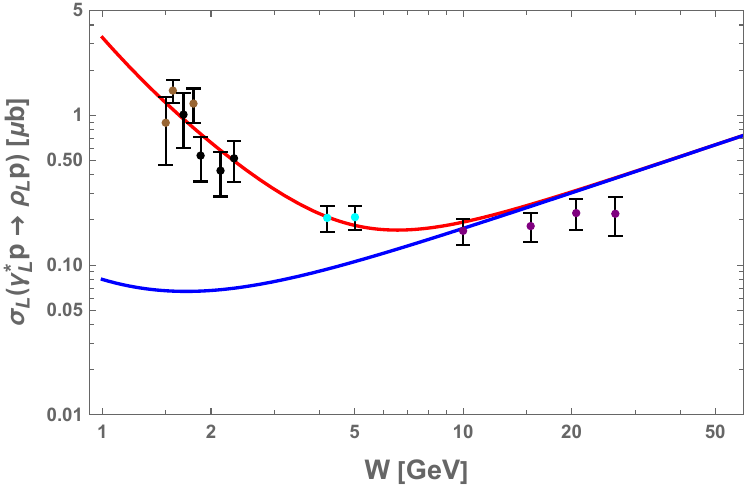}%
}\hfill
\subfloat[\label{fig_2pt5} $Q^2=2.5-2.8$~GeV$^2$]{%
  \includegraphics[height=4cm,width=.49\linewidth]{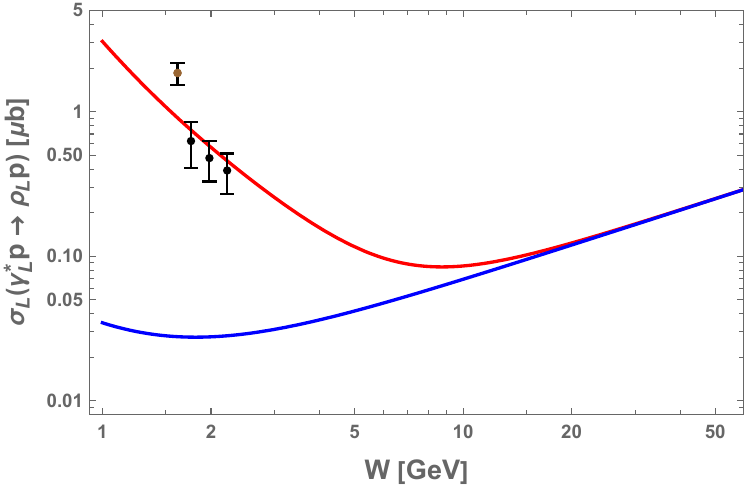}%
}\hfill
\subfloat[\label{fig_2pt8} $Q^2=2.8-3.1$~GeV$^2$]{%
  \includegraphics[height=4cm,width=.49\linewidth]{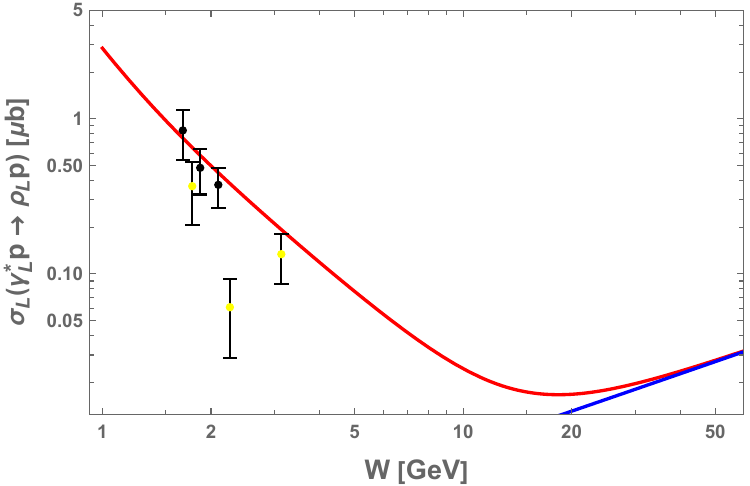}%
}\hfill
\subfloat[\label{fig_3pt1} $Q^2=3.1-3.6$~GeV$^2$]{%
  \includegraphics[height=4cm,width=.49\linewidth]{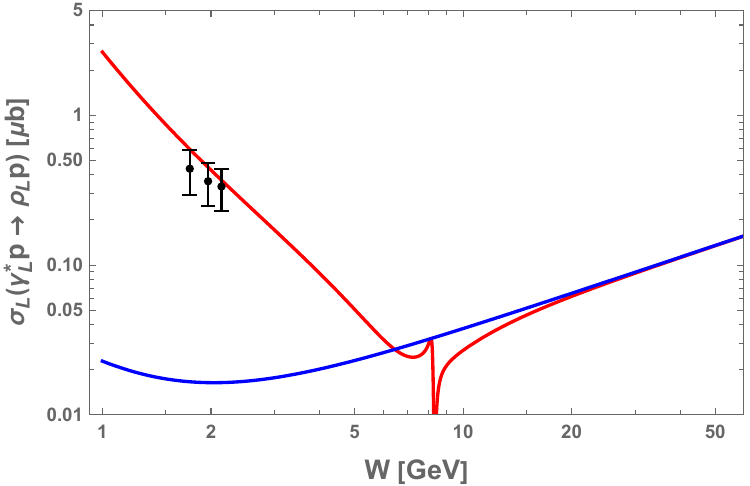}%
}\hfill
\subfloat[\label{fig_3pt6} $Q^2=3.6-4.1$~GeV$^2$]{%
  \includegraphics[height=4cm,width=.49\linewidth]{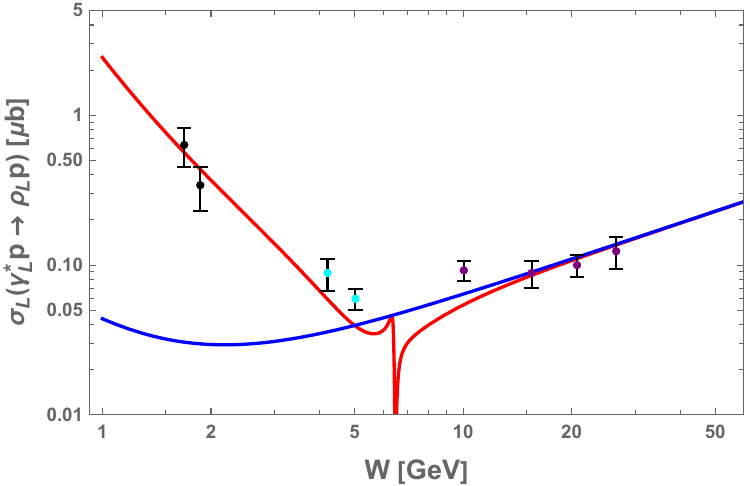}%
}\hfill
\subfloat[\label{fig_4pt1} $Q^2=4.1-4.6$~GeV$^2$]{%
  \includegraphics[height=4cm,width=.49\linewidth]{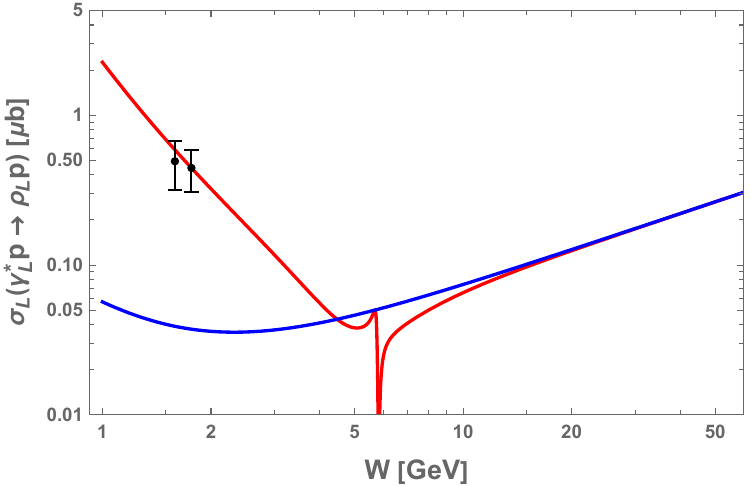}%
}\hfill
\subfloat[\label{fig_4pt6} $Q^2=4.6-5.1$~GeV$^2$]{%
  \includegraphics[height=4cm,width=.49\linewidth]{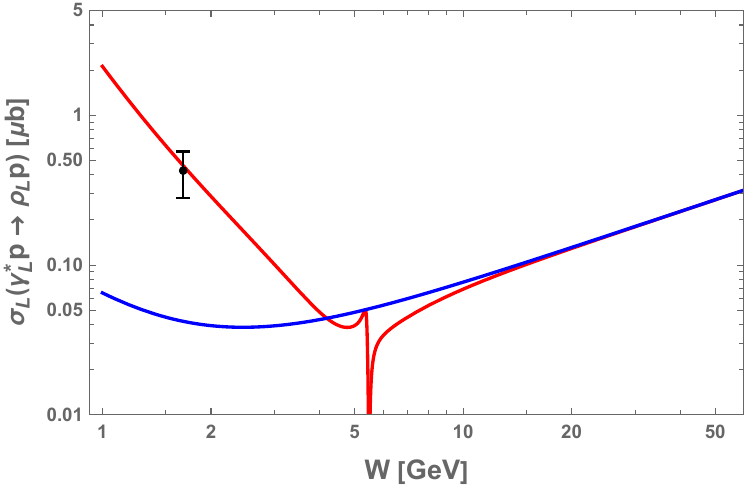}%
}\hfill
\subfloat[\label{fig_5pt1} $Q^2=5.1-5.6$~GeV$^2$]{%
  \includegraphics[height=4cm,width=.49\linewidth]{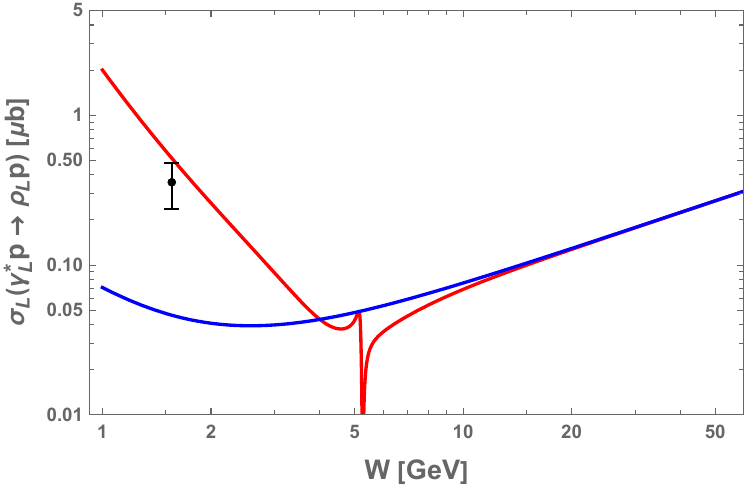}%
}
\caption{Longitudinal cross section for the electroproduction of neutral rho mesons versus $W=\sqrt{s}$ in GeV. 
}
\label{fig_FIVEE}
\end{figure*}

\begin{figure*}
\subfloat[\label{fig_2pt5}]{%
  \includegraphics[height=6cm,width=.7\linewidth]{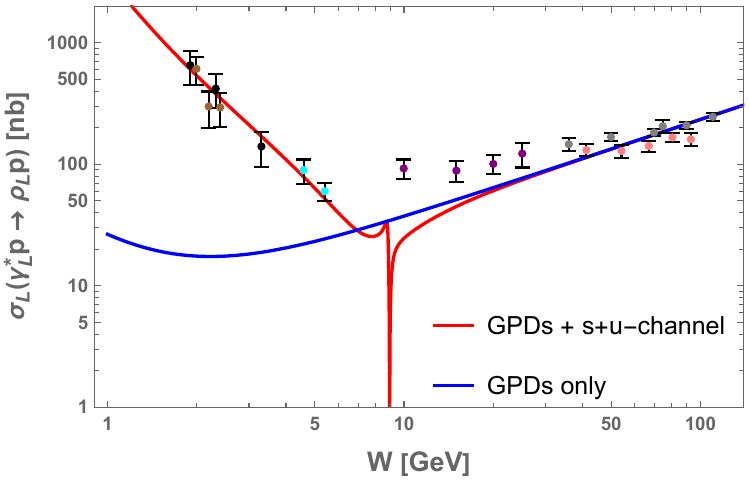}%
}
\caption{Longitudinal cross section for the electroproduction of neutral rho mesons versus $W=\sqrt{s}$ in GeV, for 
$Q^2=4$ GeV$^2$. The {solid-blue curves} are with the evolved up and down singlet quark, and gluon GPDs. 
The red curves are the sum of the s-channel background, and the evolved up and down singlet quark, and gluon GPD contribution at amplitude level (i.e., (\ref{totalcs})). Note that here we have also re-scaled our total cross section by an overall factor of $10^3$ since the unit of the cross section here is in nb. The brown data points are from 4.2 GeV CLAS \cite{CLAS:2004cri}.  The black data points are from 5.754 GeV CLAS \cite{CLAS:2008rpm}. The cyan data points are from HERMES \cite{HERMES:2000jnb}. The purple data points are from E665 \cite{E665:1997qph}.
The grey data points are from ZEUS \cite{ZEUS:1998xpo}.  The pink data points are from H1 \cite{H1:1999pji}. }
\label{fig_THIRD}
\end{figure*}

\begin{figure*}
\subfloat[\label{fig_0pt16} $x_B=0.16-0.22$]{%
  \includegraphics[height=4cm,width=.49\linewidth]{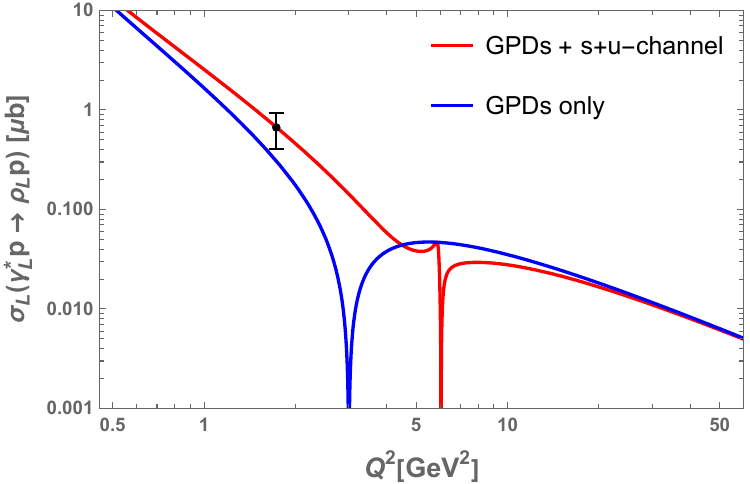}%
}\hfill
\subfloat[\label{fig_0pt22} $x_B=0.22-0.28$]{%
  \includegraphics[height=4cm,width=.49\linewidth]{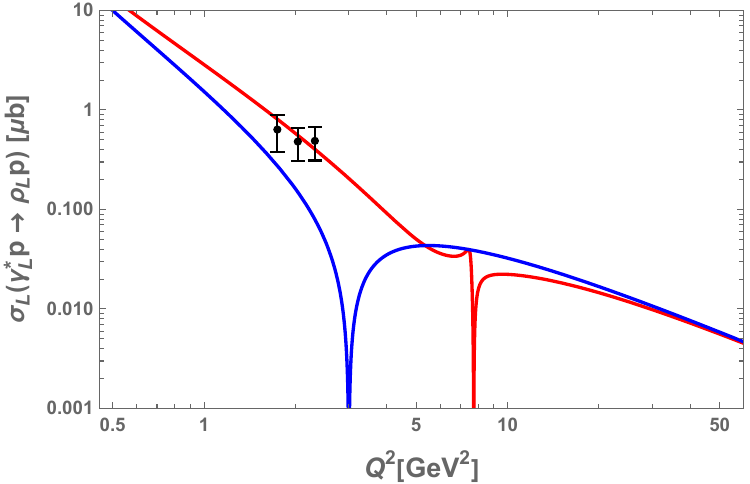}%
}\hfill
\subfloat[\label{fig_0pt28} $x_B=0.28-0.34$]{%
  \includegraphics[height=4cm,width=.49\linewidth]{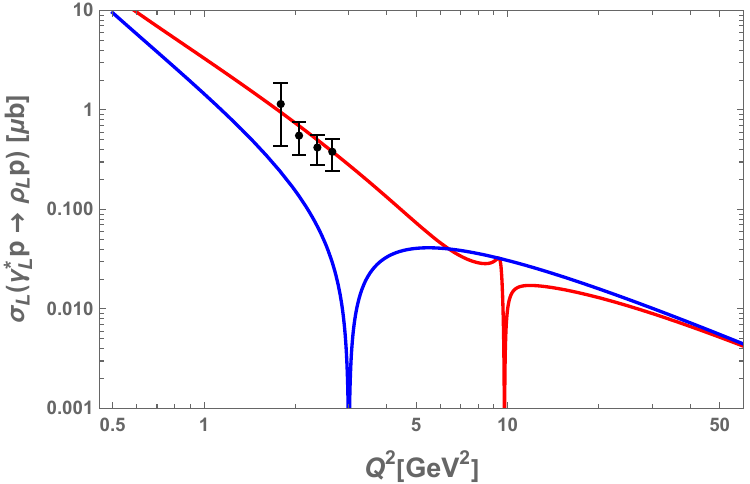}%
}\hfill
\subfloat[\label{fig_0pt34} $x_B=0.34-0.4$]{%
  \includegraphics[height=4cm,width=.49\linewidth]{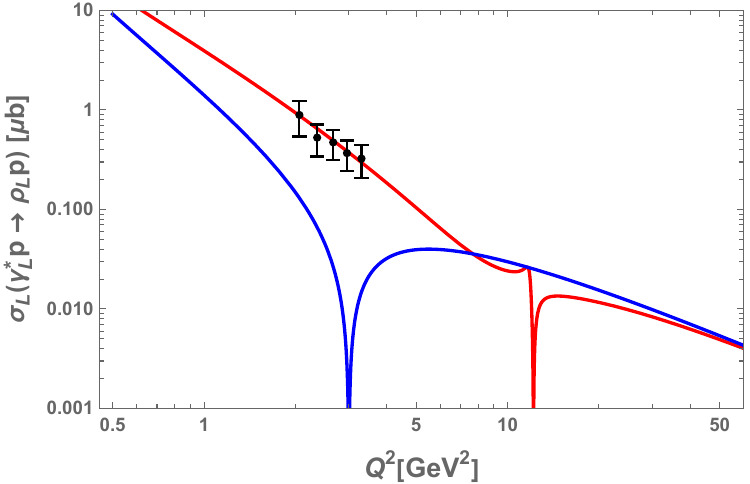}%
}\hfill
\subfloat[\label{fig_0pt4} $x_B=0.40-0.46$]{%
  \includegraphics[height=4cm,width=.49\linewidth]{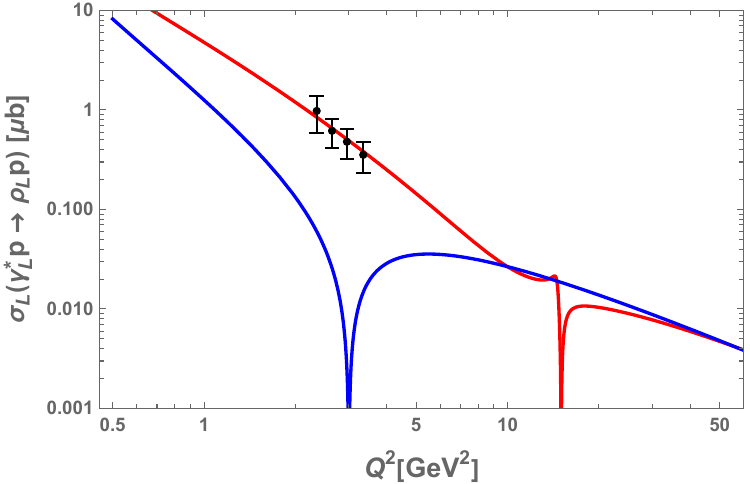}%
}\hfill
\subfloat[\label{fig_0pt46} $x_B=0.46-0.52$]{%
  \includegraphics[height=4cm,width=.49\linewidth]{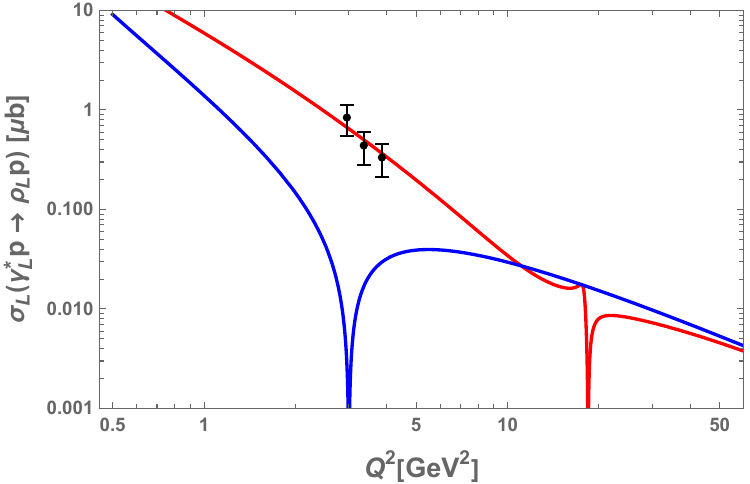}%
}\hfill
\subfloat[\label{fig_0pt52} $x_B=0.52-0.58$]{%
  \includegraphics[height=4cm,width=.49\linewidth]{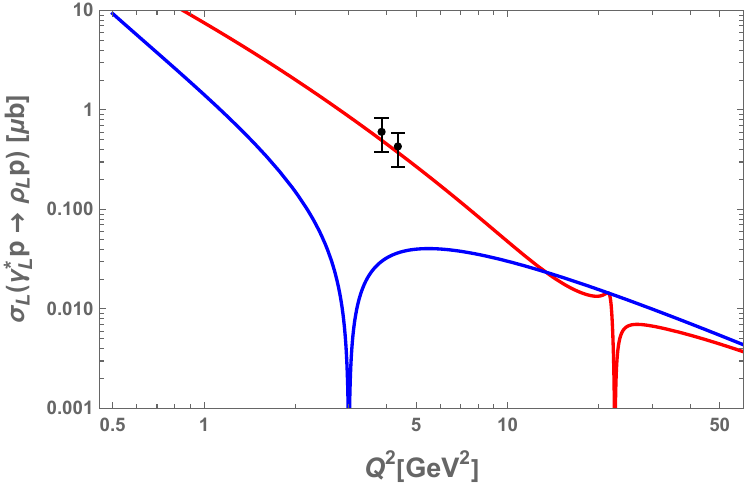}%
}\hfill
\subfloat[\label{fig_75} $W=\sqrt{s}=75$ GeV]{%
  \includegraphics[height=5cm,width=.5\linewidth]{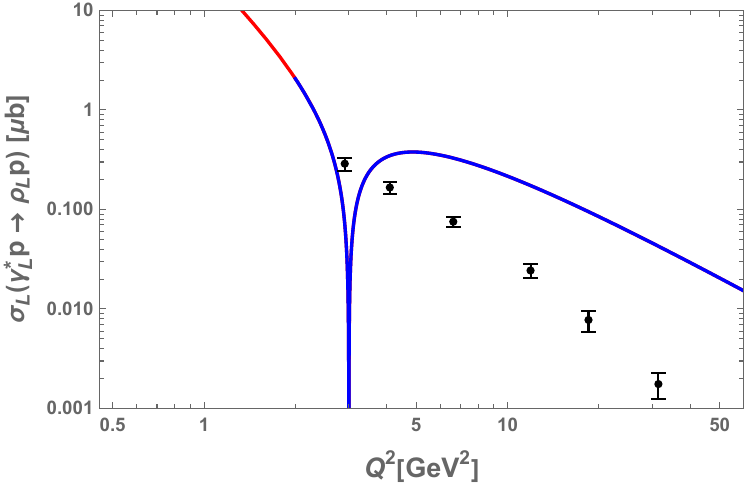}%
}
\caption{Longitudinal cross section for the electroproduction of neutral rho mesons versus $Q^2$ in GeV$^2$, for different values of Bjorken $x_B$ as indicated in the alphabet label of each subfigure, except for the last subfigure where the value of $W$ is added. The {solid-blue curves} are with the evolved up and down singlet quark, and gluon GPD contributions. The {red curves} are the sum of the s-channel background, and the evolved up and down singlet quark, and gluon GPD contributions at amplitude level. The black data points are from 5.754 GeV CLAS \cite{CLAS:2008rpm}.}
\label{fig_EIGHT}
\end{figure*}

\begin{figure*}
\subfloat[\label{fig_ELEVENTHa} $Q^2=0-1.9$ GeV$^2$, $\mathbf{x_B}=\mathbf{0.16-0.22}$]{%
  \includegraphics[height=4cm,width=.49\linewidth]{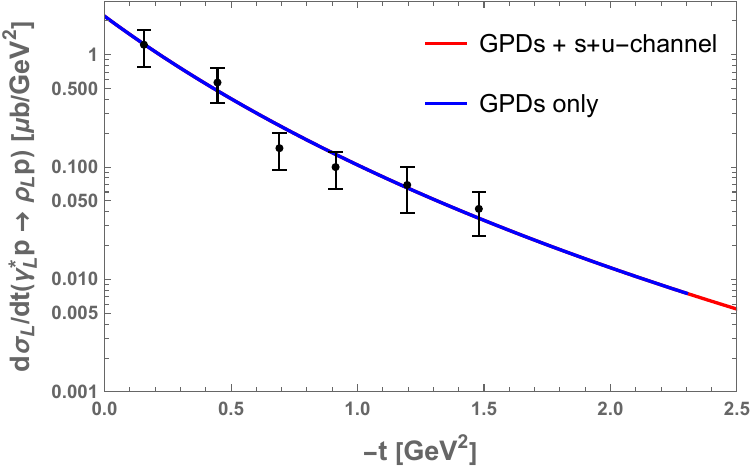}%
}\hfill
\subfloat[\label{fig_ELEVENTHb} $Q^2=2.2-2.5$ GeV$^2$, $\mathbf{x_B}=\mathbf{0.22-0.28}$]{%
  \includegraphics[height=4cm,width=.49\linewidth]{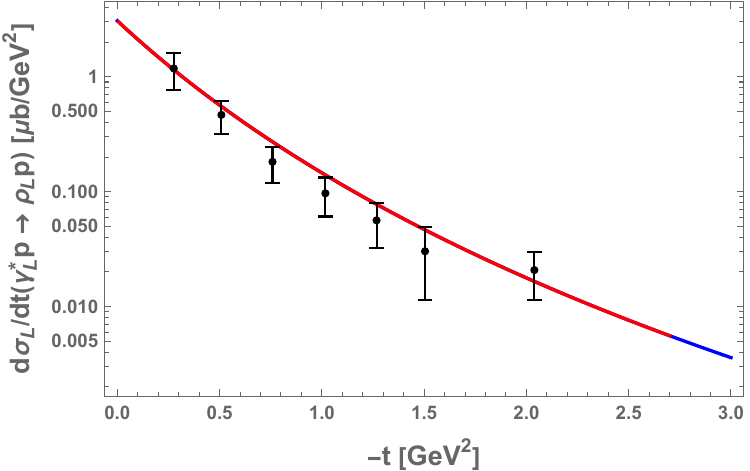}%
}\hfill
\subfloat[\label{fig_ELEVENTHc} $Q^2=1.9-2.2$ GeV$^2$, $\mathbf{x_B}=\mathbf{0.34-0.40}$]{%
  \includegraphics[height=4cm,width=.49\linewidth]{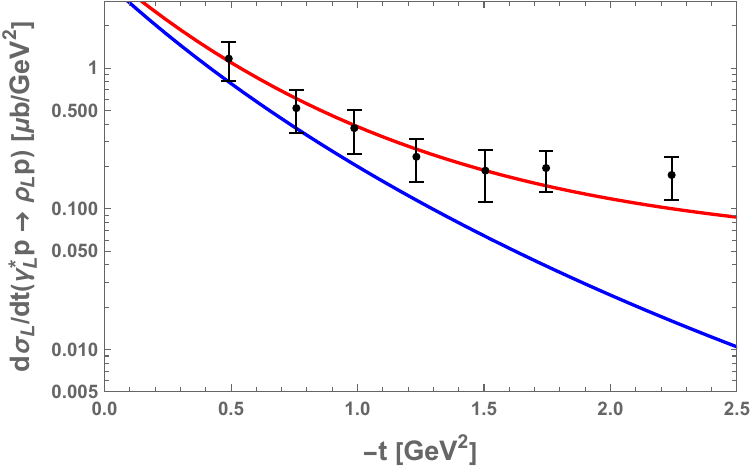}%
}\hfill
\subfloat[\label{fig_ELEVENTHd} $Q^2=3.1-3.6$ GeV$^2$, $\mathbf{x_B}=\mathbf{0.40-0.46}$]{%
  \includegraphics[height=4cm,width=.49\linewidth]{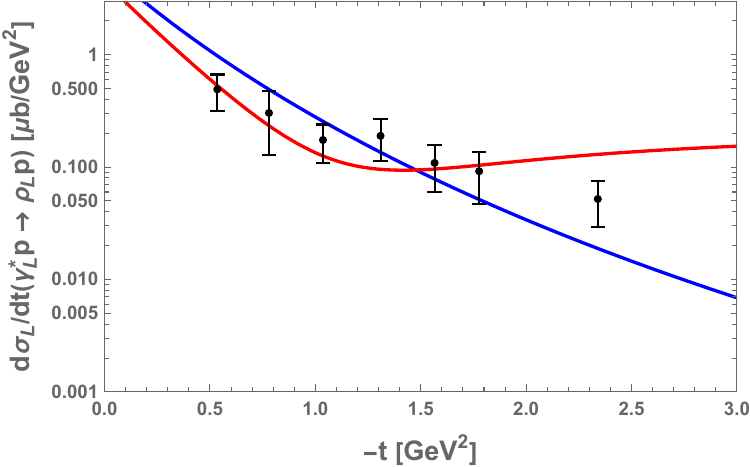}%
}\hfill
\subfloat[\label{fig_ELEVENTHe} $Q^2=4.1-4.6$ GeV$^2$, $\mathbf{x_B}=\mathbf{0.58-0.64}$]{%
  \includegraphics[height=4cm,width=.49\linewidth]{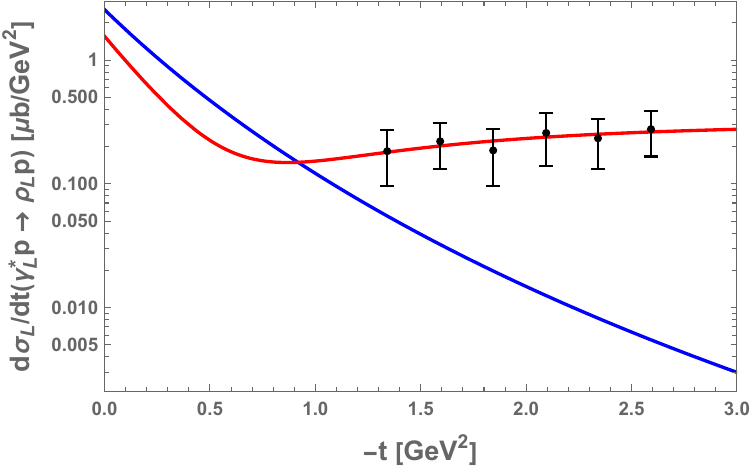}%
}
\caption{Longitudinal differential cross section for the electroproduction of neutral rho mesons versus $-t$ in GeV$^2$, for different values of $Q^2$ and $x_B$ as indicated in the alphabet label of each subfigure. The {solid-blue curves} are with the evolved up and down singlet quark, and gluon GPD contributions. The {red curves} are the sum of the s-channel, and the evolved up and down singlet quark, and gluon GPDs at amplitude level (i.e., (\ref{totaldcs})). Note that with increasing $x_{B}\geq 0.34$, that is, in (c) to (e), the contribution from the $s+u$-channel starts to control the $t$ dependence in the large $t$ regime. The black data points are from 5.754 GeV CLAS \cite{CLAS:2008rpm}, see Fig. 26 in \cite{CLAS:2008rpm}.}
\label{fig_ELEVENTH}
\end{figure*}

\subsection{Comparison to experiment for $\rho^{0}$}

Here, we present the comparison of our results with experimental measurements. In Table~\ref{table11}, we provide the values of various parameters such as $\lambda$, $\tau$, $\tilde{\kappa}_{T,S}$, and $\tilde{\kappa}_{N,V}$, which have been obtained previously by constraining the electromagnetic and gravitational form factors and the holographic $J/\psi$ photoproduction in the high energy limit~\cite{Mamo:2021jhj}. These parameters are fixed and predetermined for the GPDs under consideration here. 

However, the normalization constants $\mathcal{N}_{t,s}$ have been determined empirically by fitting to the analyzed cross sections shown in Figs.~\ref{fig_FIVEE}\,-\,\ref{fig_ELEVENTH}, as summarized in Table~\ref{table11}. It is important to note that the same normalization constants $\mathcal{N}_{t}=1.687$ and $\mathcal{N}_{s}=-131.503$ have been used for all the total cross section figures shown in Figs.~\ref{fig_FIVEE}\,-\,\ref{fig_EIGHT}. 

However, different normalization constants $\mathcal{N}_{t,s}$ have been used for each bin in the differential cross section vs. $-t$ figures shown in Figs.~\ref{fig_ELEVENTHa}\,-\,\ref{fig_ELEVENTHe}. This is because the kinematic factors of order $\mathcal{O}(-t/Q^2)$ have been ignored for both the $s+u$-channel and GPD contributions, which do not affect the total cross sections (shown in Figs.~\ref{fig_FIVEE}\,-\,\ref{fig_EIGHT}) computed at $-t=0$ using the optical theorem (\ref{totalcs}). 

Furthermore, the values of the physical parameters $m_{N}=0.94~\text{GeV}$, $m_{\rho^{0,+}}=0.775~\text{GeV}$, and $f^{\prime}_{\rho^{0,+}}=f_{\rho^{0,+}}=0.216~\text{GeV}$ have been used throughout the analysis. In Figs.~\ref{fig_FIVEE}, the {solid-blue curves} are using the evolved singlet quark (up and down) and gluon GPDs,
the {red curves} are the sum of the s-channel background, and the evolved singlet quark plus gluon GPDs (i.e., (\ref{totalcs})), the brown data points are from 4.2 GeV CLAS \cite{CLAS:2004cri}, the yellow data points are from CORNELL \cite{Cassel:1981sx}, the cyan data points are from HERMES \cite{HERMES:2000jnb}, the purple data points are from E665 \cite{E665:1997qph}, the black data points are from 5.754 GeV CLAS \cite{CLAS:2008rpm}.

\section{Electroproduction of longitudinal $\phi$ meson with evolved gluon GPD: a comparison to experiment}~\label{sec_PHI}


The electroproduction of $\phi$ meson with the evolved gluon GPD is similar to the gluon contribution to $\rho^{0}$ meson (\ref{FULLgluon2}), and is given by
\bea\label{FULLgluon222}
{\cal A}^{LL(gluon)}_{\gamma^{*} p\rightarrow  \phi p} (s,t,Q,\epsilon_{L},\epsilon_{L}^{\prime})&\propto&\widehat{\widehat{\mathbb{F}}}^g_{j_{0g}} (\xi, t ; \mu^2)\times f_V^{\prime}\,\alpha_s(\mu)\,{1\over {Q}}\,\frac{1}{2\sqrt{2} m_{N}}\,\bar u(p_2)u(p_1)\nonumber\\  
&\times &\left[\frac{(\sqrt{\lambda}/\pi)}{\xi^{2-2/\sqrt{\lambda}}}+\frac{i}{\xi^{2-2/\sqrt{\lambda}}}\right]\nonumber\\
&\propto&{\cal A}^{gluon}_{\phi} (s,t,Q)\times\frac{1}{2\sqrt{2} m_{N}}\,\bar u(p_2)u(p_1)\,,
\eea
with $j_{0g}=2-2/\sqrt{\lambda}$, $\tilde{\tau}=\text{log}(1/\xi)$, and $f_V^{\prime}\approx f_{V}$. In terms of (\ref{FULLgluon222}), 
the total longitudinal differential cross section for electroproduction of $\phi$ meson reads
\bea\label{totaldcsPhi}
\frac{d\sigma_{L}(\phi)}{dt}&=&\frac{1}{16\pi s^2}\times \frac{1}{2}\sum_{\rm spin} \Big|\mathcal{N}_t\times\mathcal{A}_{\gamma^{*} p\rightarrow  \phi p}^{LL(gluon)}\Big|^2\nonumber\\
&=&\frac{1}{16\pi s^2}\times \frac{1}{2}\ \Big|\mathcal{N}_t\times\mathcal{A}_{\phi }^{gluon}\Big|^2\,,
\eea
where the total amplitude $\mathcal{A}_{\phi }^{gluon}$ is defined in (\ref{FULLgluon222}). We also compute the total cross section using the optical theorem as
\be\label{totalcsPhi}
\sigma_{L}(\phi)=\left(\frac{16\pi\,\frac{d\sigma_{L}(\phi)}{dt}}{1+\rho}\right)^{1/2}_{t=0}\,,
\ee
where $\rho$ is the ratio of the real and imaginary part of the total amplitude. We have compared our total cross section (\ref{totalcsPhi}) to experimental data in Fig. \ref{fig_12phi} and \ref{fig_13phi}. Note that these relations carry to heavier mesons $J/\Psi$ and $\Upsilon$ with slight changes.

\begin{figure*}
\subfloat[\label{fig_2pt1-phi}]{%
  \includegraphics[height=6cm,width=.6\linewidth]{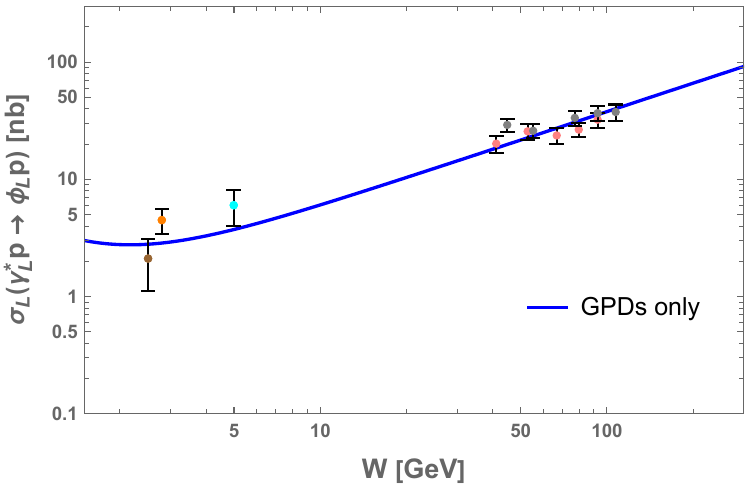}%
}
\caption{Longitudinal cross section for the electroproduction of phi mesons versus $W=\sqrt{s}$ in GeV, for 
$Q^2=3.8$ GeV$^2$ ({solid-blue curves}) in (\ref{totalcsPhi}) with $\mathcal{N}_t=0.2$.
The {solid-blue curves} are only with the evolved gluon GPD. The orange data point is from CLAS 2008 \cite{CLAS:2008cms}. The brown data point is from CLAS 2001 \cite{CLAS:2001zwd}. The cyan data point is from HERMES 2000. The grey data points are from ZEUS \cite{ZEUS:2005bhf}.
 The pink data points are from H1 \cite{H1:2000hps}.
}
\label{fig_12phi}
\end{figure*}

\begin{figure*}
\subfloat[\label{fig_75-phi}]{%
  \includegraphics[height=5.5cm,width=.55\linewidth]{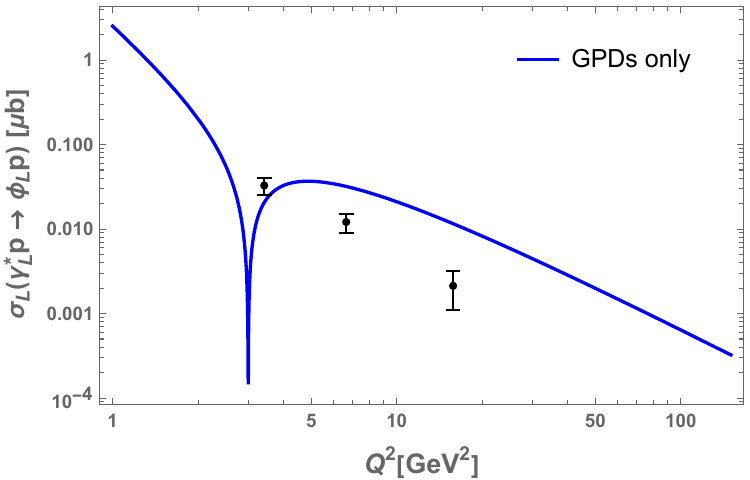}%
}
\caption{Longitudinal cross section for the electroproduction of phi mesons versus $Q^2$ in GeV$^2$, for 
$W=\sqrt{s}=75$ GeV ({solid-blue curve}) in (\ref{totalcsPhi}) with $\mathcal{N}_t=0.2$.  The {solid-blue curve} is only with the evolved gluon GPD.  The black data points are from H1, ZEUS \cite{H1:2009pze}.
}
\label{fig_13phi}
\end{figure*}

\subsection{Comparison to experiment for $\phi$}

In Table~\ref{table11}, we present the parameters for the AdS construction with a soft wall: $\lambda=11.243$, $\tau=3$, $\tilde{\kappa}_{T,S}=0.388~\text{GeV}$, and $\tilde{\kappa}_{N,V}=0.402~\text{GeV}$, as well as the normalization constants $\mathcal{N}_{t}=0.200$, and $\mathcal{N}_{s}=0$. We then illustrate the longitudinal cross section for the electroproduction of phi mesons ($\mathcal{N}_t=0.200$ in Eq.~(\ref{totalcsPhi})) as a function of $W=\sqrt{s}$ and $Q^2$ in Fig.~\ref{fig_12phi} and Fig.~\ref{fig_13phi}, respectively. The values of the physical parameters used are $m_{N}=0.94~\text{GeV}$, $m_{\phi}=1.019~\text{GeV}$, and $f^{\prime}_{\phi}=f_{\phi}=0.233~\text{GeV}$.

It is important to note that the solid-blue curve represents the evolved gluon GPD only, as the strange singlet quark GPD has been set to zero. Additionally, we assume the phi meson's coupling to the proton ($g_{\phi pp}=0$) to be zero, effectively setting the s-channel background contribution to zero. Experimental data points are included from various sources: CLAS 2008 (orange) \cite{CLAS:2008cms}, CLAS 2001 (brown) \cite{CLAS:2001zwd}, HERMES 2000 (cyan) \cite{Kroll:2014tma}, ZEUS (grey) \cite{ZEUS:2005bhf}, H1 (pink) \cite{H1:2000hps}, and H1, ZEUS (black) \cite{H1:2009pze}. Figures 5 and 7 from \cite{Kroll:2014tma} and \cite{Favart:2015umi}, respectively, also display these data points.

\section{Electroproduction of longitudinal $\rho^+$ meson with evolved valence quark GPDs: a comparison to experiment}~\label{sec_RHO+}
The electroproduction of longitudinal charged rho meson ($\rho^+$) in terms of quark DAs and quark GPDs is given by (see Eq.219 with Eq.222 and Eq.223 in \cite{Goeke:2001tz}) 
\begin{eqnarray}\label{rhoplusGPD11}
&&{\cal A}^{LL}_{\gamma^{*} p\rightarrow  \rho^{+} n} (s,t,Q,\epsilon_{L},\epsilon_{L}^{\prime}) =\nonumber\\
&& -e\times {C_{F}\over N_c}\times{1 \over 2}\times (4\pi \alpha _{s}(\mu^2))\times{1\over {Q}}\times\left[ \, \int _{0}^{1}dz{{\Phi_{\rho^{+}}(z)}\over z}\right] 
\times
\frac{1}{p^{+}}\left\{ A_{\rho^+\,n} 
\,h^{+}
\,+\, B_{\rho^+\,n} 
\,e^{+} 
\right\} \, , 
\end{eqnarray}
where $\Phi_{\rho^+}(z)$ is the charged rho meson ($\rho^+$) distribution amplitude (DA),
and
\bea\label{rhoplusGPD112}
A_{\rho^+ \, n} \,&=&\, \int_{-1}^1 dx \; 
{\left(H^u(x,\eta, t;\mu^2) \,-\, H^d(x,\eta, t;\mu^2)\right)} 
\; \left\{ {{Q_u} \over {\xi - x - i \epsilon}}
- {{Q_d} \over {\xi + x -i \epsilon}} \right\}\nonumber\\
B_{\rho^+ \, n} \,&=&\, \int_{-1}^1 dx \;
{\left( E^u(x,\eta, t;\mu^2) \,-\, E^d(x,\eta, t;\mu^2)\right)} 
\;\left\{ {{Q_u} \over {\xi - x - i \epsilon}}
- {{Q_d} \over {\xi + x - i \epsilon}} \right\}\,.
\eea
Using $C_{F}=\frac{N_c^2-1}{2N_c}$, and $Q_u=+2/3\,(Q_d=-1/3)$, we can rewrite (\ref{rhoplusGPD11}-\ref{rhoplusGPD112}) as 
\begin{eqnarray}\label{rhoplusGPD222}
{\cal A}^{LL}_{\gamma^{*} p\rightarrow  \rho^{+} n} (s,t,Q,\epsilon_{L},\epsilon_{L}^{\prime}) 
&\,=\,& -e\times {1\over 2}\times\left(1-\frac{1}{N_c^2}\right)\times 4\pi\times\alpha_s(\mu^2)\nonumber\\
&\times&{1\over {Q}}\times\left[ \, \int _{0}^{1}dz{{\Phi_{\rho^{+}}(z)}\over z}\right] \times
\frac{1}{p^{+}}\times u(p_2)\gamma^{+} u(p_1)\nonumber\\
&\times & \Bigg[\int_{0}^1 dx \; 
{\left(H_{valence}^u(x,\eta, t;\mu^2) \,-\, H_{valence}^d(x,\eta, t;\mu^2)\right)} 
\; \left\{ \frac{\xi}{2x}\times\frac{2x}{\xi^2-x^2}\right\}\nonumber\\
&+&\int_{0}^1 dx \; 
{\left(H_{singlet}^u(x,\eta, t;\mu^2) \,-\, H_{singlet}^d(x,\eta, t;\mu^2)\right)} 
\; \left\{ \frac{1}{6}\times\frac{2x}{\xi^2-x^2}\right\}\nonumber\\
&+& i\pi\times Q_u\left(H^u(\xi,\eta, t;\mu^2) \,-\, H^d(\xi,\eta, t;\mu^2)\right)\nonumber\\
&-i&\pi\times Q_d\left(H^u(-\xi,\eta, t;\mu^2) \,-\, H^d(-\xi,\eta, t;\mu^2)\right)\Bigg]\,,
\end{eqnarray}
where we have defined $H_{valence}^q(x,\eta, t;\mu^2)=H^q(x,\eta, t;\mu^2)+H^q(-x,\eta, t;\mu^2)$, $H_{singlet}^q(x,\eta, t;\mu^2)=H^q(x,\eta, t;\mu^2)-H^q(-x,\eta, t;\mu^2)$, and we have ignored the Pauli contribution for $-t\ll 4m_N^2$.

Ignoring the singlet quark GPDs contribution in (\ref{rhoplusGPD222}) by assuming $H^{u}_{singlet}=H^{d}_{singlet}$, the longitudinal electroproduction of $\rho^{+}$ meson with only evolved valence quark GPDs is given by
\begin{eqnarray}\label{rhoplusGPD22EVOLVEDHEnergy}
&&{\cal A}^{LL(valence)}_{\gamma^{*} p\rightarrow  \rho^{+} n} (s,t,Q,\epsilon_{L},\epsilon_{L}^{\prime})\nonumber\\
&=& -e\times {1\over 2}\times{1 \over 2}\times\left(1-\frac{1}{N_c^2}\right)\times 4\pi\alpha_s(\mu)\times{1\over {Q}}\times\left[ \, \int _{0}^{1}dz{{\Phi_{\rho^{+}}(z)}\over z}\right] \times
\frac{1}{2m_{N}}\times \bar u(p_2)u(p_1)\nonumber\\
&\times & \Bigg[\int_{0}^1 dx\,\,\frac{\xi}{2x}\times\frac{2x}{\xi^2 - x^2}\times\left(H_{valence}^u(x,\eta, t;\mu^2) \,-\, H_{valence}^d(x,\eta, t;\mu^2)\right)\nonumber\\
&+&i\pi\times Q_u\left(H^u(\xi,\eta, t;\mu^2) \,-\, H^d(\xi,\eta, t;\mu^2)\right)-i\pi\times Q_d\left(H^u(-\xi,\eta, t;\mu^2) \,-\, H^d(-\xi,\eta, t;\mu^2)\right)\Bigg]\,,\nonumber\\
&\,=\,& -e\times {1\over 2}\times{1 \over 2}\times\left(1-\frac{1}{N_c^2}\right)\times 4\pi\alpha_s(\mu)\times{1\over {Q}}\times\left[ \, \int _{0}^{1}dz{{\Phi_{\rho^{+}}(z)}\over z}\right] \times
\frac{1}{2m_N}\times \bar u(p_2)u(p_1)\nonumber\\
&\times &\Bigg[\frac{1}{\xi}\times\sum_{j = 1}^{\infty}
\frac{1}{\eta^{j-1}}\times\frac{1}{N_{j-1} ( \frac{3}{2})}
\times\left[\int_{0}^\eta \frac{dx}{\eta}\,\frac{\left(1 - \frac{x^2}{\eta^2}\right)}{1 - \frac{x^2}{\xi^2}}\times C_{j-1}^{3/2}\left(\frac{x}{\eta}\right)\right]\times\left(\mathbb{F}^u_{j(valence)} (\eta, t ; \mu^2) \,-\, \mathbb{F}^d_{j(valence)} (\eta, t ; \mu^2)\right)\nonumber\\
&+&i\pi\times Q_u\left(H^u(\xi,\eta, t;\mu^2) \,-\, H^d(\xi,\eta, t;\mu^2)\right)-i\pi\times Q_d\left(H^u(-\xi,\eta, t;\mu^2) \,-\, H^d(-\xi,\eta, t;\mu^2)\right)\Bigg]\,,\nonumber\\
&\,\approx\,& -e\times {1\over 2}\times\left(1-\frac{1}{N_c^2}\right)\times 4\pi\alpha_s(\mu)\times{1\over {Q}}\times\left[ \, \int _{0}^{1}dz{{\Phi_{\rho^{+}}(z)}\over z}\right] \times
\frac{1}{2m_{N}}\times \bar u(p_2)u(p_1)\nonumber\\
&\times &
\sum_{j=1}^{\infty}\frac{1}{\xi^{j}}\times\frac{1}{N_{j-1} ( \frac{3}{2})}
\times\left[\int_{0}^\xi \frac{dx}{\xi}\,C_{j-1}^{3/2}\left(\frac{x}{\xi}\right)\right]\times\left(\mathbb{F}^u_{j(valence)} (\eta\sim\xi, t ; \mu^2) \,-\, \mathbb{F}^d_{j(valence)} (\eta\sim\xi, t ; \mu^2)\right)\,,
\end{eqnarray}
with odd $j=1,3,...$, and in the last line we have used $\eta\sim\xi$. Again, the $i\pi H^{u,d}$ contributions vanish
for $\xi=\pm\eta$. We can also rewrite (\ref{rhoplusGPD22EVOLVEDHEnergy}) more compactly as 
\bea
\label{rhoplusGPD22EVOLVEDHEnergy2}
&&{\cal A}^{LL(valence)}_{\gamma^{*} p\rightarrow  \rho^{+} n} (s,t,Q,\epsilon_{L},\epsilon_{L}^{\prime})\approx  -e\times f_V^{+}\times\alpha_s(\mu)\times{1\over {Q}}\nonumber\\
&&\times \sum_{j=1}^{\infty}\frac{1}{\xi^{j}}\times\mathcal{N}_{q(valence)}(j)\times \left(\mathbb{F}^u_{j(valence)} (\xi, t ; \mu^2) \,-\, \mathbb{F}^d_{j(valence)} (\xi, t ; \mu^2)\right)
\times  \frac{1}{2m_{N}}\times \bar u(p_2)u(p_1)\,,\nonumber\\
\eea
where we have defined 
\bea
\mathcal{N}_{q(valence)}(j) &\equiv &  {1\over 2}\times\left(1-\frac{1}{N_c^2}\right)\times 4\pi\times
\frac{1}{N_{j-1} ( \frac{3}{2})}
\times I_{q(valence)}(j)\,,
\eea
and the integrals (with $\tilde{x}=x/\xi$)
\bea
f_V^{+} =\int _{0}^{1}dz{{\Phi_{\rho^{+}}(z)}\over z}\,,\qquad\qquad
I_{q(valence)}(j)=\left[\int_{0}^1 d\tilde{x}\,C_{j-1}^{3/2}\left(\tilde{x}\right)\right]\,,
\eea

\subsection{Final amplitude for $\rho^{+}$}

\begin{figure*}
  \includegraphics[height=6cm,width=.6\linewidth]{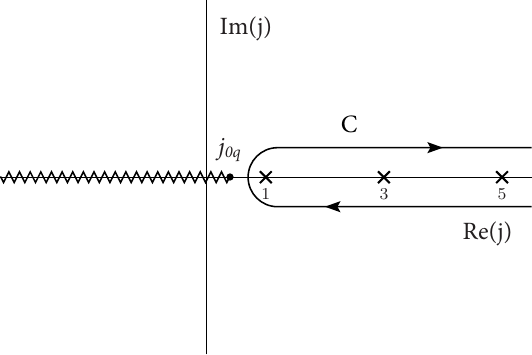}%
\caption{Illustration of the integration contour $\mathbb C$ in the complex j-plane used for computing the contour integral (\ref{rhoplusGPD22EVOLVEDHEnergy3INPUTvalence}), with a branch cut displayed for $\text{Re(j)}\leq j_{0q}=1-1/\sqrt{\lambda}$.}
\label{contourvalencequark}
\end{figure*}

The resummation by j-contour for the valence quark GPDs contribution (\ref{rhoplusGPD22EVOLVEDHEnergy2}) is similar to the resummation for the singlet quark and gluon GPDs (albeit using a different integration contour shown in Fig.~\ref{contourvalencequark}), and the detailed derivation is given in Appendix~\ref{valencequarksum}. Here we only provide the final answer.

After the resummation of (\ref{rhoplusGPD22EVOLVEDHEnergy2}) using the contour integral (\ref{rhoplusGPD22EVOLVEDHEnergy3INPUTvalence}) with the integration contour $\mathbb{C}$ in the complex j-plane, shown in Fig.~\ref{contourvalencequark}, and
using the result for $\text{Re}\left[{\cal A}^{LL(valence)}\right]$ given in (\ref{RErhoplusGPD22EVOLVEDHEnergy4}) 
and $\text{Im}\left[{\cal A}^{LL(valence)}\right]$ given in (\ref{IMrhoplusGPD22EVOLVEDHEnergy4}), the full complex amplitude with the evolved valence quark GPDs is given by
\bea\label{fullvalenceEVOLVEDvalencekINPUT}
&&{\cal A}^{LL(valence)}_{\gamma^{*} p\rightarrow  \rho^{+} n} (s,t,Q,\epsilon_{L},\epsilon_{L}^{\prime})\nonumber\\
&&\propto  \Bigg(\widehat{\mathbb{F}}^u_{j_{0q}(valence)} (\xi, t ; \mu^2)- \widehat{\mathbb{F}}^d_{j_{0q}(valence)} (\xi, t ; \mu^2)\Bigg) \times \Bigg[(\sqrt{\lambda}/4\pi)^{1/2}\times \frac{1}{\xi}\times(\sqrt{\lambda}/2\pi)\times e^{-\tilde{\tau}/\sqrt{\lambda}}\times\frac{e^{-\sqrt{\lambda}/4\tilde{\tau}}}{\tilde{\tau}^{3/2}}\nonumber\\
&& + i\times \frac{1}{\xi^{j_{0q}}}\times (\sqrt{\lambda}/4\pi)^{1/2}\times \frac{e^{-\sqrt{\lambda}/4\tilde{\tau}}}{\tilde{\tau}^{3/2}}\Bigg]\times\mathcal{N}_{q(valence)}(j_{0q})\times (-1)\times e\,f_V^{+}\,\alpha_s(\mu)\,{1\over {Q}}\,\frac{1}{2m_{N}}\,\bar u(p_2)u(p_1)\,,\nonumber\\  
&&\propto  \Bigg(\widehat{\widehat{\mathbb{F}}}^u_{j_{0q}(valence)} (\xi, t ; \mu^2)-\widehat{\widehat{\mathbb{F}}}^d_{j_{0q}(valence)} (\xi, t ; \mu^2)\Bigg)\times (-1)\times f_V^{+}\,\alpha_s(\mu)\,{1\over {Q}}\,\frac{1}{2m_{N}}\,\bar u(p_2)u(p_1)\nonumber\\ 
&&\times \left[\frac{(\sqrt{\lambda}/2\pi)}{\xi^{1-1/\sqrt{\lambda}}}+\frac{i}{\xi^{1-1/\sqrt{\lambda}}}\right]\,,\nonumber\\  
&&\propto  {\cal A}^{valence}_{\rho^{+}} (s,t,Q)\times\frac{1}{2\sqrt{2}m_{N}}\,\bar u(p_2)u(p_1)\,, 
\eea
where we have absorbed $e\times\mathcal{N}_{q(valence)}(j_{0q})\times(\sqrt{\lambda}/4\pi)^{1/2}\times \frac{e^{-\sqrt{\lambda}/4\tilde{\tau}}}{\tilde{\tau}^{3/2}}$ into the normalization constant as it varies slowly with $s$. We have also defined the normalized and flavor separated quark conformal moments as
\bea
\widehat{\widehat{\mathbb{F}}}^{u+d}_{j(valence)}(\xi, t ; \mu^2)&=&0.151\times \frac{\widehat{\mathbb{F}}^q_{j(valence)}(\xi, t ; \mu^2)}{\widehat{\mathbb{F}}^q_{j=3(valence)}(\xi=0, t=0 ; \mu^2=4~\text{GeV}^2)}\,,\nonumber\\
\widehat{\widehat{\mathbb{F}}}^{u-d}_{j(valence)}(\xi, t ; \mu^2)&=&0.070\times\frac{\widehat{\mathbb{F}}^q_{j(valence)}(\xi, t ; \mu^2)}{\widehat{\mathbb{F}}^q_{j=3(valence)}(\xi=0, t=0 ; \mu^2=4~\text{GeV}^2)}\,,\nonumber\\
\widehat{\widehat{\mathbb{F}}}^{u}_{j(valence)}(\xi, t ; \mu^2)&=&\frac{1}{2}\times\left(\widehat{\widehat{\mathbb{F}}}^{u+d}_{j(valence)}(\xi, t ; \mu^2)+\widehat{\widehat{\mathbb{F}}}^{u-d}_{j(valence)}(\xi, t ; \mu^2)\right)\,,\nonumber\\
\widehat{\widehat{\mathbb{F}}}^{d}_{j(valence)}(\xi, t ; \mu^2)&=&\frac{1}{2}\times\left(\widehat{\widehat{\mathbb{F}}}^{u+d}_{j(valence)}(\xi, t ; \mu^2)-\widehat{\widehat{\mathbb{F}}}^{u-d}_{j(valence)}(\xi, t ; \mu^2)\right)\,,
\eea
where the normalization coefficients are fixed using the lattice data in \cite{LHPC:2007blg}, and  $\widehat{\mathbb{F}}^{q}_{j(valence)} (\xi, t ; \mu^2)=\Gamma(\Delta_{q}(j)-2)\times\mathbb{F}^{q}_{j(valence)} (\xi, t ; \mu^2)$ is given in (\ref{EvolutionConfMomGPDsValence}). Note that we have used the average value of $A_3^{u+d}(-t=0)=\widehat{\widehat{\mathbb{F}}}^{u+d}_{j=3(valence)} (\xi=0, t=0 ; \mu^2=4~\text{GeV}^2)=0.151$ from Table XV and Table XVI of \cite{LHPC:2007blg}. We have also used the value of $A_3^{u-d}(-t=0)=\widehat{\widehat{\mathbb{F}}}^{u-d}_{j=3(valence)} (\xi=0, t=0 ; \mu^2=4~\text{GeV}^2=0.070$ from Table IX of \cite{LHPC:2007blg}.

\subsection{Differential and total cross sections for $\rho^{+}$}

Finally, using the valence quark GPDs contribution (\ref{fullvalenceEVOLVEDvalencekINPUT}), we have the total longitudinal differential cross section for electroproduction of $\rho^{+}$ meson 
\bea\label{totaldcsrhoplus}
\frac{d\sigma_{L}(\rho^+)}{dt}&=&\frac{1}{16\pi s^2}\times \frac{1}{2}\sum_{\rm spin} \Big|\mathcal{N}_{t}\times\mathcal{A}_{\gamma^{*} p\rightarrow  \rho^{+} n}^{LL(valence)}\Big|^2 \nonumber\\
&=&\frac{1}{16\pi s^2}\times \frac{1}{2} \Big|\mathcal{N}_{t}\times\mathcal{A}_{\rho^{+}}^{valence}\Big|^2\,.
\eea
We also calculate the total cross section using the optical theorem as
\be\label{totalcsrhoplus}
\sigma_{L}(\rho^+)=\left(\frac{16\pi\,\frac{d\sigma_{L}(\rho^+)}{dt}}{1+\rho}\right)^{1/2}_{t=0}\,,
\ee
where $\rho$ is the ratio of the real and imaginary part of our total complex amplitude. We have compared our total cross section (\ref{totalcsrhoplus}) to the experimental data in Fig. \ref{fig_14rhoplus}. We fix the value of the normalization parameter $\mathcal{N}_t$ by the charged rho meson data as detailed in the next subsection below.

\begin{figure*}
\subfloat[\label{fig_1}$Q^2=1-1.5$ GeV$^2$]{%
  \includegraphics[height=4cm,width=.49\linewidth]{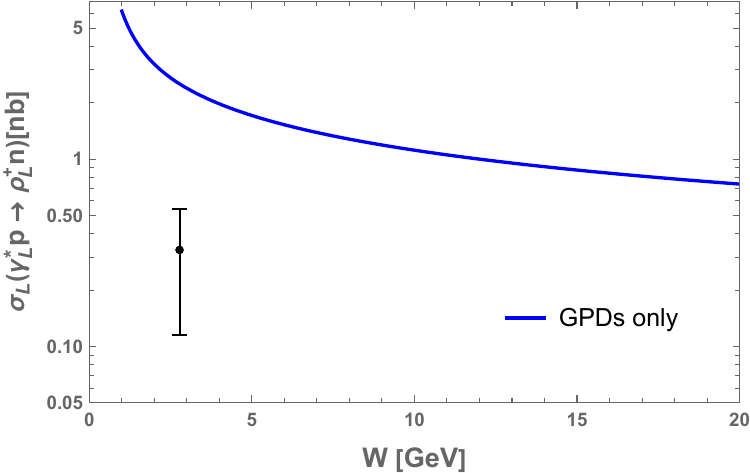}%
}\hfill
\subfloat[\label{fig_1pt5}$Q^2=1.5-2.0$ GeV$^2$]{%
  \includegraphics[height=4cm,width=.49\linewidth]{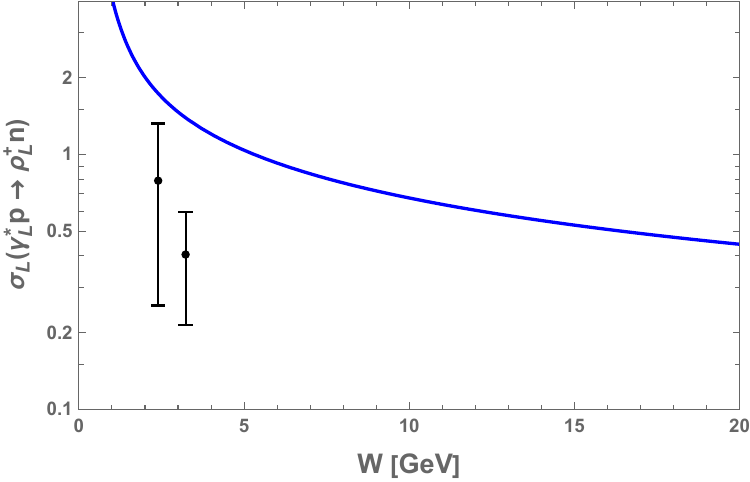}%
}\hfill
\subfloat[\label{fig_2}$Q^2=2.0-2.5$ GeV$^2$]{%
  \includegraphics[height=4cm,width=.49\linewidth]{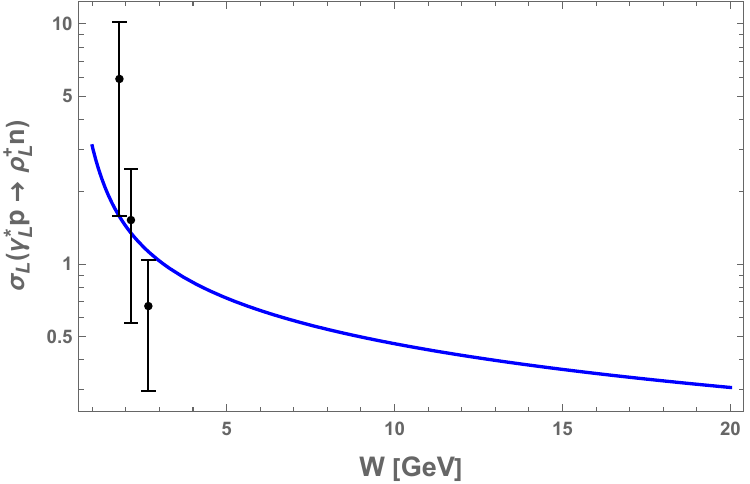}%
}\hfill
\subfloat[\label{fig_2pt5}$Q^2=2.5-3.0$]{%
  \includegraphics[height=4cm,width=.49\linewidth]{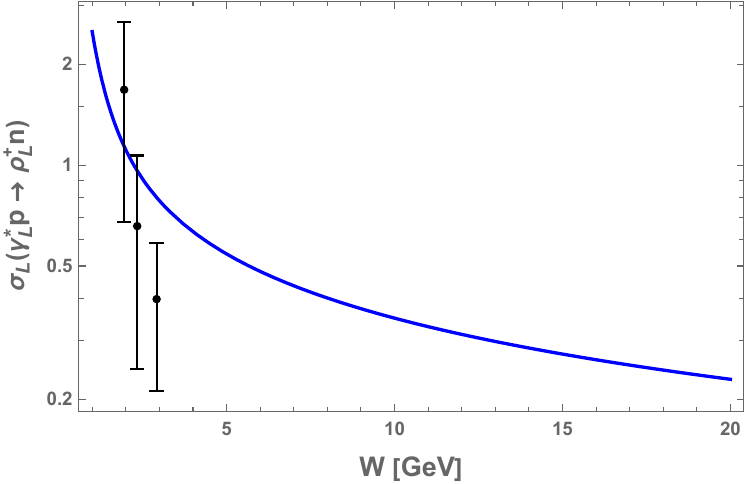}%
}\hfill
\subfloat[\label{fig_3}$Q^2=3.0-3.5$ GeV$^2$]{%
  \includegraphics[height=4cm,width=.49\linewidth]{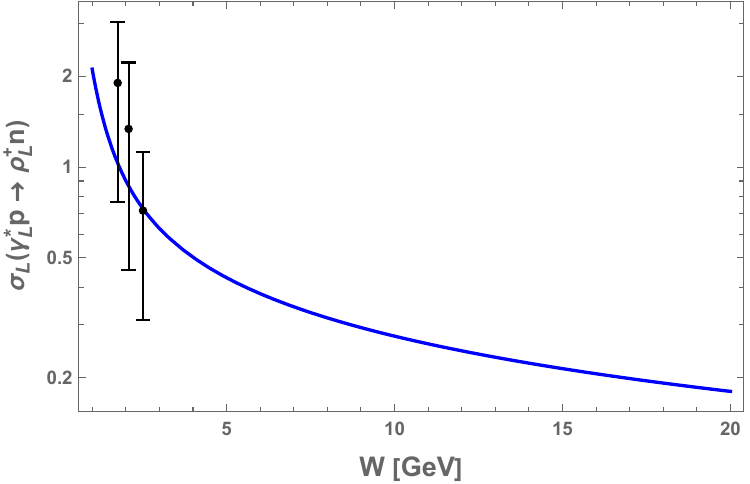}%
}\hfill
\subfloat[\label{fig_3pt5}$Q^2=3.5-4.0$]{%
  \includegraphics[height=4cm,width=.49\linewidth]{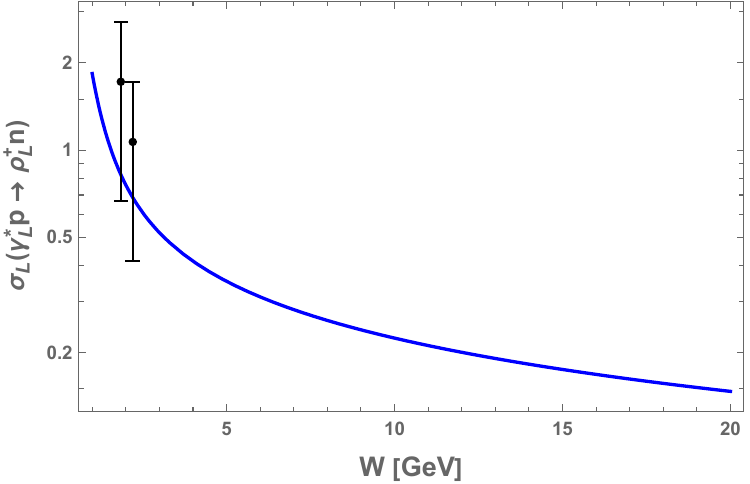}%
}\hfill
\subfloat[\label{fig_4}$Q^2=4.0-4.5$ GeV$^2$]{%
  \includegraphics[height=4cm,width=.49\linewidth]{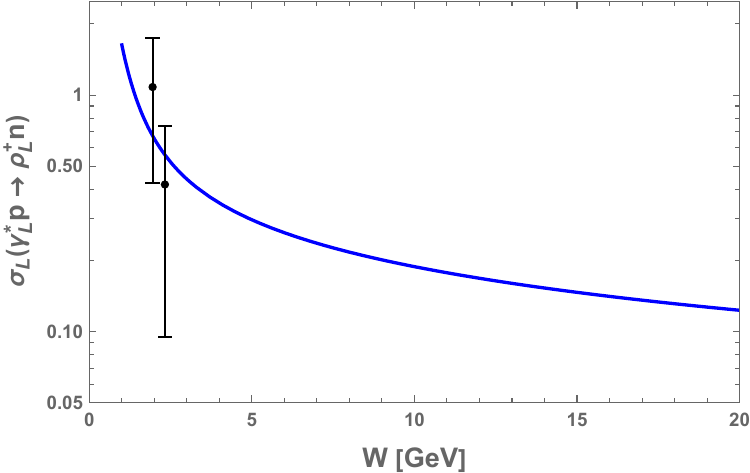}%
}
\caption{Longitudinal cross section for the electroproduction of charged rho mesons ($\rho^{+}$) versus $W=\sqrt{s}$ in GeV.
The {blue curve} is with the evolved valence up and down quark GPDs (i.e., (\ref{totalcsrhoplus}).
 The black data points are from 5.776 GeV CLAS \cite{Fradi:2010ew}.}
\label{fig_14rhoplus}
\end{figure*}


\subsection{Comparison to experiment for $\rho^{+}$}

In this section, we present our result for the electroproduction of charged rho mesons ($\rho^{+}$) and its comparison to experimental data. First, we list the parameters used in our AdS construction with a soft wall, namely $\lambda=11.243$, $\tau=3$, $\tilde{\kappa}_{T,S}=0.388~\text{GeV}$, $\tilde{\kappa}_{N,V}=0.402~\text{GeV}$, and the normalization constants $\mathcal{N}_{t}=724.513$ and $\mathcal{N}_{s}=0$, which are given in Table~\ref{table11}.

Next, we show the longitudinal cross section for the electroproduction of $\rho^{+}$ mesons versus $W=\sqrt{s}$ in GeV in Fig.~\ref{fig_14rhoplus}. In these figures, we have used the evolved valence up and down quark GPDs (i.e., (\ref{totalcsrhoplus})) with $\mathcal{N}_{t}=724.513$. We have also used the values of the physical parameters $m_{N}=0.94~\text{GeV}$, $m_{\rho^{0,+}}=0.775~\text{GeV}$, and $f^{\prime}_{\rho^{0,+}}=f_{\rho^{0,+}}=0.216~\text{GeV}$. It is important to note that we have assumed that the coupling of the $\rho^{+}$ meson to the proton and neutron ($g_{\rho^{+}pn}$) is zero and we have set the s-channel contribution to zero, that is, $\mathcal{N}_{s}=0$. 

In the comparison of our results with experimental data, in Fig.~\ref{fig_14rhoplus}, the black data points are from the 5.776 GeV CLAS experiment \cite{Fradi:2010ew} (see Fig.4 of \cite{Fradi:2010ew}).

\section{Summary and Conclusion}~\label{sec_CON}

In this paper, we have presented a framework for constructing General Parton Distributions (GPDs) at low-x and finite skewness using holographic QCD in the double limit of large $N_c$ and strong gauge coupling. We have first constructed holographic amplitudes for exclusive electroproduction processes, dominated by bulk spin-j gravitons and vector or axial meson exchanges in AdS, which are dual to Pomerons and Reggeons in QCD, respectively. By comparing the holographic exclusive amplitudes with the corresponding exclusive amplitudes in QCD, which are based on factorization theorems, we have extracted the spin-j conformal moments of GPDs and constructed the GPDs.

Using the holographically fixed conformal moments of GPDs at the initial or low renormalization scale $\mu\sim \mu_0\sim 1.8$ GeV, we have used one-loop pQCD evolution equations to obtain their counterparts at a higher renormalization scale $\mu>\mu_0$, which enhances the perturbative partonic content at low-x in addition to the nonperturbative one captured by the holographic construction at the low renormalization scale.

We have used our evolved GPDs to specifically determine the exclusive longitudinal electroproduction of neutral rho ($\rho^0$), phi ($\phi$), and charged rho ($\rho^+$) mesons, using the factorization theorems in QCD, and compare them to the existing data from CLAS, HERMES, and H1 experiments. Specifically, in the data analysis of the longitudinal electroproduction of $\rho^0$, we have found that the holographic GPDs (similar to previous GPD models) alone cannot explain the low $\sqrt{s}$ data, and we have accounted for the non-perturbative physics coming from the s-channel. We have successfully determined such non-perturbative contribution in the s-channel using holographic QCD and found that its contribution at low $\sqrt{s}$ is in good agreement with the experimental data. We expect the holographic s-channel contribution to play a key role in the future experimental extraction of GPDs at JLab and future EIC.

Our extracted quark and gluon GPDs for fixed skewness $\eta$ provide explicit and detailed dependence on the kinematical variables $t$ and $\eta$, with their moments satisfying all the polynomiality conditions in $\eta$. They should prove useful for extracting the GPDs using global data analysis and carry detailed tomographic information about the partonic distributions inside a nucleon after Fourier transforming to coordinate space with respect to $t$.

Additionally, our framework can be extended to longitudinal heavier vector mesons such as $J/\Psi$ and $\Upsilon$, and yields detailed predictions for exclusive processes such as the electroproduction of photons in DVCS or the longitudinal electroproduction of single and double axial mesons.

In conclusion, this paper provides a powerful framework for constructing GPDs at low-x and finite skewness using holographic QCD in the large $N_c$ limit. We have demonstrated the potential of this approach for analyzing exclusive electroproduction processes and provided useful results for future experimental studies and global data analyses.

\vskip 1cm
\centerline{\bf Acknowledgments}
\vskip 0.5cm
K.M. is supported by the U.S.~Department of Energy, Office of Science, Office of Nuclear Physics, contract no.~DE-AC02-06CH11357, and an LDRD initiative at Argonne National Laboratory under Project~No.~2020-0020. I.Z. is supported by the Office of Science, U.S. Department of Energy under Contract No. DE-FG-88ER40388.

\appendix



\section{Notations}\label{sec:notations}
Here, we summarize the key notations used for the GPDs and their conformal (Gegenbauer) moments:
\begin{align*}
    H^q(x) &\equiv \text{quark GPDs}\,,\\
    \widetilde{H}^q(x) &\equiv \text{axial quark GPDs}\,,\\
    H^g(x) &\equiv \text{gluon GPDs}\,,\\
    H_{\text{valence}}^{q}(x) &\equiv \text{valence quark GPDs, which are } H^{q}(x)+H^{q}(-x)\,,\\
    H_{\text{singlet}}^{q}(x) &\equiv \text{singlet quark GPDs, which are } H^{q}(x)-H^{q}(-x)\,,\\
    \widetilde{H}_{\text{valence}}^{q}(x) &\equiv \text{valence axial quark GPDs, which are } \widetilde{H}^{q}(x)-\widetilde{H}^{q}(-x)\,,\\
    \widetilde{H}_{\text{singlet}}^{q}(x) &\equiv\text{singlet axial quark GPDs, which are } \widetilde{H}^{q}(x)+\widetilde{H}^{q}(-x)\,,\\
    \mathbb{F}^q_{j(\text{valence})} &\equiv \text{conformal (Gegenbauer) moments of valence quark GPDs}\,,\\
    \mathbb{F}^q_{j(\text{singlet})} &\equiv \text{conformal (Gegenbauer) moments of singlet quark GPDs}\,,\\
    \widetilde{\mathbb{F}}^q_{j(\text{valence})} &\equiv \text{conformal (Gegenbauer) moments of valence axial quark GPDs}\,,\\
    \widetilde{\mathbb{F}}^q_{j(\text{singlet})} &\equiv \text{conformal (Gegenbauer) moments of singlet axial quark GPDs}\,,\\
    \mathbb{F}^g_{j} &\equiv \text{conformal (Gegenbauer) moments of gluon GPD}.\\
\end{align*}

\section{Kinematics}~\label{KINEMATICS}

\begin{figure*}
  \includegraphics[height=6cm,width=.6\linewidth]{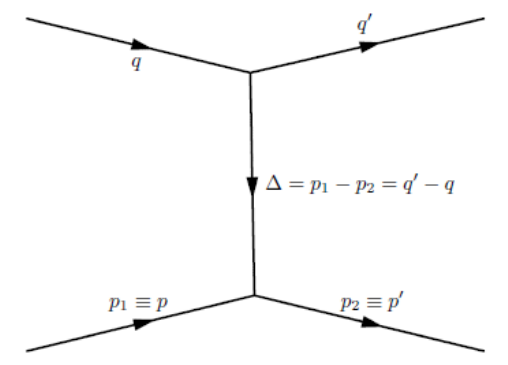}%
\caption{Illustration of the kinematic variables defined in (\ref{B1})\,-\,(\ref{B7}).}
\label{kinematics}
\end{figure*}

Here, we summarize the pertinent kinematics for the GPDs, see Fig.~\ref{kinematics}.
Following \cite{Belitsky:2005qn}, we define 
\begin{equation}\label{B1}
\tilde{q} = \frac{1}{2} (q + q^{\prime})
\, , \qquad
p = \frac{1}{2}(p_1 + p_2)
\, , \qquad
\Delta = p_1 - p_2 = q^{\prime}-q\, .
\end{equation}
\begin{equation}
\label{SymmetricVariables}
\tilde {q}^2 = - \tilde{Q}^2
\, , \qquad
\xi = \frac{\tilde{Q}^2}{2p \cdot \tilde{q}}
\, , \qquad
\eta = \frac{\Delta \cdot \tilde{q}}{2p \cdot \tilde{q}}
\, .
\end{equation}

\begin{equation}
\label{ExperVariables}
Q^2 = - q^2
\, , \qquad
Q^{\prime 2} = -q^{\prime 2}
\, , \qquad
x_{\rm B} = \frac{{Q}^2}{2 p_1 \cdot q}
\, ,
\end{equation}
\begin{equation}
\label{ResolutionScale}
\tilde{Q}^2 = \frac{1}{2}
\left(
Q^2 + Q^{\prime 2} + \frac{\Delta^2}{2}
\right) \, ,
\end{equation}

\begin{equation}
\label{XitoEta}
\xi =
\frac{Q^2 +Q^{\prime 2} + \Delta^2/2}{2Q^2/ x_{\rm B} - Q^2 + Q^{\prime 2} + \Delta^2}
\, , \qquad
\eta =
\frac{Q^2 -Q^{\prime 2}}{2Q^2/ x_{\rm B} - Q^2 + Q^{\prime 2} + \Delta^2}
\, ,
\end{equation}

\begin{equation}
Q^{\prime 2}
= \left( 1 - \frac{\eta}{\xi} \right) \tilde{Q}^2 - \frac{\Delta^2}{4}
\, , \qquad
Q^2
=
\left( 1 + \frac{\eta}{\xi} \right) \tilde{Q}^2 - \frac{\Delta^2}{4}
\, ,
\end{equation}
and
\begin{equation}\label{B7}
x_{\rm B} =
\frac{(\xi + \eta) \tilde{Q}^2 - \xi \Delta^2/4}{(1 + \eta) \tilde{Q}^2 - \xi \Delta^2/2}
\, .
\end{equation}

We also introduce a pair of
light-cone vectors such that $n^2 = n^{\ast 2} = 0$ and $n \cdot n^\ast \equiv
n^\mu n^\ast_\mu = 1$. They can be chosen as
\begin{equation}
\label{LightLikeVectors}
n^\mu \equiv \frac{1}{\sqrt{2}} (1, 0, 0, -1)
\, , \qquad
n^{\ast \mu} \equiv \frac{1}{\sqrt{2}} (1, 0, 0, 1)
\, .
\end{equation}
Any four-vector $a^\mu$ can be decomposed into its light-cone components as
\begin{equation}
a^\mu = a^+ n^{\ast \mu} + a^- n^\mu + a^\mu_\perp
\, ,
\end{equation}
with
\begin{equation}
a^+ \equiv a \cdot n = \frac{1}{\sqrt{2}} (a^0 + a^z)
\, , \qquad
a^- \equiv a \cdot n^\ast = \frac{1}{\sqrt{2}} (a^0 - a^z)
\, .
\end{equation}
Scalar products are written as

\begin{equation}
a \cdot y
\equiv
a_\mu y^\mu
=
a^+ y^- + a^- y^+ - {a}_{\perp} \cdot {y}_{\perp}
\, .
\end{equation}
The light-cone derivatives are defined as follows

\begin{equation}
\partial^+ = \frac{\partial}{\partial a^-}
\, , \qquad
\partial^- = \frac{\partial}{\partial a^+}
\, , \qquad
\partial^\mu_\perp = \frac{\partial}{\partial a_\mu^\perp}
\end{equation}

\section{Quark and Gluon GPDs from their Conformal (Gegenbauer) Moments}~\label{sec_PQCD}

The framework for analysing the evolution equations for the GPDs, consists of defining the operators in the OPE in the
representations of the conformal group.  The conformal operators transform
multiplicatively under renormalization. As a result, the anomalous dimensions of conformal operators are only 
labeled by spin-j, in reference to the conformal spin of the composite operator~\cite{Belitsky:2005qn} (and references therein).
In this section, we will  briefly review the relevant quark and gluon spin-j operators, which enter the leading GPDs at twist-2.
The spin-j operators are the Gegenbauer moments of the various quark and gluon GPDs.


\subsection{Quark and gluon operators}\label{quark operators}

{\bf Quarks:}

We define leading-twist bilocal quark operators as \cite{Belitsky:2005qn}:
\begin{eqnarray}
\label{QuarkLightRayOperators}
{\cal O}^{qq} (w^-_1, w^-_2)
\!\!\!&=&\!\!\!
\bar\psi (w^-_1)
[w^-_1 , w^-_2] \gamma^+
\psi (w^-_2)
\, , \\
\label{QuarkLightRayOperatorsOdd}
\widetilde {\cal O}^{qq} (w^-_1, w^-_2)
\!\!\!&=&\!\!\!
\bar\psi (w^-_1)
[w^-_1 , w^-_2] \gamma^+ \gamma^5
\psi (w^-_2)
\, , 
\end{eqnarray}
where have added the Wilson line in the adjoint representation
$$
[w^-_1 , w^-_2]^{ab}
=
P \exp
\left(
g \int_{w^-_2}^{w^-_1} d w^{\prime -} f^{abc} A_c^+ (w^{\prime -} , {\bf 0}_\perp)
\right)
\, ,
$$
to respect gauge invariance. 
\\
\\
{\bf Gluons:}
\\
We define the two-dimensional tensors \cite{Belitsky:2005qn}
\begin{eqnarray}
g^\perp_{\mu\nu}
\!\!\!&\equiv&\!\!\!
g_{\mu\nu} - n_\mu n^\ast_\nu
- n_\nu n^\ast_\mu
\, , \\
\varepsilon^\perp_{\mu\nu}
\!\!\!&\equiv&\!\!\!
\varepsilon^{\alpha\beta\rho\sigma}
g^\perp_{\alpha\mu} g^\perp_{\beta\nu} n^\ast_\rho n_\sigma
\, ,
\end{eqnarray}
where the totally antisymmetric tensor is normalized as $\varepsilon^{0123} = 1$. 

We also define the gauge potential in the light-cone gauge
\begin{equation}
A_\perp \equiv A^x + i A^y \, , \qquad \bar A_\perp \equiv A^x - i A^y
\, ,
\end{equation}
and the gluon field-strength tensors
\begin{eqnarray*}
\partial^+ A_\perp
=
F^{+ x} - i \widetilde F^{+ x}
=
i \left( F^{+ y} - i \widetilde F^{+ y} \right)
\, , \qquad
\partial^+ \bar A_\perp
=
F^{+ x} + i \widetilde F^{+ x}
=
- i \left( F^{+ y} + i \widetilde F^{+ y} \right)
\, ,
\end{eqnarray*}

\begin{equation}
F^{\ + \mu}_{[-] \, \perp}
=
F^{+ \mu}_{\phantom{+} \, \perp}
-
i \widetilde F^{+ \mu}_{\phantom{+} \, \perp}
\, , \qquad
F^{\ + \mu}_{[+] \, \perp}
=
F^{+ \mu}_{\phantom{+} \, \perp}
+
i \widetilde F^{+ \mu}_{\phantom{+} \, \perp}
\, ,
\end{equation}
with the dual gluon field strength defined as $\widetilde F^{\mu\nu} \equiv 12 \varepsilon^{\mu\nu\rho\sigma}
F_{\rho\sigma}$, so that $\widetilde F^+{}^\mu_\perp = \varepsilon^{\mu\nu}_\perp
F^+{}^\perp_\nu$ for two-dimensional transverse indices. 

Using these definitions we introduce the leading-twist gluonic
operators \cite{Belitsky:2005qn}:
\begin{eqnarray}
\label{GluonLightRayOperators}
{\cal O}^{gg} (w^-_1, w^-_2)
\!\!\!&=&\!\!\!
F^{+ \mu}_a (w^-_1)
[w^-_1 , w^-_2]^{ab} g_{\mu\nu}
F^{\nu +}_b (w^-_2)
\, , \\
\widetilde {\cal O}^{gg} (w^-_1, w^-_2)
\!\!\!&=&\!\!\!
F^{+ \mu}_a (w^-_1)
[w^-_1 , w^-_2]^{ab} i \varepsilon^\perp_{\mu\nu}
F^{\nu +}_b (w^-_2)
\, , 
\label{GluonLightRayOperatorMaximalHelicity}
\end{eqnarray}

\subsection{Operator matrix elements and GPDs}
\label{GPDsSpinOneHalf}
We define the spinor bilinears \cite{Belitsky:2005qn}
\begin{equation}
\label{Def-ForFac}
\begin{array}{ll}
b = \bar u (p_2) u (p_1)
\, , \qquad
&
\widetilde b = \bar u (p_2) \gamma^5 u (p_1)
\, , \\
h^{\mu} = \bar u (p_2) \gamma^\mu u (p_1)
\, , \qquad
&
\widetilde{h}^\mu = \bar u (p_2) \gamma^\mu \gamma^5 u (p_1)
\, , \\
t^{\mu\nu} = \bar u (p_2) i \sigma^{\mu\nu} u (p_1)
\, , \qquad
&
\widetilde{t}^{\mu\nu} = \bar u (p_2) i \sigma^{\mu\nu} \gamma^5 u (p_1)
\, .
\end{array}
\end{equation}

The matrix elements for
quark operators can be decomposed in terms of the above spinor bilinears as
\begin{eqnarray}
\label{vectorGPDnucleon}
\langle p_2 | {\cal O}^{qq} (- w^-, w^-) | p_1 \rangle
\!\!\!&=&\!\!\!
\int_{- 1}^{1} d x \ {\rm e}^{- i x p^+ w^- }
\left\{
h^+ H^q (x, \eta, \Delta^2) + e^+ E^q (x, \eta, \Delta^2)
\right\}
\nonumber\\
\!\!\!&\equiv&\!\!\!
p^+
\int_{-1}^1 dx \, {\rm e}^{- i x p^+ w^-} F^q (x, \eta, \Delta^2)
\, , \\
\label{axialGPDnucleon}
\langle p_2 | \widetilde {\cal O}^{qq} (- w^-, w^-) | p_1 \rangle
\!\!\!&=&\!\!\!
\int_{- 1}^{1} d x \ {\rm e}^{- i x p^+ w^- }
\left\{
\tilde h^+ \widetilde H^q (x, \eta, \Delta^2)
+
\tilde e^+ \widetilde E^q (x, \eta, \Delta^2)
\right\}
\nonumber\\
\!\!\!&\equiv&\!\!\!
p^+
\int_{-1}^1 dx \, {\rm e}^{- i x p^+ w^-} \widetilde{F}^q (x, \eta, \Delta^2)
\, . 
\end{eqnarray}

Similarly, the matrix elements for
gluon operators can be decomposed in terms of the spinor bilinears as
\begin{eqnarray}
\label{GluonGPDnucleon}
\langle p_2 | {\cal O}^{gg} (- w^-, w^-) | p_1 \rangle
\!\!\!&=&\!\!\!
\frac{1}{4} p^+
\int_{- 1}^{1} d x \
{\rm e}^{- i x p^+ w^-}
\left\{
h^+ H^g (x, \eta, \Delta^2) + e^+ E^g (x, \eta, \Delta^2)
\right\}\nonumber\\
\!\!\!&\equiv&\!\!\!
\frac{(p^+)^2}{4}
\int_{-1}^1 dx \, {\rm e}^{- i x p^+ w^-} F^g (x, \eta, \Delta^2)
\, ,\\
\langle p_2 |
\widetilde {\cal O}^{gg} (- w^-, w^-)
| p_1 \rangle
\!\!\!&=&\!\!\!
\frac{1}{4} p^+
\int_{- 1}^{1} d x \
{\rm e}^{- i x p^+ w^-}
\left\{
\tilde h^+ \widetilde H^g (x, \eta, \Delta^2)
+
\tilde e^+ \widetilde E^g (x, \eta, \Delta^2)
\right\}
\nonumber\\
\!\!\!&\equiv&\!\!\!
\frac{(p^+)^2}{4}
\int_{-1}^1 dx \, {\rm e}^{- i x p^+ w^-} \widetilde{F}^g (x, \eta, \Delta^2)
\, .
\end{eqnarray}

\subsection{Spin-j quark and gluon operators}
Classically, massless 4D QCD is invariant under the conformal group $SO(4,2)$ with 15 generators (1-dilatation, 4-conformal, 4-translation, 6-Lorentz). 
The OPE expansion on the light cone involves Wilsonian operators which transform  covariantly under the Lorentz group $SO(3,1)$, 
but not under the conformal group $SO(4,2)$. Since $SL(2)$ on the light front,  is a subgroup of the conformal group, a way 
to re-organize the Wilsonian expansion  is in terms of conformal tensors of spin $j=\frac 12 (d+s)$
(the eigenvalue of the collinear subgroup  of the conformal operator)  where the twist is $\tau=d-s$~\cite{Belitsky:2005qn}. 
More specifically,
the twist-2 vector quark and  gluon operators are (see Eq.4.138 in \cite{Belitsky:2005qn})
\begin{equation}
\label{QuarkAndGluonConformalOperators}
\mathbb{O}^{qq}_{j}
=
\bar\psi
(i \partial^+ )^{j-1}
\gamma^+
C_{j-1}^{3/2}
\left(
\stackrel{{}_\leftrightarrow}{\cal D}{\!}^+ \! / \partial^+
\right)
\psi
\, , \qquad
\mathbb{O}^{gg}_{j}
=
F^+{}_\mu
(i \partial^+ )^{j - 2}
C_{j - 2}^{5/2}
\left(
\stackrel{{}_\leftrightarrow}{\cal D}{\!}^+ \! / \partial^+
\right)
F^{\mu +}
\, ,
\end{equation}
and the twist-2 axial-vector quark operator is
\begin{equation}
\label{QuarkAndGluonConformalOperatorsAxial}
\tilde{\mathbb{O}}^{qq}_{j}
=
\bar\psi
(i \partial^+ )^{j-1}
\gamma^+
\gamma_5 C_{j-1}^{3/2}
\left(
\stackrel{{}_\leftrightarrow}{\cal D}{\!}^+ \! / \partial^+
\right)
\psi
\, , 
\end{equation}
Here $\stackrel{{}_\leftrightarrow}{\cal D}{\!}^+$ is the symmetrized long derivative on LF$^+$. $C_n^{\nu}(x)$ are the Gegenbauer polynomials, 
a generalization of the Legendre polynomials to $(2\nu+2)$-D space. Their explicit form in terms of hypergeometric functions is
\bea
C_n^\nu(x) =
\begin{pmatrix}
n+2\nu +1\\
n
\end{pmatrix} \bigg(\frac {x+1}2\bigg)^2  {}_2F_1\bigg(-n,-n-\nu+\frac 12; \nu+\frac 12; \frac{x-1}{x+1}\bigg)
\eea
with support in $[-1, +1]$, and the ortho-normalization with weight factor $w (x | \nu) = (1 - x^2)^{\nu - 1/2}$,
\bea
\label{ORTHO}
\int_{-1}^{1}dx\,(1-x^2)^{\nu -\frac 12}\,C_n^\nu(x)C_m^\nu (x) =\frac{\pi 2^{1-2\nu}\Gamma(n+2\nu}{(n+\nu)\gamma^2(\nu)\Gamma(n+1)}\,\delta_{mn}
\eea

\subsection{Conformal Moments of Quark and Gluon GPDs}

\noindent{\bf Valence vector quark:}
\\
The conformal (Gegenbauer) moments of the valence vector quark GPDs, are directly related to the matrix elements 
of the spin-j conformal operators in QCD  (see Eq.4.243 in \cite{Belitsky:2005qn})
\bea
\label{QuarkConfMoments}
\mathbb{F}^q_{j(valence)} (\eta, t  ; \mu_0^2)
=\eta^{j-1}\,\int_{0}^1 dx \  C_{j-1}^{3/2}\left(\frac{x}{\eta}\right)
F_{valence}^q (x, \eta, t  ; \mu_0^2)
=
( p^+ )^{- (j-1) - 1}
\langle p_2 | \mathbb{O}^{qq}_{j} (0) | p_1 \rangle
\, ,\nonumber\\
\eea
for odd $j=1,3,...$. The valence and x-even combination  $$F_{valence}^{q}(x, \eta ; \mu^2)=F^{q}(x, \eta ; \mu^2)+F^{q}(-x, \eta ; \mu^2)$$  stands for   either $H_{valence}^{q}(x, \eta ; \mu^2)$ or $E_{valence}^{q}(x, \eta ; \mu^2)$.  
 \\
 \\
 \noindent{\bf Singlet vector quark:}
 \\
The conformal (Gegenbauer) moments for the vector-singlet combinations,  are also related to the spin-j conformal operators by (see Eq.4.243 in \cite{Belitsky:2005qn})
\bea
\label{QuarkConfMomentsSinglet}
\mathbb{F}^q_{j(singlet)} (\eta, t  ; \mu_0^2)
=\eta^{j-1}\,\int_{0}^1 dx \  C_{j-1}^{3/2}\left(\frac{x}{\eta}\right)
F_{singlet}^q (x, \eta, t  ; \mu_0^2)
=
( p^+ )^{- (j-1) - 1}
\langle p_2 | \mathbb{O}^{qq}_{j} (0) | p_1 \rangle
\,,\nonumber\\
\eea
for even $j=2,4,...$. The valence and x-odd combination
$$F_{singlet}^{q}(x, \eta ; \mu^2)=F^{q}(x, \eta ; \mu^2)-F^{q}(-x, \eta ; \mu^2)$$ is either $H_{singlet}^{q}(x, \eta ; \mu^2)$ or $E_{singlet}^{q}(x, \eta ; \mu^2)$.  
 \\
 \\
 \noindent{\bf Singlet axial quark:}
 \\
The conformal (Gegenbauer) moments of the singlet axial quark GPDs are related to the conformal operators in QCD via (straightforward generalization of Eq.4.243 in \cite{Belitsky:2005qn} for the axial GPDs)
\bea
\label{AxialQuarkConfMoments}
\widetilde{\mathbb{F}}^q_{j(singlet)} (\eta, t  ; \mu_0^2)
=\eta^{j-1}\,\int_{0}^1 dx \  C_{j-1}^{3/2}\left(\frac{x}{\eta}\right)
\widetilde{F}_{singlet}^q (x, \eta, t  ; \mu_0^2)
=
( p^+ )^{- (j-1) - 1}
\langle p_2 | \widetilde{\mathbb{O}}^{qq}_{j} (0) | p_1 \rangle
\, ,\nonumber\\
\eea
for odd $j=1,3,...$. Again, the x-even singlet combination
$$\widetilde{F}_{singlet}^{q}(x, \eta, t ; \mu^2)=\widetilde{F}^{q}(x, \eta, t ; \mu^2)+\widetilde{F}^{q}(-x, \eta, t ; \mu^2)$$ will refer to  either $\widetilde{H}_{singlet}^{q}(x, \eta, t ; \mu^2)$ or $\widetilde{E}_{singlet}^{q}(x, \eta, t ; \mu^2)$.  
  \\
 \\
 \noindent{\bf Valence axial  quark:}
 \\
The conformal (Gegenbauer) moments for the axial-valence combinations,  are tied to the spin-j twist-2 axial operator
(straightforward generalization of Eq.4.243 in \cite{Belitsky:2005qn} for the axial GPDs) 
\bea
\label{AxialQuarkConfMomentsValence}
\tilde{\mathbb{F}}^q_{j(valence)} (\eta, t  ; \mu_0^2)
=\eta^{j-1}\,\int_{0}^1 dx \  C_{j-1}^{3/2}\left(\frac{x}{\eta}\right)
\widetilde{F}_{valence}^q (x, \eta, t  ; \mu_0^2)
=
( p^+ )^{- (j-1) - 1}
\langle p_2 | \widetilde{\mathbb{O}}^{qq}_{j} (0) | p_1 \rangle
\, ,\nonumber\\
\eea
for even $j=2,4,...$. The x-odd valence axial combination
$$\widetilde{F}_{valence}^{q}(x, \eta, t ; \mu^2)=\widetilde{F}^{q}(x, \eta, t ; \mu^2)-\widetilde{F}^{q}(-x, \eta, t ; \mu^2)$$ 
will refer to  either $\widetilde{H}_{valence}^{q}(x, \eta, t ; \mu^2)$ or $\widetilde{E}_{valence}^{q}(x, \eta, t ; \mu^2)$.
  \\
 \\
 \noindent{\bf Gluons:}
 \\
For gluon operators, we have (see Eq.4.284 in \cite{Belitsky:2005qn})
\bea
\label{GluonConfMoments}
\mathbb{F}^g_{j} (\eta, t  ; \mu_0^2)
=2\eta^{j-2}\,\int_{0}^1 dx \  C_{j-2}^{5/2}\left(\frac{x}{\eta}\right)
F^g (x, \eta, t  ; \mu_0^2)
=
4( p^+ )^{- (j-1) - 1}
\langle p_2 | \mathbb{O}^{gg}_{j} (0) | p_1 \rangle
\, ,\nonumber\\
\eea
for even $j=2,4,...$.

\subsection{Reconstruction of Quark and Gluon GPDs from their Conformal Moments}~\label{INVERT}
The conformal (Gegenbauer) moments can be inverted using their orthonormality (\ref{ORTHO}), to give the quark and gluon GPDs at a given resolution  in explicit form.
We now give the inversion with explicit support in $|x|\leq \eta$, and briefly comment on the inversion in the full support $|x|\le 1$. 
  \\
 \\
 \noindent{\bf Valence vector quark:}
 \\
 The valence vector quark GPD,  follows by inversion(see Eq.4.242 in \cite{Belitsky:2005qn})
\bea
\label{ExpansionGPDeigenfunctionsH222}
H_{valence}^q (x, \eta, t ; \mu_0^2)
&=&
\frac{1}{\eta} \sum_{j = 1}^{\infty}
\frac{w \big( \frac{x}{\eta} | \frac{3}{2} \big)}{\eta^{j-1} \, N_{j-1} ( \frac{3}{2})}
C_{j-1}^{3/2}\left(\frac{x}{\eta}\right)
\mathbb{F}^q_{j(valence)} (\eta, t ; \mu_0^2)
\, ,\nonumber\\
&=&
\frac{1}{\eta}\times \left(1-\frac{x^2}{\eta^2}\right)\times \sum_{j = 1}^{\infty}\frac{1}{\eta^{j-1}}
\times\frac{1}{N_{j-1}(\frac{3}{2})}
\times C_{j-1}^{3/2}\left(\frac{x}{\eta}\right)
\times\mathbb{F}^q_{j(valence)} (\eta, t ; \mu_0^2)
\, ,\nonumber\\
\eea
for odd $j=1,3,...$, with 
\bea
N_{j} (\nu) = 2^{1 - 2 \nu}
\frac{
{\mit\Gamma}^2 (1/2) {\mit\Gamma} (2 \nu + j)
}{
{\mit\Gamma}^2 (\nu) {\mit\Gamma} (j + 1) (\nu + j)
}
\eea
  \\
 \\
 \noindent{\bf Singlet vector   quark:}
 \\
Similarly, the singlet vector quark GPD, is (see Eq.4.264 in \cite{Belitsky:2005qn}) 
\bea
\label{ExpansionGPDsinglet222}
H_{singlet}^q (x, \eta, t ; \mu_0^2)
&=&
\frac{1}{\eta} \sum_{j = 2}^{\infty}
\frac{w \big( \frac{x}{\eta} | \frac{3}{2} \big)}{\eta^{j-1} \, N_{j-1} ( \frac{3}{2})}
C_{j-1}^{3/2}\left(\frac{x}{\eta}\right)
\mathbb{F}^q_{j(singlet)} (\eta, t ; \mu_0^2)\nonumber\\
&=&\frac{1}{\eta}\times\left(1-\frac{x^2}{\eta^2}\right) \times\sum_{j = 2}^{\infty}
\frac{1}{\eta^{j-1}}\times\frac{1}{N_{j-1} ( \frac{3}{2})}
\times C_{j-1}^{3/2}\left(\frac{x}{\eta}\right)
\times\mathbb{F}^q_{j(singlet)} (\eta, t ; \mu_0^2)
\eea
for even $j=2,4,...$.
  \\
 \\
 \noindent{\bf Valence axial  quark:}
 \\
The  valence axial quark GPD, is
\bea
\label{ExpansionAxialGPDvalence222}
\widetilde{H}_{valence}^q (x, \eta, t ; \mu_0^2)
&=&
\frac{1}{\eta} \sum_{j = 2}^{\infty}
\frac{w \big( \frac{x}{\eta} | \frac{3}{2} \big)}{\eta^{j-1} \, N_{j-1} ( \frac{3}{2})}
C_{j-1}^{3/2}\left(\frac{x}{\eta}\right)
\widetilde{\mathbb{F}}^q_{j(valence)} (\eta, t ; \mu_0^2)\nonumber\\
&=&\frac{1}{\eta}\times\left(1-\frac{x^2}{\eta^2}\right) \times\sum_{j = 2}^{\infty}
\frac{1}{\eta^{j-1}}\times\frac{1}{N_{j-1} ( \frac{3}{2})}
\times C_{j-1}^{3/2}\left(\frac{x}{\eta}\right)
\times\widetilde{\mathbb{F}}^q_{j(valence)} (\eta, t ; \mu_0^2)
\eea
for even $j=2,4,...$.
  \\
 \\
 \noindent{\bf Singlet axial  quark:}
 \\
The  singlet axial quark GPD, is
\bea
\label{singletAxialGPDinput222}
\tilde{H}_{singlet}^q (x, \eta, t ; \mu_0^2)&=&\frac{1}{\eta} \sum_{j = 2}^{\infty}
\frac{w \big( \frac{x}{\eta} | \frac{3}{2} \big)}{\eta^{j-1} \, N_{j-1} ( \frac{3}{2})}C_{j-1}^{3/2}\left(\frac{x}{\eta}\right)\times\tilde{\mathbb{F}}^q_{j(singlet)} (\eta, t ; \mu_0^2\nonumber\\
&=&
\frac{1}{\eta}\times\left(1-\frac{x^2}{\eta^2}\right)\times\sum_{j=1}^{\infty}\frac{1}{N_{j-1}(\frac{3}{2})}\times\frac{1}{\eta^{j-1}} 
\times C_{j-1}^{3/2}\left(\frac{x}{\eta}\right)\times\tilde{\mathbb{F}}^q_{j(singlet)} (\eta, t ; \mu_0^2)
\, ,
\eea
for odd $j=1,3,...$.
  \\
 \\
 \noindent{\bf Gluons:}
 \\
The  gluon GPD is (see Eq.4.264 in \cite{Belitsky:2005qn})
\bea
\label{ExpansionGPDgluon22}
H^g (x, \eta, t ; \mu_0^2)
&=&
\frac{1}{\eta} \sum_{j = 2}^{\infty}
\frac{w \big( \frac{x}{\eta} | \frac{5}{2} \big)}{\eta^{j-2} \, N_{j-2} ( \frac{5}{2})}
C_{j-2}^{5/2}\left(\frac{x}{\eta}\right)
\mathbb{F}^g_{j} (\eta, t ; \mu_0^2)
\, ,\nonumber\\
&=&
\frac{1}{\eta}\times \left(1-\frac{x^2}{\eta^2}\right)^2 \times\sum_{j = 2}^{\infty}
\frac{1}{\eta^{j-2}}\times\frac{1}{N_{j-2} ( \frac{5}{2})}
\times C_{j-2}^{5/2}\left(\frac{x}{\eta}\right)
\times\mathbb{F}^g_{j} (\eta, t ; \mu_0^2)
\, ,\nonumber\\
\eea
for even $j=2,4,...$.

\section{RG evolution of quark and gluon GPDs in QCD}~\label{RGEVOLUTION}

\subsection{RG evolution of non-singlet (valence) quark GPDs}

The leading-order RG evolution of Gegenbauer moments of valence (non-singlet) vector quark GPDs is given by (see Eq.4.244 in \cite{Belitsky:2005qn})
\begin{equation}
\label{EvolutionConfMomGPDsValence1}
\mathbb{F}^q_{j(valence)} (\eta, t ; \mu^2)
=
\mathbb{F}^q_{j(valence)} (\eta, t  ; \mu_0^2)
\times\left(
\frac{\alpha_s (\mu_0^2)}{\alpha_s (\mu^2)}
\right)^{\gamma_{(0)j-1}^{qq;\text{V};\text{NS}}/\beta_0}
\end{equation}
for odd $j=1,3,...$, where the input valence quark GPDs $\mathbb{F}^q_{j(valence)} (\eta, t  ; \mu_0^2)$ are given in (\ref{valenceGgbrMoments2updown}), and the vector non-singlet anomalous dimensions are
\bea
\gamma^{qq; \text{V}}_{(0)j}
\!\!\!&=&\!\!\!
- C_F
\left(
- 4 \psi( j + 2 ) + 4 \psi(1) + \frac{2}{( j + 1 )( j + 2 )} + 3
\right)
\, .
\eea
The  Euler $\psi$-function is given 
in terms of the harmonic numbers 
\begin{equation}
\psi (n)
=
\frac{d}{dn} \ln {\mit\Gamma} (n)
=
\psi (1)
+
\sum_{k = 1}^{n - 1} \frac{1}{k}
\, .
\end{equation}
with the Euler constant $\gamma_{\scriptscriptstyle\rm E} = - \psi (1) \approx 0.577216$.

The leading and one-loop  RG evolution equation for the valence (non-singlet) vector quark GPDs is given by (see Eq.4.240 in \cite{Belitsky:2005qn}) 
\begin{equation}
\label{LOGenerEvEqforGPDs}
\frac{d}{d \ln \mu} H_{valence}^q (x, \eta, t ; \mu^2)
=
- \frac{\alpha_s (\mu^2)}{2 \pi} \int_{-1}^1 \frac{d y}{\eta}
k^{qq}_{(0)} \left( \frac{x}{\eta} , \frac{y}{\eta} \right)
H_{valence}^q (y, \eta, t ; \mu^2) + {\cal O} (\alpha_s^2)
\, ,
\end{equation}
$k^{qq}_{(0)} \left( {x}/{\eta} , {y}/{\eta} \right)$ is  the leading-order quark-quark generalized exclusive kernel
(see Eq.4.38 in \cite{Belitsky:2005qn} for its explicit form), with  Gegenbauer polynomials $C_{j-1}^{3/2} (x)$
as eigenfunctions,
\begin{equation}
\int_{-1}^1 \frac{d x}{\eta} \, C_{j-1}^{3/2}\left(\frac{x}{\eta}\right)
k^{qq}_{(0)} \left( \frac{x}{\eta} , \frac{y}{\eta} \right)
=
\gamma_{(0)j-1}^{qq;\text{V}} C_{j-1}^{3/2}\left(\frac{y}{\eta}\right)
\, .
\end{equation}
Conversely, since the exclusive kernel is diagonal in the Gegenbauer basis which is complete in $L^2[-1,1]$,
the explicit solution to (\ref{LOGenerEvEqforGPDs}) is (see Eq.4.242 in \cite{Belitsky:2005qn}) 
\begin{equation}
\label{2ExpansionGPDeigenfunctionsH1}
H_{valence}^q (x, \eta, t ; \mu^2)
=
\frac{1}{\eta} \sum_{j = 1}^\infty
\frac{w \big( \frac{x}{\eta} | \frac{3}{2} \big)}{\eta^{j-1} \, N_{j-1} ( \frac{3}{2})}
C_{j-1}^{3/2}\left(\frac{x}{\eta}\right)
\mathbb{F}^q_{j(valence)} (\eta, t ; \mu^2)
\, ,
\end{equation}
for odd $j=1,3,...$. Here the weight and normalization factors are
$$
w (x | \nu) = (1 - x^2)^{\nu - 1/2}
\, , \qquad
N_{j-1} (\nu) = 2^{1 - 2 \nu}
\frac{
{\mit\Gamma}^2 (1/2) {\mit\Gamma} (2 \nu + (j-1))
}{
{\mit\Gamma}^2 (\nu) {\mit\Gamma} ((j-1) + 1) (\nu + (j-1))
}
\, .
$$

\subsection{RG evolution of singlet quark and gluon GPDs}
The combined singlet vector quark and gluon GPDs
\begin{eqnarray}
\label{def-SPD-sing}
\mathbf{F} (x, \eta, t ;\mu^2)
\equiv
\left(
{
\sum_q F_{singlet}^q (x, \eta, t ;\mu^2) 
\atop
F^g (x, \eta, t ;\mu^2)
}
\right)
\, ,
\end{eqnarray}
evolve in leading order as (see Eq.4.264 in \cite{Belitsky:2005qn}) 
\begin{equation}
\label{JacobiMom}
\mathbf{F} (x,\eta, t; \mu^2 )
=
\frac{1}{\eta}\sum_{j = 2}^{\infty} {\bf E}_{j}  \left(\frac{x}{\eta}\right) \mathbf{F}_j (\eta , t; \mu^2)
\, ,
\end{equation}
for even $j=2,4,...$, with the matrix valued coeffcients 
\begin{equation}
\label{E-matrix}
{\mbox{\bf E}}_{j} (x)
=
\left(
\begin{array}{cc}
E_{j-1} (x | \frac{3}{2}) & 0 \\
0 & E_{j-2} ( x | \frac{5}{2} )
\end{array}
\right)
\, ,\qquad\qquad 
E_{j} ( x | \nu )
=
\frac{w \left(x | \nu \right)}{\eta^j N_j (\nu)}
C_j^\nu(x)
\, .
\end{equation}
Note that we do not need to re-expand the GPDs using additional orthogonal polynomials, since our input GPDs are non-zero only for $-\eta<x<\eta$.  
(\ref{JacobiMom}) is invertible
\begin{equation}
\label{SingletGegenMomentsGPDs}
\mathbf{F}_j (\eta, t ; \mu^2)
=
\int_{0}^1 dx \,
\left(
\begin{array}{cc}
\eta^{j-1} \, C_{j-1}^{3/2} \left( x / \eta \right) & 0 \\
0 & \eta^{j - 2} \, C_{j - 2}^{5/2} \left( x / \eta \right)
\end{array}
\right)
\,
\mathbf{F} ( x, \eta, t ; \mu^2 )
\, ,
\end{equation}
thanks to the ortogonality and completeness of the Gegenbauer polynomials. 

(\ref{SingletGegenMomentsGPDs}) evolves as (see Eq.4.271 in \cite{Belitsky:2005qn})
\begin{equation}\label{SingletGegenMomentsGPDs2}
\mathbf{F}_j (\eta, t ; \mu^2)
=
{\mathbf{\cal E}}_{j-1} (\mu^2, \mu_0^2)
\mathbf{F}_j (\eta, t ; \mu^2_0)
\, ,
\end{equation}
with the evolution operator 
\begin{eqnarray}
\label{SingletDiagonalEvOp}
\mathbf{\cal E}_j \left( \mu^2 , \mu_0^2 \right)
\!\!\!&=&\!\!\!
{P}^+_j
\left(
\frac{\alpha_s (\mu^2_0)}{\alpha_s (\mu^2)}
\right)^{\gamma^+_j / \beta_0}
+
{P}^-_j
\left(
\frac{\alpha_s (\mu^2_0)}{\alpha_s (\mu^2)}
\right)^{\gamma^-_j / \beta_0}
\, ,
\end{eqnarray}
given in terms of the projection operators
\begin{equation}
{P}^\pm_j
= \frac{\pm 1}{\gamma^+_j - \gamma^-_j}
\left(
{\gamma}_{(0)j} - \gamma^\mp_j \,
\right)
\, ,
\end{equation}
constructed from the eigenvalues of the leading-order anomalous dimension matrix
\begin{equation}
\label{TwoByTwoMatrixAnomalousDim}
{\gamma}_{(0)j}
=
\left(
\begin{array}{cc}
\gamma^{qq}_{(0)j} & \gamma^{qg}_{(0)j} \\
\gamma^{gq}_{(0)j} & \gamma^{gg}_{(0)j}
\end{array}
\right)
\, ,
\end{equation}
as
\begin{equation}
\label{LOsingletEigenvalues}
\gamma^{\pm}_j
= \frac{1}{2}
\left(
\gamma^{qq}_{(0)j}
+
\gamma^{gg}_{(0)j}
\pm
\sqrt{\left(
\gamma^{qq}_{(0)j}
-
\gamma^{gg}_{(0)j}
\right)^2
+
4
\gamma^{gq}_{(0)j}
\gamma^{qg}_{(0)j}
}
\right)
\, .
\end{equation}
The projection operators have the properties
\begin{equation}
( {P}^\pm_j )^2 = {P}^\pm_j
\, , \qquad
{P}^+_j {P}^-_j = 0
\, , \qquad
{P}^+_j + {P}^-_j = {1}
\, ,
\end{equation}
and
\begin{equation}
{\gamma}_{(0)j} = \gamma_j^+ {P}^+_j + \gamma_j^- {P}^-_j
\, .
\end{equation}

Note that, at leading-order, the evolution operator satisfies the
equation (see Eq.4.272 in \cite{Belitsky:2005qn})
\begin{equation}
\label{DiagEvolOperMatrix}
\frac{d}{d\ln \alpha_s (\mu^2)}
\mathbf{\cal E}_j \left( \mu^2, \mu_0^2 \right)
= - \frac{1}{\beta_0}
 {\gamma}_{(0)j}
\mathbf{\cal E}_j \left( \mu^2, \mu_0^2 \right)
\, ,
\end{equation}
with the boundary condition
$$
\mathbf{\cal E}_j \left( \mu_0^2 , \mu_0^2 \right)
=
{1}
=
\left( {1 \ \, 0 \atop 0 \ \, 1} \right)
\, .
$$

In addition, the leading-order diagonal anomalous dimensions are given by (see Eq.4.152-4.159 in \cite{Belitsky:2005qn})
\begin{eqnarray}
\label{even-anomalous-dimensions}
\gamma^{qq; V}_{(0)j}
\!\!\!&=&\!\!\!
- C_F
\left(
- 4 \psi( j + 2 ) + 4 \psi(1) + \frac{2}{( j + 1 )( j + 2 )} + 3
\right)
\, , \\
\gamma^{qg; V}_{(0)j}
\!\!\!&=&\!\!\!
- 24 N_f T_F
\frac{j^2 + 3 j + 4}{j( j + 1 )( j + 2 )( j + 3 )}
\, , \\
\gamma^{gq; V}_{(0)j}
\!\!\!&=&\!\!\!
- C_F
\frac{j^2 + 3 j + 4}{3( j + 1 )( j + 2 )}
\, , \\
\label{gg-V-ad}
\gamma^{gg; V}_{(0)j}
\!\!\!&=&\!\!\!
- C_A \left(
- 4 \psi( j + 2 ) + 4 \psi(1)
+
8 \frac{j^2 + 3 j + 3}{j( j + 1 )( j + 2 )( j + 3 )}
- \frac{\beta_0}{C_A}
\right)
\, ,
\end{eqnarray}
and
\begin{eqnarray}
\label{odd-anomalous-dimensions}
\gamma^{qq; A}_{(0)j}
\!\!\!&=&\!\!\!
- C_F
\left(
- 4 \psi( j + 2 ) + 4 \psi(1) + \frac{2}{( j + 1 )( j + 2 )} + 3
\right)
\, , \\
\gamma^{qg; A}_{(0)j}
\!\!\!&=&\!\!\!
- 24 N_f T_F\frac{1}{( j + 1 )( j + 2 )}
\, , \\
\gamma^{gq; A}_{(0)j}
\!\!\!&=&\!\!\!
- C_F \frac{j ( j + 3 )}{3( j + 1 )( j + 2 )}
\, , \\
\gamma^{gg; A}_{(0)j}
\!\!\!&=&\!\!\!
- C_A \left(
- 4 \psi( j + 2 ) + 4 \psi(1)
+
\frac{8}{( j + 1 )( j + 2 )}
- \frac{\beta_0}{C_A}
\right)
\, .
\label{odd-anomalous-dimensions-last}
\end{eqnarray}

We can rewrite the RG evolution of the singlet Gegenbauer moments (\ref{SingletGegenMomentsGPDs2}) more explicitly in terms of the projection operators as 
\begin{equation}\label{SingletGegenMomentsGPDs3}
\mathbf{F}_j (\eta, t ; \mu^2)
=
{P}^+_{j-1}
\mathbf{F}_j (\eta, t ; \mu_0^2)\times\left(
\frac{\alpha_s (\mu^2_0)}{\alpha_s (\mu^2)}
\right)^{\gamma^+_{j-1} / \beta_0}
+
{P}^-_{j-1}
\mathbf{F}_j (\eta, t ; \mu_0^2)\times\left(
\frac{\alpha_s (\mu^2_0)}{\alpha_s (\mu^2)}
\right)^{\gamma^-_{j-1} / \beta_0}
\, .
\end{equation}
And, using
\begin{equation}
{P}^\pm_{j-1} {F}_j(\eta, t ; \mu_0^2)
=
\frac{\pm 2}{\gamma^+_{j-1} - \gamma^-_{j-1}}
\left(
\begin{array}{c}
\gamma^{qq}_{(0)j-1} - \gamma^\mp_{j-1}
\\
\gamma^{gq}_{(0)j-1}
\end{array}
\right)
F^\pm_j(\eta, t ; \mu_0^2)
\, ,
\end{equation}
with
\begin{equation}\label{cqcg}
F^{\pm}_j(\eta, t ; \mu_0^2) = 
\frac{1}{4}
\frac{\gamma^{qg}_{(0)j-1}}{\gamma^{qq}_{(0)j-1} - \gamma^{\mp}_{j-1}} \mathbb{F}^g_j(\eta, t ; \mu_0^2)
\, ,
\end{equation}
we find the the evolved singlet quark, and gluon GPDs in matrix form as
\bea
\label{SingletGegenMomentsGPDs4}
&&\mathbf{F}_j (\eta, t ; \mu^2)
\nonumber\\
&=&\frac{2}{\gamma^+_{j-1} - \gamma^-_{j-1}}
\left(
\begin{array}{c}
\gamma^{qq}_{(0)j-1} - \gamma^-_{j-1}
\\
\gamma^{gq}_{(0)j-1}
\end{array}
\right)
F^+_j(\eta, t;\mu_0^2)\times\left(
\frac{\alpha_s (\mu^2_0)}{\alpha_s (\mu^2)}
\right)^{\gamma^+_{j-1} / \beta_0}
\nonumber\\
&+& \frac{-2}{\gamma^+_{j-1} - \gamma^-_{j-1}}
\left(
\begin{array}{c}
\gamma^{qq}_{(0)j-1} - \gamma^+_{j-1}
\\
\gamma^{gq}_{(0)j-1}
\end{array}
\right)
F^-_j(\eta, t;\mu_0^2)\times\left(
\frac{\alpha_s (\mu^2_0)}{\alpha_s (\mu^2)}
\right)^{\gamma^-_{j-1} / \beta_0}
\,,\nonumber\\
&=&
\left(
\begin{array}{c}
\frac{2\left(\gamma^{qq}_{(0)j-1} - \gamma^-_{j-1}\right)}{\gamma^+_{j-1} - \gamma^-_{j-1}}
\\
\frac{2\left(\gamma^{gq}_{(0)j-1}\right)}{\gamma^+_{j-1} - \gamma^-_{j-1}}
\end{array}
\right)
F^+_j(\eta, t;\mu_0^2)\left(
\frac{\alpha_s (\mu^2_0)}{\alpha_s (\mu^2)}
\right)^{\gamma^+_{j-1} / \beta_0}
+
\left(
\begin{array}{c}
\frac{-2\left(\gamma^{qq}_{(0)j-1} - \gamma^+_{j-1}\right)}{\gamma^+_{j-1} - \gamma^-_{j-1}}
\\
\frac{-2\left(\gamma^{gq}_{(0)j-1}\right)}{\gamma^+_{j-1} - \gamma^-_{j-1}}
\end{array}
\right)
F^-_j(\eta, t;\mu_0^2)\left(
\frac{\alpha_s (\mu^2_0)}{\alpha_s (\mu^2)}
\right)^{\gamma^-_{j-1} / \beta_0}\,,\nonumber\\
&=&
\left(
\begin{array}{c}
\frac{2\left(\gamma^{qq}_{(0)j-1} - \gamma^-_{j-1}\right)}{\gamma^+_{j-1} - \gamma^-_{j-1}}F^+_j(\eta, t;\mu_0^2)\left(
\frac{\alpha_s (\mu^2_0)}{\alpha_s (\mu^2)}
\right)^{\gamma^+_{j-1} / \beta_0}
\\
\frac{2\left(\gamma^{gq}_{(0)j-1}\right)}{\gamma^+_{j-1} - \gamma^-_{j-1}}F^+_j(\eta, t;\mu_0^2)\left(
\frac{\alpha_s (\mu^2_0)}{\alpha_s (\mu^2)}
\right)^{\gamma^+_{j-1} / \beta_0}
\end{array}
\right)
+\left(
\begin{array}{c}
\frac{-2\left(\gamma^{qq}_{(0)j-1} - \gamma^+_{j-1}\right)}{\gamma^+_{j-1} - \gamma^-_{j-1}}F^-_j(\eta, t;\mu_0^2)\left(
\frac{\alpha_s (\mu^2_0)}{\alpha_s (\mu^2)}
\right)^{\gamma^-_{j-1} / \beta_0}
\\
\frac{-2\left(\gamma^{gq}_{(0)j-1}\right)}{\gamma^+_{j-1} - \gamma^-_{j-1}}F^-_j(\eta, t;\mu_0^2)\left(
\frac{\alpha_s (\mu^2_0)}{\alpha_s (\mu^2)}
\right)^{\gamma^-_{j-1} / \beta_0}
\end{array}
\right)
\,,\nonumber\\
&=&\left(
\begin{array}{c}
\frac{2\left(\gamma^{qq}_{(0)j-1} - \gamma^-_{j-1}\right)}{\gamma^+_{j-1} - \gamma^-_{j-1}}F^+_j(\eta, t;\mu_0^2)\left(
\frac{\alpha_s (\mu^2_0)}{\alpha_s (\mu^2)}
\right)^{\gamma^+_{j-1} / \beta_0}+\frac{-2\left(\gamma^{qq}_{(0)j-1} - \gamma^+_{j-1}\right)}{\gamma^+_{j-1} - \gamma^-_{j-1}} F^-_j(\eta, t;\mu_0^2)\left(
\frac{\alpha_s (\mu^2_0)}{\alpha_s (\mu^2)}
\right)^{\gamma^-_{j-1} / \beta_0}
\\
\frac{2\left(\gamma^{gq}_{(0)j-1}\right)}{\gamma^+_{j-1} - \gamma^-_{j-1}} F^+_j(\eta, t;\mu_0^2)\left(
\frac{\alpha_s (\mu^2_0)}{\alpha_s (\mu^2)}
\right)^{\gamma^+_{j-1} / \beta_0}+\frac{-2\left(\gamma^{gq}_{(0)j-1}\right)}{\gamma^+_{j-1} - \gamma^-_{j-1}} F^-_j(\eta, t;\mu_0^2)\left(
\frac{\alpha_s (\mu^2_0)}{\alpha_s (\mu^2)}
\right)^{\gamma^-_{j-1} / \beta_0}
\end{array}
\right)\,.\nonumber\\
\eea

Therefore, we explicitly find
\bea
\label{SingletGegenMomentsGPDs5}
\mathbf{F}_j (\eta , t; \mu^2)
&=&
\left(
\begin{array}{c}
\sum_q \mathbb{F}^q_{j(singlet)}(\eta , t; \mu^2)
\\
\mathbb{F}^g_j(\eta , t; \mu^2)
\end{array}
\right)\,,
\eea
where
\bea\label{evolvedGgnbrMomentsSingletQuark1}
&&\sum_q \mathbb{F}^q_{j(singlet)}(\eta, t ; \mu^2)\nonumber\\
&\propto&
\frac{2\left(\gamma^{qq}_{(0)j-1} - \gamma^-_{j-1}\right)}{\gamma^+_{j-1} - \gamma^-_{j-1}}\times F^+_j(\eta, t;\mu_0^2)\times\left(
\frac{\alpha_s (\mu^2_0)}{\alpha_s (\mu^2)}
\right)^{\gamma^+_{j-1} / \beta_0} \nonumber\\
&+& \frac{-2\left(\gamma^{qq}_{(0)j-1} - \gamma^+_{j-1}\right)}{\gamma^+_{j-1} - \gamma^-_{j-1}}\times F^-_j(\eta, t;\mu_0^2)\times\left(
\frac{\alpha_s (\mu^2_0)}{\alpha_s (\mu^2)}
\right)^{\gamma^-_{j-1} / \beta_0}\,,
\eea
\bea\label{evolvedGgnbrMomentsGluon1}
&&\mathbb{F}^g_j(\eta, t ; \mu^2)\nonumber\\
&\propto&\frac{2\left(\gamma^{gq}_{(0)j-1}\right)}{\gamma^+_{j-1} - \gamma^-_{j-1}}\times F^+_j(\eta, t;\mu_0^2)\times\left(
\frac{\alpha_s (\mu^2_0)}{\alpha_s (\mu^2)}
\right)^{\gamma^+_{j-1} / \beta_0} \nonumber\\ 
&+& \frac{-2\left(\gamma^{gq}_{(0)j-1}\right)}{\gamma^+_{j-1} - \gamma^-_{j-1}}\times F^-_j(\eta, t;\mu_0^2)\times\left(
\frac{\alpha_s (\mu^2_0)}{\alpha_s (\mu^2)}
\right)^{\gamma^-_{j-1} / \beta_0}\,,
\eea
for even $j=2,4,...$. Note that we fix the proportionality constants in the above equations (\ref{evolvedGgnbrMomentsSingletQuark1}) and (\ref{evolvedGgnbrMomentsGluon1}) by demanding that $\sum_q \mathbb{F}^q_{j(singlet)}(\eta, t ; \mu^2=\mu_0^2)=0$, and $\mathbb{F}^g_j(\eta, t ; \mu^2=\mu_0^2)=\mathbb{F}^g_j(\eta, t ; \mu_0^2)$, i.e.,  
\bea\label{evolvedGgnbrMomentsSingletQuark2}
&&\sum_q \mathbb{F}^q_{j(singlet)}(\eta, t ; \mu^2)\nonumber\\
&=&
\frac{1}{3}\times\frac{2\left(\gamma^{qq}_{(0)j-1} - \gamma^-_{j-1}\right)}{\gamma^+_{j-1} - \gamma^-_{j-1}}\times \mathbb{F}^+_j(\eta, t;\mu_0^2)\times\left(
\frac{\alpha_s (\mu^2_0)}{\alpha_s (\mu^2)}
\right)^{\gamma^+_{j-1} / \beta_0} \nonumber\\
&+& \frac{1}{3}\times\frac{-2\left(\gamma^{qq}_{(0)j-1} - \gamma^+_{j-1}\right)}{\gamma^+_{j-1} - \gamma^-_{j-1}}\times \mathbb{F}^-_j(\eta, t;\mu_0^2)\times\left(
\frac{\alpha_s (\mu^2_0)}{\alpha_s (\mu^2)}
\right)^{\gamma^-_{j-1} / \beta_0}\,,
\eea
\bea\label{evolvedGgnbrMomentsGluon2}
&&\mathbb{F}^g_j(\eta, t ; \mu^2)\nonumber\\
&=&2\times\frac{2\left(\gamma^{gq}_{(0)j-1}\right)}{\gamma^+_{j-1} - \gamma^-_{j-1}}\times \mathbb{F}^+_j(\eta, t;\mu_0^2)\times\left(
\frac{\alpha_s (\mu^2_0)}{\alpha_s (\mu^2)}
\right)^{\gamma^+_{j-1} / \beta_0} \nonumber\\ 
&+& 2\times\frac{-2\left(\gamma^{gq}_{(0)j-1}\right)}{\gamma^+_{j-1} - \gamma^-_{j-1}}\times \mathbb{F}^-_j(\eta, t;\mu_0^2)\times\left(
\frac{\alpha_s (\mu^2_0)}{\alpha_s (\mu^2)}
\right)^{\gamma^-_{j-1} / \beta_0}\,.
\eea

Finally, in terms of the
evolved conformal (Gegenbauer) moments (\ref{evolvedGgnbrMomentsSingletQuark}-\ref{evolvedGgnbrMomentsGluon}),  the evolved singlet quark and gluon GPDs are given by
\bea
\sum_q F_{singlet}^q (x, \eta, t;\mu^2) &=& 
\frac{1}{\eta}\sum_{j=2}^{\infty}\frac{w \left(\frac{x}{\eta} | \frac{3}{2} \right)}{\eta^{j-1} N_{j-1} \left(\frac{3}{2}\right)}
C_{j-1}^{3/2}\left(\frac{x}{\eta}\right)\times\sum_q \mathbb{F}^q_{j(singlet)}(\eta, t ; \mu^2)\,,
\nonumber\\
&=&\frac{1}{\eta}\times
\left(1-\frac{x^2}{\eta^2}\right)\times\sum_{j=2}^{\infty}\frac{1}{\eta^{j-1}}\times\frac{1}{N_{j-1}\left(\frac{3}{2}\right)}
\times C_{j-1}^{3/2}\left(\frac{x}{\eta}\right)\times\sum_q \mathbb{F}^q_{j(singlet)}(\eta, t ; \mu^2)\,,\nonumber\\
\eea
and
\bea
F^g (x, \eta, t;\mu^2)&=&\frac{1}{\eta}\sum_{j=2}^{\infty}\frac{w \left(\frac{x}{\eta} | \frac{5}{2} \right)}{\eta^{j-2} N_{j-2} \left(\frac{5}{2}\right)}
C_{j-2}^{5/2}\left(\frac{x}{\eta}\right)\times\mathbb{F}^g_{j}(\eta, t ; \mu^2)\,,\nonumber\\
&=&\frac{1}{\eta}\times \left(1-\frac{x^2}{\eta^2}\right)^2\times \sum_{j=2}^{\infty}\frac{1}{\eta^{j-2}}\times \frac{1}{N_{j-2} \left(\frac{5}{2}\right)}
\times C_{j-2}^{5/2}\left(\frac{x}{\eta}\right)\times\mathbb{F}^g_{j}(\eta, t ; \mu^2)\,.
\eea

\subsection{Evolution of quark and gluon GPDs to the asymptotic regime $\mu\rightarrow\infty$}~\label{ASYMPTOTICX}
For completeness,
we summarise the asymptotic forms of the GPDs at
infinite resolution, which are fixed by the free parton model, and are dominated by the leading spin-1 or spin-2 Gegenbauer moments. 
More explicitly, the asymptotic GPDs for non-singlet (valence) quark GPDs, are (see Eq.4.245 in \cite{Belitsky:2005qn}) 
\bea
\label{AsyQuarkGPDns}
H_{valence}^q (x, \eta, t; \mu^2 \to \infty)
&=&
\frac{3}{4}\times\frac{1}{\eta}
\times \left( 1 - \frac{x^2}{\eta^2} \right)\times F^q_1(t) \times \theta (\eta - |x|) \,,\nonumber\\
E_{valence}^q (x, \eta, t; \mu^2 \to \infty)
&=&
\frac{3}{4}\times\frac{1}{\eta}
\times \left( 1 - \frac{x^2}{\eta^2} \right)\times F^q_2(t) \times \theta (\eta - |x|) \,.\nonumber\\
\eea
For singlet quark GPDs, they are (see Eq.4.289 and Eq.4.290 in \cite{Belitsky:2005qn}) 
\begin{eqnarray}
\label{AsyFormSingletQuarkGPDs}
\sum_q H_{singlet}^q(x, \eta, t; \mu^2 \to \infty)
\!\!\!&=&\!\!\!
\frac{15}{4} \frac{N_f}{4 C_F + N_f} \times\frac{1}{\eta}\times \left( 1 - \frac{x^2}{\eta^2} \right)\times\frac{1}{\eta}\times \frac{x}{\eta}\times\sum_q
\bigg\{
A_q(t) + \eta^2 D_q(t)
\bigg\}\times\theta (\eta - |x|) 
\, ,\nonumber\\
\sum_q E_{singlet}^q(x, \eta, t; \mu^2 \to \infty)
\!\!\!&=&\!\!\!
\frac{15}{4} \frac{N_f}{4 C_F + N_f} \times\frac{1}{\eta}\times \left( 1 - \frac{x^2}{\eta^2} \right)\times\frac{1}{\eta}\times \frac{x}{\eta}\times\sum_q
\bigg\{
E_q(t) - \eta^2 D_q(t)
\bigg\}\times\theta (\eta - |x|) 
\, .\nonumber\\
\end{eqnarray}
And, for gluon GPDs, they are (see Eq.4.291 and Eq.4.292 in \cite{Belitsky:2005qn})  
\begin{eqnarray}
\label{AsyFormGluonGPDs}
H^g(x, \eta, t; \mu^2 \to \infty)
\!\!\!&=&\!\!\!
\frac{15}{8} \frac{4C_F}{4 C_F + N_f} \times\frac{1}{\eta}\times
\left ( 1 - \frac{x^2}{\eta^2} \right)^2 \times\bigg\{
A_g(t) + \eta^2 D_g(t)
\bigg\}\times\theta (\eta - |x|)\, ,\nonumber\\
E^g(x, \eta, t; \mu^2 \to \infty)
\!\!\!&=&\!\!\!
\frac{15}{8} \frac{4C_F}{4 C_F + N_f} \times\frac{1}{\eta}
\times \left ( 1 - \frac{x^2}{\eta^2} \right)^2 \times\bigg\{
E_g(t) - \eta^2 D_g(t)
\bigg\}\times\theta (\eta - |x|)\,.\nonumber\\
\end{eqnarray}

\subsection{Matching $\lambda$ in soft-wall holographic QCD to $\lambda_s(\mu_0^2)$ in pQCD}~\label{sec_MATCH}
In our previous work \cite{Mamo:2019mka}, we have fixed the $^\prime$t Hooft coupling $\lambda=11.243$ in the soft-wall holographic QCD,
by using the photoproduction data for $J/\Psi$. Now, we  suggest to map this fixed and large  $^\prime$t Hooft coupling in holographic QCD, to the running 
$^\prime$t Hooft coupling in pQCD, which will allow us to determine the value of the initial scale $\mu_0$. More specifically, the running 
 $^\prime$t Hooft coupling  $\lambda_{s}(\mu_0^2)=g_s^2(\mu_0^2)N_c=\alpha_{s}(\mu_0^2)\times 4\pi\times N_c$ at $\mu=\mu_0$, maps onto
\be\label{coupling-mapping}
\lambda_s(\mu_0^2)=\lambda=11.243\,. 
\ee
which translates to
\be
\alpha_{s}(\mu_0^2)=\frac{\lambda_s(\mu_0^2)}{4\pi N_c}=\frac{\lambda}{4\pi N_c}=\frac{11.243}{4\pi N_c}=0.298\,. 
\ee
This result is remarkably close to the value of $\frac 13$ commonly used to initialize pQCD evolutions.
This is welcome, since our holographic description is set to describe the non-perturbative physics at 
$\mu_0$. Therefore, using $\alpha_{s}(\mu_0^2)=0.298$ in
the two-loop exact transcendental
equation, for the QCD running coupling constant $\alpha_{s}(\mu^2)$, \cite{Belitsky:2005qn} 
\begin{equation}
\label{ExactTwoLoopCoupling}
- \beta_0 \ln \frac{\mu^2}{\Lambda^2_{\overline{\rm MS}}}
= \frac{4 \pi}{\alpha_s (\mu^2)}
- \frac{\beta_1}{\beta_0}
\ln \left( - \frac{4 \pi}{\beta_0 \alpha_s (\mu^2)}
- \frac{\beta_1}{\beta_0^2} \right)
\, ,
\end{equation}
with 
\begin{eqnarray}
\label{BetaFunction}
\beta_0
\!\!\!&=&\!\!\!
\frac{4}{3} T_F N_f - \frac{11}{3} C_A
\, , \\
\beta_1
\!\!\!&=&\!\!\!
\frac{10}{3} C_A N_f + 2 C_F N_f - \frac{34}{3} C_A^2
\, ,
\end{eqnarray}   
and 
\begin{equation}
T_F = \frac{1}{2}
\, , \qquad
C_F = \frac{N_c^2 - 1}{2 N_c}
\, , \qquad
C_A = N_c
\  ,
\end{equation}
we find the input scale $\mu_0=1.808~\text{GeV}$ (or $\mu_0^2=3.269~\text{GeV}^2$ ), using $N_c=N_f=3$, and $\Lambda_{\overline{\rm MS}}=0.34~\text{GeV}$.

\section{Exclusive electroproduction of mesons and photon with twist-2 quark and gluon GPDs in QCD}~\label{PQCDAMPLITUDES}
In this section of the appendix, we review the various amplitudes for electroproduction of mesons and photon with twist-2 quark and gluon GPDs in QCD.

\subsection{Electroproduction of vector mesons with twist-2 gluon GPD in QCD}

In pQCD, the twist-2 gluon GPD drives the near-forward electromagnetic production of heavy vector mesons such as 
charmonia and bottomonia, in the Regge limit, as illustrated in Fig.~\ref{fig_HVM2} (left).
More specifically, the gluon GPDs contribution $F^g(x,\eta, t;\mu^2)$ to the longitudinal 
vector meson electroproduction amplitude is given by (see Eq.282 in \cite{Diehl:2003ny})

\begin{eqnarray}
  \label{meson-2}
&&\epsilon_{L\mu}\epsilon_{L\nu}^{\prime*} T_{t(\text{with gluon GPDs})}=\nonumber\\
\mathcal{A}_{\rm gluon}^{LL}&=& e\times\frac{1}{8}\times\frac{1}{N_c^2}\times g_s^2(\mu^2)N_c \times \frac{1}{Q}\nonumber\\
&&\times \left[\int_0^1 dz\, \sum_{q} e_q \frac{\Phi^q(z)}{z(1-z)}\right] \times
\int_{-1}^1 dx\, \frac{F^g(x,\eta, t;\mu^2)}{x}  
\left[ \frac{1}{\xi - x -i\epsilon} -
                        \frac{1}{\xi + x -i\epsilon} \right]\nonumber\\
\end{eqnarray}
In leading order in $1/Q$ and $\alpha_s$, the meson structure function enters as a constant,  
through the integral of the pertinent quark flavor combination in the meson DA $\Phi^q(z)$.
On the light front, 
$$h^+=\bar u(p_2)\gamma^{+}u(p_1)\approx \frac{p^+}{2m_N}\times \bar u(p_2)u(p_1),$$
and ignoring the Pauli contribution $E^g(x,\eta, t;\mu^2)$ to $F^g(x,\eta, t;\mu^2)$ for $Q\gg\sqrt{-t}$, we have
\bea
F^g(x,\eta, t;\mu^2)\approx\frac{h^{+}}{p^{+}}H^g(x,\eta, t;\mu^2)
= H^g(x,\eta, t;\mu^2)\times \frac{1}{2m_N}\times \bar u(p_2)u(p_1)\,,
\eea
which simplifies to
\bea\label{gluonFinal2211}
\mathcal{A}_{\rm gluon}^{LL}(\xi,t,\mu^2)=&& e\times\frac{1}{4}\times\frac{1}{N_c^2}\times \frac{1}{Q}
\times \lambda_{s}(\mu^2)\nonumber\\
&&\times\left[\int_0^1 dz\, \sum_{q} e_q \frac{\Phi^q(z)}{z(1-z)}\right]\times \frac{1}{\xi^2} \times
\int_{0}^1 dx\, \frac{H^g(x,\eta, t;\mu^2)}{(1-\frac{x^2}{\xi^2})}  
\times \frac{1}{m_N}
\times\bar u(p_2)u(p_1)\,,\nonumber\\
\eea
where we have defined the running 't Hooft coupling constant in QCD as $\lambda_{s}(\mu^2)\equiv g_s^2(\mu^2)N_c$. 

The gluon GPD at $\mu=\mu_0$ can be written in terms of their Gegenbauer (conformal) moments $\mathbb{F}^g_{j-2} (\eta, t ; \mu_0^2)$ as 
\bea
\label{ExpansionGPDgluon222111}
H^g (x, \eta, t ; \mu_0^2)
&=&
\frac{1}{\eta} \sum_{j = 2}^{\infty}
\frac{w \big( \frac{x}{\eta} | \frac{5}{2} \big)}{\eta^{j-2} \, N_{j-2} ( \frac{5}{2})}
C_{j-2}^{5/2}\left(\frac{x}{\eta}\right)
\mathbb{F}^g_{j} (\eta, t ; \mu_0^2)\times\theta(\eta - |x|)
\, ,\nonumber\\
&=&
\frac{1}{\eta}\times \left(1-\frac{x^2}{\eta^2}\right)^2 \times\sum_{j = 2}^{\infty}
\frac{1}{\eta^{j-2}}\times\frac{1}{N_{j-2} ( \frac{5}{2})}
\times C_{j-2}^{5/2}\left(\frac{x}{\eta}\right)
\times\mathbb{F}^g_{j} (\eta, t ; \mu_0^2)\times\theta(\eta - |x|)
\, ,\nonumber\\
\eea
for even $j=2,4,...$, where we used the weight and normalization factors as
$$
w (x | \nu) = (1 - x^2)^{\nu - 1/2}
\, , \qquad
N_{j-2} (\nu) = 2^{1 - 2 \nu}
\frac{
{\mit\Gamma}^2 (1/2) {\mit\Gamma} (2 \nu + (j-2))
}{
{\mit\Gamma}^2 (\nu) {\mit\Gamma} ((j-2) + 1) (\nu + (j-2))
}
\,.
$$
And, using the conformal expansion of gluon GPDs (\ref{ExpansionGPDgluon222111}) in the amplitude  (\ref{gluonFinal2211}), we find
\bea\label{gluonFinal2}
\mathcal{A}_{\rm gluon}^{LL}(\xi,t,\mu_0^2)&=& e\times\frac{1}{4}\times\frac{1}{N_c^2}\times \frac{1}{Q_0}
\times \lambda_{s}(\mu^2)\times\left[\int_0^1 dz\, \sum_{q} e_q \frac{\Phi^q(z)}{z(1-z)}\right]\times \Bigg[\frac{1}{\xi^2} \times
\int_{0}^1 dx\, \frac{H^g(x,\eta, t;\mu_0^2)}{(1-\frac{x^2}{\xi^2})}\nonumber\\  
&+&i\pi H^g(\xi,\eta, t;\mu_0^2)+i\pi H^g(-\xi,\eta, t;\mu_0^2)\Bigg]\times \frac{1}{m_N}
\times\bar u(p_2)u(p_1)\,,\nonumber\\
&=&e\times\frac{1}{4}\times\frac{1}{N_c^2}\times \frac{1}{Q_0}
\times \lambda_{s}(\mu^2)\times\left[\int_0^1 dz\, \sum_{q} e_q \frac{\Phi^q(z)}{z(1-z)}\right]\times \frac{1}{m_N}
\times\bar u(p_2)u(p_1)\nonumber\\
&\times & \Bigg[\frac{1}{\xi^2}\times\sum_{j = 2}^{\infty}
\frac{1}{\eta^{j-2}}\times\frac{1}{N_{j-2} ( \frac{5}{2})}\times
\left[\int_{0}^{\eta}\frac{dx}{\eta}\, \frac{\left(1-\frac{x^2}{\eta^2}\right)^2}{(1-\frac{x^2}{\xi^2})}\times C_{j-2}^{5/2}\left(\frac{x}{\eta}\right)\right]\times\mathbb{F}^g_{j} (\eta, t ; \mu_0^2)\nonumber\\ 
&+&i\pi H^g(\xi,\eta, t;\mu_0^2)+i\pi H^g(-\xi,\eta, t;\mu_0^2)\Bigg]\nonumber\\
&\approx &e\times\frac{1}{4}\times\frac{1}{N_c^2}\times \frac{1}{Q_0}
\times \lambda_{s}(\mu^2)\times\left[\int_0^1 dz\, \sum_{q} e_q \frac{\Phi^q(z)}{z(1-z)}\right]\times \frac{1}{m_N}
\times\bar u(p_2)u(p_1)\nonumber\\
&\times &\sum_{j = 2}^{\infty}
\frac{1}{\xi^{j}}\times\frac{1}{N_{j-2} ( \frac{5}{2})}\times
\left[\int_{0}^{\xi}\frac{dx}{\xi}\, \left(1-\frac{x^2}{\xi^2}\right)\times C_{j-2}^{5/2}\left(\frac{x}{\xi}\right)\right]\times\mathbb{F}^g_{j} (\xi, t ; \mu_0^2)\,,
\eea
where in the last line we have used $\eta\sim\xi$, and dropped the $i\pi H^g$ contributions as they vanish in the Regge kinematics.

\subsection{DVCS with twist-2 quark GPDs in QCD}


The leading order pQCD deep virtual Compton scattering (DVCS) amplitude written in terms of quark GPDs in the twist-2 approximation, is given by~\cite{Belitsky:2005qn} (see Eq. 5.32 there)

\begin{eqnarray}
\label{Tw2Parton}
T^{\mu\nu}_{\text{with quark GPDs}}
&=&\!\!\!
\frac{1}{2 p \cdot \tilde q}
\int_{-1}^{1} d x \,
\sum_q
C^{q[-]}_{(0)} (x, \xi)
\left\{
(p \cdot \tilde q) S^{\mu \nu; \rho \sigma} n_\sigma F^q_\rho (x, \eta, \Delta)
+
i \varepsilon^{\mu \nu \rho \sigma} p_\sigma
\left(
x \, \widetilde F^q_\rho (x, \eta, \Delta)
\right)
\right\}
\nonumber
\\
\!\!\!&+&\!\!\!
\frac{1}{2 p \cdot \tilde q}
\int_{-1}^{1} d x \,
\sum_q
C^{q[+]}_{(0)} (x, \xi)
\left\{2 i \varepsilon^{\mu \nu \rho \sigma} \tilde q_\sigma
\widetilde F^q_\rho (x, \eta, \Delta)
\right\}
\, .
\end{eqnarray}
The tree-level coefficient
functions are

\begin{equation}
\label{LOquarkCoeffFunct}
C^{q[\pm]}_{(0)} (x, \xi)
=
\frac{Q_q^2}{\xi - x - i\epsilon} \pm \frac{Q_q^2}{\xi + x - i\epsilon}\equiv Q_q^2\times C^{[\pm]}_{(0)} (x, \xi)
\, ,
\end{equation}
and the distributions in the twist-2 approximation are given by

\begin{eqnarray}
\label{VectorTw2Parton}
F^q_\rho (x, \eta, \Delta)
\!\!\!&=&\!\!\!
p_\rho\left( \frac{h^+}{p^+} H^q (x, \eta, \Delta) + \frac{e^+}{p^+} E^q (x, \eta, \Delta) \right)=p_\rho\times F^q(x, \eta, \Delta)\,,
\end{eqnarray}
\begin{eqnarray}
\label{AxialTw2Parton}
\widetilde{F}^q_\rho (x, \eta, \Delta)
\!\!\!&=&\!\!\!
 p_\rho\left(
\frac{\widetilde{h}^+}{p^+} \widetilde{H}^q (x, \eta, \Delta)
+
\frac{\widetilde{e}^+}{p^+} \widetilde{E}^q (x, \eta, \Delta) \right)=p_\rho\times\widetilde{F}^q(x, \eta, \Delta)\, .
\end{eqnarray}
\newline
\newline
\newline
{\bf Vector contribution:\nonumber}
\newline
\newline
By contracting (\ref{Tw2Parton}) with the transverse $TT$ polarization tensors, we can separate the vector contribution to DVCS as 

\begin{eqnarray}
\label{cont1Tw2Parton}
\epsilon_{T\mu}\epsilon_{T\nu}^{\prime*}T^{\mu\nu}_{\text{with quark GPDs}}
&=&\!\!\!
-\epsilon_{T}\cdot\epsilon_{T}^{\prime}\times \frac{1}{2}
\int_{-1}^{1} d x \,
\sum_q
C^{q[-]}_{(0)} (x, \xi)
\times p^{+}\times F^q(x, \eta, \Delta)
\, .
\end{eqnarray}

The singlet vector quark GPDs at $\mu=\mu_0$ can be written in terms of their conformal (Gegenbauer) moments $\mathbb{F}^q_{j(singlet)} (\eta, t ; \mu_0^2)$ as 
\bea
\label{ExpansionGPDsinglet2211}
H_{singlet}^q (x, \eta, t ; \mu_0^2)
&=&
\frac{1}{\eta} \sum_{j = 2}^{\infty}
\frac{w \big( \frac{x}{\eta} | \frac{3}{2} \big)}{\eta^{j-1} \, N_{j-1} ( \frac{3}{2})}
C_{j-1}^{3/2}\left(\frac{x}{\eta}\right)
\mathbb{F}^q_{j(singlet)} (\eta, t ; \mu_0^2)\nonumber\\
&=&\frac{1}{\eta}\times\left(1-\frac{x^2}{\eta^2}\right) \times\sum_{j = 2}^{\infty}
\frac{1}{\eta^{j-1}}\times\frac{1}{N_{j-1} ( \frac{3}{2})}
\times C_{j-1}^{3/2}\left(\frac{x}{\eta}\right)
\times\mathbb{F}^q_{j(singlet)} (\eta, t ; \mu_0^2)\,,\nonumber\\
\eea
for even $j=2,4,...$, where we used the weight and normalization factors as
$$
w (x | \nu) = (1 - x^2)^{\nu - 1/2}
\, , \qquad
N_{j-1} (\nu) = 2^{1 - 2 \nu}
\frac{
{\mit\Gamma}^2 (1/2) {\mit\Gamma} (2 \nu + (j-1))
}{
{\mit\Gamma}^2 (\nu) {\mit\Gamma} ((j-1) + 1) (\nu + (j-1))
}
\,.
$$
And, using the conformal expansion of singlet quark GPDs (\ref{ExpansionGPDsinglet2211}) in the vector part of the amplitude (\ref{cont1Tw2Parton}), we find 
\begin{eqnarray}
\label{cont1Tw2Parton22}
&&\epsilon_{T\mu}\epsilon_{T\nu}^{\prime*}T^{\mu\nu}_{\text{with quark GPDs}}
\nonumber\\
&=&-\epsilon_{T}\cdot\epsilon_{T}^{\prime}\times N_c^0\times\frac{1}{2m_N}\times\bar u(p_2)u(p_1)\times\frac{1}{2}
\Bigg[\int_{0}^{1} d x \,
\sum_q Q_q^2\times\frac{2x}{\xi^2-x^2}
\times H_{singlet}^q(x, \eta, t; \mu_0^2)\nonumber\\
&+&i\pi\sum_q Q_q^2\times H_{singlet}^q(\xi, \eta, t; \mu_0^2)+i\pi\sum_q Q_q^2\times H_{singlet}^q(-\xi, \eta, t; \mu_0^2)\Bigg]\,,\nonumber\\
&=&\!\!\!
-\epsilon_{T}\cdot\epsilon_{T}^{\prime}\times N_c^0\times\frac{1}{2m_N}\times\bar u(p_2)u(p_1)\nonumber\\
&\times & \frac{1}{2}\Bigg[\frac{1}{\xi}\times\sum_{j = 2}^{\infty}
\frac{1}{\eta^{j-1}}\times\frac{1}{N_{j-1} ( \frac{3}{2})}\times 2\left[\int_{0}^{\eta} \frac{dx}{\eta}\,\frac{\frac{x}{\xi}\left(1-\frac{x^2}{\eta^2}\right)}{1-\frac{x^2}{\xi^2}}
\times C_{j-1}^{3/2}\left(\frac{x}{\eta}\right)
\right]\nonumber\\
&\times & \sum_q Q_q^2\times\mathbb{F}^q_{j(singlet)} (\eta, t ; \mu_0^2)+i\pi\sum_q Q_q^2\times H_{singlet}^q(\xi, \eta, t; \mu_0^2)\nonumber\\
&+&i\pi\sum_q Q_q^2\times H_{singlet}^q(-\xi, \eta, t; \mu_0^2)\Bigg]\nonumber\\
&\approx &\!\!\!
-\epsilon_{T}\cdot\epsilon_{T}^{\prime}\times N_c^0\times\frac{1}{2m_N}\times\bar u(p_2)u(p_1)\nonumber\\
&\times &\sum_{j = 2}^{\infty}
\frac{1}{\xi^{j}}\times\frac{1}{N_{j-1} ( \frac{3}{2})}\times \left[\int_{0}^{\xi} \frac{dx}{\xi}\times\frac{x}{\xi}
\times C_{j-1}^{3/2}\left(\frac{x}{\xi}\right)
\right]\times \sum_q Q_q^2\times\mathbb{F}^q_{j(singlet)} (\xi, t ; \mu_0^2)\,,\nonumber\\
\end{eqnarray}
for even $j=2,4,...$, and, in the last line, we have used $\eta\sim\xi$. The $i\pi H^q$ contribution vanishes at $\xi=\pm \eta$.
\\
\\
{\bf Axial contribution:}
\\
By contracting (\ref{Tw2Parton}) with the longitudinal-transverse $LT$ polarization tensors, we can separate the axial contribution to DVCS as  
\begin{eqnarray}
\label{cont2Tw2Parton1122}
\epsilon_{L\mu}\epsilon_{T\nu}^{\prime*}T^{\mu\nu}_{\text{with quark GPDs}}
&=&\frac{1}{2 p \cdot \tilde q}
\int_{-1}^{1} d x \,
\sum_q
C^{q[+]}_{(0)} (x, \xi)
\left\{\epsilon_{L\mu}\epsilon_{T\nu}^{\prime*}\times 2 i \varepsilon^{\mu \nu \rho \sigma} \tilde q_\sigma p_\rho
\times\widetilde F^q(x, \eta, \Delta)
\right\}
\, .
\end{eqnarray}

The singlet axial-vectror quark GPDs at $\mu=\mu_0$ can be written in terms of their Gegenbauer (conformal) moments $\tilde{\mathbb{F}}^q_{j(singlet)} (\eta, t ; \mu_0^2)$ as
\bea
\label{singletAxialGPDinput22}
\sum_{q}Q_q^2\,\tilde{H}_{singlet}^q (x, \eta, t ; \mu_0^2)
&=&
\frac{1}{\eta}\times\left(1-\frac{x^2}{\eta^2}\right)\times\sum_{j=1}^{\infty}\frac{1}{N_{j-1}(\frac{3}{2})}\times\frac{1}{\eta^{j-1}} 
\times C_{j-1}^{3/2}\left(\frac{x}{\eta}\right)\times\sum_{q}Q_q^2\,\tilde{\mathbb{F}}^q_{j(singlet)} (\eta, t ; \mu_0^2)\nonumber\\
\eea
for odd $j=1,3,...$. And, using the conformal expansion of the axial-vector singlet quark GPDs (\ref{singletAxialGPDinput22}) in the axial part of the DVCS amplitude (\ref{cont2Tw2Parton1122}), we find 
\begin{eqnarray}
\label{cont2Tw2Parton222}
&&\epsilon_{L\mu}\epsilon_{T\nu}^{\prime*}T^{\mu\nu}_{\text{with quark GPDs}}
\nonumber\\
&=&
\epsilon_{L\mu}\epsilon_{T\nu}^{\prime*}\times N_c^0\times i \varepsilon^{\mu \nu \rho \sigma} \tilde q_\sigma p_\rho
\times \frac{1}{p^{+}}
\times \bar u(p_2)\gamma^{+}\gamma_5 u(p_1)\nonumber\\
&\times &\frac{2}{Q^2}\times\xi
\times\Bigg[\int_{0}^{1} d x \,
\frac{2\xi}{\xi^2-x^2}
\times\sum_q Q_q^2\,\widetilde H_{singlet}^q(x, \eta,t;\mu_0^2)+i\pi\sum_q Q_q^2\,\widetilde H_{singlet}^q(\xi, \eta,t;\mu_0^2)\nonumber\\
&+&i\pi\sum_q Q_q^2\,\widetilde H_{singlet}^q(-\xi, \eta,t;\mu_0^2)\Bigg]\,,\nonumber\\
&=&\!\!\!
\epsilon_{L\mu}\epsilon_{T\nu}^{\prime*}\times N_c^0\times i \varepsilon^{\mu \nu \rho \sigma} \tilde q_\sigma p_\rho
\times \frac{1}{p^{+}}
\times \bar u(p_2)\gamma^{+}\gamma_5 u(p_1)\nonumber\\
&\times &\frac{2}{Q^2}\times \Bigg[\sum_{j=1}^{\infty}\,\frac{1}{\eta^{j-1}} 
\times\frac{2}{N_{j-1}(\frac{3}{2})}\times\left[\int_{0}^{\eta}\frac{dx}{\eta} \,
\frac{1-\frac{x^2}{\eta^2}}{1-\frac{x^2}{\xi^2}}
\times C_{j-1}^{3/2}\left(\frac{x}{\eta}\right)\right]\times\sum_{q}Q_q^2\,\tilde{\mathbb{F}}^q_{j(singlet)} (\eta, t ; \mu_0^2)\nonumber\\
&+&i\pi\sum_q Q_q^2\,\widetilde H_{singlet}^q(\xi, \eta,t;\mu_0^2)+i\pi\sum_q Q_q^2\,\widetilde H_{singlet}^q(-\xi, \eta,t;\mu_0^2)\Bigg]\,,\nonumber\\
&\approx &\!\!\!
\epsilon_{L\mu}\epsilon_{T\nu}^{\prime*}\times N_c^0\times i \varepsilon^{\mu \nu \rho \sigma} \tilde q_\sigma p_\rho
\times \frac{1}{p^{+}}
\times \bar u(p_2)\gamma^{+}\gamma_5 u(p_1)\nonumber\\
&\times &\frac{4}{Q^2}\times\sum_{j=1}^{\infty}\,\frac{1}{\xi^{j-1}} 
\times\frac{1}{N_{j-1}(\frac{3}{2})}\times\left[\int_{0}^{\xi}\frac{dx}{\xi}\,C_{j-1}^{3/2}\left(\frac{x}{\xi}\right)\right]\times\sum_{q}Q_q^2\,\tilde{\mathbb{F}}^q_{j(singlet)} (\eta\sim\xi, t ; \mu_0^2)\,,\nonumber\\
\end{eqnarray}
for odd $j=1,3,...$. In  the last line we have used $\eta\sim\xi$, and dropped the $i\pi\widetilde{H}^q$ in the Regge limit.

\subsection{Pair meson electroproduction with non-singlet (valence) vector quark GPDs in QCD}


In leading order in $\alpha_s$, the 
 amplitude for electroproduction of pair of pions (or any other pair of mesons) in pQCD
 is illustrated in Fig.~\ref{fig_PIPI2} left. 
 It is given in terms of gluon DA of mesons, and quark GPDs as (see Eq.283 in \cite{Diehl:2003ny}, and Eq.10 in \cite{Lehmann-Dronke:2000hlo})
\bea\label{pairGPD}
{\cal A}^{LL}_{\gamma^* p\rightarrow  \pi\pi p} (s,t,Q,\epsilon_{L};m_{\pi\pi})=&&
e\times \frac{2\times 4\pi \alpha_s}{N_c} \frac{1}{Q}
\int_0^1 dz\, \frac{\Phi^g(z)}{z(1-z)} \,\nonumber\\
&&\times\int_{-1}^1 dx\, \sum_{q} Q_q\,F^q(x,\eta, t;\mu^2)  
\left[ \frac{1}{\xi - x -i\epsilon} +
                     \frac{1}{\xi + x -i\epsilon} \right]\,.\nonumber\\
\eea
In leading order, $\Phi^g(z)$ is gluon distribution amplitude for the emission of 2 pions, with $z$ a gluon-parton fraction of the 
2 pion longitudinal momentum in the final state. $F^q(x,\eta, t;\mu^2)$ is the unpolarized and skewed quark parton distribution 
in the nucleon, with both skewed Dirac and Pauli form factors. In the Regge limit with $s\gg -t$, we can ignore 
 the Pauli contribution and rewrite (\ref{pairGPD}) as
\bea\label{pairGPD222}
{\cal A}^{LL}_{\gamma^* p\rightarrow  \pi\pi p} (s,t,Q,\epsilon_{L};m_{\pi\pi})&=&
e\times\frac{2g_s^2N_c}{N_c^2} \frac{1}{Q}
\left[\int_0^1 dz\, \frac{\Phi^g(z)}{z(1-z)}\right]\times\frac{1}{2m_N}\times\bar u(p_2)u(p_1) \,\nonumber\\
&\times &\int_{0}^1 dx\, \sum_{q} Q_q \,H_{valence}^q(x,\eta, t;\mu^2)  
\left[ \frac{1}{\xi - x -i\epsilon} +
                     \frac{1}{\xi + x -i\epsilon} \right]\,.\nonumber\\
\eea

The non-singlet (valence) vector quark GPDs at $\mu=\mu_0$ can be written in terms of their conformal (Gegenbauer) moments $\mathbb{F}^q_{j(valence)} (\eta, t ; \mu_0^2)$ as
\bea
\label{ExpansionGPDeigenfunctionsH2211}
H_{valence}^q (x, \eta, t ; \mu_0^2)
&=&
\frac{1}{\eta} \sum_{j = 1}^{\infty}
\frac{w \big( \frac{x}{\eta} | \frac{3}{2} \big)}{\eta^{j-1} \, N_{j-1} ( \frac{3}{2})}
C_{j-1}^{3/2}\left(\frac{x}{\eta}\right)
\mathbb{F}^q_{j(valence)} (\eta, t ; \mu_0^2)\times\theta(\eta - |x|)
\, ,\nonumber\\
&=&
\frac{1}{\eta}\times \left(1-\frac{x^2}{\eta^2}\right)\times \sum_{j = 1}^{\infty}\frac{1}{\eta^{j-1}}
\times\frac{1}{N_{j-1}(\frac{3}{2})}
\times C_{j-1}^{3/2}\left(\frac{x}{\eta}\right)
\times\mathbb{F}^q_{j(valence)} (\eta, t ; \mu_0^2)\times\theta(\eta - |x|)
\, ,\nonumber\\
\eea
for odd $j=1,3,...$, where we have used the weight and normalization factors as
$$
w (x | \nu) = (1 - x^2)^{\nu - 1/2}
\, , \qquad
N_{j-1} (\nu) = 2^{1 - 2 \nu}
\frac{
{\mit\Gamma}^2 (1/2) {\mit\Gamma} (2 \nu + (j-1))
}{
{\mit\Gamma}^2 (\nu) {\mit\Gamma} ((j-1) + 1) (\nu + (j-1))
}
\,.
$$
And, using the conformal expansion (\ref{ExpansionGPDeigenfunctionsH2211}) in the amplitude (\ref{pairGPD222}), we find
\bea\label{pairGPD22}
&&{\cal A}^{LL}_{\gamma^* p\rightarrow  \pi\pi p} (s,t,Q_0,\epsilon_{L};m_{\pi\pi})=
e\times\lambda_{s}(\mu_0)\times \frac{2}{N_c^2}\times \frac{1}{Q_0}
\times\left[\int_0^1 dz\, \frac{\Phi^g(z)}{z(1-z)}\right]\times\frac{1}{2m_N}\times\bar u(p_2)u(p_1) \,\nonumber\\
&&\times \Bigg[\int_{0}^1 dx\,\frac{2\xi}{\xi^2 - x^2}\times \sum_{q} Q_q \,H_{valence}^q(x,\eta, t;\mu_0^2)+i\pi\sum_{q} Q_q \,H_{valence}^q(\xi,\eta, t;\mu_0^2)\nonumber\\
&+&i\pi\sum_{q} Q_q \,H_{valence}^q(-\xi,\eta, t;\mu_0^2)\Bigg]\,,\nonumber\\
&=&
e\times\lambda_{s}(\mu_0)\times \frac{2}{N_c^2}\times \frac{1}{Q_0}
\times\left[\int_0^1 dz\, \frac{\Phi^g(z)}{z(1-z)}\right]\times\frac{1}{2m_N}\times\bar u(p_2)u(p_1) \nonumber\\
&\times &\Bigg[\frac{2}{\xi}\times\sum_{j = 1}^{\infty}\frac{1}{\eta^{j-1}}
\times\frac{1}{N_{j-1}(\frac{3}{2})}
\times\left[\int_{0}^{\eta}\frac{dx}{\eta}\times\frac{1 - \frac{x^2}{\eta^2}}{1 - \frac{x^2}{\xi^2}}\times C_{j-1}^{3/2}\left(\frac{x}{\eta}\right)
\right]\times\sum_{q} Q_q\,\mathbb{F}^q_{j(valence)} (\eta, t ; \mu_0^2) \nonumber\\
&+&i\pi\sum_{q} Q_q \,H_{valence}^q(\xi,\eta, t;\mu_0^2)+i\pi\sum_{q} Q_q \,H_{valence}^q(-\xi,\eta, t;\mu_0^2)\Bigg]\,,\nonumber\\
&\approx &
e\times\lambda_{s}(\mu_0)\times \frac{4}{N_c^2}\times \frac{1}{Q_0}
\times\left[\int_0^1 dz\, \frac{\Phi^g(z)}{z(1-z)}\right]\times\frac{1}{2m_N}\times\bar u(p_2)u(p_1) \nonumber\\
&\times &\sum_{j = 1}^{\infty}\frac{1}{\xi^{j}}
\times\frac{1}{N_{j-1}(\frac{3}{2})}
\times\left[\int_{0}^{\xi}\frac{dx}{\xi}\,C_{j-1}^{3/2}\left(\frac{x}{\xi}\right)
\right]\times\sum_{q} Q_q\,\mathbb{F}^q_{j(valence)} (\eta\sim\xi, t ; \mu_0^2)  
\,,\nonumber\\
\eea
for odd $j=1,3,...$. In the last line, we have used $\eta\sim\xi$, and dropped the $i\pi H^q$ contributions as they
vanish in the Regge limit.

\subsection{Neutral pion electroproduction with non-singlet (valence) axial quark GPDs in QCD}


In the leading twist-expansion, the pQCD electroproduction of neutral pion ($\pi^0$) in Fig.~\ref{fig_PI2} left,  can be written
in terms of the meson DAs and the nucleon axial quark GPDs (see Eq.230 with Eq.231 and Eq.232 in \cite{Goeke:2001tz})

\begin{eqnarray}\label{pi0GPD}
{\cal A}^{LL}_{\gamma^{*} p\rightarrow  \pi^{0} p} (s,t,Q,\epsilon_{L},\epsilon_{L}^{\prime}) &\,=\,& e\times {C_{F}\over N_c}\times{1 \over 2}\times (4\pi \alpha _{s})\times{1\over {Q}}\times\left[ \, \int _{0}^{1}dz{{\Phi_{\pi}(z)}\over z}\right] \nonumber\\
&\times &
\frac{1}{p^{+}}\left\{ A_{\pi ^0N} 
\,\widetilde{h}^{+}
\,+\, B_{\pi^0 N} 
\,\widetilde{e}^{+} 
\right\} \, , 
\end{eqnarray}
where $\Phi_{\pi^{0}}(z)$ is the pion distribution amplitude (DA), 
\bea
A_{\pi^0 \, p} \,&=&\, \int_{-1}^1 dx \; 
{-1 \over {\sqrt 2}} \, 
{\left(Q_u \ \widetilde H^u(x,\eta, t;\mu^2) \,-\, Q_d \ \widetilde H^d(x,\eta, t;\mu^2)\right)} 
\; \left\{ {{1} \over {\xi - x - i \epsilon}}
- {{1} \over {\xi + x -i \epsilon}} \right\}\nonumber\\
B_{\pi^0 \, p} \,&=&\, \int_{-1}^1 dx \;
{-1 \over {\sqrt 2}} \, 
{\left(Q_u \ \widetilde E^u(x,\eta, t;\mu^2) \,-\, Q_d \ \widetilde E^d(x,\eta, t;\mu^2)\right)} 
\;\left\{ {{1} \over {\xi - x - i \epsilon}}
- {{1} \over {\xi + x - i \epsilon}} \right\}\,.
\eea
Using the Casimir $C_{F}=\frac{N_c^2-1}{2N_c}\approx \frac{N_c}{2}$ in the large-$N_c$ limit, and defining the 't Hooft coupling constant as $g_s^2N_c =\lambda_s$, we can re-write (\ref{pi0GPD}) as 
\begin{eqnarray}\label{pi0GPD211}
&&{\cal A}^{LL}_{\gamma^{*} p\rightarrow  \pi^{0} p} (s,t,Q,\epsilon_{L},\epsilon_{L}^{\prime}) \,=\, e\times {1\over 2}\times{1 \over 2}\times \frac{1}{N_c}\times\lambda_s\times{1\over {Q}}\times\left[ \, \int _{0}^{1}dz{{\Phi_{\pi}(z)}\over z}\right] \times
\frac{1}{p^{+}}\times u(p_2)\gamma^{+}\,\gamma_5 u(p_1)\times{-1 \over {\sqrt 2}}\nonumber\\
&&\times  \int_{0}^1 dx \; 
{\left(Q_u \ \widetilde H_{valence}^u(x,\eta, t;\mu^2) \,-\, Q_d \ \widetilde H_{valence}^d(x,\eta, t;\mu^2)\right)} 
\; \left\{ {{1} \over {\xi - x - i \epsilon}}
- {{1} \over {\xi + x -i \epsilon}} \right\}\,,\nonumber\\
\end{eqnarray}
where we have ignored the Pauli contribution for $s\gg -t$.

The non-singlet (valence) axial quark GPDs at $\mu=\mu_0$ can be written in terms of their Gegenbauer (conformal) moments $\tilde{\mathbb{F}}^q_{j(valence)} (\eta, t ; \mu_0^2)$ as
\bea
\label{ExpansionAxialGPDvalence2211}
\widetilde{H}_{valence}^q (x, \eta, t ; \mu_0^2)
&=&
\frac{1}{\eta} \sum_{j = 2}^{\infty}
\frac{w \big( \frac{x}{\eta} | \frac{3}{2} \big)}{\eta^{j-1} \, N_{j-1} ( \frac{3}{2})}
C_{j-1}^{3/2}\left(\frac{x}{\eta}\right)
\widetilde{\mathbb{F}}^q_{j(valence)} (\eta, t ; \mu_0^2)\nonumber\\
&=&\frac{1}{\eta}\times\left(1-\frac{x^2}{\eta^2}\right) \times\sum_{j = 2}^{\infty}
\frac{1}{\eta^{j-1}}\times\frac{1}{N_{j-1} ( \frac{3}{2})}
\times C_{j-1}^{3/2}\left(\frac{x}{\eta}\right)
\times\widetilde{\mathbb{F}}^q_{j(valence)} (\eta, t ; \mu_0^2)\,,\nonumber\\
\eea
for even $j=2,4,...$. And, using the conformal expansion of axial-vector valence quark GPDs (\ref{ExpansionAxialGPDvalence2211}) in the $\pi^0$ production amplitude (\ref{pi0GPD211}), we find
\begin{eqnarray}\label{pi0GPD22}
&&{\cal A}^{LL}_{\gamma^{*} p\rightarrow  \pi^{0} p} (s,t,Q_0,\epsilon_{L},\epsilon_{L}^{\prime}) \nonumber\\
&=& e\times {1\over 2}\times{1 \over 2}\times \frac{1}{N_c}\times\lambda_s(\mu_0)\times{1\over {Q_0}}\times\left[ \, \int _{0}^{1}dz{{\Phi_{\pi}(z)}\over z}\right] \times
\frac{1}{p^{+}}\times u(p_2)\gamma^{+}\,\gamma_5 u(p_1)\times{-1 \over {\sqrt 2}}\nonumber\\
&\times & \Bigg[\int_{0}^1 dx\,\frac{2x}{\xi^2 - x^2}\times\left(Q_u \ \widetilde H_{valence}^u(x,\eta, t;\mu_0^2) \,-\, Q_d \ \widetilde H_{valence}^d(x,\eta, t;\mu_0^2)\right)\nonumber\\
&+&i\pi\left(Q_u \ \widetilde H_{valence}^u(\xi,\eta, t;\mu_0^2) \,-\, Q_d \ \widetilde H_{valence}^d(\xi,\eta, t;\mu_0^2)\right)\nonumber\\
&+&i\pi\left(Q_u \ \widetilde H_{valence}^u(-\xi,\eta, t;\mu_0^2) \,-\, Q_d \ \widetilde H_{valence}^d(-\xi,\eta, t;\mu_0^2)\right)\Bigg]\,,\nonumber\\
&\,=\,& e\times {1\over 2}\times{1 \over 2}\times \frac{1}{N_c}\times\lambda_s(\mu_0)\times{1\over {Q_0}}\times\left[ \, \int _{0}^{1}dz{{\Phi_{\pi}(z)}\over z}\right] \times
\frac{1}{p^{+}}\times u(p_2)\gamma^{+}\,\gamma_5 u(p_1)\times{-1 \over {\sqrt 2}}\nonumber\\
&\times &\Bigg[\frac{2}{\xi}\times\sum_{j = 2}^{\infty}
\frac{1}{\eta^{j-1}}\times\frac{1}{N_{j-1} ( \frac{3}{2})}
\times\left[\int_{0}^\eta \frac{dx}{\eta}\,\frac{\frac{x}{\xi}\left(1 - \frac{x^2}{\eta^2}\right)}{1 - \frac{x^2}{\xi^2}}\times C_{j-1}^{3/2}\left(\frac{x}{\eta}\right)\right]\nonumber\\
&\times &\left(Q_u \ \widetilde{\mathbb{F}}^u_{j(valence)} (\eta, t ; \mu_0^2) \,-\, Q_d \ \widetilde{\mathbb{F}}^d_{j(valence)} (\eta, t ; \mu_0^2)\right)\nonumber\\
&+&i\pi\left(Q_u \ \widetilde H_{valence}^u(\xi,\eta, t;\mu_0^2) \,-\, Q_d \ \widetilde H_{valence}^d(\xi,\eta, t;\mu_0^2)\right)\nonumber\\
&+&i\pi\left(Q_u \ \widetilde H_{valence}^u(-\xi,\eta, t;\mu_0^2) \,-\, Q_d \ \widetilde H_{valence}^d(-\xi,\eta, t;\mu_0^2)\right)\Bigg]\,,\nonumber\\
&\,\approx\,& e\times {1\over 2}\times \frac{1}{N_c}\times\lambda_s(\mu_0)\times{1\over {Q_0}}\times\left[ \, \int _{0}^{1}dz{{\Phi_{\pi}(z)}\over z}\right] \times
\frac{1}{p^{+}}\times u(p_2)\gamma^{+}\,\gamma_5 u(p_1)\times{-1 \over {\sqrt 2}}\nonumber\\
&\times &\sum_{j = 2}^{\infty}
\frac{1}{\xi^{j}}\times\frac{1}{N_{j-1} ( \frac{3}{2})}
\times\left[\int_{0}^\xi \frac{dx}{\xi}\,\frac{x}{\xi}\,C_{j-1}^{3/2}\left(\frac{x}{\xi}\right)\right]\nonumber\\
&\times &\left(Q_u \ \widetilde{\mathbb{F}}^u_{j(valence)} (\eta\sim\xi, t ; \mu_0^2) \,-\, Q_d \ \widetilde{\mathbb{F}}^d_{j(valence)} (\eta\sim\xi, t ; \mu_0^2)\right)\,,\nonumber\\
\end{eqnarray}
for even $j=2,4,...$. Again, in  the last line, we have used $\eta\sim\xi$, and dropped the $i\pi\widetilde{H}$ contributions.

\section{Details on the resummation by the j-contour for gluon GPD contribution in the electroproduction of $\rho^0$}~\label{gluonsum}

{\bf Resummation by j-contour}

We first rewrite the sum over even $j=2,4,...$ in (\ref{gluonFinal2EVOLVEDHEnergy2}) as 
\bea
\label{gluonFinal2EVOLVEDHEnergy33INPUTgluon}
{\cal A}^{LL(gluon)}_{\gamma^{*} p\rightarrow  \rho^{0} p} (s,t,Q,\epsilon_{L},\epsilon_{L}^{\prime})&=& -\int_{\mathbb C}\frac{dj}{4i}\frac{1+e^{-i\pi j}}{{\rm sin}\pi j}
\times\frac{1}{\xi^{j}}\times\frac{1}{\Gamma(\Delta_{g}(j)-2)}\times\mathcal{N}_{g}(j)\times\widehat{\mathbb{F}}^g_{j} (\xi, t ; \mu^2)\nonumber\\
&\times&  e\times f_V^{\prime}\times\alpha_s(\mu)\times{1\over {Q}}\times\frac{1}{2\sqrt{2} m_{N}}\times \bar u(p_2)u(p_1)\,,
\eea
where the contour $\mathbb C $ is at the right most of the branch point of $$\frac{1}{\Gamma(\Delta_{g}(j)-2)}=\frac{1}{\Gamma\left(\sqrt{2}\lambda^{1/4}(j-j_{0g})^{1/2}\right)}$$, and leftmost of the poles at even $j=2,4,...$. Here we have defined
\bea
\widehat{\mathbb{F}}^g_{j} (\xi, t ; \mu^2)=\Gamma(\Delta_{g}(j)-2)\times\mathbb{F}^g_{j} (\xi, t ; \mu^2) 
\eea
to make explicit the  branch point at $j=j_{0g}=2-2/\sqrt{\lambda}$. originating from the holographic input conformal (Gegenbauer) moments of gluon GPD at $\mu=\mu_0\sim Q_0$.

We now evaluate the input gluon GPD part of the evolved gluon GPD
\bea
\label{gluonFinal2EVOLVEDHEnergy3INPUTgluon}
{\cal A}^{LL(gluon)}_{\gamma^{*} p\rightarrow  \rho^{0} p} (s,t,Q,\epsilon_{L},\epsilon_{L}^{\prime})&=&-\int_{\mathbb C}\frac{dj}{4i}\frac{1+e^{-i\pi j}}{{\rm sin}\pi j}
\times\frac{1}{\xi^{j}}\times\frac{1}{\Gamma(\Delta_{g}(j)-2)}\times\mathcal{N}_{g}(j)\times\widehat{\mathbb{F}}^g_{j} (\xi, t ; \mu^2)\nonumber\\
&\times&  e\times f_V^{\prime}\times\alpha_s(\mu)\times{1\over {Q}}\times\frac{1}{2\sqrt{2} m_{N}}\times \bar u(p_2)u(p_1)\,.
\eea
 by wrapping the j-plane contour $\mathbb C $ to the left
\bea
\label{gluonFinal2EVOLVEDHEnergy4INPUTgluon}
{\cal A}^{LL(gluon)}_{\gamma^{*} p\rightarrow  \rho^{0} p} (s,t,Q,\epsilon_{L},\epsilon_{L}^{\prime})&\approx& -\int_{-\infty}^{j_0}\frac{dj}{2i}\frac{1+e^{-i\pi j}}{{\rm sin}\pi j}
\times\frac{1}{\xi^{j}}\times\text{Im}\left[\frac{1}{\Gamma(iy)}\right]\times\mathcal{N}_{g}(j)\times\widehat{\mathbb{F}}^g_{j} (\xi, t ; \mu^2)\nonumber\\
&\times&  e\times f_V^{\prime}\times\alpha_s(\mu)\times{1\over {Q}}\times\frac{1}{2\sqrt{2} m_{N}}\times \bar u(p_2)u(p_1)\,,
\eea
where we have defined $$iy = i\sqrt{2\sqrt{\lambda}(j_{0g}-j)}\,.$$ For, $y\rightarrow 0$, we may approximate $\frac{1}{\Gamma(iy)}\approx iy\,e^{i\gamma_E y}$ with the Euler constant $\gamma_E\approx 0.577216$, and write $\text{Im}\left[\frac{1}{\Gamma(iy)}\right]\approx y\,\text{cos}(\gamma_E y)\approx \text{sin}(y)$ for $y\rightarrow 0$ or $j\rightarrow j_{0g}$. Therefore, for $j\rightarrow j_{0g}$, we find
\bea
\label{gluonFinal2EVOLVEDHEnergy5INPUTgluon}
{\cal A}^{LL(gluon)}_{\gamma^{*} p\rightarrow  \rho^{0} p} (s,t,Q,\epsilon_{L},\epsilon_{L}^{\prime})&\approx&  \mathcal{N}_{g}(j_{0g})\,\widehat{\mathbb{F}}^g_{j_{0g}} (\xi, t ; \mu^2)\times e\,f_V^{\prime}\,\alpha_s(\mu)\,{1\over {Q}}\,\frac{1}{2\sqrt{2} m_{N}}\,\bar u(p_2)u(p_1)\nonumber\\
&\times& -\int_{-\infty}^{j_{0g}}\frac{dj}{2}\frac{1+e^{-i\pi j}}{{\rm sin}\pi j}
\times\frac{1}{\xi^{j}}\times \text{sin}\left(\sqrt{2\sqrt{\lambda}(j_{0g}-j)}\right)\,.
\eea
\\
\\
{\bf Real and Imaginary parts}
\\
We separate the amplitude (\ref{gluonFinal2EVOLVEDHEnergy5INPUTgluon}) into its real and imaginary parts $${\cal A}^{LL(gluon)}_{\gamma^{*} p\rightarrow  \rho^{0} p} (s,t,Q,\epsilon_{L},\epsilon_{L}^{\prime})\equiv{\cal A}^{LL(gluon)}=\text{Re}\left[{\cal A}^{LL(gluon)}\right]+i\text{Im}\left[{\cal A}^{LL(gluon)}\right]$$,
\bea\label{REIMgluonFinal2EVOLVEDHEnergy5INPUTgluon}
\text{Re}\left[{\cal A}^{LL(gluon)}\right]&=& \mathcal{N}_{g}(j_{0g})\times\widehat{\mathbb{F}}^g_{j_{0g}} (\xi, t ; \mu^2)\times e\,f_V^{\prime}\,\alpha_s(\mu)\,{1\over {Q}}\,\frac{1}{2\sqrt{2} m_{N}}\,\bar u(p_2)u(p_1) \nonumber\\
&\times &-\int_{-\infty}^{j_{0g}}\frac{dj}{2}\frac{1+\text{cos}\pi j}{{\rm sin}\pi j}
\times\frac{1}{\xi^{j}}\times \text{sin}\left(\sqrt{2\sqrt{\lambda}(j_{0g}-j)}\right)\,,\nonumber\\
\text{Im}\left[{\cal A}^{LL(gluon)}\right] &=&
\mathcal{N}_{g}(j_{0g})\times \widehat{\mathbb{F}}^g_{j_{0g}} (\xi, t ; \mu^2)\times e\,f_V^{\prime}\,\alpha_s(\mu)\,{1\over {Q}}\,\frac{1}{2\sqrt{2} m_{N}}\,\bar u(p_2)u(p_1) \nonumber\\
&\times &\int_{-\infty}^{j_{0g}}\frac{dj}{2}\,\frac{1}{\xi^{j}}\times \text{sin}\left(\sqrt{2\sqrt{\lambda}(j_{0g}-j)}\right)\,.
\eea
With the change of integration variable $y^2=2\sqrt{\lambda}(j_{0g}-j)$, we turn the imaginary part into a Gaussian integral that can be evaluated exactly (with $\tilde{\tau}=\text{log}(1/\xi)$)
\be\label{IMgluonFinal2EVOLVEDHEnergy5INPUTgluon}
\text{Im}\left[{\cal A}^{LL(gluon)}\right] &=&
\mathcal{N}_{g}(j_{0g})\times\widehat{\mathbb{F}}^g_{j_{0g}} (\xi, t ; \mu^2)\times e\,f_V^{\prime}\,\alpha_s(\mu)\,{1\over {Q}}\,\frac{1}{2\sqrt{2} m_{N}}\,\bar u(p_2)u(p_1) \nonumber\\
&\times &\frac{1}{\xi^{j_{0g}}}\times\int_{-\infty}^{\infty}\frac{dy}{4\pi i\sqrt{\lambda}}\,ye^{-\tilde{\tau}y^2/2\sqrt{\lambda}}e^{iy}\,,\nonumber\\
&=&\mathcal{N}_{q}(j_{0g})\times\widehat{\mathbb{F}}^g_{j_{0g}} (\xi, t ; \mu^2)\times e\,f_V^{\prime}\,\alpha_s(\mu)\,{1\over {Q}}\,\frac{1}{2\sqrt{2} m_{N}}\,\bar u(p_2)u(p_1) \nonumber\\
&\times &\frac{1}{\xi^{j_{0g}}}\times (\sqrt{\lambda}/2\pi)^{1/2}\times \frac{e^{-2\sqrt{\lambda}/4\tilde{\tau}}}{\tilde{\tau}^{3/2}}\,,
\ee

We compute the real part of the amplitude $\text{Re}\left[{\cal A}^{LL(gluon)}\right]$ by first expanding its prefactor near $j=2$ 
\bea
\frac{1+\text{cos}\pi j}{{\rm sin}\pi j}\simeq \frac{2}{\pi(j-2)}+\mathcal{O}(j-2)\,,
\eea
and establishing the identity $\partial_{\tilde{\tau}}\left[e^{-2\tilde{\tau}}\text{Re}\left[{\cal A}^{LL(gluon)}\right]\right]=-(2/\pi)e^{-2\tilde{\tau}}\text{Im}\left[{\cal A}^{LL(gluon)}\right]$ which gives us an approximation for the real part
\bea\label{REgluonFinal2EVOLVEDHEnergy5INPUTgluon}
\text{Re}\left[{\cal A}^{LL(gluon)}\right]&\simeq &\mathcal{N}_{g}(j_{0g})\times\widehat{\mathbb{F}}^g_{j_{0g}} (\xi, t ; \mu^2)\times e\,f_V^{\prime}\,\alpha_s(\mu)\,{1\over {Q}}\,\frac{1}{2\sqrt{2} m_{N}}\,\bar u(p_2)u(p_1)\nonumber\\ 
&\times & (\sqrt{\lambda}/2\pi)^{1/2}\times \frac{1}{\xi^{2}}\times\int_{\tilde{\tau}}^{\infty}d\tilde{\tau}^{\prime}\,\frac{2e^{-2\tilde{\tau}^{\prime}/\sqrt{\lambda}-\sqrt{\lambda}/2\tilde{\tau}^{\prime}}}{\pi\tilde{\tau}^{\prime 3/2}}\,.\nonumber\\
\eea
Small corrections in the order of $\mathcal{O}(\text{Im}\left[{\cal A}^{LL(gluon)}\right]/\text{log}(1/\xi))$ and $\mathcal{O}(\text{Im}\left[{\cal A}^{LL(gluon)}\right]/\sqrt{\lambda})$ to (\ref{REgluonFinal2EVOLVEDHEnergy5INPUTgluon}), can be computed in a standard perturbation series but can be ignored for $\xi\rightarrow 0$ and fixed large $\sqrt{\lambda}$. 
\\
\\
{\bf Small-$\xi$ regime}
\\
We then compute the integral in (\ref{REgluonFinal2EVOLVEDHEnergy5INPUTgluon}) in the small-$\xi$ regime, i.e., $\tilde{\tau}=\text{log}(1/\xi)\rightarrow\infty$ at fixed large $\sqrt{\lambda}$ where the integral is dominated by its end point, and can be approximated by
\be
\int_{\tilde{\tau}}^{\infty}d\tilde{\tau}^{\prime}\,\frac{2e^{-2\tilde{\tau}^{\prime}/\sqrt{\lambda}-\sqrt{\lambda}/2\tilde{\tau}^{\prime}}}{\pi\tilde{\tau}^{\prime 3/2}}=(\sqrt{\lambda}/\pi)\times e^{-2\tilde{\tau}/\sqrt{\lambda}}\times\frac{e^{-\sqrt{\lambda}/2\tilde{\tau}}}{\tilde{\tau}^{3/2}}\left(1+\mathcal{O}(\sqrt{\lambda}/\tilde{\tau})\right)\,.
\ee 
Finally, combining this approximation for $\text{Re}\left[{\cal A}^{LL(gluon)}\right]$ with the exact result for $\text{Im}\left[{\cal A}^{LL(gluon)}\right]$ (\ref{IMgluonFinal2EVOLVEDHEnergy5INPUTgluon}), we find the full complex amplitude with evolved gluon GPD (for input gluon GPD part)  
\bea
\label{REIMgluonFinal2EVOLVEDHEnergy6INPUTgluon}
{\cal A}^{LL(gluon)}_{\gamma^{*} p\rightarrow  \rho^{0} p} (s,t,Q,\epsilon_{L},\epsilon_{L}^{\prime}) &\simeq&  \mathcal{N}_{g}(j_{0g})\, \widehat{\mathbb{F}}^g_{j_{0g}} (\xi, t ; \mu^2)\times e\,f_V^{\prime}\,\alpha_s(\mu)\,{1\over {Q}}\,\frac{1}{2\sqrt{2} m_{N}}\,\bar u(p_2)u(p_1)\nonumber\\  
&\times& \Bigg[(\sqrt{\lambda}/2\pi)^{1/2}\times \frac{1}{\xi^{2}}\times(\sqrt{\lambda}/\pi)\times e^{-2\tilde{\tau}/\sqrt{\lambda}}\times\frac{e^{-\sqrt{\lambda}/2\tilde{\tau}}}{\tilde{\tau}^{3/2}}
\nonumber\\
&+& i\times \frac{1}{\xi^{j_{0g}}}\times (\sqrt{\lambda}/2\pi)^{1/2}\times \frac{e^{-\sqrt{\lambda}/2\tilde{\tau}}}{\tilde{\tau}^{3/2}}\Bigg]\,,\nonumber\\  
&\simeq&  \mathcal{N}_{g}(j_{0g})\,\widehat{\mathbb{F}}^g_{j_{0g}} (\xi, t ; \mu^2)\times e\,f_V^{\prime}\,\alpha_s(\mu)\,{1\over {Q}}\,\frac{1}{2\sqrt{2} m_{N}}\,\bar u(p_2)u(p_1)\nonumber\\  
&\times& \frac{1}{\xi^{2-2/\sqrt{\lambda}}}\times\left[(\sqrt{\lambda}/\pi)+i\right]\times(\sqrt{\lambda}/2\pi)^{1/2}\times \frac{e^{-\sqrt{\lambda}/2\tilde{\tau}}}{\tilde{\tau}^{3/2}}\,, 
\eea
with $j_{0g}=2-2/\sqrt{\lambda}$, and $\tilde{\tau}=\text{log}(1/\xi)$.

\section{Details on the resummation by the j-contour for valence quark GPDs contribution in the electroproduction of $\rho^+$}~\label{valencequarksum}
\\
\\
{\bf Resummation by j-contour}
\\
We rewrite the sum over odd $j=1,3,...$ in (\ref{rhoplusGPD22EVOLVEDHEnergy2}) as contour integrals in the complex j-plane 
\bea
\label{rhoplusGPD22EVOLVEDHEnergy3INPUTvalence}
&&{\cal A}^{LL(valence)}_{\gamma^{*} p\rightarrow  \rho^{+} n} (s,t,Q,\epsilon_{L},\epsilon_{L}^{\prime})\approx \int_{\mathbb C}\frac{dj}{4i}\frac{1-e^{-i\pi j}}{{\rm sin}\pi j}
\times\frac{1}{\xi^{j}}\times\frac{1}{\Gamma(\Delta_{q}(j)-2)}\times\mathcal{N}_{q(valence)}(j)\nonumber\\
&\times &\left(\widehat{\mathbb{F}}^u_{j(valence)} (\xi, t ; \mu^2) \,-\, \widehat{\mathbb{F}}^d_{j(valence)} (\xi, t ; \mu^2)\right) \times (-1)\times e\times f_V^{+}\times\alpha_s(\mu)\times{1\over {Q}}\times \frac{1}{2m_{N}}\times \bar u(p_2)u(p_1)\,,\nonumber\\
\eea
where the contour $\mathbb C $ is at the right most of the branch point of $$\frac{1}{\Gamma(\Delta_{q}(j)-2)}=\frac{1}{\Gamma\left(\lambda^{1/4}(j-j_{0q})^{1/2}\right)}$$ and leftmost of the poles at odd $j=1,3,...$. Here we have defined
\bea
\widehat{\mathbb{F}}^q_{j(valence)} (\xi, t ; \mu^2 )=\Gamma(\Delta_{q}(j)-2)\times\mathbb{F}^q_{j(valence)} (\xi, t ; \mu^2) 
\eea
in order to reveal the branch point at $j=j_{0q}=1-1/\sqrt{\lambda}$ coming from our holographic input conformal (Gegenbauer) moments of valence quark GPDs at $\mu=\mu_0\sim Q_0$.

We evaluate the amplitude (\ref{rhoplusGPD22EVOLVEDHEnergy3INPUTvalence}) by wrapping the j-plane contour $\mathbb C $ to the left
\bea
\label{rhoplusGPD22EVOLVEDHEnergy4INPUTvalence}
{\cal A}^{LL(valence)}_{\gamma^{*} p\rightarrow  \rho^{+} n} (s,t,Q,\epsilon_{L},\epsilon_{L}^{\prime})&\approx& \int_{-\infty}^{j_{0q}}\frac{dj}{2i}\frac{1-e^{-i\pi j}}{{\rm sin}\pi j}
\times\frac{1}{\xi^{j}}\times\text{Im}\left[\frac{1}{\Gamma(iy)}\right]\nonumber\\
&\times & \left(\widehat{\mathbb{F}}^u_{j(valence)} (\xi, t ; \mu^2) \,-\, \widehat{\mathbb{F}}^d_{j(valence)} (\xi, t ; \mu^2)\right)\nonumber\\
&\times&\mathcal{N}_{q(valence)}(j)\times (-1)\times e\times f_V^{+}\times\alpha_s(\mu)\times{1\over {Q}}\times\frac{1}{2m_{N}}\times \bar u(p_2)u(p_1)\,,\nonumber\\
\eea
where we have defined $$iy = i\sqrt{\sqrt{\lambda}(j_{0q}-j)}\,.$$ For, $y\rightarrow 0$, we may approximate $\frac{1}{\Gamma(iy)}\approx iy\,e^{i\gamma_E y}$ with the Euler constant $\gamma_E\approx 0.577216$, and write $\text{Im}\left[\frac{1}{\Gamma(iy)}\right]\approx y\,\text{cos}(\gamma_E y)\approx \text{sin}(y)$ for $y\rightarrow 0$ or $j\rightarrow j_{0q}$. Therefore, for $j\rightarrow j_{0q}$, we find
\bea
\label{rhoplusGPD22EVOLVEDHEnergy5INPUTvalence}
&&{\cal A}^{LL(valence)}_{\gamma^{*} p\rightarrow  \rho^{+} n} (s,t,Q,\epsilon_{L},\epsilon_{L}^{\prime})\approx\mathcal{N}_{q(valence)}(j_{0q})\times (-1)\times e\times f_V^{+}\times\alpha_s(\mu)\times {1\over {Q}}\times\frac{1}{2m_{N}}\times\bar u(p_2)u(p_1)\nonumber\\
&\times & \left(\widehat{\mathbb{F}}^u_{j_{0q}(valence)} (\xi, t ; \mu^2)-\widehat{\mathbb{F}}^d_{j_{0q}(valence)} (\xi, t ; \mu^2)\right)\times\int_{-\infty}^{j_{0q}}\frac{dj}{2}\frac{1-e^{-i\pi j}}{{\rm sin}\pi j}
\times\frac{1}{\xi^{j}}\times \text{sin}\left(\sqrt{\sqrt{\lambda}(j_{0q}-j)}\right)\,.\nonumber\\
\eea
\\
\\
{\bf Real and imaginary parts}
\\
We separate the amplitude (\ref{rhoplusGPD22EVOLVEDHEnergy5INPUTvalence}) into its real and imaginary parts $${\cal A}^{LL(valence)}_{\gamma^{*} p\rightarrow  \rho^{+} n} (s,t,Q,\epsilon_{L},\epsilon_{L}^{\prime})\equiv{\cal A}^{LL(valence)}=\text{Re}\left[{\cal A}^{LL(valence)}\right]+i\text{Im}\left[{\cal A}^{LL(valence)}\right]\,,$$ with
\bea\label{REIMrhoplusGPD22EVOLVEDHEnergy4}
\text{Re}\left[{\cal A}^{LL(valence)}\right]&=&\mathcal{N}_{q(valence)}(j_{0q})\times \left(\widehat{\mathbb{F}}^u_{j_{0q}(valence)} (\xi, t ; \mu^2) \,-\, \widehat{\mathbb{F}}^d_{j_{0q}(valence)} (\xi, t ; \mu^2)\right)\nonumber\\
&\times & (-1)\times e\,f_V^{+}\,\alpha_s(\mu)\,{1\over {Q}}\,\frac{1}{2 m_{N}}\,\bar u(p_2)u(p_1) \times \int_{-\infty}^{j_{0q}}\frac{dj}{2}\frac{1-\text{cos}\pi j}{{\rm sin}\pi j}
\times\frac{1}{\xi^{j}}\times \text{sin}\left(\sqrt{\sqrt{\lambda}(j_{0q}-j)}\right)\,,\nonumber\\
\text{Im}\left[{\cal A}^{LL(valence)}\right]&=&\mathcal{N}_{q(valence)}(j_{0q})\times \left(\widehat{\mathbb{F}}^u_{j_{0q}(valence)} (\xi, t ; \mu^2) \,-\, \widehat{\mathbb{F}}^d_{j_{0q}(valence)} (\xi, t ; \mu^2)\right)\nonumber\\
&\times & (-1)\times e\,f_V^{+}\,\alpha_s(\mu)\,{1\over {Q}}\,\frac{1}{2 m_{N}}\,\bar u(p_2)u(p_1) \times\int_{-\infty}^{j_{0q}}\frac{dj}{2}\,\frac{1}{\xi^{j}}\times \text{sin}\left(\sqrt{\sqrt{\lambda}(j_{0q}-j)}\right)\,.
\eea
With the change of integration variable $y^2=\sqrt{\lambda}(j_{0q}-j)$, we turn the imaginary part into a Gaussian integral that can be evaluated exactly
\be\label{IMrhoplusGPD22EVOLVEDHEnergy4}
\text{Im}\left[{\cal A}^{LL(valence)}\right]&=&\mathcal{N}_{q(valence)}(j_{0q})\times \left(\widehat{\mathbb{F}}^u_{j_{0q}(valence)} (\xi, t ; \mu^2) \,-\, \widehat{\mathbb{F}}^d_{j_{0q}(valence)} (\xi, t ; \mu^2)\right)\nonumber\\
&\times & (-1)\times e\,f_V^{+}\,\alpha_s(\mu)\,{1\over {Q}}\,\frac{1}{2 m_{N}}\,\bar u(p_2)u(p_1) \times\frac{1}{\xi^{j_{0q}}}\times\int_{-\infty}^{\infty}\frac{dy}{2\pi i\sqrt{\lambda}}\,y\,e^{-\tilde{\tau}y^2/\sqrt{\lambda}}\,e^{iy}\,,\nonumber\\
&=&\mathcal{N}_{q(valence)}(j_{0q})\times \left(\widehat{\mathbb{F}}^u_{j_{0q}(valence)} (\xi, t ; \mu^2) \,-\, \widehat{\mathbb{F}}^d_{j_{0q}(valence)} (\xi, t ; \mu^2)\right)\nonumber\\
&\times & (-1)\times e\,f_V^{+}\,\alpha_s(\mu)\,{1\over {Q}}\,\frac{1}{2 m_{N}}\,\bar u(p_2)u(p_1) \times\frac{1}{\xi^{j_{0q}}}\times (\sqrt{\lambda}/4\pi)^{1/2}\times \frac{e^{-\sqrt{\lambda}/4\tilde{\tau}}}{\tilde{\tau}^{3/2}}\,,
\ee
with $\tilde{\tau}=\text{log}(1/\xi)$.

We compute the real part of the amplitude $\text{Re}\left[{\cal A}^{LL(valence)}\right]$ by first expanding its prefactor near $j=1$ 
\bea
\frac{1-\text{cos}\pi j}{{\rm sin}\pi j}\simeq -\frac{2}{\pi(j-1)}+\mathcal{O}(j-1)\,,
\eea
and establishing the identity $\partial_{\tilde{\tau}}\left[e^{-\tilde{\tau}}\text{Re}\left[{\cal A}^{LL(valence)}\right]\right]=-(2/\pi)e^{-\tilde{\tau}}\text{Im}\left[{\cal A}^{LL(valence)}\right]$ which gives us an approximation for the real part
\bea\label{RErhoplusGPD22EVOLVEDHEnergy4}
\text{Re}\left[{\cal A}^{LL(valence)}\right]&\simeq&\mathcal{N}_{q(valence)}(j_{0q})\times \left(\widehat{\mathbb{F}}^u_{j_{0q}(valence)} (\xi, t ; \mu^2) \,-\, \widehat{\mathbb{F}}^d_{j_{0q}(valence)} (\xi, t ; \mu^2)\right)\nonumber\\
&\times & (-1)\times e\,f_V^{+}\,\alpha_s(\mu)\,{1\over {Q}}\,\frac{1}{2m_{N}}\,\bar u(p_2)u(p_1)\times (\sqrt{\lambda}/4\pi)^{1/2}\times \frac{1}{\xi}\times\int_{\tilde{\tau}}^{\infty}d\tilde{\tau}^{\prime}\,\frac{2e^{-\tilde{\tau}^{\prime}/\sqrt{\lambda}-\sqrt{\lambda}/4\tilde{\tau}^{\prime}}}{\pi\tilde{\tau}^{\prime 3/2}}\,.\nonumber\\
\eea
Small corrections in the order of $\mathcal{O}(\text{Im}\left[{\cal A}^{LL(valence)}\right]/\text{log}(1/\xi))$ and $\mathcal{O}(\text{Im}\left[{\cal A}^{LL(valence)}\right]/\sqrt{\lambda})$ to (\ref{RErhoplusGPD22EVOLVEDHEnergy4}) can be computed in a standard perturbation series but can be ignored for $\xi\rightarrow 0$ and fixed large $\sqrt{\lambda}$. 
\\
\\
{\bf Small-$\xi$ regime}
\\
We then compute the integral in (\ref{RErhoplusGPD22EVOLVEDHEnergy4}) in the small-$\xi$ regime, i.e., $\tilde{\tau}=\text{log}(1/\xi)\rightarrow\infty$ at fixed large $\sqrt{\lambda}$ where the integral is dominated by its end point, and can be approximated by
\be
\int_{\tilde{\tau}}^{\infty}d\tilde{\tau}^{\prime}\,\frac{2e^{-\tilde{\tau}^{\prime}/\sqrt{\lambda}-\sqrt{\lambda}/4\tilde{\tau}^{\prime}}}{\pi\tilde{\tau}^{\prime 3/2}}=(\sqrt{\lambda}/2\pi)\times e^{-4\tilde{\tau}/\sqrt{\lambda}}\times\frac{e^{-\sqrt{\lambda}/4\tilde{\tau}}}{\tilde{\tau}^{3/2}}\left(1+\mathcal{O}(\sqrt{\lambda}/\tilde{\tau})\right)\,.
\ee

\section{Details of the holographic calculations}~\label{ACTION11}

A simple way to capture AdS/CFT duality in the non-conformal limit is  to model it using a slice of AdS$_5$ with various bulk fields with assigned 
anomalous dimensions and pertinent boundary values, in the so-called bottom-up approach which we will follow here using the conventions in our
recent  work in DIS scattering~\cite{Mamo:2019mka,Mamo:2021cle}.
 We consider AdS$_5$ with a soft wall with a background metric  $g_{MN}=(\eta_{\mu\nu},-1)R^2/z^2$ with the flat metric  $\eta_{\mu\nu}=(1,-1,-1,-1)$
 at the boundary.  Confinement will be described by  a harmonic background dilaton $\phi={\tilde{\kappa}_N}^2z^2$.

\subsection{Bulk vector mesons}

The vector mesons fields $L,R$  are  described by the bulk  effective action~\cite{Hirn:2005nr,Domokos:2009cq}

\begin{eqnarray}
\label{01}
S_{M}=&&-\frac 1{4g_5^2}\int d^5x e^{-\phi(z)}\sqrt{g}\,g^{MP}g^{NQ}{\rm Tr}\bigg({\cal F}^L_{MN}{\cal F}^L_{PQ}+{\cal F}^R_{MN}{\cal F}^R_{PQ}\bigg)
+\int d^5x\,\bigg(\omega^L_5({\cal A})-\omega^R_5({\cal A})\bigg)
\end{eqnarray}
with the Chern-Simons contribution

\be
\omega_5({\cal A})=
\frac{N_c}{24\pi^2}\int d^5x \,{\rm Tr}\left({\cal AF}^2+\frac 12 {\cal A}^3{\cal F}-\frac 1{10}{\cal A}^5\right)
\ee
Here ${\cal F}=d{\cal A}-i{\cal A}^2$ and ${\cal A}={\cal A}^aT^a$  with $T^0=\frac{3}{4}{\bf 1_2}$ and $T^i=\frac{1}{4}\tau^i$, with the form notation subsumed. 
Also the vector fields are given by $V=(R+L)/2$ and the  axial-vector  fields are given by $A=(R-L)/2$.
The coupling $g_5$ in (\ref{01})  is fixed by the brane embeddings in bulk, or phenomenologically as
${1}/{g_{5}^2}\equiv{N_c}/(12\pi^2)$~\cite{Cherman:2008eh}.

The flavor gauge fields solve

\begin{eqnarray}
\label{Solutionsgauge}
\Box  V^\mu  +z e^{\tilde{\kappa}_N^2 z^2}  \partial_z \Big( e^{-\tilde{\kappa}_N^2 z^2} \frac{1}{z}
\partial_z V^\mu \Big) \,=\, 0\qquad 
\Box V_z -  \partial_z \Big( \partial_\mu V^\mu \Big) \,=\, 0\,.
\end{eqnarray} 
subject to  the gauge condition

\begin{equation}
\label{gaugechoice}
 \partial_\mu V^\mu \,+\,
z e^{ \kappa^2 z^2} \partial_z \Big( e^{-\kappa^2 z^2} \frac{1}{z}
 V_z \Big) \,=\,0 \,,
\end{equation}
with the boundary condition $V_\mu (z, y) \vert_{z\to 0} \,=\, \epsilon_{\mu} (q)\, e^{-iq\cdot y}$.
The non-normalizable solutions are

\begin{eqnarray}
V_\mu (z, y) &=& \epsilon_\mu (q)\, e^{-iq\cdot y} \,\Gamma \bigg(1+\frac{Q^2}{4\tilde{\kappa}_V^2} \bigg)\,\,\tilde{\kappa}_N^2z^2 
\,\,{\cal U} \bigg(1+\frac{Q^2}{4 \tilde{\kappa}_N^2} ; 2 ; \tilde{\kappa}_N^2 z^2 \bigg)
\nonumber\\
V_z (z, y)  &=& \frac{i}{2} \,  \epsilon(q) \cdot q \,  e^{-iq\cdot y} \,\, \Gamma \bigg(1+\frac{Q^2}{4\tilde{\kappa}_V^2 } \bigg)\,\, z \,\, 
{\cal U} \bigg(1+\frac{Q^2}{4 \tilde{\kappa}_V^2} ; 1 ; \tilde{\kappa}_V^2 z^2 \bigg)\,,
\label{Gauge}
\end{eqnarray} 
with  $\,{\cal U} (a;b;w) \,$ the confluent hypergeometric functions of the second kind.  


 \subsection{Bulk Dirac fermions}

The bulk Dirac fermion action in a sliced of AdS$_5$ is

\be
S_F=\frac 1{2g_5^2}\int d^{5} x \,e^{-\phi(z)}\,\sqrt{g}\,\Big(\mathcal{L}_{F1}+\mathcal{L}_{F2}\Big)+\frac 1{2g_5^2}\int d^4 x \sqrt{-g^{(4)}}\,\Big(\mathcal{L}_{UV1}+\mathcal{L}_{UV2}\Big)\,,\nonumber\\
\label{Action}
\ee
The Dirac and Pauli  contributions to ${\cal L}_{F1,2}$ are respectively


\bea
\label{fermionAction}
\mathcal{L}_{\rm Dirac1,2}&=&\bigg( \frac{i}{2} \overline{\Psi}_{1,2} e^N_A \Gamma^A\big(\overrightarrow{D}_N^{L,R}-\overleftarrow{D}_N^{L,R}\big)\Psi_{1,2}-(\pm M+V(z))\bar{\Psi}_{1,2}\Psi_{1,2}\bigg)\,,\nonumber\\
\mathcal{L}_{\rm Pauli1,2}&=&\pm 2g_5^2\times\eta\, \bar{\Psi}_{1,2} e^M_Ae^N_B\sigma^{AB}{\cal F}^{L,R}_{MN}\Psi_{1,2}\,,\nonumber\\
\eea
with  $V(z)={\tilde \kappa}^2z^2$,  $e^N_A=z \delta^N_A$, $\sigma^{AB}=\frac i2  [\Gamma^A,\Gamma^B]$, 
and  $\omega_{\mu z\nu}=-\omega_{\mu\nu z}=\frac{1}{z}\eta_{\mu\nu}$. The Dirac gamma matrices  
$\Gamma^A=(\gamma^\mu, -i\gamma^5)$ are chosen in the chiral representation.  They  satisfy the flat
anti-commutation relation $\{\Gamma^A,\Gamma^B\}=2\eta^{AB}$. 
The left and right covariant derivatives are defined as

\bea
\overrightarrow{D}_N^{X=L,R}=&&\overrightarrow{\partial}_N +\frac{1}{8}\omega_{NAB}[\Gamma^A,\Gamma^B]-iX_N^aT^a\equiv \overrightarrow{\mathcal{D}}_N-iX_N^aT^a \nonumber\\
\overleftarrow{D}_N^{X=L,R}=&&\overleftarrow{\partial}_N +\frac{1}{8}\omega_{NAB}[\Gamma^A,\Gamma^B]+iX_N^aT^a\equiv \overleftarrow{\mathcal{D}}_N+iX_N^aT^a 
\eea
The nucleon doublet  refers to 

\be
\Psi_{1,2}\equiv 
\begin{pmatrix} 
  \Psi_{p1,2}\\ 
  \Psi_{n1,2}
\end{pmatrix}\,.
\ee
The nucleon fields in bulk form an iso-doublet $p,n$ with $1,2$ referring to their  boundary chirality  $1,2=\pm=R,L$~\cite{Hong:2006ta}.  They are dual to the boundary sources
$\Psi_{p1,2}\leftrightarrow {\cal O}_{p,\pm}$  and $\Psi_{n1,2}\leftrightarrow {\cal O}_{n,\pm}$ with anomalous dimensions
$\pm M=\pm (\Delta-2)=\pm (\tau-3/2)$. 

The equation of motions for the bulk Dirac chiral doublet  is

\bea
\label{EOM12}
&&\bigg(i e^N_A \Gamma^A D_N^{L,R} -\frac{i}{2}(\partial_N\phi)\, e^N_A \Gamma^A- (\pm M+V(z))\bigg)\Psi_{1,2}=0\,,
\eea
The normalizable solution to (\ref{EOM12}) are

\bea\label{SolutionFermions}
\Psi_1(p,z;n)&=&\psi_R(z;n)\Psi^0_{R}(p)+ \psi_L(z;n)\Psi^0_{L}(p)\nonumber\\
\Psi_2(p,z;n)&=&\psi_R(z;n)\Psi^0_{L}(p)+ \psi_L(z;n)\Psi^0_{R}(p)
\eea
with the normalized bulk wave functions

\be
&&\psi_R(z;n)=z^{\Delta}\times\tilde{\psi}_R(z;n)=n_R {\tilde{\xi}_N^{\tau-\frac{3}{2}}}
			L^{(\tau-2)}_n(\tilde{\xi}_N)\times z^{\Delta}\,,\nonumber\\
&&\psi_L(z;n)=z^{\Delta}\times\tilde{\psi}_L(z;n)=\tilde{n}_L{\tilde{\xi}_N^{\tau-1}} 
			L^{(\tau-1)}_n(\tilde{\xi}_N)\times z^{\Delta}\,,\nonumber\\
\ee
Here $\Delta=\tau + \frac{1}{2}$, $\tilde{\xi}_N=\tilde{\kappa}_N^2z^2$,  $L_n^{(\alpha)}(\tilde{\xi})$ are the generalized Laguerre, and $n_R = n_L \sqrt{\tau-1}$
and $n_L=\tilde{\kappa}^{-(\tau-2)}\sqrt{{2}/{\Gamma(\tau)}}$.  The free Weyl spinors $\Psi^0_{R/L}(p)= P_{\pm}u(p)$ and $\bar\Psi^0_{R/L}(p)=\bar u(p)P_{\mp}$,  and the free boundary spinors satisfy

\be
\bar u(p)u(p)=2m_N\qquad\qquad 
2m_N\times\bar u(p')\gamma^{\mu}u(p)=\bar u(p')(p'+p)^{\mu}u(p)\,.
\ee
The fermionic spectrum Reggeizes  $m_n^2=4\tilde{\kappa}_N^2(n+\tau-1)$.
The assignments $1=+$ and $2=-$ at the boundary are commensurate with the substitutions
$\psi_{R,L}\leftrightarrow \mp \psi_{L,R}$ by parity.

Using the  Dirac 1-form currents

\begin{eqnarray}
J^{aN}_L=&&\frac{\partial\mathcal{L}_{\rm Dirac1}}{\partial L^a_N}=\overline{\Psi}_1 e^N_A \Gamma^A T^a\Psi_1\,,\nonumber\\
J^{aN}_R=&&\frac{\partial\mathcal{L}_{\rm Dirac2}}{\partial R^a_N}=\overline{\Psi}_2 e^N_A \Gamma^A T^a\Psi_2\,,
\end{eqnarray}
and Pauli 2-form currents

\begin{eqnarray}
J^{aMN}_L=&&\frac{\partial\mathcal{L}_{\rm Pauli1}}{\partial L^a_{MN}}=+2g_5^2\times\eta^{a}\overline{\Psi}_1 e^M_Ae^N_B \sigma^{AB}T^a\Psi_1\,,\nonumber\\
J^{aMN}_R=&&\frac{\partial\mathcal{L}_{\rm Pauli2}}{\partial R^a_{MN}}=-2g_5^2\times\eta^{a}\overline{\Psi}_2 e^M_Ae^N_B\sigma^{AB}T^a\Psi_2\,.
\end{eqnarray}
we can rewrite  (\ref{fermionAction}) with the explicit isoscalar ($a=0$) and isovector ($a=3$) contributions

\bea
\label{fermionAction2}
&&\mathcal{L}_{F1}+\mathcal{L}_{F2}\supset \nonumber\\
&&\frac{i}{2} \bar{\Psi}_1 e^N_A \Gamma^A\big(\overrightarrow{\mathcal{D}}_N-\overleftarrow{\mathcal{D}}_N\big)\Psi_1-(M+V(z))\bar{\Psi}_1\Psi_1 +\frac{i}{2} \bar{\Psi}_2 e^N_A \Gamma^A\big(\overrightarrow{\mathcal{D}}_N-\overleftarrow{\mathcal{D}}_N\big)\Psi_2-(-M+V(z))\bar{\Psi}_2\Psi_2\nonumber\\
&&+V_{N}^{0}J_{V}^{0N}+A_{N}^{0}J_{A}^{0N}+V_{N}^{3}J_{V}^{3N}+A_{N}^{3}J_{A}^{3N}+V^{0}_{MN}J_{V}^{0MN}+A^{0}_{MN}J_{A}^{0MN}+V^{3}_{MN}J_{V}^{3MN}+A^{3}_{MN}J_{A}^{3MN}\nonumber\\
\eea
with $J_{V,A}^{aN}=J_{L}^{aN}\pm J_{R}^{aN}$ and $J_{V,A}^{aMN}=J_{L}^{aMN}\mp J_{R}^{aMN}$.

\subsection{Bulk glueballs}

The graviton in the bulk of AdS space is dual to a glueball on the boundary. The graviton tensor $h_{\mu\nu}$ can be decomposed into its transverse and traceless part, $h$, and its trace-full part, $f$, using \cite{Kanitscheider:2008kd}. The decomposition is given by
\be
\label{exp}
h_{\mu\nu}=\epsilon_{\mu\nu}^{TT}\,h+\epsilon_{\mu\nu}^{T}\,f-k_\mu k_\nu H+ k_\mu A_{\nu}^{\perp}+ k_\nu A_{\mu}^{\perp}\,\nonumber\\
\ee
where $k^\mu\epsilon_{\mu\nu}^{TT}= \eta^{\mu\nu}\epsilon_{\mu\nu}^{TT}=0$, and $\epsilon_{\mu\nu}^{T}=\frac{1}{4}\eta_{\mu\nu}$. In a gauge where $A_{\mu}^{\perp}=0$, the equation of motion for $h$ decouples, but the equations for $f$, $H$, and $\varphi$ are coupled (see Eqs.7.16-20 in \cite{Kanitscheider:2008kd}). Diagonalizing the equations shows that $f$ satisfies the same equation of motion as $h$ \cite{Kanitscheider:2008kd}. It's important to note that $f$ couples to $T^{\mu}_{\mu}$ of the gauge theory, while $H$ couples to $k^{\mu}k^{\nu}T_{\mu\nu}\equiv 0$ (see Eq.7.6 of \cite{Kanitscheider:2008kd}). 

The effective action for the graviton  ($\eta_{\mu\nu}\rightarrow\eta_{\mu\nu}+h_{\mu\nu}$) and dilaton fluctuations ($\phi\rightarrow\phi+\varphi$) follows from the Einstein-Hilbert action plus dilaton by expanding to quadratic order, and after adding the  background de-Donder gauge fixing term. The result is

\be
S=\int d^{5} x \sqrt{g}\,e^{-2\phi}\big(\mathcal{L}_{h+f}+\mathcal{L}_\varphi \big)\,,\nonumber\\
\label{Action2}
\ee
with

\bea
\label{kinetic}
\mathcal{L}_{h+f} =&& -\frac{1}{4\tilde{g}_5^2}\,g^{\mu\nu}\,\eta^{\lambda\rho}\eta^{\sigma\tau}\partial_{\mu}h_{\lambda\sigma}\partial_{\nu}h_{\rho\tau}+\frac{1}{8\tilde{g}_5^2}\,g^{\mu\nu}\eta^{\alpha\beta}\eta^{\gamma\sigma}\,\partial_{\mu}h_{\alpha\beta}\,\partial_{\nu}h_{\gamma\sigma}\,,\nonumber\\
\mathcal{L}_\varphi=&&+\frac{1}{2\tilde{g}_5^2}\,g^{\mu\nu}\,\partial_{\mu}\varphi\,\partial_{\nu}\varphi\,,
\ee
and $\tilde{g}_5^2=2\kappa^2=16\pi G_N={8\pi^2}/{N_c^2}$. Here $g_{\mu\nu}$ is the AdS metric.


For the graviton in the axial gauge $h_{\mu z}=h_{zz}=0$.
The pertinent couplings follow from linearizing the action (\ref{Action}) by replacing $\eta_{\mu\nu}\rightarrow\eta_{\mu\nu}+h_{\mu\nu}$, are

  \be
 h\overline\Psi\Psi:\quad &&-\frac{\sqrt{2\kappa^2}}{2}\int d^5x\,\sqrt{g}\,h_{\mu\nu}T_F^{\mu\nu}\nonumber\\
  h AA:\quad && -\frac{\sqrt{2\kappa^2}}{2}\int d^5x\,\sqrt{g}\,h_{\mu\nu}T_V^{\mu\nu}\nonumber\\
  \label{vertices1}
 \ee
with the energy-momentum tensors

 \bea
T_F^{\mu\nu}&=&e^{-\phi}\frac{i}{2}\,z\,\overline\Psi\gamma^\mu\overset{\leftrightarrow}{\partial^\nu}\Psi-\eta^{\mu\nu}\mathcal{L}_F\,,\nonumber\\
T_V^{\mu\nu} &=&-e^{-\phi}\Big(z^4\eta^{\rho\sigma}\eta^{\mu\beta}\eta^{\nu\gamma}\,F^V_{\beta\rho}F^V_{\gamma\sigma}-z^4\,\eta^{\mu\beta}\eta^{\nu\gamma}\,F^V_{\beta z}F^V_{\gamma z}\Big)-\eta^{\mu\nu}\mathcal{L}_V\,.
  \label{EMT}
 \eea

\subsection{t-channel spin-2 glueball exchange}

In the soft-wall model the normalized wave function for spin-2 glueballs is given by \cite{BallonBayona:2007qr} (note that the discussion in \cite{BallonBayona:2007qr} is for general massive bulk scalar fluctuation but can be used for spin-2 glueball which has an effective bulk action similar to massless bulk scalar fluctuation)
\bea
 \label{wfSW}
&&J_{h}(m_n,z)\equiv \psi_{n}(z)=c_n\,z^{4}L_{n}^{2}(2\tilde{\xi}_T)\,,\nonumber\\
\eea
with $\tilde{\xi}_T=\tilde{\kappa}_T^2z^2$, and
\be
c_n=\Bigg(\frac{2^{4}\tilde{\kappa}_{T}^{6}\Gamma(n+1)}{\Gamma(n+3)}\Bigg)^{\frac 12}=\frac{4\tilde{\kappa}_{T}^{3}}{\sqrt{(n+2)(n+1)}}\,,
\ee
which is determined from the normalization condition (for soft-wall model with background dilaton $\phi=\tilde{\kappa}_T^2z^2$)
\be
\int dz\,\sqrt{g}e^{-\phi}\,\abs{g^{xx}}\,\psi_n(z)\psi_m(z)=\delta_{nm}\,.\nonumber\\
\ee
Therefore we have
\be
F_n=\frac{1}{\sqrt{2}\kappa}\bigg(-\frac{1}{z^{\prime 3}}\partial_{z^\prime}\psi_n(z^\prime)\bigg)_{z^\prime=\epsilon}=-\frac{4}{\sqrt{2}\kappa}c_n L_{n}^{2}(0)\,,\nonumber\\
\ee
with $\psi_n(z\rightarrow 0)\approx c_n\,z^{4}L_{n}^{2}(0)$ and with $L_{n}^{2}(0)=\binom{2+n}{n}$. We can also re-write the normalized wave function of glueballs (\ref{wfSW}) in terms of $F_n$ as
\be\label{psigluon}
\psi_{n}(z)=-\sqrt{2}\kappa\times F_{n}\times\frac{1}{4}\times\frac{1}{\tilde{\kappa}_T^4}\times\frac{1}{L_{n}^{2}(0)}\times\tilde{\xi}_T^2 L_{n}^{2}(2\tilde{\xi}_T)\,,
\ee
 Also note that $m_n^2=8\tilde{\kappa}_T^2(n+2)$,
\be\label{FN2}
F_n^2=\frac{16}{2\kappa^2}\times \frac{16\tilde{\kappa}_T^6}{(n+2)(n+1)}\times \left(L_{n}^{2}(0)\right)^2\,.
\ee

For space-like momenta ($t=k^2=-K^2$), we have the bulk-to-bulk propagator near the boundary
\be
G(z\rightarrow 0,z^{\prime})\approx \frac{z^4}{4}\sum_n \frac{\sqrt{2}\kappa F_n\psi_n(z^{\prime})}{K^2+m_n^2}=\frac{z^4}{4}\mathcal{H}(K,z^{\prime}) , \nonumber\\\label{hbbt3SW2}
\ee
where, for the soft-wall model~\cite{Abidin:2008ku,BallonBayona:2007qr}

\bea
\mathcal{H}(K,z)&=&\sum_n \frac{\sqrt{2}\kappa F_n\psi_n(z^{\prime})}{K^2+m_n^2}\,,\nonumber\\
&=&4z^{4}\Gamma(\frac{a_K}{2} +2)U\Big(\frac{a_K}{2}+2,3;2\tilde{\xi}_T\Big)
=\Gamma(\frac{a_K}{2}+2)U\Big(\frac{a_K}{2},-1;2\tilde{\xi}_T\Big)\nonumber\\
&=&\frac{\Gamma(\frac{a_K}{2}+2)}{\Gamma(\frac{a_K}{2})}
\int_{0}^{1}dx\,x^{\frac{a_K}{2}-1}(1-x){\rm exp}\Big(-\frac{x}{1-x}(2\tilde{\xi}_T)\Big)\,,
\label{BBSWj2}
\eea
with $\tilde{\xi}_T=\tilde{\kappa}_T^2z^2$, $a_K={K^2}/{4\tilde{\kappa}_T^2}$, and we have used the transformation $U(m,n;y)=y^{1-n}U(1+m-n,2-n,y)$. (\ref{BBSWj2}) satisfies the normalization condition ${\cal H}(0,z)={\cal H}(K,0)=1$.

For example, the t-channel spin-2 glueball exchange contribution to electroproduction of vector mesons, is given by
\be
\label{Amph}
&&i{\cal A}^{h}_{\gamma^{*}_{T/L}p\rightarrow  Vp} (s,t)=\sum_n i\tilde{{\cal A}}^{h}_{\gamma^{*}_{T/L}p\rightarrow  Vp} (m_n,s,t)\nonumber\\
&&i\tilde{{\cal A}}^{h}_{\gamma_{T/L}^{*}p\rightarrow  Vp} (m_n,s,t)=\frac{1}{g_5}\times(-i)V_{h\gamma^{*}_{T/L}V}^{\mu\nu}(q,q^{\prime},k,m_n)\times \tilde{G}_{\mu\nu\alpha\beta}(m_n,k)\times
 (-i)V_{h\bar\Psi\Psi}^{\alpha\beta}(p_1,p_2,k,m_n)\,,	\nonumber\\
\ee
with the bulk vertices (defining $t=k^2=\Delta^2$, and $k=\Delta=p_2-p_1=q-q^{\prime}$)

\be
&&V_{h\gamma^{*}_{T/L}V}^{\mu\nu}(q,q^{\prime},k,m_n)\equiv \left(\frac{\delta S_{h\gamma^{*}_{T/L}V}^k}{\delta (\epsilon_{\mu\nu}h(k,z))}\right)\,J_{h}(m_n,z)=\sqrt{2\kappa^2}\times\frac{1}{2}\int dz\sqrt{g}\,e^{-\phi}z^4K_{T/L}^{\mu\nu}(q,q^{\prime},\epsilon,\epsilon^{\prime},z)J_{h}(m_n,z)\,,\nonumber\\
&&V_{h\bar\Psi\Psi}^{\alpha\beta}(p_1,p_2,k,m_n)\equiv \left(\frac{\delta S_{h\bar\Psi\Psi}^k}{\delta (\epsilon_{\alpha\beta}h(k,z))}\right)\,J_{h}(m_n,z)=-\sqrt{2\kappa^2}\times\frac{1}{2}\int dz\sqrt{g}\,e^{-\phi}z\bar\Psi(p_2,z)\gamma^\alpha p^\beta\Psi(p_1,z)J_{h}(m_n,z)\,,\nonumber\\ \label{vh}
\ee
where we have defined the kinematic factors (in the high energy limit $s\gg -t$ with $p=({p_1+p_2})/{2}\approx p_1$, and $q^{\mu}\approx q^{\prime\mu}$) as
\bea
&&K_{T}^{\mu\nu}(q,q^{\prime},\epsilon_T,\epsilon^{\prime}_T,z)\approx -q^\mu q^{\nu}\times\mathcal{V}_{\gamma^{*}}(Q,z)\mathcal{V}_{V}(M_{V},z)\equiv q^\mu q^{\nu}\times K_{T}(Q,M_V,z) 
\,,\nonumber\\
&&K_{L}^{\mu\nu}(q,q^{\prime},\epsilon_L,\epsilon^{\prime}_L,z)\approx - q^\mu q^{\nu}\times\frac{1}{Q\,M_V}\times \partial_z\mathcal{V}_{\gamma^{*}}(Q,z)\partial_z\mathcal{V}_{V}(M_V,z)\equiv q^\mu q^{\nu}\times K_{L}(Q,M_V,z) \,,\nonumber\\
\label{BKhighenergy}
\eea
using
\bea
\mathcal{V}_{\gamma^{*}}(Q,z)
&=&\tilde\xi \,\Gamma(1+a_Q)\,\,{\cal U} (1+a_Q; 2 ; \tilde\xi)
=\tilde\xi\int_{0}^{1}\frac{dx}{(1-x)^2}x^{a_Q}{\rm exp}\Big(-\frac{x\,\tilde\xi}{1-x}\Big)\,,\nonumber\\
\mathcal{V}_{V}(M_V,z)&\equiv&\phi_0(z)=g_5\times\frac{f_V}{M_V}\times 2\tilde\xi L_0^1(\tilde\xi)\,,
\label{vps2swQMV}
\eea
with $a_Q=Q^2/(4\tilde{\kappa}_V^2)$ and $\tilde\xi=\tilde{\kappa}_V^2z^2$.

The full bulk-to-bulk spin-2 glueball (graviton) propagator $G_{\mu\nu\alpha\beta}(m_n,k,z,z^{\prime})$ is given by \cite{Raju:2011mp,DHoker:1999bve}
\be
 \label{Gh}
&&G_{\mu\nu\alpha\beta}(m_n,k,z,z^{\prime})=J_{h}(m_n,z)\times \tilde{G}_{\mu\nu\alpha\beta}(m_n,k)\times J_{h}(m_n,z^{\prime})\,,\nonumber\\
\ee
where the massive spin-2 boundary propagator $\tilde{G}_{\mu\nu\alpha\beta}(m_n,k)$ is given by
\be\label{spin2boundary}
&&\tilde{G}_{\mu\nu\alpha\beta}(m_n,k)=
P_{\mu\nu;\alpha\beta}(m_n,k)\times\frac{-i} {k^2-m_n^2+i\epsilon}\,,
\ee
with the massive spin-2 projection operator $P_{\mu\nu;\alpha\beta}(m_n,k)$ defined as
\be
P_{\mu\nu;\alpha\beta}(m_n,k)={1 \over 2} \left(P_{\mu;\alpha} P_{\nu;\beta} + P_{\mu;\beta} P_{\nu;\alpha} -
\frac 23 P_{\mu;\nu} P_{\alpha;\beta}\right)\,,
\ee
which is written in terms of the massive spin-1 projection operator
\be
P_{\mu;\alpha}(m_n,k)= \eta_{\mu\alpha} - k_{\mu} k_{\alpha}/m_n^2\,.
\ee

For $z\rightarrow 0$, $t=-K^2$, and focusing on the transverse part of the boundary propagator, i.e., $$\tilde{G}_{\mu\nu\alpha\beta}(m_n,k)\approx\frac{1}{2}\eta_{\mu\alpha}\eta_{\nu\beta}\times \frac{-i} {k^2-m_n^2+i\epsilon}$$ (as we did in~\cite{Mamo:2019mka}), we can simplify (\ref{Amph}) as

\be
\label{nAmph}
&&i{\cal A}^{h}_{\gamma^{*}_{T/L}p\rightarrow  Vp} (s,t)\approx\frac{1}{g_5}\times(-i)\mathcal{V}^{\mu\nu}_{h\gamma^{*}_{T/L}V}(q,q^{\prime},k)\times \bigg(\frac{-i}{2}\eta_{\mu\alpha}\eta_{\nu\beta}\bigg)\times(-i)\mathcal{V}^{\alpha\beta}_{h\bar\Psi\Psi}(p_1,p_2,k)\,,\nonumber\\
\ee
with
\be
&&\mathcal{V}^{\mu\nu}_{h\gamma^{*}_{T/L}V}(q,q^{\prime},k_)=\sqrt{2\kappa^2}\times\frac{1}{2}\int dz\sqrt{g}\,e^{-\phi}z^4K_{T/L}^{\mu\nu}(q,q^{\prime},\epsilon,\epsilon^{\prime},z)\frac{z^4}{4}\,,\nonumber\\
&&\mathcal{V}^{\alpha\beta}_{h\bar\Psi\Psi}(p_1,p_2,k_z)=-\sqrt{2\kappa^2}\times\frac{1}{2}\int dz\,\sqrt{g}\,e^{-\phi}z\,\bar\Psi(p_2,z)\gamma^\mu p^\nu\,\Psi(p_1,z)\mathcal{H}(K,z)\,.
\ee


However, if we instead use the full massive spin-2 boundary propagator $\tilde{G}_{\mu\nu\alpha\beta}(m_n,k)$ (\ref{spin2boundary}), we find (particularly for longitudinal vector meson production)
\be
\label{nAmphfull}
{\cal A}^{h}_{\gamma_{L}^{*}p\rightarrow Vp} (s,t) &=& \frac{1}{g_5}\times 2\kappa^2\times\frac{1}{QM_V}\times\mathcal{V}_{h\gamma_L^{*}V}(Q,M_V)\times\frac{1}{m_N}\times\bar u(p_2)u(p_1)\nonumber\\
&\times &\sum_{n=0}^{\infty}\,\left[q^{\mu}q^{\nu}\,P_{\mu\nu;\alpha\beta}(m_n,k) \,p_1^{\alpha}p_1^{\beta}\right]\times\mathcal{V}_{h\bar\Psi\Psi}(K,m_n)\,,\nonumber\\
\ee
with
\be
&&\mathcal{V}_{h\gamma^{*}_LV}(Q,M_V)=-\frac{1}{2}\int dz\sqrt{g}\,e^{-\phi}z^4\partial_z\mathcal{V}_{\gamma^{*}}(Q,z)\partial_z\mathcal{V}_{V}(M_V,z)\frac{z^4}{4}\,,\nonumber\\
&&\mathcal{V}_{h\bar\Psi\Psi}(K,m_n)=-\frac{1}{2}\int dz\,\sqrt{g}\,e^{-\phi}z\,\left(\psi_R^2(z)+\psi_L^2(z)\right)\times\frac{\sqrt{2}\kappa F_{n}\psi_n(z)}{K^2+m_n^2}\,.\nonumber\\
\ee

Following~\cite{Nishio:2014eua}, we can evaluate  
\bea\label{spin2contraction}
q^{\mu}q^{\nu}\,P_{\mu\nu;\alpha\beta}(m_n,k) \,p_1^{\alpha}p_1^{\beta}=\left(p_1\cdot q\right)^{2}\times \left(1+\frac{4}{3}\frac{p_1^2}{m_{n}^2}\times\eta^2\right)\,,
\eea
using the general result (see Eq.279 in~\cite{Nishio:2014eua})
\be\label{spinjcontraction}
q^{\mu_1}q^{\mu_2}...q^{\mu_j}\,P_{\mu_1\mu_2...\mu_j;\nu_1\nu_2...\nu_j}(m_n,k) \,p_1^{\nu_1}p_1^{\nu_2}...p_1^{\nu_j}=\left(p_1\cdot q\right)^{j}\times \hat{d}_{j}(\eta,m_n^2)
\ee
where $\hat{d}_{j}(\eta,m_n^2)$, for even $j=2,4,...$, is a polynomial of skewness $\eta$ of degree $j$ which can be written explicitly in terms of the hypergeometric function $_2F_1(a,b,c,d)$ as 
\be\label{etapoly}
\hat{d}_{j}(\eta,m_n^2)=\, _2F_1\left(-\frac{j}{2},\frac{1-j}{2};\frac{1}{2}-j;-\frac{4p_1^2}{m_n^2}\times\eta ^2\right)\,.
\ee
Note that we have also replaced $p^2\sim p_1^2$ by $-p_1^2$, and $-\Delta^2$ by $m_n^2$ in Eq.279 of~\cite{Nishio:2014eua}, since we are using the massive spin-1 projection operator $P_{\mu;\alpha}(m_n,k)= -\eta_{\mu\alpha} + k_{\mu} k_{\alpha}/m_n^2$ instead of the massless one $P_{\mu;\alpha}(\Delta=k)= \eta_{\mu\alpha} - k_{\mu} k_{\alpha}/\Delta^2$ used in~\cite{Mamo:2019mka} for the conformal case with $\eta_{\mu\alpha}=(-,+,+,+)$ signature.

Therefore, using (\ref{spin2contraction}) in (\ref{nAmphfull}), we find 
\be
\label{nAmphfull2}
{\cal A}^{h}_{\gamma_{L}^{*}p\rightarrow Vp} (s,t) &=& -\frac{1}{g_5^2}\times 2\kappa^2\times\frac{1}{QM_V}\times\mathcal{V}_{h\gamma_L^{*}V}(Q,M_V)\times\frac{1}{m_N}\times\bar u(p_2)u(p_1)\nonumber\\
&\times & \left(p_1\cdot q\right)^{2}\times\left[A(K,\tilde{\kappa}_T)+\eta^2 D(K,\tilde{\kappa}_T,\tilde{\kappa}_S)\right]\,,\nonumber\\
\ee
with the A and D gravitational form factors~\cite{Mamo:2019mka,Mamo:2021tzd,Mamo:2022eui}
\bea\label{AK1}
A(K,\tilde{\kappa}_T)&\equiv&\frac{1}{2}\int dz\,\sqrt{g}\,e^{-\phi}z\,\left(\psi_R^2(z)+\psi_L^2(z)\right)\times\sum_{n=0}^{\infty}\frac{\sqrt{2}\kappa F_{n}\psi_n(z)}{K^2+m_n^2}\,,\nonumber\\
&=&\frac{1}{2}\int dz\,\sqrt{g}\,e^{-\phi}z\,\left(\psi_R^2(z)+\psi_L^2(z)\right)\times\mathcal{H}(K,z)\,,\nonumber\\
&=&A(0)\times 6\times\frac{\Gamma \left(2+\frac{a_K}{2}\right)}{\Gamma \left(4+\frac{a_K}{2}\right)}\times \, _2F_1\left(3,\frac{a_K}{2};\frac{a_K}{2}+4;-1\right)\,,
\eea
and
\bea\label{DK1}
D(K,\tilde{\kappa}_T)&\equiv&\frac{1}{2}\int dz\,\sqrt{g}\,e^{-\phi}z\,\left(\psi_R^2(z)+\psi_L^2(z)\right)\times\sum_{n=0}^{\infty}\,\frac{4}{3}\frac{p_1^2}{m_n^2}\times\frac{\sqrt{2}\kappa F_{n}\psi_n(z)}{K^2+m_n^2}\,,\nonumber\\
&=&\frac{1}{2}\int dz\,\sqrt{g}\,e^{-\phi}z\,\left(\psi_R^2(z)+\psi_L^2(z)\right)\times\sum_{n=0}^{\infty}\,\frac{4}{3}\frac{p_1^2}{k^2}\times\frac{\sqrt{2}\kappa F_{n}\psi_n(z)}{-k^2+m_n^2}\,,\nonumber\\
&=&-\frac{4}{3}\frac{m_N^2}{K^2}\times A(K,\tilde{\kappa}_T)\,.
\eea
Following our arguments in~\cite{Mamo:2021krl,Mamo:2022eui}, we also replace
\bea\label{DK2}
D(K, \tilde{\kappa}_T)\rightarrow D(K,\tilde{\kappa}_T,\tilde{\kappa}_S)&=&-\frac{4}{3}\frac{m_N^2}{K^2}\times \left[A(K,\tilde{\kappa}_T)-A_{S}(K,\tilde{\kappa}_S)\right]\,,
\eea
where
\bea
\label{AS11}
A_{S}(K,\tilde{\kappa}_S)&=&A(0)\times 6\times\frac{\Gamma \left(2+\frac{\tilde{a}_K}{2}\right)}{\Gamma \left(4+\frac{\tilde{a}_K}{2}\right)}\times \, _2F_1\left(3,\frac{\tilde{a}_K}{2};\frac{\tilde{a}_K}{2}+4;-1\right)\,,
\eea
with $\tilde{a}_K=K^2/4\tilde{\kappa}_S^2$.

\subsection{t-channel spin-j glueball exchange}


In the soft-wall model, the spin-j glueballs' normalized wave functions are given in terms of the generalized Laguerre polynomials as~\cite{Mamo:2019mka,Mamo:2021tzd}
\be
 \label{wfSW}
&&\psi_{n}(j,z)=c_n(j)\,z^{\Delta}L_{n}^{\Delta_g(j)-2}(2\tilde{\xi}_T)\,,
\ee
where $\tilde{\xi}_T=\tilde{\kappa}_{T}^2z^2$, and the normalization coefficients are
\be
c_n(j)=\Big(\frac{2^{\Delta_g(j)}\tilde{\kappa}_{T}^{2(\Delta_g(j)-1)}\Gamma(n+1)}{\Gamma(n+\Delta_g(j)-1)}\Big)^{\frac 12}\,,
\ee
with
\bea
\Delta_g(j)
&=&2+\sqrt{2\sqrt{\lambda}(j-j_{0g})}\,,
\eea
where $j_{0g}=2-2/\sqrt{\lambda}$ for closed strings.

The non-normalized bulk-to-boundary propagators for spin-j glueballs are given in terms of Kummer's (confluent hypergeometric) function of the second kind and its integral representation as (for space-like momenta $k^2=-K^2$)
\bea
\mathcal{H}(j,K,z)
&&=z^{\Delta}U\Big(\frac{a_K}{2}+\frac{\Delta_g(j)}{2},\Delta_g(j)-1;2\tilde{\xi}_T\Big)
=z^{\Delta_g(j)}(2\tilde{\xi}_T)^{2-\Delta_g(j)}U\Big(\tilde{a}(j),\tilde{b}(j);2\tilde{\xi}_T\Big)\nonumber\\
&&=z^{\Delta_g(j)}(2\tilde{\xi}_T)^{2-\Delta_g(j)}\frac{1}{\Gamma(\tilde{a}(j))}
\int_{0}^{1}dx\,x^{\tilde{a}(j)-1}(1-x)^{-\tilde{b}(j)}{\rm exp}\Big(-\frac{x}{1-x}(2\tilde{\xi}_T)\Big)\,,
\label{BBSWj}
\eea
where
\be
a_K=\frac{K^2}{4\tilde{\kappa}_T^2}\qquad
\tilde{a}(j)=\frac{a_K}{2}+2-\frac{\Delta_g(j)}{2}\qquad
\tilde{b}(j)=3-\Delta_g(j)
\ee
and we have used the transformation $U(m,n;y)=y^{1-n}U(1+m-n,2-n,y)$.
The bulk-to-bulk propagator can also be approximated as (for space-like momenta $k^2=-K^2$)
\be
G(j,z\rightarrow 0,z^{\prime})&\approx& \frac{\psi_{n}(j,z\rightarrow 0)}{(\sqrt{2}\kappa)\mathcal{F}_n(j)}\times(\sqrt{2}\kappa)\sum_n \frac{ \mathcal{F}_n(j)\psi_n(j,z^{\prime})}{K^2+m_n^2(j)}\nonumber\\
&=&\frac{2^{\Delta_g(j)-2}\Gamma(\frac{a_K}{2}+\frac{\Delta_g(j)}{2})}{\Gamma(\Delta_g(j)-2)}\times \tilde{\kappa}_{T}^{2\Delta_g(j)-4}\times \frac{z^{\Delta_g(j)}}{\Delta_g(j)}\times\mathcal{H}(j,K,z^{\prime}) , \nonumber\\\label{hbbt2jSW}
\ee
where we used
\bea
\mathcal{H}(j,K,z^{\prime})=&&\sum_n \frac{\sqrt{2}\kappa\mathcal{F}_n(j)\psi_n(j,z^{\prime})}{K^2+m_n^2(j)}\,,\nonumber\\
\mathcal{F}_n(j)=&&\frac{\mathcal{C}(j,K,\epsilon)}{\sqrt{2}\kappa}\times (-\sqrt{g}\,e^{-\phi}\,\abs{g^{xx}}\,\partial_{z^{\prime}}\psi_n(j,z^{\prime}))|_{z^{\prime}=\epsilon}\,,\nonumber\\
\mathcal{C}(j,K,\epsilon)=&&\mathcal{H}(j,K,\epsilon)\,,
\eea
with $\psi_n(j,z\rightarrow 0)\approx c_n(j)\,z^{\Delta}L_{n}^{\Delta_g(j)-2}(0)$ for the soft-wall model.

For example, the electroproduction of vector mesons (using the spin-j glueball bulk-to-boundary propagator (\ref{hbbt2jSW})) is given by~\cite{Mamo:2021tzd}
\be\label{spinjAMP}
{\cal A}^{h}_{\gamma^{*}_{L/T} p\rightarrow V p} (j,s,t)&=&-\frac{1}{g_5}\times 2\kappa^2\times\mathcal{V}^{\mu\nu}_{h\gamma^{*}_{L/T}V}(j,Q,M_V)\times\frac{1}{m_N}\times\bar u(p_2)u(p_1)\nonumber\\
&\times &\sum_{n=0}^{\infty}\,\left[q^{\mu_1}q^{\mu_2}...q^{\mu_j}\,P_{\mu_1\mu_2...\mu_j;\nu_1\nu_2...\nu_j}(m_n,k) \,p_1^{\nu_1}p_1^{\nu_2}...p_1^{\nu_j}\right]\times\mathcal{V}_{h\bar\Psi\Psi}(j,K,m_n)\,,	\nonumber\\
\ee
where
\bea
\label{VSWjSW}
\mathcal{V}_{h\gamma^{*}_{L/T}V}(j,Q,M_V)&=&\frac{1}{2}\int dz\sqrt{g}\,e^{-\phi}\,z^{4+2(j-2)}\nonumber\\
&&\times K_{T/L}(Q,M_V)\times \frac{z^{\Delta_g(j)-(j-2)}}{\Delta_g(j)}\times \tilde{\kappa}_V^{2\Delta_g(j)-4}\times \frac{1}{\Gamma(\Delta_g(j)-2)}\nonumber\\
\mathcal{V}_{h\bar\Psi\Psi}(j,K,m_n)
&=&\frac{1}{2}\int dz\,\sqrt{g}\,e^{-\phi}\,z^{1+2(j-2)}\,\big(\psi_R^2(z)+\psi_L^2(z)\big)\,z^{-(j-2)}\nonumber\\
&&\times\frac{\sqrt{2}\kappa\mathcal{F}_n(j)\psi_n(j,z)}{K^2+m_n^2(j)}\times 2^{\Delta_g(j)-2}\,\Gamma\left(\frac{a_K}{2}+\frac{\Delta_g(j)}{2}\right)\times\tilde{\kappa}_T^{j-2+\Delta_g(j)}\,,
\eea
with
\bea
\label{PARA}
&&\Delta_g(j)=2+\sqrt{2\sqrt{\lambda}(j-j_{0g})}\,,\qquad \frac {a_K}{2}=\frac{K^2}{8{\tilde\kappa}_T^2}\,,\qquad j_{0g}=2-\frac{2}{\sqrt{\lambda}}\,,
\eea
and
\bea
&&K_{T}(Q,M_V,z)\equiv -\mathcal{V}_{\gamma^{*}}(Q,z)\mathcal{V}_{V}(M_{V},z)\,,\nonumber\\
&&K_{L}(Q,M_V,z)\equiv -\frac{1}{Q\,M_V}\times\partial_z\mathcal{V}_{\gamma^{*}}(Q,z)\partial_z\mathcal{V}_{V}(M_V,z) \,.\nonumber\\
\label{BKhighenergy}
\eea

Finally using (\ref{spinjcontraction}) 
\be\label{spinjcontraction2}
q^{\mu_1}q^{\mu_2}...q^{\mu_j}\,P_{\mu_1\mu_2...\mu_j;\nu_1\nu_2...\nu_j}(m_n,k) \,p_1^{\nu_1}p_1^{\nu_2}...p_1^{\nu_j}=\left(p_1\cdot q\right)^{j}\times \hat{d}_{j}(\eta,m_n^2)
\ee
in (\ref{spinjAMP}), we find
\be\label{spinjAMP2}
{\cal A}^{h}_{\gamma^{*}_{L/T} p\rightarrow V p} (j,s,t)&=&-\frac{1}{g_5}\times 2\kappa^2\times\mathcal{V}^{\mu\nu}_{h\gamma^{*}_{L/T}V}(j,Q,M_V)\times\left(p_1\cdot q\right)^{j}\times\frac{1}{m_N}\times\bar u(p_2)u(p_1)\nonumber\\
&\times &\left[\mathcal{A}(j,\tau,\Delta_g(j),K,\tilde{\kappa}_T)+\mathcal{D}_{\eta}(j,\tau,\Delta_g(j),K,\tilde{\kappa}_T,\tilde{\kappa}_S)\right]\,,	\nonumber\\
\ee
where we have defined the spin-j moments (or spin-j form factors), for even $j=2,4,..$, as
\bea\label{spinjFFA}
\mathcal{A}(j,\tau,\Delta_g(j),K,\tilde{\kappa}_T)&=&\frac{1}{2}\int dz\,\sqrt{g}\,e^{-\phi}\,z^{1+2(j-2)}\,\big(\psi_R^2(z)+\psi_L^2(z)\big)\,z^{-(j-2)}\nonumber\\
&&\times\sum_n \frac{\sqrt{2}\kappa\mathcal{F}_n(j)\psi_n(j,z)}{K^2+m_n^2(j)}\times 2^{\Delta_g(j)-2}\,\Gamma\left(\frac{a_K}{2}+\frac{\Delta_g(j)}{2}\right)\times\tilde{\kappa}_T^{j-2+\Delta_g(j)}\,,\nonumber\\
&=&\frac{1}{2}\int dz\,\sqrt{g}\,e^{-\phi}\,z^{1+2(j-2)}\,\big(\psi_R^2(z)+\psi_L^2(z)\big)\,z^{-(j-2)}\nonumber\\
&&\times\mathcal{H}(j,K,z)\times 2^{\Delta_g(j)-2}\times\Gamma\left(\frac{a_K}{2}+\frac{\Delta_g(j)}{2}\right)
\times \tilde{\kappa}_T^{j-2+\Delta_g(j)}\,,\nonumber\\
&=&\frac{2^{1-\Delta_{g}(j) }}{\Gamma(\tau)} \Bigg( (\tau -1) \Gamma \left(\frac{j-2}{2}+\tau -\frac{\Delta_{g}(j) }{2}+1\right) \Gamma \left(\frac{j-2}{2}+\frac{\Delta_{g}(j) }{2}+\tau -1\right)\nonumber\\
&\times & \, _2\tilde{F}_1\left(\frac{a_K}{2}-\frac{\Delta_{g}(j) }{2}+2,\frac{j-2}{2}+\tau -\frac{\Delta_{g}(j) }{2}+1;\frac{a_K}{2}+\frac{j-2}{2}+\tau +1;-1\right)\nonumber\\
&&+\Gamma \left(\frac{j-2}{2}+\tau -\frac{\Delta_{g}(j) }{2}+2\right) \Gamma \left(\frac{j-2}{2}+\frac{\Delta_{g}(j) }{2}+\tau \right)\nonumber\\
&\times & _2\tilde{F}_1\left(\frac{a_K}{2}-\frac{\Delta_{g}(j) }{2}+2,\frac{j-2}{2}+\tau -\frac{\Delta_{g}(j) }{2}+2;\frac{a_K}{2}+\frac{j-2}{2}+\tau +2;-1\right)\Bigg)\,,\nonumber\\
\eea
where $_2\tilde{F}_1$ is the regularized hypergeometric function, 
and we have defined the generalized $\mathcal{D}_{\eta}$-terms at finite skewness $\eta\neq 0$ and even $j=2,4,...$ as
\bea\label{spinjFFD}
\mathcal{D}_{\eta}(j,\tau,\Delta_g(j),K,\tilde{\kappa}_T)
&=&\frac{1}{2}\int dz\,\sqrt{g}\,e^{-\phi}\,z^{1+2(j-2)}\,\big(\psi_R^2(z)+\psi_L^2(z)\big)\,z^{-(j-2)}\nonumber\\
&&\times\sum_{n=0}^{\infty}\,\left(\hat{d}_{j}(\eta,m_n^2)-1\right)\times\frac{\sqrt{2}\kappa\mathcal{F}_n(j)\psi_n(j,z)}{K^2+m_n^2(j)}\nonumber\\
&&\times  2^{\Delta_g(j)-2}\times\Gamma\left(\frac{a_K}{2}+\frac{\Delta_g(j)}{2}\right) \times \tilde{\kappa}_T^{j-2+\Delta_g(j)}\,,\nonumber\\
&=&\frac{1}{2}\int dz\,\sqrt{g}\,e^{-\phi}\,z^{1+2(j-2)}\,\big(\psi_R^2(z)+\psi_L^2(z)\big)\,z^{-(j-2)}\nonumber\\
&&\times\left(\hat{d}_{j}(\eta,k^2)-1\right)\times\sum_{n=0}^{\infty}\,\frac{\sqrt{2}\kappa\mathcal{F}_n(j)\psi_n(j,z)}{-k^2+m_n^2(j)}\nonumber\\
&&\times  2^{\Delta_g(j)-2}\times\Gamma\left(\frac{a_K}{2}+\frac{\Delta_g(j)}{2}\right) \times \tilde{\kappa}_T^{j-2+\Delta_g(j)}\,,\nonumber\\
&=&\left(\hat{d}_{j}(\eta,-K^2)-1\right)\times\mathcal{A}(j,\tau,\Delta_g(j),K,\tilde{\kappa}_T) \,.
\eea

Generalizing our arguments in~\cite{Mamo:2021tzd} for $j=2$ to an arbitrary even $j\geq 2$, we replace
\bea\label{DKj}
\mathcal{D}_{\eta}(j,\tau,\Delta_g(j),K,\tilde{\kappa}_T)&\rightarrow& \mathcal{D}_{\eta}(j,\tau,\Delta_g(j),K,\tilde{\kappa}_T,\tilde{\kappa}_S)\nonumber\\
&=&\left(\hat{d}_{j}(\eta,-K^2)-1\right)\times \left[\mathcal{A}(j,\tau,\Delta_g(j),K,\tilde{\kappa}_T)-\mathcal{A}_S(j,\tau,\Delta_g(j),K,\tilde{\kappa}_S)\right]\,,
\eea
with
\bea
\label{ASj}
\mathcal{A}_S(j,\tau,\Delta_g(j),K,\tilde{\kappa}_S)&\equiv&\mathcal{A}(j,\tau,\Delta_g(j),K,\tilde{\kappa}_T\rightarrow \tilde{\kappa}_S)\,.
\eea

\subsection{t-channel spin-1 meson exchange}

In the soft-wall model the normalized wave function for spin-1 (vector) mesons is given by \cite{Grigoryan:2007my}

\be
\phi_n(z)=c_{n}\tilde{\kappa}_V^2z^2 L_n^1( \tilde{\kappa}_V^2z^2)\equiv J_{V}(m_n,z)  \,,\nonumber\\
\ee
with $c_{n}=\sqrt{{2}/{(n+1)}}$ which is determined from the normalization condition (for the soft-wall model with background dilaton $\phi=\tilde{\kappa}_V^2z^2$)
\be
\int dz\,\sqrt{g}e^{-\phi}\,(g^{xx})^2\,\phi_n(z)\phi_m(z)=\delta_{nm}\,.\nonumber\\
\ee
Therefore, we have

\be
F_n=\frac 1{g_5}\bigg(-e^{-\phi}\frac{1}{z^\prime}\partial_{z^\prime}\phi_n(z^\prime)\bigg)_{z^\prime=\epsilon}=-\frac{2}{g_5}c_n(n+1)\tilde{\kappa}_V^2\,,\nonumber\\
\ee
with $\phi_n(z\rightarrow 0)\approx c_n\tilde{\kappa}_V^2z^2(n+1)$. If we define the decay constant as $f_n=-{F_n}/{m_n}$,  we have

\be
\phi_n(z)=g_5\times\frac{f_n}{m_n}\times 2\tilde{\kappa}_V^2z^2 L_n^1( \tilde{\kappa}_V^2z^2)\,.
\ee

Note that for $z\rightarrow 0$, we can write the bulk-to-bulk propagator as

\be
G(z\rightarrow 0,z^{\prime})&&\approx \frac{\phi_n(z\rightarrow 0)}{-g_5F_n}\sum_n \frac{-g_5F_n\phi_n(z^{\prime})}{k^2-m_n^2}=\frac{z^2}{2}\sum_n \frac{-g_5F_n\phi_n(z^{\prime})}{k^2-m_n^2}=\frac{z^2}{2}V(k,z^{\prime}) . \nonumber\\\label{vbbt2sw}
\ee
For space-like momenta ($t=k^2=-K^2$), we have the bulk-to-bulk propagator near the boundary

\be
G(z\rightarrow 0,z^{\prime})\approx \frac{z^2}{2}\sum_n \frac{g_5F_n\phi_n(z^{\prime})}{K^2+m_n^2}=\frac{z^2}{2}\mathcal{V}(K,z^{\prime})\,, \nonumber\\\label{vbbt2sw}
\ee
where~\cite{Grigoryan:2007my}

\bea
\mathcal{V}(K,z)=g_5\sum_n \frac{F_n\phi_n(z)}{K^2+m_n^2}
=\tilde{\kappa}_V^2z^2\int_{0}^{1}\frac{dx}{(1-x)^2}x^{a_{K}}{\rm exp}\Big[-\frac{x}{1-x}\tilde{\kappa}_V^2z^2\Big]\,,
\label{vps2sw}
\ee
with the normalization ${\cal V}(0,z)={\cal V}(K,0)=1$, and we have defined $a_K=\frac{K^2}{4\tilde{\kappa}_V^2}$.

\subsection{t-channel spin-j meson exchange}

The spin-1 transverse bulk gauge field defined as $zV_\mu(m_n,z)$ obeys the same bulk equation of motion as a bulk massive scalar field $\tilde\phi_n(j=1,z)$ with $m^2R^2=-3$. Therefore, the spin-j normalized meson wave functions $J_V(m_n(j),z)$ can be expressed in terms of the wave functions of massive scalar 
fields $\tilde\phi_n(j,z)$ which are given, for the soft wall model, in terms of the generalized Laguerre polynomials as \cite{BallonBayona:2007qr}

\be
 \label{wfSW}
&&zJ_V(m_n(j),z)\equiv \tilde\phi_{n}(j,z)=c_n(j)\,z^{\Delta_{q}(j)}L_{n}^{\Delta_{q}(j)-2}(\tilde{\xi})\,,
\ee
with $\tilde{\xi}=\tilde{\kappa}_{V}^2z^2$.  The normalization coefficients are

\be
c_n(j)=\Big(\frac{2\tilde{\kappa}_{V}^{2(\Delta_{q}(j)-1)}\Gamma(n+1)}{\Gamma(n+\Delta_{q}(j)-1)}\Big)^{\frac 12}\,,
\ee
and the dimension of the massive scalar fields (with an additional mass coming from the massive open string states attached to the  D9 or D7-branes) $\Delta_{q}(j)$ is given by
\bea
\Delta_{q}(j)=2+\sqrt{4+m^2R^2+\frac{R^2}{\alpha^\prime}(j-1)}=2+\sqrt{\sqrt{\lambda}(j-j_{0q})}\,,
\eea
where, in the last line, we have used the fact that $m^2R^2=-3$, the open string quantized mass spectrum $m_j^2R^2=(j-1)({R^2}/{\alpha^\prime})=\sqrt{\lambda}(j-1)$ for open strings attached to the D9-brane in bulk, and we have defined $j_{0q}=1-{1}/{\sqrt{\lambda}}$.

We now recall that the non-normalized bulk-to-boundary propagators of massive scalar fields are given in terms of Kummer's (confluent hypergeometric) function of the second kind,  and their integral representations  are (for space-like momenta $k^2=-K^2$)

\bea
\mathcal{\tilde V}(j,K,z)
=&&z^{\Delta_{q}(j)}U\Big(a_K+\frac{\Delta_{q}(j)}{2},\Delta_{q}(j)-1;\tilde{\xi}\Big)
=z^{\Delta_{q}(j)}\tilde{\xi}^{2-\Delta_{q}(j)}U\Big(\hat{a}(j),\hat{b}(j);\tilde{\xi}\Big)\nonumber\\
=&&z^{\Delta_{q}(j)}\tilde{\xi}^{2-\Delta_{q}(j)}\frac{1}{\Gamma(\tilde{a}(j))}
\int_{0}^{1}dx\,x^{\hat{a}(j)-1}(1-x)^{-\hat{b}(j)}{\rm exp}\Big(-\frac{x}{1-x}\tilde{\xi}\Big)\,,
\label{BBSWj}
\eea
with $\tilde{\xi}=\tilde{\kappa}_{V}^2z^2$
\be
a_K=\frac{K^2}{4\tilde{\kappa}_{V}^2}\qquad
\hat{a}(j)=a_K+2-\frac{\Delta_{q}(j)}{2}\qquad
\hat{b}(j)=3-\Delta_{q}(j)
\ee
after using the identity  $U(m,n;y)=y^{1-n}U(1+m-n,2-n,y)$.
Therefore, the bulk-to-bulk propagator of spin-j mesons (defined as massive bulk scalar fields) can be approximated at the boundary as (for space-like momenta $k^2=-K^2$)

\be
 \tilde{G}(j,z\rightarrow 0,z^{\prime})\approx &&-\bigg[\frac{\tilde{\phi}_{n}(j,z\rightarrow 0)}{g_5\tilde{\mathcal{F}}_n(j)}\bigg]\times\sum_n \frac{g_5\tilde{\mathcal{F}}_n(j)\tilde{\phi}_n(j,z^{\prime})}{K^2+m_n^2(j)}\nonumber\\
 =&&(\tilde{\kappa}_V^2)^{\Delta_{q}(j)-2}\frac{z^{\Delta_{q}(j)-1}}{\Delta_{q}(j)-1}\frac{\Gamma(\Delta_{q}(j)-2+a_K)}{\Gamma(\Delta_{q}(j)-2)}\,\tilde{\mathcal{V}}(j,K,z^{\prime})\nonumber\\
\ee
where we have defined the non-normalized bulk-to-boundary propagator of spin-j mesons (which are defined as massive spin-j scalar fields)
\be\label{spinjBB1}
\tilde{\mathcal{V}}(j,K,z^{\prime})=\sum_n \frac{g_5\tilde{\mathcal{F}}_n(j)\phi_n(j,z^{\prime})}{K^2+m_n^2(j)}\,,\nonumber\\
\ee
with the mass spectrum of massive spin-j scalar fields $m_n^2(j)=4\tilde{\kappa}_V^2(n+\frac{\Delta_{q}(j)}{2})$, and we have also defined
\bea 
\tilde{\mathcal{F}}_n(j)=&&\frac{\tilde{\mathcal{C}}(j,K,\epsilon)}{g_5}\bigg(-\sqrt{g}\,e^{-\phi}\,\big(g^{xx}\big)^2\,\partial_{z^{\prime}}\tilde{\phi}_n(j,z^\prime)\bigg)_{z^\prime=\epsilon}\,,\nonumber\\
\tilde{\mathcal{C}}(j,K,\epsilon)=&&\tilde{\mathcal{V}}(j,K,\epsilon)
\eea
with $\tilde{\phi}_n(j,z\rightarrow 0)\approx c_n(j)\,z^{\Delta_{q}(j)-2}L_{n}^{\Delta_{q}(j)-2}(0)$ for the soft wall model.

Note that the bulk-to-boundary propagator of spin-j mesons (\ref{spinjBB1}) is equivalent to the spin-j glueball bulk-to-boundary propagator (\ref{BBSWj}) with the replacements: $j\rightarrow j+1, \Delta_{g}\rightarrow\Delta_{q}, \tilde{\kappa}_T\rightarrow\frac{1}{2}\times\tilde{\kappa}_V$ in (\ref{BBSWj}), i.e.,
\be\label{spinjBB2}
\tilde{\mathcal{V}}(j,K,z;\Delta_{q};\tilde{\kappa_V})\equiv\mathcal{H}(j\rightarrow j+1,K,z;\Delta_{g}\rightarrow\Delta_{q};\tilde{\kappa}_T\rightarrow\frac{1}{2}\times\tilde{\kappa}_V)\,.\nonumber\\
\ee

\subsection{Electroproduction of vector mesons with t-channel spin-j closed string exchange in AdS}

\noindent{\bf Spin-2 glueball t-exchange\nonumber}
\newline
\newline
The diffractive production vector mesons in holography is mediated by bulk spin-2 gravitons near threshold, and their Reggeized
contribution to the Pomeron away from treshold. 
The t-channel spin-2 graviton (spin-2 glueball resonances) contribution to the transverse and longitudinal vector meson production amplitude at finite skewness, $\eta\neq 0$, is illustrated in Fig.~\ref{fig_HVM2} (right)  with the result
\bea
\label{DCTT2}
{\cal A}^{TT}_{\gamma^* p\rightarrow  V p} (s,t,Q,M_{V},\epsilon_{T},\epsilon^\prime_{T})&=&e\times\frac{1}{g_5}\times 2\kappa^2\times \left(\frac{s}{\tilde{\kappa}_T^2}\right)^{2}
\times  \mathcal{I}(Q,M_{V})\times\frac{\tilde{\kappa}_T^4}{\tilde{\kappa}_{V}^4}\nonumber\\
&&\times\frac{1}{4}\times \left[A(K,\tilde{\kappa}_T)+\eta^2 D(K,\tilde{\kappa}_T,\tilde{\kappa}_S)\right]\times\frac{1}{m_N}
\times\bar u(p_2)u(p_1)\,,\nonumber\\
{\cal A}^{LL}_{\gamma^* p\rightarrow  V p} (s,t,Q,M_{V},\epsilon_{L},\epsilon^\prime_{L})&=&e\times\frac{1}{g_5}\times 2\kappa^2\times \left(\frac{s}{\tilde{\kappa}_T^2}\right)^{2}
\times  \mathcal{I}(Q,M_{V})\times\frac{\tilde{\kappa}_T^4}{\tilde{\kappa}_{V}^4}\nonumber\\
&&\times\frac{1}{4}\times \frac{1}{3}\times \frac{Q}{M_V}\times \left[A(K,\tilde{\kappa}_T)+\eta^2 D(K,\tilde{\kappa}_T,\tilde{\kappa}_S)\right]\times\frac{1}{m_N}
\times\bar u(p_2)u(p_1)\,,\nonumber\\
\eea
A detailed account of this result can be found in~\cite{Mamo:2021tzd} to which we refer for completeness.
The  spin-2 gravitational form factors $A(K,\tilde{\kappa}_T)$ and $D(K,\tilde{\kappa}_T,\tilde{\kappa}_S)$ are given by
\bea
A(K,\tilde{\kappa}_T)&\equiv&\frac{1}{2}\int dz\,\sqrt{g}\,e^{-\phi}z\,\left(\psi_R^2(z)+\psi_L^2(z)\right)\times\sum_{n=0}^{\infty}\frac{\sqrt{2}\kappa F_{n}\psi_n(z)}{K^2+m_n^2}\,,\nonumber\\
&=&\frac{1}{2}\int dz\,\sqrt{g}\,e^{-\phi}z\,\left(\psi_R^2(z)+\psi_L^2(z)\right)\times\mathcal{H}(K,z)\,,\nonumber\\
&=&6\times\frac{\Gamma \left(2+\frac{a_K}{2}\right)}{\Gamma \left(4+\frac{a_K}{2}\right)}\times \, _2F_1\left(3,\frac{a_K}{2};\frac{a_K}{2}+4;-1\right)\,,
\nonumber\\
D(K,\tilde{\kappa}_T,\tilde{\kappa}_S)&\equiv&-\frac{4}{3}\frac{m_N^2}{K^2}\times \left[A(K,\tilde{\kappa}_T)-A_{S}(K,\tilde{\kappa}_S)\right]\,,\nonumber\\
\eea
with
\bea
\label{AS112}
A_{S}(K,\tilde{\kappa}_S)&=&A(0)\times 6\times\frac{\Gamma \left(2+\frac{\tilde{a}_K}{2}\right)}{\Gamma \left(4+\frac{\tilde{a}_K}{2}\right)}\times \, _2F_1\left(3,\frac{\tilde{a}_K}{2};\frac{\tilde{a}_K}{2}+4;-1\right)\,,
\eea
with $\tilde{a}_K=K^2/4\tilde{\kappa}_S^2$.

The  bulk-to-boundary vector propagators associated to the incoming virtual photon, and outgoing vector meson
in Fig.~\ref{fig_HVM2}, set the  scale factors in (\ref{DCTT2}) namely ( $\hat{\xi}\equiv Qz$)
\bea
 \label{IJ}
\mathcal{I}(Q,\tilde{\kappa}_{V})
=&&\Big(\frac{\tilde{\kappa}_{V}}{Q}\Big)^4\times\frac{1}{2}\int_{0}^{\infty} d\hat{\xi}\,e^{-\hat{\xi}^2\frac{\tilde{\kappa}_{V}^2}{Q^2}}\,\hat{\xi}^{-1}\times \mathcal{V}_{\gamma^*}(\hat{\xi})\mathcal{V}_{V}(\hat{\xi} M_{V}/Q)\times \frac{\hat{\xi}^{4}}{4}\nonumber\\
=&&\frac{f_{V}}{M_{V}}\times g_5\times\Bigg(\frac{3}{\frac{1}{32}\frac{Q^6}{\tilde{\kappa}_{V}^6}+\frac{3}{4}\frac{Q^4}{\tilde{\kappa}_{V}^4}+\frac{11}{2}\frac{Q^2}{\tilde{\kappa}_{V}^2}+12}\Bigg)\nonumber\\
=&&\frac{1}{2}\frac{f_{V}}{M_{V}}\times g_5\times\left(\frac{3}{\left(\frac{Q^2}{4 \tilde{\kappa}_{V}^2}+3\right)\left(\frac{Q^2}{4 \tilde{\kappa}_{V}^2}+2\right)\left(\frac{Q^2}{4 \tilde{\kappa}_{V}^2}+1\right)}\right)
\,,\nonumber\\
\mathcal{J}(Q,\tilde{\kappa}_{V})
=&&\Big(\frac{\tilde{\kappa}_{V}}{Q}\Big)^4\times\frac{1}{2}\int_{0}^{\infty} d\hat{\xi}\,e^{-\hat{\xi}^2\frac{\tilde{\kappa}_{V}^2}{Q^2}}\,\hat{\xi}^{-1}\times\partial_{\hat{\xi}}\mathcal{V}_{\gamma^*}(\hat{\xi})\times\partial_{\hat{\xi}}\mathcal{V}_{V}(\hat{\xi} M_{V}/Q)\times \frac{\hat{\xi}^{4}}{4}\nonumber\\
=&&-\frac{f_{V}}{M_{V}}\times g_5\times\Bigg(\frac{1}{\frac{1}{32}\frac{Q^6}{\tilde{\kappa}_{V}^6}+\frac{3}{4}\frac{Q^4}{\tilde{\kappa}_{V}^4}+\frac{11}{2}\frac{Q^2}{\tilde{\kappa}_{V}^6}+12}\Bigg)\nonumber\\
=&&-\frac{1}{3}\times\mathcal{I}(Q,M_{V})\,,
\eea
\newline
\newline
\newline
\noindent{\bf Spin-j glueball t-exchange\nonumber}
\newline
\newline
Away from threshold, the diffractive electroproduction of vector mesons involve heavier spin-j
gravitons which are the holographic dual of spin-j glueballs at the boundary. More specifically, 
the spin-j glueball exchange contribution to Fig.~\ref{fig_HVM2} (right) can be written as
\bea
\label{IJ2}
&&{\cal A}^{TT}_{\gamma^* p\rightarrow  V p} (j,s,t,Q,M_{V},\epsilon_{T},\epsilon^\prime_{T})=\frac{2\kappa^2}{g_5}\times\sum_{j=2}^{\infty}\,\Big(\frac{\tilde{\kappa}_T}{\tilde{\kappa}_V}\Big)^{8-3\Delta_{g}(j)+j-2}\times\frac{4}{\Delta_{g}(j)}\times\frac{1}{\Gamma(\Delta_{g}(j)-2)}\nonumber\\
&&\times\frac{1}{4^{j/2}}\times\mathcal{I}(j,Q,\tilde{\kappa}_{V})\times\left[\mathcal{A}(j,\tau,\Delta_{g}(j), K)+\mathcal{D}_{\eta}(j,\tau,\Delta_g(j),K)\right]\times\frac{1}{m_N}\times\bar u(p_2)u(p_1)\,,\nonumber\\ 
&&{\cal A}^{LL}_{\gamma^* p\rightarrow  V p} (j,s,t,Q,M_{V},\epsilon_{L},\epsilon^\prime_{L})=\frac{2\kappa^2}{g_5}\times\sum_{j=2}^{\infty}\,\Big(\frac{\tilde{\kappa}_T}{\tilde{\kappa}_V}\Big)^{8-3\Delta_{g}(j)+j-2}\times\frac{4}{\Delta_{g}(j)}\times\frac{1}{\Gamma(\Delta_{g}(j)-2)}\nonumber\\
&&\times\frac{1}{4^{j/2}}\times\frac{2}{j+\Delta_{g}(j)}\times\frac{Q}{M_V}\times\mathcal{I}(j,Q,\tilde{\kappa}_{V})\times\left[\mathcal{A}(j,\tau,\Delta_{g}(j), K)+\mathcal{D}_{\eta}(j,\tau,\Delta_g(j),K)\right]\times\frac{1}{m_N}\times\bar u(p_2)u(p_1)\,,\nonumber\\ 
\eea
for even $j=2,4...$, with the spin-j form factors
\bea\label{Ajglueball}
&&\mathcal{A}(j,\tau,\Delta_{g}(j), K,\tilde{\kappa}_T) \equiv\nonumber\\ 
&&\frac{2^{1-\Delta_{g}(j) }}{\Gamma(\tau)} \Bigg( (\tau -1) \Gamma \left(\frac{j-2}{2}+\tau -\frac{\Delta_{g}(j) }{2}+1\right) \Gamma \left(\frac{j-2}{2}+\frac{\Delta_{g}(j) }{2}+\tau -1\right)\nonumber\\
&\times & \, _2\tilde{F}_1\left(\frac{a_K}{2}-\frac{\Delta_{g}(j) }{2}+2,\frac{j-2}{2}+\tau -\frac{\Delta_{g}(j) }{2}+1;\frac{a_K}{2}+\frac{j-2}{2}+\tau +1;-1\right)\nonumber\\
&&+\Gamma \left(\frac{j-2}{2}+\tau -\frac{\Delta_{g}(j) }{2}+2\right) \Gamma \left(\frac{j-2}{2}+\frac{\Delta_{g}(j) }{2}+\tau \right)\nonumber\\
&\times & _2\tilde{F}_1\left(\frac{a_K}{2}-\frac{\Delta_{g}(j) }{2}+2,\frac{j-2}{2}+\tau -\frac{\Delta_{g}(j) }{2}+2;\frac{a_K}{2}+\frac{j-2}{2}+\tau +2;-1\right)\Bigg)\,,\nonumber\\
\eea
where $_2\tilde{F}_1$ is the regularized hypergeometric function,  the skewness $\eta$ dependent spin-j $\mathcal{D}_{\eta}$-terms are also given by
\bea\label{DKj2}
\mathcal{D}_{\eta}(j,\tau,\Delta_g(j),K,\tilde{\kappa}_T,\tilde{\kappa}_S)&=&\left(\hat{d}_{j}(\eta,-K^2)-1\right)\times \left[\mathcal{A}(j,\tau,\Delta_g(j),K,\tilde{\kappa}_T)-\mathcal{A}_S(j,\tau,\Delta_g(j),K,\tilde{\kappa}_S)\right]\,,\nonumber\\
\eea
where
\bea
\label{ASj}
\mathcal{A}_S(j,\tau,\Delta_g(j),K,\tilde{\kappa}_S)&\equiv&\mathcal{A}(j,\tau,\Delta_g(j),K,\tilde{\kappa}_T\rightarrow \tilde{\kappa}_S)\,.
\eea
and
\be\label{etapoly23}
\hat{d}_{j}(\eta,-K^2)=\, _2F_1\left(-\frac{j}{2},\frac{1-j}{2};\frac{1}{2}-j;\frac{4 m_N^2}{K^2}\times\eta ^2\right)\,.
\ee
We have also defined the dimensionless scale functions
\bea
\mathcal{I}(j,Q,\tilde{\kappa}_{V}) &\equiv & \Big(\frac{\tilde{\kappa}_{V}^2}{Q^2}\Big)^{-\frac{1}{2} (-\Delta_{g}(j)-j+2+4)}\times\Big(\frac{\tilde{\kappa}_{V}}{Q}\Big)^4\nonumber\\
&&\times\frac{1}{2}\int_{0}^{\infty} d\hat{\xi}\,e^{-\hat{\xi}^2\frac{\tilde{\kappa}_{V}^2}{Q^2}}\,\frac{\hat{\xi}^{\Delta_{g}(j)+j+2-5}}{4}\times \mathcal{V}_{\gamma^*}(\hat{\xi})\mathcal{V}_{V}(\hat{\xi} M_{V}/Q)\nonumber\\
&=&\frac{1}{2}\frac{f_{V}}{M_{V}}\times g_5\times\frac{\Gamma \left(\frac{Q^2}{4 \tilde{\kappa}_{V}^2}+1\right)}{\Gamma\left(\frac{Q^2}{4\tilde{\kappa}_{V}^2}+\frac{1}{2}(j+\Delta_{g} (j)+2)\right)}\times\left(\frac{j+\Delta_{g}(j)}{2}\right)\times\frac{1}{4}\Gamma^2 \left(\frac{j+\Delta_{g}(j)}{2}\right)\nonumber\\
&=&\frac{1}{2}\frac{f_{V}}{M_{V}}\times g_5\times\frac{\Gamma \left(\frac{Q^2}{4 \tilde{\kappa}_{V}^2}+1\right)}{\Gamma\left(\frac{Q^2}{4\tilde{\kappa}_{V}^2}+\frac{1}{2}(j+\Delta_{g} (j))-2\right)}\times\left(\frac{j+\Delta_{g}(j)}{2}\right)\times\frac{1}{4}\Gamma^2 \left(\frac{j+\Delta_{g}(j)}{2}\right)\nonumber\\
&&\times \frac{1}{\left(\frac{Q^2}{4\tilde{\kappa}_{V}^2}+\frac{1}{2}(j+\Delta_{g} (j))\right)\left(\frac{Q^2}{4\tilde{\kappa}_{V}^2}+\frac{1}{2}(j+\Delta_{g} (j))-1\right)\left(\frac{Q^2}{4\tilde{\kappa}_{V}^2}+\frac{1}{2}(j+\Delta_{g} (j))-2\right)}
\,,\nonumber\\\nonumber\\\nonumber\\
\mathcal{J}(j,Q,\tilde{\kappa}_{V}) &\equiv & \Big(\frac{\tilde{\kappa}_{V}^2}{Q^2}\Big)^{-\frac{1}{2} (-\Delta_{g}(j)-j+2+4)}\times\Big(\frac{\tilde{\kappa}_{V}}{Q}\Big)^4\nonumber\\
&&\times\frac{1}{2}\int_{0}^{\infty} d\hat{\xi}\,e^{-\hat{\xi}^2\frac{\tilde{\kappa}_{V}^2}{Q^2}}\,\frac{\hat{\xi}^{\Delta_{g}(j)+j+2-5}}{4}\times\partial_{\hat{\xi}}\mathcal{V}_{\gamma^*}(\hat{\xi})\times\partial_{\hat{\xi}}\mathcal{V}_{V}(\hat{\xi} M_{V}/Q)\,,\nonumber\\
&=&-\frac{1}{2}\frac{f_{V}}{M_{V}}\times g_5\times\frac{\Gamma \left(\frac{Q^2}{4 \tilde{\kappa}_{V}^2}+1\right)}{\Gamma\left(\frac{Q^2}{4\tilde{\kappa}_{V}^2}+\frac{1}{2}(j+\Delta_{g} (j)+2)\right)}\times\frac{1}{4}\Gamma^2 \left(\frac{j+\Delta_{g}(j)}{2}\right)\nonumber\\
&=&-\left(\frac{2}{j+\Delta_{g}(j)}\right)\times\mathcal{I}(j,Q,M_{V})\,.
\eea
We also have
\bea
\Delta_{g}(j)
&=&2+\sqrt{2\sqrt{\lambda}(j-j_{0g})}\,,
\eea
with $j_{0g}=2-2/\sqrt{\lambda}$ for closed strings.


\subsection{DVCS with t-channel open string exchanges in AdS}

\noindent{\bf Spin-2 vector contribution}
\newline
\newline
The spin-2 vector meson contribution to the transverse holographic DVCS amplitude at finite skewness, $\eta\neq 0$, 
is illustrated in Fig.~\ref{fig_DVCS2} right.
Its explicit contribution to the $TT$ holographic amplitude is
\bea 
\label{DCTT2meson}
{\cal A}^{TT}_{\gamma^* p\rightarrow  \gamma p} (s,t,Q,\epsilon_{T},\epsilon^\prime_{T})&=&-\epsilon_{T}\cdot\epsilon^\prime_{T}\times\frac{1}{g_5^2}\times g_5^2\times \left(\frac{s}{\tilde{\kappa}_N^2}\right)^{2}
\times \frac{1}{4}\mathcal{I}_{singlet}^{vector}(Q,\tilde{\kappa}_{N})\times\frac{1}{4}\nonumber\\
&\times & \left[A_{q}(K)+\eta^2D_{q}(K)\right]\times\frac{1}{2m_N}
\times\bar u(p_2)u(p_1)\times\mathcal{O}(1/N_c)\,,
\eea
where we have defined
\bea
 \label{AqIqmeson}
A_{q}(K)&=&\mathcal{A}_q(j=2,\tau,\Delta_{q}(j=2), K)\,,\nonumber\\
\eta^2 D_{q}(K)&=&\mathcal{D}_{q\eta}(j=2,\tau,\Delta_{q}(j=2), K)\,,\nonumber\\
\mathcal{I}_{singlet}^{vector}(Q,\tilde{\kappa}_{N})
&=&\mathcal{O}(1/N_c)\,,\nonumber\\
\eea
with $\mathcal{A}_q(j,\tau,\Delta_{q}(j), K)$ defined below in~(\ref{Aqq}).
\newline
\newline
\newline
\noindent{\bf Spin-j vector contribution}
\newline
\newline
The spin-j vector mesons contribution is given by (for even $j=2,4,...$). 
\bea
\label{IJ2meson}
{\cal A}^{TT}_{\gamma^* p\rightarrow  \gamma p} (s,t,Q,\epsilon_{T},\epsilon^\prime_{T})&=& \epsilon_{T}\cdot\epsilon^\prime_{T}\times\frac{1}{g_5^2}\times g_5^2\times\frac{1}{2m_N}\times\bar u(p_2)u(p_1)\times\mathcal{O}(1/N_c)\nonumber\\
&\times & \sum_{j=2}^{\infty}\,\frac{1}{4^{j/2}}\times \left(\frac{s}{\tilde{\kappa}_{N}^2}\right)^{j}\times\frac{1}{4}\mathcal{I}_{singlet}^{vector}(j,Q,\tilde{\kappa}_{N})\times\frac{1}{\Gamma(\Delta_{q}(j)-2)}\nonumber\\
&\times & \left[\mathcal{A}_q(j,\tau,\Delta_{q}(j), K)+\mathcal{D}_{q\eta}(j,\tau,\Delta_{q}(j), K)\right]\,,\nonumber\\  
\eea
We have defined the quark spin-j form factors, for even $j=2,4,...$, as
\bea\label{Aqq}
\mathcal{A}_q(j,\tau,\Delta_{q}(j),K;\tilde{\kappa}_V) &\equiv &\mathcal{A}(j\rightarrow j+1,\tau,\Delta_{g}\rightarrow\Delta_{q},K;\tilde{\kappa}_T\rightarrow\frac{1}{2}\times\tilde{\kappa}_V)\,,
\eea
and the quark spin-j skewness dependent $\mathcal{D}_{q\eta}$-terms as 
\bea\label{spinjFFD2quark}
\mathcal{D}_{q\eta}(j,\tau,\Delta_{q}(j),K)
&\equiv& \mathcal{D}_{\eta}(j\rightarrow j+1,\tau,\Delta_{g}\rightarrow\Delta_{q},K;\tilde{\kappa}_T\rightarrow\frac{1}{2}\times\tilde{\kappa}_V)\,,
\eea
where $\mathcal{A}$ and $\mathcal{D}_{\eta}$ are given by (\ref{Ajglueball}) and (\ref{DKj2}), respectively. The polynomial $\hat{d}_{j}(\eta,m_n^2)$ is also the same as for the spin-j glueballs (for even $j=2,4,..$) (\ref{etapoly23}).
We have also defined the dimensionless scale factors as
\bea
\mathcal{I}_{singlet}^{vector}(j,Q,\tilde{\kappa}_{N}) & = &\mathcal{O}(1/N_c)\,,
\eea
and
\bea
\Delta_{q}(j)=2+\sqrt{4+m^2R^2+\frac{R^2}{\alpha^\prime}(j-1)}=2+\sqrt{\sqrt{\lambda}(j-j_{0q})}\,,
\eea
for even $j=2,4,...$ with $m^2R^2=-3$. Note that $j_{0q}=1-1/\sqrt{\lambda}$ refers to open strings attached to $N_f$\,D9-branes (Reggeons),
which is to be contrasted with $j_{0g}=2-2/\sqrt{\lambda}$ for the closed strings (Pomerons).
\newline
\newline
\newline
\noindent{\bf Spin-j axial-vector contribution}
\newline
\newline
The arguments for the vector meson exchange, can be extended to the axial vector meson exchange in bulk.
In particular, the 
 spin-j axial meson exchange (for odd $j=1,3,...$) contribution to  the DVCS amplitude is

\bea 
\label{LTDVCSaxialH}
{\cal A}^{LT}_{\gamma^{*} p\rightarrow  \gamma p} (s,t,Q,\epsilon_{L},\epsilon^{\prime}_{T})&=&\frac{1}{g_5}\times\frac{1}{g_5}\times g_5\times g_5^3\kappa_{cs}\times\frac{1}{\tilde{\kappa}_{N}^2}\epsilon_{L\mu}\epsilon_{T\nu}^{\prime*}\times i \varepsilon^{\mu \nu \rho \sigma} \tilde q_\sigma
\times\bar u(p_2)\gamma_{\rho}\gamma_5 u(p_1)\nonumber\\
&\times &\sum_{j=1}^{\infty}\,\frac{1}{\tilde{\kappa}_{N}^{2(j-1)}}s^{j-1}\times\mathcal{I}_{singlet}^{axial}(j,Q,\tilde{\kappa}_{N})\times\frac{1}{\Gamma(\Delta_{q}(j)-2)}\nonumber\\
&\times&\left[\mathcal{F}_{A}(j,\tau,\Delta_{q}(j), K)+\mathcal{D}_{A\eta}(j,\tau,\Delta_{q}(j),K)\right]\,,\nonumber\\
\eea
with the Chern-Simons coupling $\kappa_{cs}=\frac{N_c}{24\pi^2}$. We have defined the spin-j axial form factors as
\bea\label{FAj1}
\mathcal{F}_A(j,\tau,\Delta_{q}(j),t) &\equiv& \widetilde{\mathcal{A}}(j\rightarrow j+1,\tau,\Delta_{g}\rightarrow\Delta_{q},K;\tilde{\kappa}_T\rightarrow\frac{1}{2}\times\tilde{\kappa}_V)\,,\nonumber\\
\eea
with 
\bea\label{Ajjaxial}
&&\widetilde{\mathcal{A}}(j,\tau,\Delta_{g}(j),K) = \nonumber\\ 
&&\frac{2^{1-\Delta_{g}(j) }}{\Gamma(\tau)} \Bigg( (\tau -1) \Gamma \left(\frac{j-2}{2}+\tau -\frac{\Delta_{g}(j) }{2}+1\right) \Gamma \left(\frac{j-2}{2}+\frac{\Delta_{g}(j) }{2}+\tau -1\right)\nonumber\\
&\times & \, _2\tilde{F}_1\left(\frac{a_K}{2}-\frac{\Delta_{g}(j) }{2}+2,\frac{j-2}{2}+\tau -\frac{\Delta_{g}(j) }{2}+1;\frac{a_K}{2}+\frac{j-2}{2}+\tau +1;-1\right)\nonumber\\
&&-\Gamma \left(\frac{j-2}{2}+\tau -\frac{\Delta_{g}(j) }{2}+2\right) \Gamma \left(\frac{j-2}{2}+\frac{\Delta_{g}(j) }{2}+\tau \right)\nonumber\\
&\times & _2\tilde{F}_1\left(\frac{a_K}{2}-\frac{\Delta_{g}(j) }{2}+2,\frac{j-2}{2}+\tau -\frac{\Delta_{g}(j) }{2}+2;\frac{a_K}{2}+\frac{j-2}{2}+\tau +2;-1\right)\Bigg)\,,\nonumber\\
\eea
and the anomalous dimension of the spin-j conformal singlet axial-vector quark operator at $\mu=\mu_0$
\be
\Delta_{q}(j)=2+\sqrt{\sqrt{\lambda}(j-j_{0q})}\,, 
\ee
with $j_{0q}=1-1/\sqrt{\lambda}$. The skewness or
$\eta$-dependent spin-j $\mathcal{D}_{A\eta}$-terms are given by
\bea\label{FAetaKj22}
\mathcal{D}_{A\eta}(j,\tau,\Delta_{q}(j),K)&=&\left(\hat{d}_{j}(\eta,t)-1\right)\times \left[\mathcal{F}_{A}(j,\tau,\Delta_{q}(j),K;\tilde{\kappa}_V)-\mathcal{F}_{AS}(j,\tau,\Delta_q(j),t;\tilde{\kappa}_S)\right]\,,\nonumber\\
\eea
where
\bea
\label{FAetaSj22}
\mathcal{F}_{AS}(j,\tau,\Delta_{q}(j),K;\tilde{\kappa}_S)&\equiv&\mathcal{F}_{A}(j,\tau,\Delta_{q}(j),K;\tilde{\kappa}_V\rightarrow \tilde{\kappa}_S)\,.
\eea

In principle, we can also calculate the scale factor $\mathcal{I}_{singlet}^{axial}(j,Q,\tilde{\kappa}_{N})$ very precisely,
 using the two virtual photon coupling to one axial meson derived in~\cite{Mamo:2021cle}. However, we do not need its precise form to extract the moments of axial spin-j (odd $j=1,3,...$) or axial singlet moments.


\subsection{Pair meson electroproduction with t-channel open string exchange in AdS}

\noindent{\bf Spin-1 vector meson t-exchange}
\newline
\newline
The t-channel spin-1 vector meson resonances contribution to the transverse holographic pair meson production amplitude at finite skewness, $\eta\neq 0$, is illustrated in Fig.~\ref{fig_PIPI2} right. The bulk t-exchange refers to a spin-1 meson, coupled in bulk to the bulk-to-boundary
virtual photon of momentum $q$, and a spin-2 bulk-to-boundary graviton of momentum $q^\prime$ decaying to a $\pi\pi$ pair at the boundary with momentum $p_{1\pi}$ and $p_{2\pi}$ respectively. Specifically, the contribution is
\bea 
\label{LLmesonPair}
{\cal A}^{LL}_{\gamma^* p\rightarrow  \pi\pi p} (s,t,Q,\epsilon_{L};m_{\pi\pi})
&=&-e\times\frac{1}{g_5}\times g_5\times 2\kappa^2
\times\frac{1}{\tilde{\kappa}_{N}^2s}\epsilon_{L\mu}p_{\nu}T^{\mu\nu}_{\pi(gluon)}(q^{\prime})
\times\mathcal{I}_{valence}^{vector}(Q,\tilde{\kappa}_{N})\times F_{1}(K)\nonumber\\
&\times &\frac{1}{2m_N}
\times\bar u(p_2)u(p_1)\,,
\eea
where $T^{\mu\nu}_{\pi(gluon)}(q^{\prime})$ is the gravitational form factor of the pion (or any other meson) in the time-like region, and we have defined the dimensionless functions 
\bea
 \label{AqIqmeson}
F_{1}(K)&=&\mathcal{F}_{1}(j=1,\tau,\Delta_{q}(j=1), K)\nonumber\\
\mathcal{I}_{valence}^{vector}(Q,\tilde{\kappa}_{N})
&=&\frac{1}{8}\times\Big(\frac{\tilde{\kappa}_{N}}{Q}\Big)^5\times\frac{1}{2}\int_{0}^{\infty} d\hat{\xi}\,e^{-\hat{\xi}^2\frac{\tilde{\kappa}_{N}^2}{Q^2}}\,\hat{\xi}^{-1}\times \partial_{\xi}\mathcal{V}_{\gamma^*}(\hat{\xi})\times \frac{\hat{\xi}^4}{4}\times \frac{\hat{\xi}^{2}}{1}\nonumber\\
&=&-\frac{1}{8}\times\frac{Q^2}{\tilde{\kappa}_{N}^2}\times\frac{225 \pi}{2048}\times\frac{\Gamma \left(\frac{Q^2}{4\tilde{\kappa}_{N}^2}+1\right)}{\Gamma \left(\frac{Q^2}{4 \tilde{\kappa}_{N}^2}+\frac{9}{2}\right)}
\,,
\eea
with $\hat{\xi}\equiv Qz$, and $\mathcal{F}_1(j,\tau,\Delta_{q}(j), K)$ defined below in~(\ref{F1jj}).
\newline
\newline
\newline
\noindent{\bf Spin-j vector meson t-exchange}
\newline
\newline
The extension of the holographic spin-1 exchange in bulk near threshold, extends to the spin-j exchange away from threshold,
using a similar Witten diagram as in Fig.~\ref{fig_PIPI2} right. Specifically, 
the  spin-j vector meson exchange in bulk contribution reads

\bea 
\label{LLmesonPairj}
{\cal A}^{LL}_{\gamma^* p\rightarrow  \pi\pi p} (s,t,Q,\epsilon_{L};m_{\pi\pi})&=&-e\times\frac{1}{g_5}\times g_5\times 2\kappa^2\times\frac{1}{\tilde{\kappa}_{N}^2s}\epsilon_{L\mu}p_{\nu}T^{\mu\nu}_{\pi(gluon)}(q^{\prime})\times \frac{1}{2m_N}
\times\bar u(p_2)u(p_1)\nonumber\\
&\times &\sum_{j=1}^{\infty}\,\frac{1}{\tilde{\kappa}_{N}^{2(j-1)}}s^{j-1}\times\mathcal{I}_{valence}^{vector}(j,Q,\tilde{\kappa}_{N})\times\frac{1}{\Gamma(\Delta_{q}(j)-2)}\nonumber\\
&\times&\left[\mathcal{F}_1(j,\tau,\Delta_{q}(j), K)+\mathcal{D}_{q\eta}\right]\,,\nonumber\\
\eea
for odd $j=1,3,...$. We have defined the spin-j vector form factors as
\bea\label{F1jj}
\mathcal{F}_1(j,\tau,\Delta_{q}(j), K) &\equiv& \mathcal{A}_q(j,\tau,\Delta_{q}(j),K;\tilde{\kappa}_V)\nonumber\\
\eea
for odd $j=1,3,...$. The spin-j vector form factors $\mathcal{A}_q$ and $\mathcal{D}_{q\eta}$ are given by (\ref{Aqq}) and (\ref{spinjFFD2quark}), respectively. We have also defined the scale factors
\bea
\mathcal{I}_{valence}^{vector}(j,Q,\tilde{\kappa}_{N}) &=&\frac{1}{8}\times\Big(\frac{\tilde{\kappa}_{N}}{Q}\Big)^{1+\Delta_{q}+j}\times\frac{1}{2}\int_{0}^{\infty} d\hat{\xi}\,e^{-\hat{\xi}^2\frac{\tilde{\kappa}_{N}^2}{Q^2}}\,\hat{\xi}^{-1+2(j-1)}\times \partial_{\xi}\mathcal{V}_{\gamma^*}(\hat{\xi})\times \frac{\hat{\xi}^4}{4}\times \frac{\hat{\xi}^{\Delta_{q} (j)-1-(j-1)}}{1}\nonumber\\
&=&-\frac{1}{8}\times\frac{Q^2}{\tilde{\kappa}_{N}^2}\times\frac{1}{32}\times\Gamma \left(\frac{1}{2} (j+\Delta_{q}(j) +3)\right)^2\times\frac{\Gamma \left(\frac{Q^2}{4 \tilde{\kappa}_{N}^2}+1\right)}{\Gamma \left(\frac{Q^2}{4\tilde{\kappa}_{N}^2+\frac{1}{2} (\Delta_{q}(j) +j+5)}\right)}
\,,
\eea
with the anomalous dimension 
\bea
\Delta_{q}(j)=2+\sqrt{4+m^2R^2+\frac{R^2}{\alpha^\prime}(j-1)}=2+\sqrt{\sqrt{\lambda}(j-j_{0q})}\,,
\eea
Here  $m^2R^2=-3$, and $j_{0q}=1-1/\sqrt{\lambda}$ for open strings attached to $N_f$\,D9 bulk filling branes.

\subsection{Neutral pion electroproduction with t-channel open string exchange in AdS}

\noindent{\bf Spin-j axial meson t-exchange}
\newline
\newline
The holographic dual of the neutral pion production as shown in in Fig.~\ref{fig_PI2} right, involves 
the bulk spin-j axial meson exchange (even $j=2,4,...$) 


\bea 
\label{LLmesonAxialj}
{\cal A}^{LL}_{\gamma^{*} p\rightarrow  \pi^{0} p} (s,t,Q,\epsilon_{L},\epsilon_{L}^{\prime})&=&e\times\frac{1}{g_5}\times g_5^2\times\frac{1}{\tilde{\kappa}_{N}^4}\times\epsilon_{L}^{\prime}\cdot p\times\bar u(p_2)\epsilon_{L}\cdot \gamma\,\gamma_5 u(p_1)\nonumber\\
&\times &\sum_{j=2}^{\infty}\,\frac{1}{\tilde{\kappa}_{N}^{2(j-2)}}s^{j-2}\times \frac{f_{\pi}}{m_{\pi}}\times g_{5}\times \mathcal{I}_{valence}^{axial}(j,Q,\tilde{\kappa}_{N})\times\frac{1}{\Gamma(\Delta_{va}(j)-2)}\nonumber\\
&\times &\left[\mathcal{F}_{A}(j,\tau,\Delta_{va}(j), K)+\mathcal{D}_{A\eta}(j,\tau,\Delta_{q}(j), K)\right]\,,
\eea
where the scale factor
\be
\mathcal{I}_{valence}^{axial}(j,Q,\tilde{\kappa}_{N})=\mathcal{O}(1/N_c)\,,
\ee
and $\mathcal{F}_A$ and $\mathcal{D}_{A\eta}$ are given by (\ref{FAj1}) and (\ref{FAetaKj22}), respectively, for even $j=2,4,...$.

\end{widetext}

\bibliography{GPDs_DVCS}

\begin{thebibliography}{45}
\expandafter\ifx\csname natexlab\endcsname\relax\def\natexlab#1{#1}\fi
\expandafter\ifx\csname bibnamefont\endcsname\relax
  \def\bibnamefont#1{#1}\fi
\expandafter\ifx\csname bibfnamefont\endcsname\relax
  \def\bibfnamefont#1{#1}\fi
\expandafter\ifx\csname citenamefont\endcsname\relax
  \def\citenamefont#1{#1}\fi
\expandafter\ifx\csname url\endcsname\relax
  \def\url#1{\texttt{#1}}\fi
\expandafter\ifx\csname urlprefix\endcsname\relax\def\urlprefix{URL }\fi
\providecommand{\bibinfo}[2]{#2}
\providecommand{\eprint}[2][]{\url{#2}}

\bibitem[{\citenamefont{Abdul~Khalek et~al.}(2021)}]{AbdulKhalek:2021gbh}
\bibinfo{author}{\bibfnamefont{R.}~\bibnamefont{Abdul~Khalek}}
  \bibnamefont{et~al.} (\bibinfo{year}{2021}), \eprint{2103.05419}.

\bibitem[{\citenamefont{Anderle et~al.}(2021)}]{Anderle:2021wcy}
\bibinfo{author}{\bibfnamefont{D.~P.} \bibnamefont{Anderle}}
  \bibnamefont{et~al.} (\bibinfo{year}{2021}), \eprint{2102.09222}.

\bibitem[{\citenamefont{Ji}(1997)}]{Ji:1996nm}
\bibinfo{author}{\bibfnamefont{X.-D.} \bibnamefont{Ji}},
  \bibinfo{journal}{Phys. Rev. D} \textbf{\bibinfo{volume}{55}},
  \bibinfo{pages}{7114} (\bibinfo{year}{1997}), \eprint{hep-ph/9609381}.

\bibitem[{\citenamefont{Radyushkin}(1997)}]{Radyushkin:1997ki}
\bibinfo{author}{\bibfnamefont{A.~V.} \bibnamefont{Radyushkin}},
  \bibinfo{journal}{Phys. Rev. D} \textbf{\bibinfo{volume}{56}},
  \bibinfo{pages}{5524} (\bibinfo{year}{1997}), \eprint{hep-ph/9704207}.

\bibitem[{\citenamefont{Bertone et~al.}(2021)\citenamefont{Bertone, Dutrieux,
  Mezrag, Moutarde, and Sznajder}}]{Bertone:2021yyz}
\bibinfo{author}{\bibfnamefont{V.}~\bibnamefont{Bertone}},
  \bibinfo{author}{\bibfnamefont{H.}~\bibnamefont{Dutrieux}},
  \bibinfo{author}{\bibfnamefont{C.}~\bibnamefont{Mezrag}},
  \bibinfo{author}{\bibfnamefont{H.}~\bibnamefont{Moutarde}}, \bibnamefont{and}
  \bibinfo{author}{\bibfnamefont{P.}~\bibnamefont{Sznajder}},
  \bibinfo{journal}{Phys. Rev. D} \textbf{\bibinfo{volume}{103}},
  \bibinfo{pages}{114019} (\bibinfo{year}{2021}), \eprint{2104.03836}.

\bibitem[{\citenamefont{Nastase}(2015)}]{Nastase:2015wjb}
\bibinfo{author}{\bibfnamefont{H.}~\bibnamefont{Nastase}},
  \emph{\bibinfo{title}{{Introduction to the ADS/CFT Correspondence}}}
  (\bibinfo{publisher}{Cambridge University Press}, \bibinfo{year}{2015}), ISBN
  \bibinfo{isbn}{978-1-107-08585-5, 978-1-316-35530-5}.

\bibitem[{\citenamefont{Nishio and
  Watari}(2014{\natexlab{a}})}]{Nishio:2014rya}
\bibinfo{author}{\bibfnamefont{R.}~\bibnamefont{Nishio}} \bibnamefont{and}
  \bibinfo{author}{\bibfnamefont{T.}~\bibnamefont{Watari}},
  \bibinfo{journal}{Phys. Rev. D} \textbf{\bibinfo{volume}{90}},
  \bibinfo{pages}{125001} (\bibinfo{year}{2014}{\natexlab{a}}).

\bibitem[{\citenamefont{Mamo and Zahed}(2020)}]{Mamo:2019mka}
\bibinfo{author}{\bibfnamefont{K.~A.} \bibnamefont{Mamo}} \bibnamefont{and}
  \bibinfo{author}{\bibfnamefont{I.}~\bibnamefont{Zahed}},
  \bibinfo{journal}{Phys. Rev. D} \textbf{\bibinfo{volume}{101}},
  \bibinfo{pages}{086003} (\bibinfo{year}{2020}), \eprint{1910.04707}.

\bibitem[{\citenamefont{Mamo and Zahed}(2021{\natexlab{a}})}]{Mamo:2021tzd}
\bibinfo{author}{\bibfnamefont{K.~A.} \bibnamefont{Mamo}} \bibnamefont{and}
  \bibinfo{author}{\bibfnamefont{I.}~\bibnamefont{Zahed}},
  \bibinfo{journal}{Phys. Rev. D} \textbf{\bibinfo{volume}{104}},
  \bibinfo{pages}{066023} (\bibinfo{year}{2021}{\natexlab{a}}),
  \eprint{2106.00722}.

\bibitem[{\citenamefont{Mamo and Zahed}(2022)}]{Mamo:2022eui}
\bibinfo{author}{\bibfnamefont{K.~A.} \bibnamefont{Mamo}} \bibnamefont{and}
  \bibinfo{author}{\bibfnamefont{I.}~\bibnamefont{Zahed}}
  (\bibinfo{year}{2022}), \eprint{2204.08857}.

\bibitem[{\citenamefont{Belitsky and Radyushkin}(2005)}]{Belitsky:2005qn}
\bibinfo{author}{\bibfnamefont{A.~V.} \bibnamefont{Belitsky}} \bibnamefont{and}
  \bibinfo{author}{\bibfnamefont{A.~V.} \bibnamefont{Radyushkin}},
  \bibinfo{journal}{Phys. Rept.} \textbf{\bibinfo{volume}{418}},
  \bibinfo{pages}{1} (\bibinfo{year}{2005}), \eprint{hep-ph/0504030}.

\bibitem[{\citenamefont{Goeke et~al.}(2001)\citenamefont{Goeke, Polyakov, and
  Vanderhaeghen}}]{Goeke:2001tz}
\bibinfo{author}{\bibfnamefont{K.}~\bibnamefont{Goeke}},
  \bibinfo{author}{\bibfnamefont{M.~V.} \bibnamefont{Polyakov}},
  \bibnamefont{and}
  \bibinfo{author}{\bibfnamefont{M.}~\bibnamefont{Vanderhaeghen}},
  \bibinfo{journal}{Prog. Part. Nucl. Phys.} \textbf{\bibinfo{volume}{47}},
  \bibinfo{pages}{401} (\bibinfo{year}{2001}), \eprint{hep-ph/0106012}.

\bibitem[{\citenamefont{Brower et~al.}(2009)\citenamefont{Brower, Strassler,
  and Tan}}]{Brower:2007xg}
\bibinfo{author}{\bibfnamefont{R.~C.} \bibnamefont{Brower}},
  \bibinfo{author}{\bibfnamefont{M.~J.} \bibnamefont{Strassler}},
  \bibnamefont{and} \bibinfo{author}{\bibfnamefont{C.-I.} \bibnamefont{Tan}},
  \bibinfo{journal}{JHEP} \textbf{\bibinfo{volume}{03}}, \bibinfo{pages}{092}
  (\bibinfo{year}{2009}), \eprint{0710.4378}.

\bibitem[{\citenamefont{Diehl}(2003)}]{Diehl:2003ny}
\bibinfo{author}{\bibfnamefont{M.}~\bibnamefont{Diehl}},
  \bibinfo{journal}{Phys. Rept.} \textbf{\bibinfo{volume}{388}},
  \bibinfo{pages}{41} (\bibinfo{year}{2003}), \eprint{hep-ph/0307382}.

\bibitem[{\citenamefont{Hagler et~al.}(2008)}]{LHPC:2007blg}
\bibinfo{author}{\bibfnamefont{P.}~\bibnamefont{Hagler}} \bibnamefont{et~al.}
  (\bibinfo{collaboration}{LHPC}), \bibinfo{journal}{Phys. Rev. D}
  \textbf{\bibinfo{volume}{77}}, \bibinfo{pages}{094502}
  (\bibinfo{year}{2008}), \eprint{0705.4295}.

\bibitem[{\citenamefont{Hadjidakis et~al.}(2005)}]{CLAS:2004cri}
\bibinfo{author}{\bibfnamefont{C.}~\bibnamefont{Hadjidakis}}
  \bibnamefont{et~al.} (\bibinfo{collaboration}{CLAS}), \bibinfo{journal}{Phys.
  Lett. B} \textbf{\bibinfo{volume}{605}}, \bibinfo{pages}{256}
  (\bibinfo{year}{2005}), \eprint{hep-ex/0408005}.

\bibitem[{\citenamefont{Morrow et~al.}(2009)}]{CLAS:2008rpm}
\bibinfo{author}{\bibfnamefont{S.~A.} \bibnamefont{Morrow}}
  \bibnamefont{et~al.} (\bibinfo{collaboration}{CLAS}), \bibinfo{journal}{Eur.
  Phys. J. A} \textbf{\bibinfo{volume}{39}}, \bibinfo{pages}{5}
  (\bibinfo{year}{2009}), \eprint{0807.3834}.

\bibitem[{\citenamefont{Airapetian et~al.}(2000)}]{HERMES:2000jnb}
\bibinfo{author}{\bibfnamefont{A.}~\bibnamefont{Airapetian}}
  \bibnamefont{et~al.} (\bibinfo{collaboration}{HERMES}),
  \bibinfo{journal}{Eur. Phys. J. C} \textbf{\bibinfo{volume}{17}},
  \bibinfo{pages}{389} (\bibinfo{year}{2000}), \eprint{hep-ex/0004023}.

\bibitem[{\citenamefont{Adams et~al.}(1997)}]{E665:1997qph}
\bibinfo{author}{\bibfnamefont{M.~R.} \bibnamefont{Adams}} \bibnamefont{et~al.}
  (\bibinfo{collaboration}{E665}), \bibinfo{journal}{Z. Phys. C}
  \textbf{\bibinfo{volume}{74}}, \bibinfo{pages}{237} (\bibinfo{year}{1997}).

\bibitem[{\citenamefont{Breitweg et~al.}(1999)}]{ZEUS:1998xpo}
\bibinfo{author}{\bibfnamefont{J.}~\bibnamefont{Breitweg}} \bibnamefont{et~al.}
  (\bibinfo{collaboration}{ZEUS}), \bibinfo{journal}{Eur. Phys. J. C}
  \textbf{\bibinfo{volume}{6}}, \bibinfo{pages}{603} (\bibinfo{year}{1999}),
  \eprint{hep-ex/9808020}.

\bibitem[{\citenamefont{Adloff et~al.}(2000{\natexlab{a}})}]{H1:1999pji}
\bibinfo{author}{\bibfnamefont{C.}~\bibnamefont{Adloff}} \bibnamefont{et~al.}
  (\bibinfo{collaboration}{H1}), \bibinfo{journal}{Eur. Phys. J. C}
  \textbf{\bibinfo{volume}{13}}, \bibinfo{pages}{371}
  (\bibinfo{year}{2000}{\natexlab{a}}), \eprint{hep-ex/9902019}.

\bibitem[{\citenamefont{Mamo and Zahed}(2021{\natexlab{b}})}]{Mamo:2021jhj}
\bibinfo{author}{\bibfnamefont{K.~A.} \bibnamefont{Mamo}} \bibnamefont{and}
  \bibinfo{author}{\bibfnamefont{I.}~\bibnamefont{Zahed}}
  (\bibinfo{year}{2021}{\natexlab{b}}), \eprint{2106.00752}.

\bibitem[{\citenamefont{Cassel et~al.}(1981)}]{Cassel:1981sx}
\bibinfo{author}{\bibfnamefont{D.~G.} \bibnamefont{Cassel}}
  \bibnamefont{et~al.}, \bibinfo{journal}{Phys. Rev. D}
  \textbf{\bibinfo{volume}{24}}, \bibinfo{pages}{2787} (\bibinfo{year}{1981}).

\bibitem[{\citenamefont{Santoro et~al.}(2008)}]{CLAS:2008cms}
\bibinfo{author}{\bibfnamefont{J.~P.} \bibnamefont{Santoro}}
  \bibnamefont{et~al.} (\bibinfo{collaboration}{CLAS}), \bibinfo{journal}{Phys.
  Rev. C} \textbf{\bibinfo{volume}{78}}, \bibinfo{pages}{025210}
  (\bibinfo{year}{2008}), \eprint{0803.3537}.

\bibitem[{\citenamefont{Lukashin et~al.}(2001)}]{CLAS:2001zwd}
\bibinfo{author}{\bibfnamefont{K.}~\bibnamefont{Lukashin}} \bibnamefont{et~al.}
  (\bibinfo{collaboration}{CLAS}), \bibinfo{journal}{Phys. Rev. C}
  \textbf{\bibinfo{volume}{64}}, \bibinfo{pages}{059901}
  (\bibinfo{year}{2001}), \eprint{hep-ex/0101030}.

\bibitem[{\citenamefont{Chekanov et~al.}(2005)}]{ZEUS:2005bhf}
\bibinfo{author}{\bibfnamefont{S.}~\bibnamefont{Chekanov}} \bibnamefont{et~al.}
  (\bibinfo{collaboration}{ZEUS}), \bibinfo{journal}{Nucl. Phys. B}
  \textbf{\bibinfo{volume}{718}}, \bibinfo{pages}{3} (\bibinfo{year}{2005}),
  \eprint{hep-ex/0504010}.

\bibitem[{\citenamefont{Adloff et~al.}(2000{\natexlab{b}})}]{H1:2000hps}
\bibinfo{author}{\bibfnamefont{C.}~\bibnamefont{Adloff}} \bibnamefont{et~al.}
  (\bibinfo{collaboration}{H1}), \bibinfo{journal}{Phys. Lett. B}
  \textbf{\bibinfo{volume}{483}}, \bibinfo{pages}{360}
  (\bibinfo{year}{2000}{\natexlab{b}}), \eprint{hep-ex/0005010}.

\bibitem[{\citenamefont{Aaron et~al.}(2010)}]{H1:2009pze}
\bibinfo{author}{\bibfnamefont{F.~D.} \bibnamefont{Aaron}} \bibnamefont{et~al.}
  (\bibinfo{collaboration}{H1, ZEUS}), \bibinfo{journal}{JHEP}
  \textbf{\bibinfo{volume}{01}}, \bibinfo{pages}{109} (\bibinfo{year}{2010}),
  \eprint{0911.0884}.

\bibitem[{\citenamefont{Kroll}(2015)}]{Kroll:2014tma}
\bibinfo{author}{\bibfnamefont{P.}~\bibnamefont{Kroll}}, \bibinfo{journal}{EPJ
  Web Conf.} \textbf{\bibinfo{volume}{85}}, \bibinfo{pages}{01005}
  (\bibinfo{year}{2015}), \eprint{1410.4450}.

\bibitem[{\citenamefont{Favart et~al.}(2016)\citenamefont{Favart, Guidal, Horn,
  and Kroll}}]{Favart:2015umi}
\bibinfo{author}{\bibfnamefont{L.}~\bibnamefont{Favart}},
  \bibinfo{author}{\bibfnamefont{M.}~\bibnamefont{Guidal}},
  \bibinfo{author}{\bibfnamefont{T.}~\bibnamefont{Horn}}, \bibnamefont{and}
  \bibinfo{author}{\bibfnamefont{P.}~\bibnamefont{Kroll}},
  \bibinfo{journal}{Eur. Phys. J. A} \textbf{\bibinfo{volume}{52}},
  \bibinfo{pages}{158} (\bibinfo{year}{2016}), \eprint{1511.04535}.

\bibitem[{\citenamefont{Fradi}(2011)}]{Fradi:2010ew}
\bibinfo{author}{\bibfnamefont{A.}~\bibnamefont{Fradi}}, \bibinfo{journal}{AIP
  Conf. Proc.} \textbf{\bibinfo{volume}{1374}}, \bibinfo{pages}{537}
  (\bibinfo{year}{2011}), \eprint{1010.1198}.

\bibitem[{\citenamefont{Lehmann-Dronke
  et~al.}(2001)\citenamefont{Lehmann-Dronke, Schaefer, Polyakov, and
  Goeke}}]{Lehmann-Dronke:2000hlo}
\bibinfo{author}{\bibfnamefont{B.}~\bibnamefont{Lehmann-Dronke}},
  \bibinfo{author}{\bibfnamefont{A.}~\bibnamefont{Schaefer}},
  \bibinfo{author}{\bibfnamefont{M.~V.} \bibnamefont{Polyakov}},
  \bibnamefont{and} \bibinfo{author}{\bibfnamefont{K.}~\bibnamefont{Goeke}},
  \bibinfo{journal}{Phys. Rev. D} \textbf{\bibinfo{volume}{63}},
  \bibinfo{pages}{114001} (\bibinfo{year}{2001}), \eprint{hep-ph/0012108}.

\bibitem[{\citenamefont{Mamo and Zahed}(2021{\natexlab{c}})}]{Mamo:2021cle}
\bibinfo{author}{\bibfnamefont{K.~A.} \bibnamefont{Mamo}} \bibnamefont{and}
  \bibinfo{author}{\bibfnamefont{I.}~\bibnamefont{Zahed}},
  \bibinfo{journal}{Phys. Rev. D} \textbf{\bibinfo{volume}{104}},
  \bibinfo{pages}{066010} (\bibinfo{year}{2021}{\natexlab{c}}),
  \eprint{2102.00608}.

\bibitem[{\citenamefont{Hirn and Sanz}(2005)}]{Hirn:2005nr}
\bibinfo{author}{\bibfnamefont{J.}~\bibnamefont{Hirn}} \bibnamefont{and}
  \bibinfo{author}{\bibfnamefont{V.}~\bibnamefont{Sanz}},
  \bibinfo{journal}{JHEP} \textbf{\bibinfo{volume}{12}}, \bibinfo{pages}{030}
  (\bibinfo{year}{2005}), \eprint{hep-ph/0507049}.

\bibitem[{\citenamefont{Domokos et~al.}(2009)\citenamefont{Domokos, Grigoryan,
  and Harvey}}]{Domokos:2009cq}
\bibinfo{author}{\bibfnamefont{S.~K.} \bibnamefont{Domokos}},
  \bibinfo{author}{\bibfnamefont{H.~R.} \bibnamefont{Grigoryan}},
  \bibnamefont{and} \bibinfo{author}{\bibfnamefont{J.~A.}
  \bibnamefont{Harvey}}, \bibinfo{journal}{Phys. Rev. D}
  \textbf{\bibinfo{volume}{80}}, \bibinfo{pages}{115018}
  (\bibinfo{year}{2009}), \eprint{0905.1949}.

\bibitem[{\citenamefont{Cherman et~al.}(2009)\citenamefont{Cherman, Cohen, and
  Werbos}}]{Cherman:2008eh}
\bibinfo{author}{\bibfnamefont{A.}~\bibnamefont{Cherman}},
  \bibinfo{author}{\bibfnamefont{T.~D.} \bibnamefont{Cohen}}, \bibnamefont{and}
  \bibinfo{author}{\bibfnamefont{E.~S.} \bibnamefont{Werbos}},
  \bibinfo{journal}{Phys. Rev. C} \textbf{\bibinfo{volume}{79}},
  \bibinfo{pages}{045203} (\bibinfo{year}{2009}), \eprint{0804.1096}.

\bibitem[{\citenamefont{Hong et~al.}(2007)\citenamefont{Hong, Inami, and
  Yee}}]{Hong:2006ta}
\bibinfo{author}{\bibfnamefont{D.~K.} \bibnamefont{Hong}},
  \bibinfo{author}{\bibfnamefont{T.}~\bibnamefont{Inami}}, \bibnamefont{and}
  \bibinfo{author}{\bibfnamefont{H.-U.} \bibnamefont{Yee}},
  \bibinfo{journal}{Phys. Lett. B} \textbf{\bibinfo{volume}{646}},
  \bibinfo{pages}{165} (\bibinfo{year}{2007}), \eprint{hep-ph/0609270}.

\bibitem[{\citenamefont{Kanitscheider et~al.}(2008)\citenamefont{Kanitscheider,
  Skenderis, and Taylor}}]{Kanitscheider:2008kd}
\bibinfo{author}{\bibfnamefont{I.}~\bibnamefont{Kanitscheider}},
  \bibinfo{author}{\bibfnamefont{K.}~\bibnamefont{Skenderis}},
  \bibnamefont{and} \bibinfo{author}{\bibfnamefont{M.}~\bibnamefont{Taylor}},
  \bibinfo{journal}{JHEP} \textbf{\bibinfo{volume}{09}}, \bibinfo{pages}{094}
  (\bibinfo{year}{2008}), \eprint{0807.3324}.

\bibitem[{\citenamefont{Ballon~Bayona et~al.}(2008)\citenamefont{Ballon~Bayona,
  Boschi-Filho, and Braga}}]{BallonBayona:2007qr}
\bibinfo{author}{\bibfnamefont{C.~A.} \bibnamefont{Ballon~Bayona}},
  \bibinfo{author}{\bibfnamefont{H.}~\bibnamefont{Boschi-Filho}},
  \bibnamefont{and} \bibinfo{author}{\bibfnamefont{N.~R.~F.}
  \bibnamefont{Braga}}, \bibinfo{journal}{JHEP} \textbf{\bibinfo{volume}{03}},
  \bibinfo{pages}{064} (\bibinfo{year}{2008}), \eprint{0711.0221}.

\bibitem[{\citenamefont{Abidin and Carlson}(2008)}]{Abidin:2008ku}
\bibinfo{author}{\bibfnamefont{Z.}~\bibnamefont{Abidin}} \bibnamefont{and}
  \bibinfo{author}{\bibfnamefont{C.~E.} \bibnamefont{Carlson}},
  \bibinfo{journal}{Phys. Rev. D} \textbf{\bibinfo{volume}{77}},
  \bibinfo{pages}{095007} (\bibinfo{year}{2008}), \eprint{0801.3839}.

\bibitem[{\citenamefont{Raju}(2011)}]{Raju:2011mp}
\bibinfo{author}{\bibfnamefont{S.}~\bibnamefont{Raju}}, \bibinfo{journal}{Phys.
  Rev. D} \textbf{\bibinfo{volume}{83}}, \bibinfo{pages}{126002}
  (\bibinfo{year}{2011}), \eprint{1102.4724}.

\bibitem[{\citenamefont{D'Hoker et~al.}(1999)\citenamefont{D'Hoker, Freedman,
  Mathur, Matusis, and Rastelli}}]{DHoker:1999bve}
\bibinfo{author}{\bibfnamefont{E.}~\bibnamefont{D'Hoker}},
  \bibinfo{author}{\bibfnamefont{D.~Z.} \bibnamefont{Freedman}},
  \bibinfo{author}{\bibfnamefont{S.~D.} \bibnamefont{Mathur}},
  \bibinfo{author}{\bibfnamefont{A.}~\bibnamefont{Matusis}}, \bibnamefont{and}
  \bibinfo{author}{\bibfnamefont{L.}~\bibnamefont{Rastelli}},
  \bibinfo{journal}{Nucl. Phys. B} \textbf{\bibinfo{volume}{562}},
  \bibinfo{pages}{330} (\bibinfo{year}{1999}), \eprint{hep-th/9902042}.

\bibitem[{\citenamefont{Nishio and
  Watari}(2014{\natexlab{b}})}]{Nishio:2014eua}
\bibinfo{author}{\bibfnamefont{R.}~\bibnamefont{Nishio}} \bibnamefont{and}
  \bibinfo{author}{\bibfnamefont{T.}~\bibnamefont{Watari}}
  (\bibinfo{year}{2014}{\natexlab{b}}), \eprint{1408.6365}.

\bibitem[{\citenamefont{Mamo and Zahed}(2021{\natexlab{d}})}]{Mamo:2021krl}
\bibinfo{author}{\bibfnamefont{K.~A.} \bibnamefont{Mamo}} \bibnamefont{and}
  \bibinfo{author}{\bibfnamefont{I.}~\bibnamefont{Zahed}},
  \bibinfo{journal}{Phys. Rev. D} \textbf{\bibinfo{volume}{103}},
  \bibinfo{pages}{094010} (\bibinfo{year}{2021}{\natexlab{d}}),
  \eprint{2103.03186}.

\bibitem[{\citenamefont{Grigoryan and Radyushkin}(2007)}]{Grigoryan:2007my}
\bibinfo{author}{\bibfnamefont{H.~R.} \bibnamefont{Grigoryan}}
  \bibnamefont{and} \bibinfo{author}{\bibfnamefont{A.~V.}
  \bibnamefont{Radyushkin}}, \bibinfo{journal}{Phys. Rev. D}
  \textbf{\bibinfo{volume}{76}}, \bibinfo{pages}{095007}
  (\bibinfo{year}{2007}), \eprint{0706.1543}.

\end{thebibliography}

\end{document}